\documentclass[12pt]{article}
\PassOptionsToPackage{hyphens}{url}
\usepackage[pagebackref=true,colorlinks]{hyperref}
\usepackage{amsmath,amssymb,amsthm,mathtools}
\usepackage[alphabetic]{amsrefs}
\usepackage{etex} 
\usepackage{authblk}
\usepackage[toc,page]{appendix}
\usepackage{stmaryrd}
\usepackage{fancyhdr}
\usepackage{soul}
\usepackage{dsfont}
\usepackage[normalem]{ulem}
\usepackage{longtable}
\usepackage[bottom]{footmisc}
\usepackage{graphicx}
\usepackage{caption}
\usepackage{subcaption}
\usepackage{pictexwd, dcpic}
\usepackage{float}
\usepackage{calligra} 
\usepackage{enumerate}
\usepackage[all]{xy} 
\hypersetup{
	colorlinks=true,
    linktoc=all,
    linkcolor=blue,  
    citecolor=red,
    }

\makeatletter
\renewcommand*\env@matrix[1][\arraystretch]{%
  \edef\arraystretch{#1}%
  \hskip -\arraycolsep
  \let\@ifnextchar\new@ifnextchar
  \array{*\c@MaxMatrixCols c}}
\makeatother

\theoremstyle{theorem}
\newtheorem{theorem}[equation]{Theorem}
\newtheorem{lemma}[equation]{Lemma}
\newtheorem{proposition}[equation]{Proposition}
\newtheorem{corollary}[equation]{Corollary}
\theoremstyle{definition}
\newtheorem{definition}[equation]{Definition}
\newtheorem{construction}[equation]{Construction}
\newtheorem{question}[equation]{Question}
\newtheorem{problem}[equation]{Problem}
\newtheorem{example}[equation]{Example}
\newtheorem{exercise}[equation]{Exercise}
\newtheorem*{answer}{Answer}
\newtheorem*{solution}{Solution}
\newtheorem{remark}[equation]{Remark}

\numberwithin{equation}{section}

\let\a=\alpha \let\b=\beta \let\g=\gamma \let\de=\delta \let\e=\epsilon
\let\z=\zeta \let\h=\eta \let\q=\theta \let\i=\iota \let\k=\kappa
\let\l=\lambda \let\r=\rho
\let\s=\sigma \let\t=\tau    
\let\w=\omega      \let\G=\Gamma \let\D=\Delta  
  \let\S=\Sigma   
\let\C=\Chi \let\W=\Omega
\def\vf{\varphi}

\newcommand{\be}{\begin{equation}}
\newcommand{\ee}{\end{equation}}
\def\ba{\begin{align}} 
\def\ea{\end{align}}
\newcommand{\bea}{\begin{eqnarray}}
\newcommand{\eea}{\end{eqnarray}}
\newcommand{\bx}{\begin{example}}
\newcommand{\ex}{\end{example}}
\newcommand{\bex}{\begin{exercise}}
\newcommand{\eex}{\end{exercise}}
\newcommand{\ban}{\begin{answer}}
\newcommand{\ean}{\end{answer}}
\newcommand{\bt}{\begin{theorem}}
\newcommand{\et}{\end{theorem}}
\newcommand{\bc}{\begin{corollary}}
\newcommand{\ec}{\end{corollary}}
\newcommand{\blem}{\begin{lemma}}
\newcommand{\elem}{\end{lemma}}
\newcommand{\bp}{\begin{problem}}
\newcommand{\ep}{\end{problem}}
\newcommand{\bn}{\begin{proposition}}
\newcommand{\en}{\end{proposition}}
\newcommand{\bd}{\begin{definition}}
\newcommand{\ed}{\end{definition}}
\newcommand{\bcon}{\begin{construction}}
\newcommand{\econ}{\end{construction}}
\newcommand{\bq}{\begin{question}}
\newcommand{\eq}{\end{question}}
\newcommand{\bprf}{\begin{proof}}
\newcommand{\eprf}{\end{proof}}
\newcommand{\br}{\begin{remark}}
\newcommand{\er}{\end{remark}}
\newcommand{\bs}{\begin{solution}}
\newcommand{\es}{\end{solution}}
\newcommand{\beqs}{\begin{eqnarray}}
\newcommand{\eeqs}{\end{eqnarray}}

\let\p=\partial \let\ov=\overline
\let\un=\underline
\newcommand{\id}{\mathrm{id}}

\newcommand{\mC}{\mathcal{C}}
\newcommand{\mD}{\mathcal{D}}
\newcommand{\vect}{\textbf{Vect}_{\K}}
\newcommand{\Hilb}{\textbf{Hilb}}
\newcommand{\isom}{\mathrm{Isom}}
\newcommand{\proj}{\mathrm{proj}}
\newcommand{\Hilbip}{\Hilb_{\isom}^{\proj}}

\newcommand{\aand}{\qquad \& \qquad}

\def\triv{{{\mathrm{triv}}}}

\def\R{{{\mathbb R}}}
\def\C{{{\mathbb C}}}

\def\K{{{\mathbb K}}}

\def\N{{{\mathbb N}}}
\def\Z{{{\mathbb Z}}}

\def\B{{{\mathbb B}}}

\def\Hi{{{\mathcal{H}}}}

\def\mK{{{\mathcal{K}}}}

\DeclareMathAlphabet{\mathcalligra}{T1}{calligra}{q}{n}
\DeclareFontShape{T1}{calligra}{q}{n}{<->s*[2.2]callig15}{}

\newcommand{\ds}{\displaystyle}
\newcommand{\ben}{\renewcommand{\theenumi}{\alph{enumi}} 
\renewcommand{\labelenumi}{(\theenumi)}\begin{enumerate}}
\newcommand{\een}{\end{enumerate}}

\title{Two-dimensional algebra in lattice gauge theory}
\author{Arthur J. Parzygnat}
\affil{\small \it Mathematics Department, University of Connecticut 
Storrs, CT 06269, USA}

\begin{document}
\maketitle

\begin{abstract}
We provide a visual and intuitive 
introduction to effectively calculating in 2-groups along with
explicit examples coming from non-abelian 1- and 2-form 
gauge theory. In particular, we utilize string diagrams, 
tools similar to tensor networks,
to compute the parallel transport along a surface using 
approximations on a lattice. 
Although this work is mainly intended as expository, 
we prove a convergence theorem for the surface transport
in the continuum limit. 
Locality is used to define infinitesimal parallel transport
and two-dimensional algebra is used to derive
finite versions along arbitrary surfaces with sufficient
orientation data. 
The correct surface ordering is dictated by two-dimensional algebra
and leads to an
interesting diagrammatic picture for gauge fields interacting with particles
and strings on a lattice. 
The surface ordering is inherently complicated, but we
prove a simplification theorem 
confirming earlier results of Schreiber and Waldorf. 
Assuming little background, 
we present a simple way
to understand some abstract concepts of higher category theory. 
In doing so, we review all the necessary categorical concepts from the tensor 
network point of view as well as many aspects of higher gauge theory. 
\end{abstract} 

\tableofcontents

\section{Introduction}
\label{sec:2dalgebraintro}
We use string diagrams to express many
concepts in gauge theory in the broader context of
two-dimensional algebra.
By two-dimensional algebra,
we mean the manipulation of algebraic quantities
along surfaces. Such manipulations are dictated
by 2-category theory and we include a thorough
and visual introduction to 2-categories based on
string diagrams. 
Such string diagrams, including their close relatives 
known as tensor networks,
have been found to provide exceptionally
clear interpretations in areas such as open quantum systems \cite{WBC15}, 
foundations of quantum mechanics \cite{AC04}, 
entanglement entropy \cite{Or14}, and
braiding statistics in topological condensed matter theory \cite{Bo17}
to name a few. 

We postulate simple rules for
associating algebraic data to surfaces with boundary
and use the rules of two-dimensional algebra to 
derive non-abelian surface transport from infinitesimal
pieces arising from a triangulation/cubulation of the surface.
One of the novelties in this work is an analytic
proof for the convergence of surface transport
together with a more direct derivation of the 
iterated surface integral than what appears in \cite{SW2} for instance. 
To be as self-contained as possible, we include discussions
on gauge transformations, orientation data on surfaces,
and a two-dimensional calculation of a Wilson cube
deriving the curvature 3-form. We also review
ordinary transport for particles to make the
transition from one-dimensional algebra to
two-dimensional algebra less mysterious. 

Ordinary algebra, matrix multiplication,
group theory, etc. are special cases of one-dimensional
algebra in the sense that they can all be described by
ordinary category theory. For example, a group is
a type of category that consists of only a single object.
Thanks to the advent of higher category theory,
beginning with the work of B\'enabou on 2-categories \cite{Be},
it has been possible to conceive of a general framework
for manipulating algebraic quantities in higher dimensions.
In particular, monoidal categories and the string diagrams
associated with them \cite{JSV} can be viewed as
2-categories with a single object. The special case of this where
all algebraic quantities have inverses are known as 2-groups,
with a simple review given in \cite{BH} and a more
thorough investigation in \cite{BL}. We do not expect the 
reader is knowledgeable of these definitions and we only
assume the reader knows about Lie groups (even a heuristic
knowledge will suffice since our formulas will be expressed 
for matrix groups). 

While there already exist several articles 
\cite{BH}, \cite{Pf}, \cite{GP}, \cite{SW4}, 
introducing the \emph{conceptual} basic ideas of higher
gauge theory and parallel transport for strings in terms of
category theory and even
a book by Schreiber describing the mathematical framework
of higher-form gauge theories \cite{Sc}, there are few articles that
provide explicit and computationally effective methods
for calculating such parallel transport \cite{Pa}. 
Although Girelli and Pfeiffer explain many ideas, most results 
useful for computations are infinitesimal and it is not clear
how to build local quantities from the infinitesimal ones \cite{GP}. 
Baez and Schreiber \cite{BS} focus on similar aspects as we do in this article,
but our presentation is significantly simplified since we 
assume certain results on path spaces without further discussion,
such as relationships between differential forms on a manifold and 
smooth functions on its path space,
and therefore do not deal with the delicate 
analytical issues on such path spaces.
Our goal is to provide tools and visualizations
to perform more intuitive calculations involving mainly calculus 
and matrix algebra. 

\subsection{Some background and history}
\label{sec:2dalgebrabg}
In 1973, Kalb and Ramond first introduced the idea of 
coupling classical abelian gauge fields to strings in \cite{KR}.
Actions for interacting charged strings were written down
together with equations of motions for both the fields
and the strings themselves. 
Furthermore, a little bit of the quantization of the theory
was discussed.  
The next big step took place in 1985 with the work
of Teitelboim (aka Bunster) and Henneaux,
who introduced higher form abelian gauge fields
that could couple to higher-dimensional manifolds 
\cite{Te}, \cite{HT}. 
In \cite{Te}, Teitelboim studied the generalization of 
parallel transport for higher dimensional surfaces
and concluded that non-abelian $p$-form gauge fields 
for $p\ge2$ cannot be coupled to $p$-dimensional manifolds
in order to construct parallel transport. The conclusion was that the 
only possibilities for string interactions involved abelian gauge fields. 
As a result, it seemed that only a
few tried to get around this in the early 1980's.
For example, the non-abelian Stoke's theorem came from
analyzing these issues in the context of Yang-Mills theories 
and confinement \cite{Ar} 
(see also for instance Section 5.3 of \cite{Ma}). 
Although 
such calculations led people to believe defining non-abelian surface 
parallel transport is possible, the expressions were not invariant 
under reparametrizations
and they did not seem well-controlled under gauge transformations.
Without a different perspective, 
interest in it seemed to fade. 

The crux of the argument of Teitelboim is related to the fact
that higher homotopy groups are abelian. This is
sometimes also known as the Eckmann-Hilton argument
\cite{BH}. 
However, J. H. C. Whitehead in 1949 realized that
higher \emph{relative} homotopy groups can be described
by \emph{non-abelian} groups \cite{Wh}. In fact, it was Whitehead
who introduced the concept of a crossed module to describe
homotopy 2-types. This work was
in the area of algebraic topology and the connection between
crossed modules and higher groups were not made until much
later. A review of this is given in \cite{BH}. Eventually, 
non-abelian generalizations of parallel transport for surfaces 
were made using category theory and ideas from homotopy theory
stressing that one should also associate differential form data
to lower-dimensional submanifolds beginning with the work of
Girelli and Pfeiffer \cite{GP}. Before this,
most of the work on non-abelian forms associated
to higher-dimensional objects did not discuss 
parallel transport but developed the combinatorial
and cocycle data \cite{At},\cite{Pf} building on the 
foundational 
work of Breen and Messing \cite{BrMe}. 
This cocycle perspective 
eventually led to the field of non-abelian differential cohomology
\cite{Sc}, \cite{Wo11}, \cite{Wal1}. 
The idea of decorating lower-dimensional manifolds
is consistent with the explicit locality exhibited in the 
extended functorial field theory approach to 
axiomatizing quantum field theories 
\cite{Se88}, \cite{Ati88}, \cite{BD}, \cite{Lu}. 
Recently, in a series of four papers, 
Schreiber and Waldorf axiomatized parallel transport
along curves and surfaces 
\cite{SW1}, \cite{SW2}, \cite{SW3}, \cite{SW4}, 
building on earlier work of Caetano and Picken \cite{CP}. 

\subsection{Motivation}
\label{sec:2dalgebramotiv}
We have already indicated one of the motivations
of pursuing an understanding of parallel transport
along surfaces, namely in the context of string theory.
Strings can be charged under non-abelian groups
and interact via non-abelian differential forms.
Just as parallel transport can be used to described
non-perturbative effects in ordinary gauge theories
for particles, parallel transport along higher-dimensional
surfaces might be used to describe non-perturbative
effects in string theory and M-theory.
Yet another use of parallel transport is in the context of lattice
gauge theory where it is used to construct Actions whose
continuum limit approaches Yang-Mills type Actions \cite{Wi74}. 

Higher form symmetries have also
been of recent interest in high energy physics
and condensed matter in the exploration
of surface operators and charges for higher-dimensional
excitations \cite{GKSW}. However, 
the forms in the latter are strictly abelian and
the proper mathematical framework for describing
them is provided by \emph{abelian} gerbes 
(aka higher bundles) \cite{MP02},\cite{TWZ} and Deligne cohomology. 
Higher non-abelian forms appear in many other contexts
in physics, such as in a stack of D-branes in string theory
\cite{My}, in the ABJM model \cite{PaSa}, and in the 
quantum field theory on the M5-brane \cite{FSS}. 
In fact, the authors of
\cite{PaSa} show how higher gauge theories 
provide a unified framework for describing certain
M-brane models and how the 3-algebras
of \cite{BagLam} can be described in this framework. 
Further work, including an explicit Action for modeling M5-branes, 
was provided recently in \cite{SS17}.

Although a description of the 
non-abelian forms themselves is described by
higher differential cohomology \cite{Sc}, parallel transport
seems to require additional flatness conditions on these
forms \cite{BH}, \cite{BS}, \cite{GP}, \cite{Pf}, \cite{Sc}, \cite{Wal2}. 
For example, in the special case of surfaces,
this condition is known as the vanishing of the fake
curvature. Some argue that this condition should
be dropped and the existence of parallel transport is not
as important for such theories \cite{Ch}. However, our perspective
is to take this condition seriously and work out
some of its consequences. Indeed, since 
higher-dimensional objects can be charged
in many physical models besides just string theory, parallel transport
might be used to study non-perturbative or effective aspects
of these theories, an important tool to understand
quantization (see the discussion at the end
of \cite{Sctalk}). 
Because it is not yet known how to avoid these
flatness conditions, further investigation is necessary, 
with some recent progress by Waldorf \cite{Wal1}, \cite{Wal2}. 

Therefore, because of the subject's infancy,
it is a good idea to devote some time to 
understanding how to \emph{calculate} surface transport
explicitly to better understand how branes
of different dimensions can be charged under
various gauge groups. Here, we focus on the
case of two-dimensional surfaces such as strings,
or D1-branes.  However, we make no explicit reference to
any known physical models. For these, we refer the reader to
other works in the literature such as \cite{SS17} and the references therein. 

Higher category theory is notoriously, and inaccurately, thought to be
too abstract of a theory to be useful for calculations or describing 
physical phenomenon. We hope to dispel this misconception in our work
and show how it can be used to expand our perspectives on
algebra, geometry, and analysis. 

\subsection{Outline}
\label{sec:2dalgebraout}
In Section \ref{sec:2dalgebracatalg}, we describe how
categorical ideas can be used to express a mix of algebraic and geometric 
concepts. Namely, in Section \ref{sec:2dalgebra1cat}, 
we review in detail ``string diagrams'' for ordinary categories
and how group theory arises as a special case of
ordinary category theory. In Section \ref{sec:2dalgebra2cat}, 
we define 2-categories and other relevant structures
providing a two-dimensional visualization of the algebraic
quantities in terms of string diagrams. In 
Section \ref{sec:2dalgebra2group}, we specialize to the case
where the algebraic data are invertible. We restrict attention
to strict 2-groups, which is sufficient for 
many interesting applications 
\cite{GKSW}, \cite{GuKa13}, \cite{PaSa}, \cite{SS17}, \cite{Sh15}. 

In Section \ref{sec:2dalgebralcgt}, we describe how
gauge theory for 0-dimensional objects (particles) 
and 1-dimensional objects (strings) can be expressed
conveniently in the language of two-dimensional algebra.
In detail, in Section \ref{sec:2dalgebra1dgt}, we review how
classical gauge theory for particles is described categorically. 
We include a review of the formula for parallel transport describing
it in terms of one-dimensional algebra as an iterated integral obtained
from a lattice discretization and a limiting procedure.
In Section \ref{sec:2dalgebra2dgt}, we include several crucial
calculations for gauge theory for 1-dimensional objects (strings)
expressing everything in terms of two-dimensional algebra. 
In particular, we derive the local infinitesimal data
of a higher gauge theory. To our knowledge, these ideas seem to have 
first been analyzed in \cite{At}, \cite{GP}, and \cite{BS}, though our
inspiration for this viewpoint came from \cite{CT}.
Furthermore, we use the rules of
two-dimensional algebra to derive an \emph{explicit} formula for
the discretized
and continuous limit versions
of the local parallel transport of non-abelian gauge fields along
a surface. Although such a formula appears in the literature \cite{BS},
\cite{SW2}, we provide a more intuitive 
derivation as well as a useful expression for lattice computations. 
We provide a picture for
the correct surface ordering needed to describe parallel
transport along surfaces with non-abelian gauge fields in 
Proposition \ref{prop:trivnconverges} and the discussion surrounding
this new result. We then proceed to prove that the surface ordering 
can be dramatically simplified in Theorem
\ref{thm:fulltoreducedsurfacetransport}. 
In Remark \ref{rmk:SW}, we show our resulting formula
agrees with the one given by Schreiber and Waldorf
that was obtained through different means \cite{SW2}. 
In Section \ref{sec:2dalgebragtst}, we study the gauge covariance
of the earlier expressions and derive the infinitesimal counterparts
in terms of differential forms. In Section \ref{sec:2dalgebraorient}, we
discuss the subtle issue of orientations of surfaces and how our
formalism incorporates them. In Section \ref{sec:2dalgebra3curv}, 
we again use two-dimensional algebra to calculate a Wilson cube
on a lattice and from it obtain the 3-form curvature. We then study how
it changes under gauge transformations showing consistency 
with the results of Girelli and Pfeiffer \cite{GP}. 

Finally, in Section
\ref{sec:2dalgebraconc} we discuss some indication
as to how these ideas might be used in physical situations
and indicate several open questions. 

\subsection{Acknowledgements}
\label{sec:2dalgebraack}
We express our sincere thanks to Urs Schreiber
and Radboud University in Nijmegen, Holland,
who hosted us for several
productive days in the summer of 2012 during which
a preliminary version of some ideas here were
prepared and presented there. 
We also thank Urs for many helpful comments and suggestions. 
We would like to thank Stefan Andronache, 
Sebastian Franco, Cheyne Miller, 
V. P. Nair, Xing Su, Steven Vayl, 
Scott O. Wilson, and Zhibai Zhang, 
for discussions, ideas, interest, and insight. 
Most of this work was done when the author was at 
the CUNY Graduate Center under the NSF 
Graduate Research Fellowship Grant No. 40017-01-04
and during a Capelloni Dissertation Fellowship. 
The present work is an updated version of a part of the author's
Ph.D. thesis \cite{Pa16}.

\section{Categorical algebra}
\label{sec:2dalgebracatalg}

\subsection{Categories as one-dimensional algebra}
\label{sec:2dalgebra1cat}
We do not assume the reader is familiar with categories
in this paper. We will present categories in terms of
what are known as ``string diagrams'' since we find
that they are simpler to manipulate and compute with
when working with 2-categories.
Therefore, we will define categories, functors, and
natural transformations in terms of string diagrams.
Afterwards, we will make a simplification and discuss
special examples of categories known as groups.
\bd
\label{defn:1dcat}
A \emph{\uline{category}}, denoted by $\mC,$ consists of 
\begin{enumerate}[i)]
\item
a collection of 1-d domains (aka objects)
\begin{figure}[H]
\centering
    \includegraphics[width=0.80\textwidth]{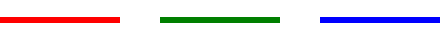}
      \begin{picture}(0,0)
\put(-345,22){$R$}
\put(-200,22){$V$}
\put(-60,22){$A$}
\end{picture}
\end{figure} 
\vspace{-3mm}
(labelled for now by some color),
\item
between any two 1-d domains, 
a collection (which could be empty) 
of 0-d defects (aka morphisms)%
\footnote{Technically, 0-d defects have
a direction/orientation. 
In this paper, the convention
is that we read the expressions from right
to left. Hence, $g$ is thought
of as ``beginning'' at $A$ and ``ending''
at $R$ or transitioning from $A$ to $R.$ 
In many cases, as in the theory
of groups, we will always be able to go back
by an inverse operation. However, in general,
$g$ will merely be a transformation from $A$ to $R.$
If at any point confusion may arise as to the direction,
we will signify with an arrow close to the 0-d defect.
See Remark 
\ref{rmk:categorydirection} for further details.}
\begin{figure}[H]
\centering
    \includegraphics[width=0.40\textwidth]{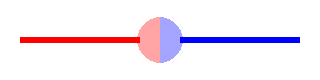}
      \begin{picture}(0,0)
\put(-155,29){$R$}
\put(-105,22){$g$}
\put(-58,29){$A$}
\end{picture}
\end{figure} 
\vspace{-3mm}
(labelled by lower-case Roman letters),
\item
an ``in series'' composition rule
\begin{figure}[H]
\centering
\begin{subfigure}{0.58\textwidth}
    \includegraphics[width=0.98\textwidth]{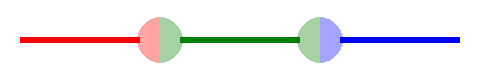}
      \begin{picture}(0,0)
\put(-195,23){$g_{2}$}
\put(-102,23){$g_{1}$}
\put(-4,21){$\to$}
\end{picture}
\end{subfigure}
\begin{subfigure}{0.38\textwidth}
\includegraphics[width=0.98\textwidth]{category_defects}
\begin{picture}(0,0)
\put(-106,22){$g_{2}g_{1}$}
\end{picture}
\end{subfigure}
\end{figure} 
\noindent
whenever 1-d domains match,
\item
and between every 1-d domain and itself, a
specified 0-d defect
\begin{figure}[H]
\centering
    \includegraphics[width=0.40\textwidth]{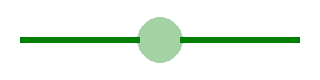}
      \begin{picture}(0,0)
\put(-104,22){$e$}
\end{picture}
\end{figure} 
\vspace{-3mm}
called the identity.
\end{enumerate}
These data must satisfy the conditions that
\begin{enumerate}[(a)]
\item
the composition rule is associative and
\item
the identity 0-d defect is a left and right identity for the
composition rule. 
\end{enumerate}
\ed
\br
\label{rmk:categorydirection}
For the reader familiar with categories, we
are defining them in terms of their Poincar\'e duals.
The relationship can be visualized by the following diagram.
\begin{figure}[H]
\centering
\includegraphics[width=0.40\textwidth]{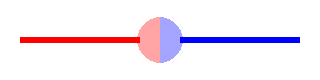}
\begin{picture}(0,0)
\put(-160,32){\xy0;/r.25pc/:
	(17,0)*+{A}="1";
	(-17,0)*+{R}="2";
	{\ar"1";"2"^{\displaystyle g}};
	\endxy
	}
\end{picture}
\end{figure} 
\vspace{-3mm}
In this article, we may occasionally use the notation
\be
\xy0;/r.25pc/:
	(17,0)*+{A}="1";
	(-17,0)*+{R}="2";
	{\ar"1";"2"_{\displaystyle g}};
	\endxy
\ee
instead and denote the 1-d domains as
``objects'' and the 0-d defects as ``morphisms.''
The motivation for using the terminology of
domains and defects comes from physics
(see Remark \ref{rmk:2catstandard} for more details). 
\er
\bx
\label{ex:BG}
Let $G$ be a group. From $G,$ one can construct
a category, denoted by $\B G,$ consisting of only
a single domain (say, red) and the collection of 0-d defects
from that domain to itself consists of all the elements of $G.$
The composition is group multiplication. The identity
at the single domain is the identity of the group.
\ex
The previous example of a category is one in which
all 0-d defects are invertible. 
\bd
\label{defn:functor}
Let $\mC$ and $\mD$ be two categories. A 
\emph{\uline{functor}}
$F:\mC\to\mD$ is an assignment sending
1-d domains in $\mC$ to 1-d domains in $\mD$ 
and 0-d defects in $\mC$ to 0-d defects in $\mD$
satisfying 
\begin{enumerate}[(a)]
\item
the source-target matching condition
\begin{figure}[H]
\centering
\begin{subfigure}{0.45\textwidth}
\centering
\includegraphics[width=0.90\textwidth]{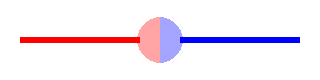}
\begin{picture}(0,0)
\put(-154,28){$R$}
\put(-106,23){$g$}
\put(-58,28){$A$}
\end{picture}
\end{subfigure}
$\xmapsto{\quad F\quad}$
\begin{subfigure}{0.45\textwidth}
\centering
\includegraphics[width=0.90\textwidth]{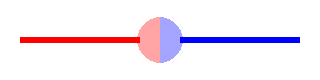}
\begin{picture}(0,0)
\put(-162,29){$F(R)$}
\put(-115,22){$F(g)$}
\put(-64,29){$F(A)$}
\end{picture}
\end{subfigure}
\end{figure} 
\item
preservation of the identity
\begin{figure}[H]
\centering
\begin{subfigure}{0.45\textwidth}
\centering
\includegraphics[width=0.97\textwidth]{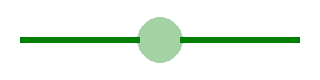}
\begin{picture}(0,0)
\put(-161,30){$V$}
\put(-117,24){$\id_{V}$}
\put(-62,30){$V$}
\end{picture}
\end{subfigure}
$\xmapsto{\quad F\quad}$
\begin{subfigure}{0.45\textwidth}
\centering
\includegraphics[width=0.97\textwidth]{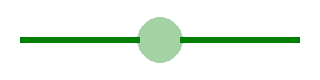}
\begin{picture}(0,0)
\put(-172,32){$F(V)$}
\put(-125,24){$\id_{F(V)}$}
\put(-70,32){$F(V)$}
\end{picture}
\end{subfigure}
\end{figure} 
\item
and preservation of the composition in series
\begin{figure}[H]
\centering
\begin{subfigure}{0.58\textwidth}
\centering
\includegraphics[width=0.96\textwidth]{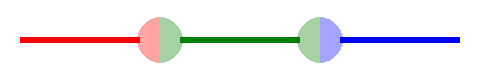}
\begin{picture}(0,0)
\put(-196,21){$\scriptstyle F(g_{2})$}
\put(-105,21){$\scriptstyle F(g_{1})$}
\end{picture}
\end{subfigure}
$=$
\begin{subfigure}{0.38\textwidth}
\centering
\includegraphics[width=0.96\textwidth]{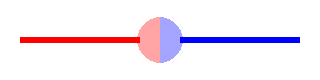}
\begin{picture}(0,0)
\put(-150,29){$F(R)$}
\put(-105,21){$\scriptstyle\!F(\!g_{1}g_{2}\!)$}
\put(-64,29){$F(A)$}
\end{picture}
\end{subfigure}
\end{figure} 
\end{enumerate}
\ed
This last condition can be expressed by saying that
the following triangle of defects commutes
\begin{figure}[H]
\centering
\includegraphics[width=0.36\textwidth]{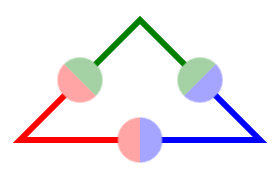}
\begin{picture}(0,0)
\put(-150,29){$F(R)$}
\put(-105,22){$\scriptstyle\!F(\!g_{1}g_{2}\!)$}
\put(-64,29){$F(A)$}
\put(-139,60){$\scriptstyle\!F(\!g_{2})$}
\put(-62,60){$\scriptstyle\!F(\!g_{1})$}
\end{picture}
\end{figure}
\noindent
meaning that going left along the top two
parts of the triangle and composing in series is the same
as going left along the bottom. 

There are several ways to think about what functors do.
On the one hand, they can be viewed as a \emph{construction}
in the sense that one begins with data and from them
constructs
new data in a consistent way. Another perspective is that
functors are invariants and give a way of associating information
that only depends on the isomorphism class of 1-d defects.
Another perspective that we will find useful in this
article is to think of a functor as attaching algebraic data to
geometric data. We will explore this last idea in Section
\ref{sec:2dalgebra1dgt} and generalize it in Section 
\ref{sec:2dalgebra2dgt}. 
Yet another perspective is to view categories more algebraically
and think of a functor as a generalization of a group homomorphism
since the third condition in Definition \ref{defn:functor}
resembles this concept. We will explore this last perspective in
in the following example. 
\bx
Let $G$ and $H$ be two groups and let 
$\B G$ and $\B H$ be their associated one-object categories
as discussed in Example \ref{ex:BG}. Then functors
$F:\B G\to\B H$ are in one-to-one correspondence
with group homomorphisms $f:G\to H.$
\ex
\bd
\label{defn:nattrans1cat}
Let $\mC$ and $\mD$ be two categories and
$F,G:\mC\to\mD$ be two functors. A
\emph{\uline{natural transformation}}
$\s:F\Rightarrow G$ is an assignment sending 1-d domains
of $\mC$
to 0-d defects of $\mD$ in such a way so that
\begin{figure}[H]
\centering
\begin{subfigure}{0.35\textwidth}
\centering
\includegraphics[width=0.70\textwidth]{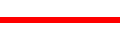}
\begin{picture}(0,0)
\put(-68,27){$R$}
\end{picture}
\end{subfigure}
$\xmapsto{\quad \s\quad}$
\begin{subfigure}{0.55\textwidth}
\centering
\reflectbox{\includegraphics[width=0.90\textwidth]{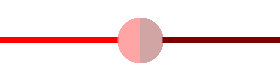}}
\begin{picture}(0,0)
\put(-202,42){$G(R)$}
\put(-136,32){$\s(R)$}
\put(-70,42){$F(R)$}
\end{picture}
\end{subfigure}
\end{figure} 
\noindent
and to every 0-d defect
\begin{figure}[H]
\centering
\includegraphics[width=0.40\textwidth]{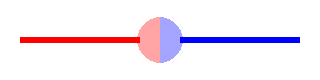}
\begin{picture}(0,0)
\put(-155,30){$R$}
\put(-105,22){$g$}
\put(-62,30){$A$}
\end{picture}
\end{figure} 
\noindent
the condition
\begin{figure}[H]
\centering
\includegraphics[width=0.60\textwidth]{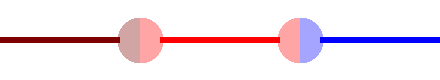}
\begin{picture}(0,0)
\put(-265,31){$G(R)$}
\put(-215,24){$\s(R)$}
\put(-163,31){$F(R)$}
\put(-110,24){$F(g)$}
\put(-60,31){$F(A)$}
\end{picture}
\end{figure} 
\vspace{-12mm}
\[
=
\]
\vspace{-9mm}
\begin{figure}[H]
\centering
\includegraphics[width=0.60\textwidth]{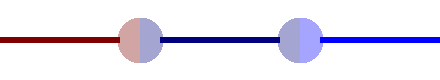}
\begin{picture}(0,0)
\put(-265,31){$G(R)$}
\put(-217,24){$G(g)$}
\put(-163,31){$G(A)$}
\put(-110,24){$\s(A)$}
\put(-60,31){$F(A)$}
\end{picture}
\end{figure} 
\noindent
must hold.
\ed
The last condition in the definition of a natural
transformation can be thought of as saying both ways
of composing in the following ``square''
\begin{figure}[H]
\centering
\includegraphics[width=0.35\textwidth]{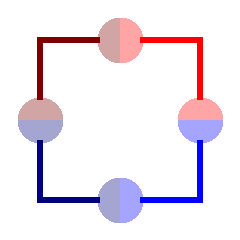}
\begin{picture}(0,0)
\put(-143,130){$G(R)$}
\put(-153,97){\rotatebox{-90}{$G(g)$}}
\put(-143,34){$G(A)$}
\put(-101,25){$\s(A)$}
\put(-62,34){$F(A)$}
\put(-101,140){$\s(R)$}
\put(-62,130){$F(R)$}
\put(-38,97){\rotatebox{-90}{$F(g)$}}
\put(-128,85){\xy0;/r.30pc/:
(-10,10)*+{}="1";
(10,10)*+{}="2";
(10,-10)*+{}="3";
(-10,-10)*+{}="4";
{\ar"2";"1"};
{\ar"3";"2"};
{\ar"4";"1"};
{\ar"3";"4"};
\endxy}
\end{picture}
\end{figure} 
\noindent
are equal (the arrows have been drawn to
be clear about the order in which one should
multiply), i.e. as an algebraic equation without
pictures
\be
\s(R)F(g)=G(g)\s(A).
\ee 
Natural transformations can be composed
though we will not need this now and 
will instead discuss this in greater
generality for 2-categories later. 

\bx
Let $G$ be a group and $\B G$ its associated
category. Let $\vect$ be the category of
vector spaces over a field $\mathbb{K}.$ Namely, 
the 1-d domains are vector spaces and the
0-d defects are $\mathbb{K}$-linear operators
between vector spaces. Let us analyze
what a functor $\rho:\B G\to\vect$ is. To the single
1-d domain of $\B G,$ $\rho$ assigns to it
some vector space, $V.$ To every group
element $g\in G,$ i.e. to every 0-d defect of $\B G,$ 
$\rho$ assigns an invertible operator $\rho(g):V\to V.$ 
This assignment satisfies $\rho(e)=\id_{V}$ and
$\rho(gh)=\rho(g)\rho(h).$ Thus, the functor $\rho$ 
encodes the data of a representation of $G.$ 
Now, let $\rho$ and $\rho'$ be two representations,
where the vector space associated to $\rho'$ is 
denoted by $V'.$ A natural transformation 
$\s:\rho\Rightarrow\rho'$ consists of a single linear 
operator
$\s:V\to V'$ satisfying the condition that
\be
\s\rho(g)=\rho'(g)\s
\ee
for all $g\in G.$
In other words, a natural transformation encodes
the data of a intertwiner of representations of $G.$%
\footnote{For the physicist not familiar with the terminology ``intertwiners,'' 
these are used to relate two different representations. 
For instance, the Fourier transform is a unitary
intertwiner between the position and momentum
representations of the Heisenberg algebra in
quantum mechanics. As another example, all tensor
operators in quantum mechanics are intertwiners
\cite{Ha13}. 
}
\ex

\subsection{2-categories as two-dimensional algebra}
\label{sec:2dalgebra2cat}
2-categories provide one realization of 
manipulating algebraic data in two dimensions.
\bd
\label{defn:2dcat}
A \emph{\uline{2-category}}, also denoted by $\mC$,
consists of 
\begin{enumerate}[i)]
\item
a collection of 2-d domains (aka objects)
\begin{figure}[H]
\centering
\begin{subfigure}{0.30\textwidth}
\centering
    \includegraphics[width=0.70\textwidth]{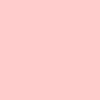}
      \begin{picture}(0,0)
\put(-60,47){$R$}
\end{picture}
\end{subfigure} 
\begin{subfigure}{0.30\textwidth}
\centering
    \includegraphics[width=0.70\textwidth]{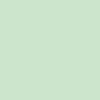}
      \begin{picture}(0,0)
\put(-60,47){$V$}
\end{picture}
\end{subfigure} 
\begin{subfigure}{0.30\textwidth}
\centering
    \includegraphics[width=0.70\textwidth]{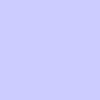}
      \begin{picture}(0,0)
\put(-60,47){$A$}
\end{picture}
\end{subfigure} 
\end{figure} 
\vspace{-3mm}
(labelled for now by some color),
\item
between any two 2-d domains, 
a collection (which could be empty) 
of 1-d defects (aka 1-morphisms or domain walls)
\begin{figure}[H]
\centering
 \includegraphics[width=0.40\textwidth]{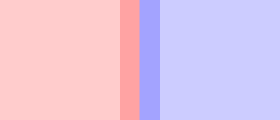}
      \begin{picture}(0,0)
\put(-105,38){$g$}
\put(-160,38){$R$}
\put(-50,38){$A$}
\end{picture}
\end{figure} 
\vspace{-3mm}
(labelled by lower-case Roman letters),
\item
between any two 1-d defects that are
themselves between the same two 2-d domains,
a collection (which could be empty)
of 0-d defects (aka 2-morphisms or excitations)%
\footnote{Technically, both 1-d defect and 0-d defects
have direction as explained later in Remark 
\ref{rmk:2catstandard}. Our convention in this
paper is that 1-d defects are read from right to left
and 0-d defects are read from top to bottom on the page. 
Occasionally, 
it will be convenient to move diagrams around
and draw them sideways or in other directions
for visual purposes. In these cases, we will
label the directionality when it might be unclear.
}
\begin{figure}[H]
\centering
\includegraphics[width=0.33\textwidth]{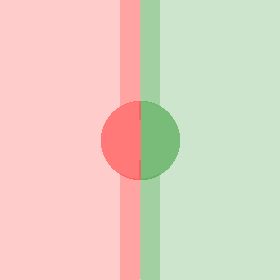}
\begin{picture}(0,0)
\put(-88,130){$g$}
\put(-88,78){$\lambda$}
\put(-88,30){$f$}
\put(-138,78){$R$}
\put(-38,78){$V$}
\end{picture}
\end{figure} 
(labelled by lower case Greek letters),
\item
an ``in parallel'' composition (aka horizontal composition) rule for 1-d defects
\begin{figure}[H]
\centering
\begin{subfigure}{0.54\textwidth}
\centering
\hspace{3mm}
\includegraphics[width=0.87\textwidth]{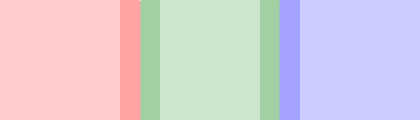}
\begin{picture}(0,0)
\put(-160,30){$f$}
\put(-83,30){$g$}
\end{picture}
\end{subfigure}
$\to$\hspace{-6mm}
\begin{subfigure}{0.435\textwidth}
\centering
\includegraphics[width=0.72\textwidth]{1ddefectfusion}
\begin{picture}(0,0)
\put(-87,30){$fg$}
\end{picture}
\end{subfigure}
\hspace{-3mm}
\end{figure} 
\item
an ``in series'' composition (aka vertical composition) rule for 0-d defects
\vspace{-1mm}
\begin{figure}[H]
\centering
\begin{subfigure}{0.45\textwidth}
\centering
\includegraphics[width=0.5\textwidth]{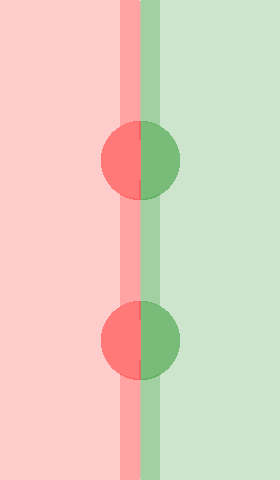}
\begin{picture}(0,0)
\put(-62,160){$h$}
\put(-62,125){$\mu$}
\put(-61,90){$g$}
\put(-61,20){$f$}
\put(-62,52){$\lambda$}
\end{picture}
\end{subfigure}
$\to$
\begin{subfigure}{0.45\textwidth}
\centering
\includegraphics[width=0.50\textwidth]{0ddefect}
\begin{picture}(0,0)
\put(-62,85){$h$}
\put(-62,52){$\begin{matrix}[0.7]\mu\\\lambda\end{matrix}$}
\put(-61,20){$f$}
\end{picture}
\end{subfigure}
\end{figure} 
\vspace{-1mm}
\item
an ``in parallel'' composition (aka horizontal composition) rule for 0-d defects
\begin{figure}[H]
\centering
\begin{subfigure}{0.45\textwidth}
\centering
\includegraphics[width=0.75\textwidth]{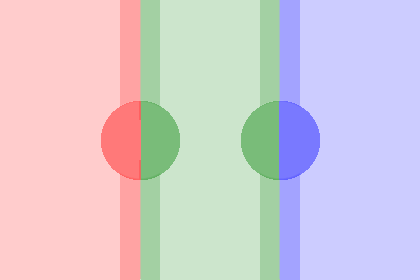}
\begin{picture}(0,0)
\put(-117,88){$g$}
\put(-117,16){$f$}
\put(-117,52){$\lambda$}
\put(-62,88){$k$}
\put(-62,16){$h$}
\put(-62,52){$\sigma$}
\end{picture}
\end{subfigure}
\hspace{4mm}$\to$\hspace{-4mm}
\begin{subfigure}{0.45\textwidth}
\centering
\includegraphics[width=0.50\textwidth]{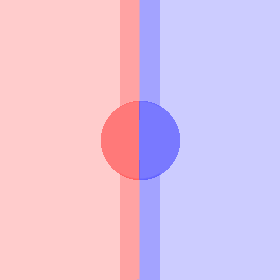}
\begin{picture}(0,0)
\put(-66,88){$gk$}
\put(-66,52){$\lambda\sigma$}
\put(-66,16){$fh$}
\end{picture}
\end{subfigure}
\end{figure} 
\item
Every 2-d domain $R$ has both an identity 1-d defect 
and an identity 0-d defect
\begin{figure}[H]
\centering
\begin{subfigure}{0.32\textwidth}
\centering
\includegraphics[width=0.775\textwidth]{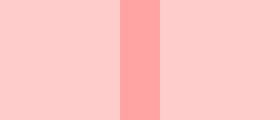}
\begin{picture}(0,0)
\put(-105,22){$R$}
\put(-72,22){$\id_{R}$}
\put(-35,22){$R$}
\end{picture}
\end{subfigure}
\begin{subfigure}{0.32\textwidth}
\centering
\includegraphics[width=0.775\textwidth]{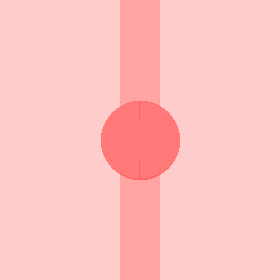}
\begin{picture}(0,0)
\put(-105,58){$R$}
\put(-72,98){$\id_{R}$}
\put(-74,58){$\id_{\id_{R}}$}
\put(-72,20){$\id_{R}$}
\put(-35,58){$R$}
\end{picture}
\end{subfigure}
\end{figure}
respectively,
and every 1-d defect has an
identity 2-d defect
\begin{figure}[H]
\centering
\includegraphics[width=0.25\textwidth]{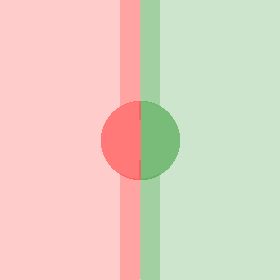}
\centering
\begin{picture}(0,0)
\put(-105,58){$R$}
\put(-68,98){$g$}
\put(-71,58){$\id_{g}$}
\put(-68,20){$g$}
\put(-32,58){$V$}
\end{picture}
.
\end{figure} 
\end{enumerate}
These data must satisfy the following conditions.
\begin{enumerate}[(a)]
\item
All composition rules are associative.%
\footnote{This will be implicit in drawing the diagrams as
we have.}
%
\item
The identities obey rules 
exhibiting them as identities for the
compositions.
\item
The composition in series and in parallel
must satisfy the ``interchange law'' 
\vspace{-3mm}
\begin{figure}[H]
\centering
\begin{subfigure}{0.45\textwidth}
\centering
\includegraphics[width=0.58\textwidth]{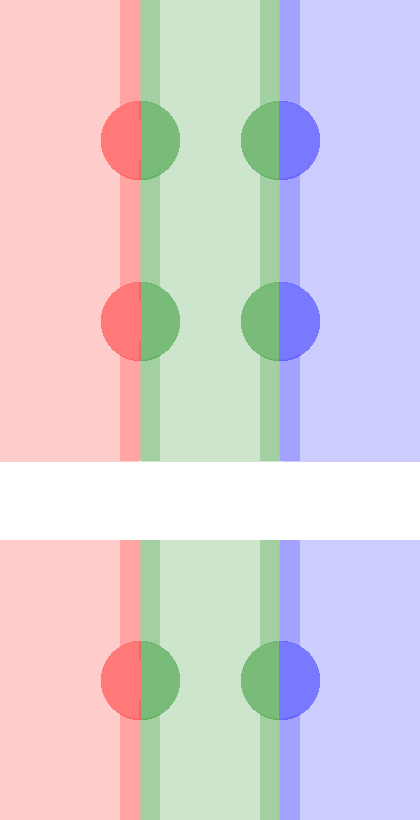}
\begin{picture}(0,0)
\put(-92,178){$g$}
\put(-92,122){$f$}
\put(-92,148){$\lambda$}
\put(-92,230){$h$}
\put(-92,205){$\mu$}
\put(-50,230){$k$}
\put(-49,122){$i$}
\put(-49,178){$j$}
\put(-49,148){$\sigma$}
\put(-50,205){$\tau$}
\put(-72,95){$\downarrow$}
\put(-92,65){$h$}
\put(-92,10){$f$}
\put(-93,39){$\begin{matrix}[0.7]\mu\\\lambda\end{matrix}$}
\put(-49,65){$k$}
\put(-48,10){$i$}
\put(-50,39){$\begin{matrix}[0.7]\tau\\\sigma\end{matrix}$}
\put(7,178){$\to$}
\put(7,40){$\to$}
\end{picture}
\end{subfigure}
\hspace{-30mm}
\begin{subfigure}{0.45\textwidth}
\centering
\includegraphics[width=0.39\textwidth]{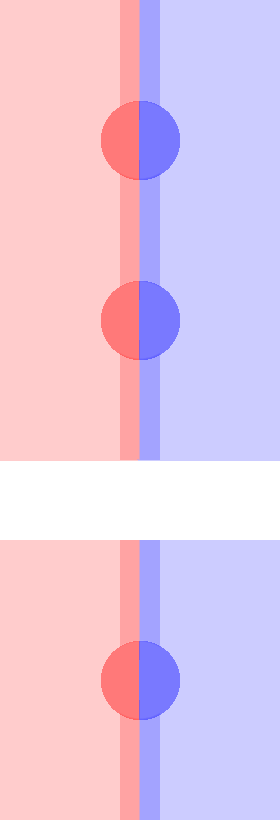}
\begin{picture}(0,0)
\put(-54,232){$hk$}
\put(-54,206){$\mu\tau$}
\put(-52,180){$gj$}
\put(-54,151){$\lambda\sigma$}
\put(-53,125){$fi$}
\put(-50,97){$\downarrow$}
\put(-54,67){$hk$}
\put(-54,40){$\begin{matrix}[0.7]\mu\tau\\\lambda\sigma\end{matrix}$}
\put(-53,13){$fi$}
\end{picture}
\end{subfigure}
\end{figure} 
meaning that the final diagram is unambiguous, i.e. 
\be
\label{eq:interchangelaw}
\begin{matrix}(\mu\tau)\\(\lambda\sigma)\end{matrix}
=
\left(\begin{matrix}\mu\\\lambda\end{matrix}\right)
\left(\begin{matrix}\tau\\\sigma\end{matrix}\right).
\ee
\end{enumerate}
\ed
These laws guarantee the well-definedness of concatenating
defects in all allowed combinations. 
\br
\label{rmk:2catstandard}
The above depiction of 2-categories
is related to the usual presentation of 2-categories via
\vspace{-5mm}
\begin{figure}[H]
\centering
\includegraphics[width=0.225\textwidth]{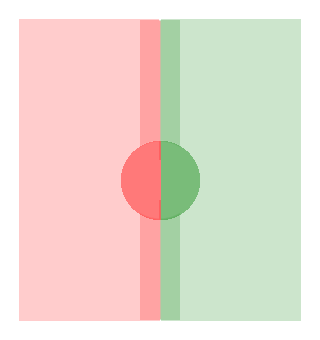}
\begin{picture}(0,0)
\put(-108,52){
$\xy0;/r.35pc/:
(-9,0)*+{R}="red";
(9,0)*+{V}="green";
(0,5.5)*{}="g";
(0,-5.5)*{}="f";
{\ar@/_2.0pc/"green";"red"_{\displaystyle g}};
{\ar@/^2.0pc/"green";"red"^{\displaystyle f}};
{\ar@{=>}"g";"f"|-{\displaystyle\lambda}};
\endxy$
}
\end{picture}
\end{figure} 
\vspace{-5mm}
\noindent
and are called ``string diagrams.''
We prefer the string diagram approach as opposed to the 
``globular'' approach because they are used in
more areas of physics such as in condensed matter
\cite{KiKo12}
and open quantum systems \cite{WBC15}. 
The terminology of domains, domain walls, defects, and excitations
comes from physics \cite{KiKo12}. 
\er
Using this definition, we can make sense of
combinations of defects such as 
\begin{figure}[H]
\centering
\includegraphics[width=0.20\textwidth]{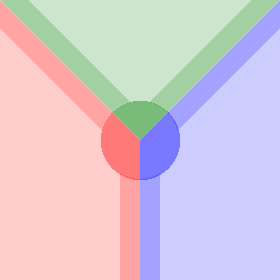}
\begin{picture}(0,0)
\put(-84,74){$h$}
\put(-28,74){$k$}
\put(-56,46){$\s$}
\put(-56,16){$g$}
\end{picture}
\end{figure} 
\noindent
interpreting it as the composition in parallel of the
top two 1-d defects along the common 2-d domain
(drawn in green)
\begin{figure}[H]
\centering
\includegraphics[width=0.20\textwidth]{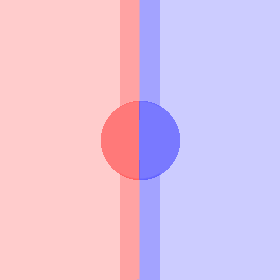}
\begin{picture}(0,0)
\put(-59,78){$hk$}
\put(-56,46){$\s$}
\put(-56,16){$g$}
\end{picture}
\end{figure} 
\noindent
In fact, a 0-d defect can have any valence with respect
to 1-d defects 
\begin{figure}[H]
\centering
\includegraphics[width=0.20\textwidth]{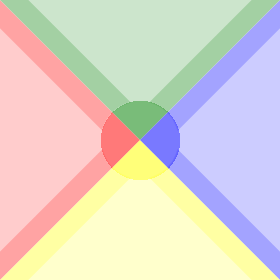}
\begin{picture}(0,0)
\put(-84,74){$h$}
\put(-28,73){$i$}
\put(-56,46){$\s$}
\put(-84,20){$g$}
\put(-28,19){$k$}
\end{picture}
\end{figure} 
\noindent
but it is important to keep in mind which 1-d defects
are incoming and outgoing. Our convention is that
all incoming 1-d defects come from above the 0-d defect and all 
outgoing 1-d defects go towards the bottom of the page. 
Occasionally,
we will go against this convention, and we will rely on the
context to be clear, or to be cautious, we may even include
arrows to indicate the direction. For example, this
last 4-valence diagram might be drawn as
\begin{figure}[H]
\centering
\includegraphics[width=0.20\textwidth]{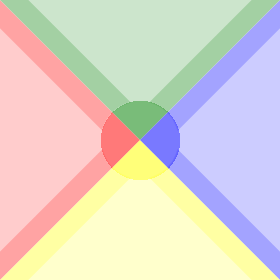}
\begin{picture}(0,0)
\put(-84,74){$h$}
\put(-28,73){$i$}
\put(-56,46){$\s$}
\put(-84,20){$g$}
\put(-28,19){$k$}
\put(-86,50){\xy0;/r.25pc/:
(10,0)*+{}="A";
(0,10)*+{}="V";
(-10,0)*+{}="R";
(0,-10)*+{}="L";
(0,0)*+{\begin{smallmatrix}\phantom{xy}\\ \phantom{xy}\end{smallmatrix}}="center";
{\ar@/_0.75pc/"A";"V"};
{\ar@/_0.75pc/"V";"R"};
{\ar@/^0.75pc/"A";"L"};
{\ar@/^0.75pc/"L";"R"};
{\ar@{=}"V";"center"};
{\ar@{=>}"center";"L"};
\endxy}
\end{picture}
\end{figure} 
\noindent
Furthermore, we can define composition in parallel
between a 1-d defect and a 0-d defect as in 
\begin{figure}[H]
\centering
\includegraphics[width=0.338\textwidth]{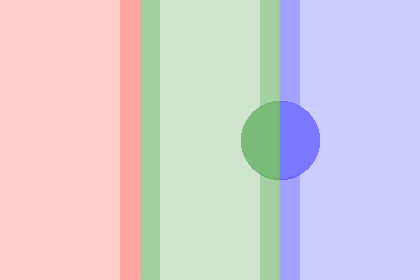}
\begin{picture}(0,0)
\put(-117,52){$g$}
\put(-62,88){$k$}
\put(-62,16){$h$}
\put(-62,52){$\sigma$}
\end{picture}
\end{figure} 
\noindent
by viewing the 1-d defect with an identity 0-d defect
and then use the already defined 
composition of 0-d defects in parallel
\begin{figure}[H]
\centering
\includegraphics[width=0.338\textwidth]{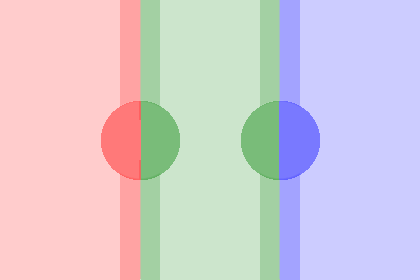}
\begin{picture}(0,0)
\put(-117,88){$g$}
\put(-117,16){$g$}
\put(-119,52){$\id_{g}$}
\put(-62,88){$k$}
\put(-62,16){$h$}
\put(-62,52){$\sigma$}
\end{picture}
\end{figure} 
\noindent
A similar idea can be used if the right side was just a 1-d defect.
Using these rules, we can make sense of diagrams such as
\begin{figure}[H]
\centering
\includegraphics[width=0.338\textwidth]{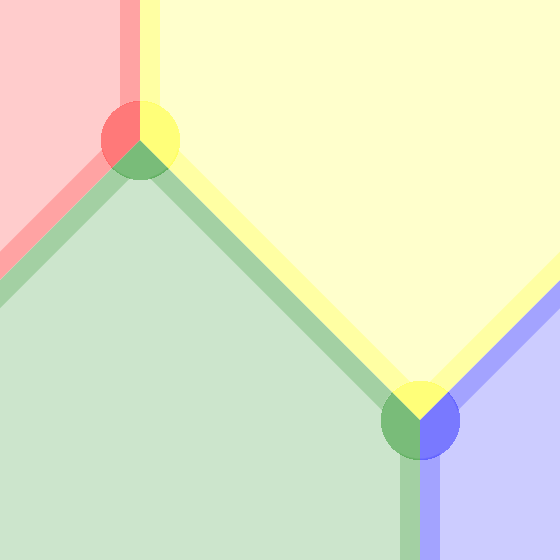}
\begin{picture}(0,0)
\put(-131,148){$h$}
\put(-131,121){$\sigma$}
\put(-156,98){$g$}
\put(-90,80){$k$}
\put(-48,13){$j$}
\put(-22,63){$i$}
\put(-48,38){$\t$}
\end{picture}
\end{figure} 
\noindent
by extending the left ``dangling'' 1-d defect to the bottom
and the right ``dangling'' 1-d defect to the top
as follows
\begin{figure}[H]
\centering
\includegraphics[width=0.676\textwidth]{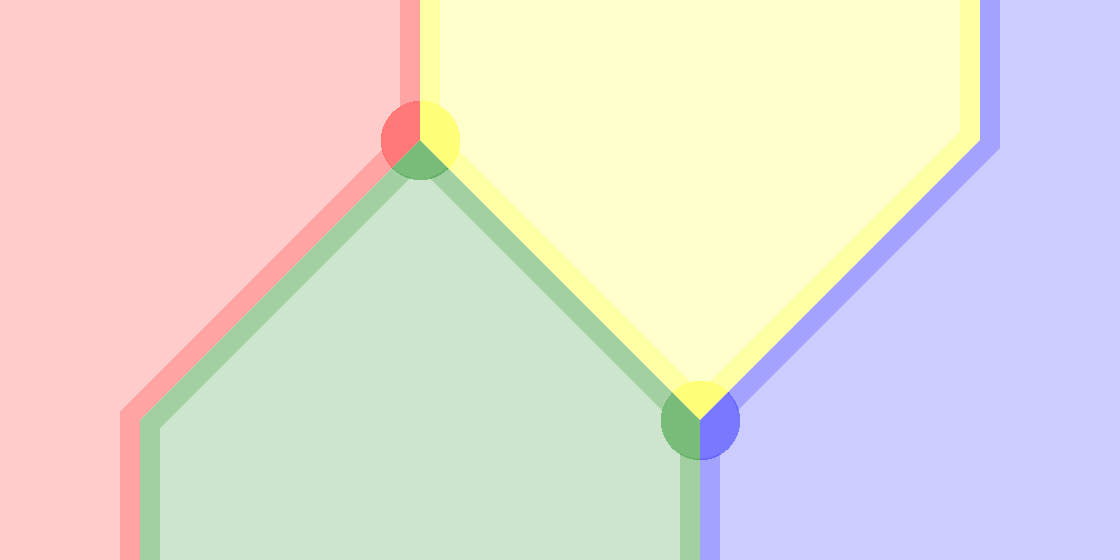}
\begin{picture}(0,0)
\put(-214,148){$h$}
\put(-214,121){$\sigma$}
\put(-257,80){$g$}
\put(-170,80){$k$}
\put(-131,13){$j$}
\put(-88,80){$i$}
\put(-131,38){$\t$}
\end{picture}
\end{figure} 
\noindent
Then we can compose in parallel to obtain
\begin{figure}[H]
\centering
\includegraphics[width=0.20\textwidth]{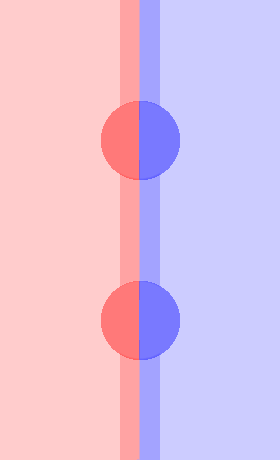}
\begin{picture}(0,0)
\put(-58,136){$hi$}
\put(-62,109){$\s\id_{i}$}
\put(-62,78){$gki$}
\put(-62,46){$\id_{g}\t$}
\put(-58,16){$gj$}
\end{picture}
\end{figure} 
\noindent
and finally compose in series
\begin{figure}[H]
\centering
\includegraphics[width=0.20\textwidth]{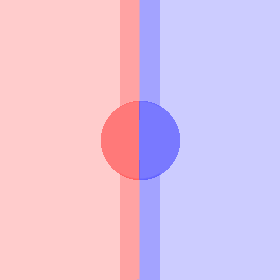}
\begin{picture}(0,0)
\put(-59,78){$hi$}
\put(-62,46){$\begin{matrix}[0.7]\s\id_{i}\\\id_{g}\t\end{matrix}$}
\put(-59,16){$gj$}
\end{picture}
\end{figure} 
\noindent
One must be cautious in such an expression. 
It does not make sense to compose $\s$ with $\id_{g}$ alone
in series because $k$ is an outgoing 1-d defect from $\s.$
Therefore, the expression 
$\begin{matrix}[0.7]\s\id_{i}\\\id_{g}\t\end{matrix}$
must be calculated by first composing in parallel
and then one can compose the results in series
as we have done. It may be less ambiguous to write this
expression as $\begin{matrix}[0.7](\s\id_{i})\\(\id_{g}\t)\end{matrix}.$
More details can be found in Joyal and Street's seminal
paper on the invariance of string diagrams under continuous
deformations \cite{JoSt91} or in many introductory accounts
of string diagrams in 2-categories. 
Examples of 2-categories
related to groups will be given in Section 
\ref{sec:2dalgebra2group}. 
\bx
\label{ex:Hilbip}
Let $\Hilb$ be the category of Hilbert spaces,
i.e. 1-d domains are Hilbert spaces and 
0-d defects are bounded linear operators. 
Let $\Hilb_{\isom}$ be the subcategory
whose 1-d domains are Hilbert spaces 
and whose 0-d defects are isometries. 
Finally, let $\Hilb_{\isom}^{\proj}$ be the 
2-category whose 2-d domains
are Hilbert spaces, 1-d defects are isometries,
and 0-d defects are elements of $U(1).$
More precisely, given two Hilbert spaces $\Hi$
and $\Hi'$ and two isometries $L,K:\Hi'\to\Hi$
a 0-d defect from $L$ to $K$ is an element $\l\in U(1)$
such that $K=\l L.$ The composition in series
is given by the product of elements in $U(1)$ 
\begin{figure}[H]
\centering
\begin{subfigure}{0.45\textwidth}
\centering
\includegraphics[width=0.4\textwidth]{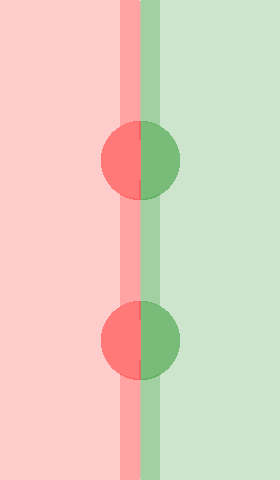}
\begin{picture}(0,0)
\put(-52,130){$L$}
\put(-52,97){$\l$}
\put(-53,70){$K$}
\put(-81,70){$\Hi$}
\put(-28,70){$\Hi'$}
\put(-52,13){$J$}
\put(-51,42){$\mu$}
\end{picture}
\end{subfigure}
$\to$
\begin{subfigure}{0.45\textwidth}
\centering
\includegraphics[width=0.40\textwidth]{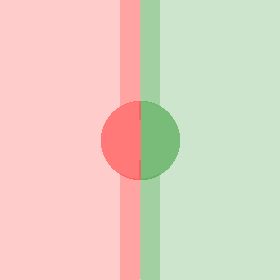}
\begin{picture}(0,0)
\put(-53,72){$L$}
\put(-55,41){$\mu\l$}
\put(-80,41){$\Hi$}
\put(-25,41){$\Hi'$}
\put(-52,12){$J$}
\end{picture}
\end{subfigure}
\end{figure} 
\noindent
and the in parallel composition is also
defined by the product of elements in $U(1)$
\begin{figure}[H]
\centering
\begin{subfigure}{0.45\textwidth}
\centering
\includegraphics[width=0.60\textwidth]{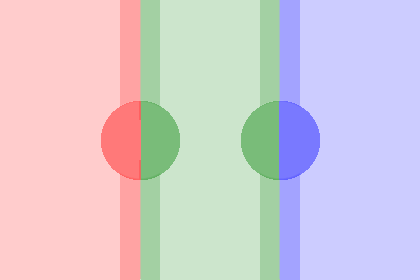}
\begin{picture}(0,0)
\put(-97,70){$L$}
\put(-97,13){$K$}
\put(-126,41){$\Hi$}
\put(-96,41){$\lambda$}
\put(-76,41){$\Hi'$}
\put(-53,70){$L'$}
\put(-54,13){$K'$}
\put(-52,41){$\lambda'$}
\put(-26,41){$\Hi''$}
\end{picture}
\end{subfigure}
\hspace{4mm}$\to$\hspace{-4mm}
\begin{subfigure}{0.45\textwidth}
\centering
\includegraphics[width=0.41\textwidth]{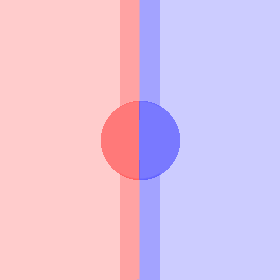}
\begin{picture}(0,0)
\put(-58,73){$LL'$}
\put(-82,42){$\Hi$}
\put(-57,42){$\lambda\lambda'$}
\put(-27,42){$\Hi''$}
\put(-59,13){$KK'$}
\end{picture}
\end{subfigure}
\end{figure} 
\noindent
The products $LL'$ and $KK'$ are given
by the composition of linear operators. 
The reader should check that this is
indeed a 2-category. 
\ex
\bx
\label{ex:tensornetworks}
A common 2-category that appears in tensor networks 
in quantum information theory is $\mathbf{Hilb}^{\otimes}$ \cite{WBC15}. 
In this 2-category, there is only a single object (2-d domain).
The 1-d defects are Hilbert spaces and 0-d defects are bounded
linear transformations. 
The parallel composition of Hilbert spaces and bounded
linear transformations is the tensor product. 
The series composition of linear transformations is the functional
composition of these operators. It is a basic property of the tensor 
product and functional composition that 
if $\mathcal{H}\xrightarrow{f}\mathcal{K}\xrightarrow{g}\mathcal{J}$
and $\mathcal{H}'\xrightarrow{f'}\mathcal{K}'\xrightarrow{g'}\mathcal{J}'$
are given, then
\be
(g\otimes g')\circ(f\otimes f')=(g\circ f)\otimes(g'\circ f').
\ee
This equality is precisely the interchange law for the compositions
in 2-categories, but writing the composition in two dimensions, namely
vertically and horizontally, makes it more clear that these expressions 
are equal. 
Note that the identity Hilbert space for the parallel composition, 
the tensor product, is the Hilbert space of complex numbers $\C.$ 
Technically, this is not an identity on the nose, nor is the tensor product
strictly associative, but one can safely ignore this issue due
to MacLane's coherence theorem on monoidal categories
\cite{Ma63}. 
\ex
Kitaev and Kong provide more examples of 2-categories in their
discussion of domains, defects, and excitations in the context
of condensed matter \cite{KiKo12}. In their language, 
we are viewing excitations as generalized defects. 
\bd
Let $\mC$ and $\mD$ be two 2-categories.
A \emph{\uline{(normalized) weak functor}} 
$F:\mC\to\mD$ 
is an assignment sending $d$-dimensional
domains/defects of $\mC$ to $d$-dimensional
domains/defects of $\mD$ 
together with an assignment $c^{F}$ that associates
to every pair of parallel composable 1-d
defects $f$ and $g$ in $\mC$ an
invertible 0-d defect in $\mD$
interpolating from $F(f)F(g)$ to
$F(fg)$ as in
\begin{figure}[H]
\centering
\begin{subfigure}{0.54\textwidth}
\centering
\hspace{3mm}
\includegraphics[width=0.87\textwidth]{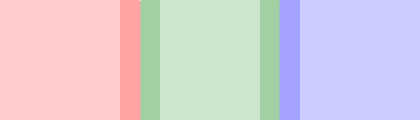}
\begin{picture}(0,0)
\put(-160,30){$f$}
\put(-83,30){$g$}
\end{picture}
\end{subfigure}
$\to$\hspace{-6mm}
\hspace{-10mm}
\begin{subfigure}{0.45\textwidth}
\centering
\includegraphics[width=0.52\textwidth]{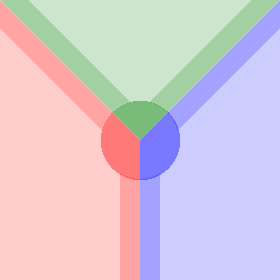}
\begin{picture}(0,0)
\put(-104,85){$F(f)$}
\put(-43,85){$F(g)$}
\put(-69,54){$c^{F}_{f,g}$}
\put(-76,18){$F(fg)$}
\end{picture}
\end{subfigure}
\end{figure} 
\noindent
These assignments must satisfy the following conditions.
\begin{enumerate}[(a)]
\item
The assignment $F$ is such that all sources and targets are respected, i.e. 
\begin{figure}[H]
\centering
\begin{subfigure}{0.45\textwidth}
\centering
\includegraphics[width=0.50\textwidth]{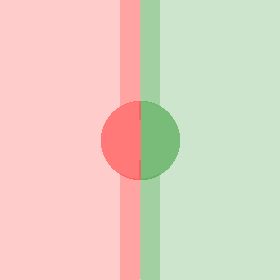}
\begin{picture}(0,0)
\put(-62,88){$g$}
\put(-62,51){$\lambda$}
\put(-63,18){$f$}
\put(-100,51){$R$}
\put(-26,51){$V$}
\end{picture}
\end{subfigure}
$\xmapsto{\quad F\quad}$
\begin{subfigure}{0.45\textwidth}
\centering
\includegraphics[width=0.50\textwidth]{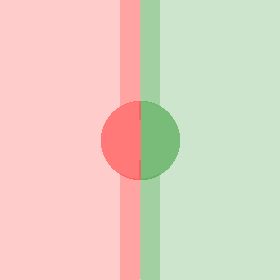}
\begin{picture}(0,0)
\put(-71,88){$F(g)$}
\put(-71,51){$F(\lambda)$}
\put(-72,18){$F(f)$}
\put(-109,51){$F(R)$}
\put(-35,51){$F(V)$}
\end{picture}
\end{subfigure}
\end{figure} 
\item
All identities are preserved (this is the ``normalized'' condition). 
\item
For any 1-d defect $f$
\begin{figure}[H]
\centering
\includegraphics[width=0.33\textwidth]{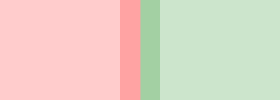}
\begin{picture}(0,0)
\put(-87,23){$f$}
\put(-131,23){$R$}
\put(-47,23){$V$}
\end{picture}
\end{figure}
the equalities
\begin{figure}[H]
\centering
\begin{subfigure}{0.30\textwidth}
\centering
    \includegraphics[width=0.75\textwidth]{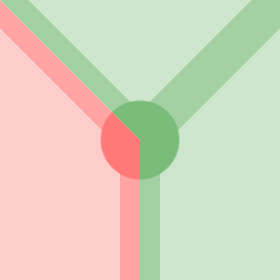}
      \begin{picture}(0,0)
\put(-104,85){$F(f)$}
\put(-43,85){$\id_{F(V)}$}
\put(-69,54){$c^{F}_{f,\id_{V}}$}
\put(-73,18){$F(f)$}
\end{picture}
\end{subfigure} 
=
\begin{subfigure}{0.30\textwidth}
\centering
    \includegraphics[width=0.75\textwidth]{0ddefect}
      \begin{picture}(0,0)
\put(-71,88){$F(f)$}
\put(-72,51){$\id_{F(f)}$}
\put(-71,18){$F(f)$}
\end{picture}
\end{subfigure} 
=
\begin{subfigure}{0.30\textwidth}
\centering
    \includegraphics[width=0.75\textwidth]{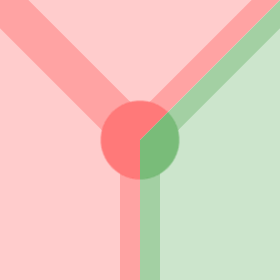}
      \begin{picture}(0,0)
\put(-104,85){$\id_{F(R)}$}
\put(-41,85){$F(f)$}
\put(-72,54){$c^{F}_{\id_{R},f}$}
\put(-72,18){$F(f)$}
\end{picture}
\end{subfigure} 
\end{figure} 
i.e.
\be
c^{F}_{f,\id_{V}}=\id_{F(f)}=c^{F}_{\id_{R},f}
\ee
must hold. 
\item
To every triple of parallel composable 1-d defects
\begin{figure}[H]
\centering
\includegraphics[width=0.75\textwidth]{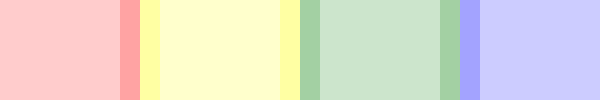}
\begin{picture}(0,0)
\put(-289,27){$f$}
\put(-190,27){$g$}
\put(-93,27){$h$}
\put(-340,27){$R$}
\put(-245,27){$L$}
\put(-141,27){$V$}
\put(-47,27){$A$}
\end{picture}
\end{figure}
the equality
\begin{figure}[H]
\centering
\includegraphics[width=0.90\textwidth]{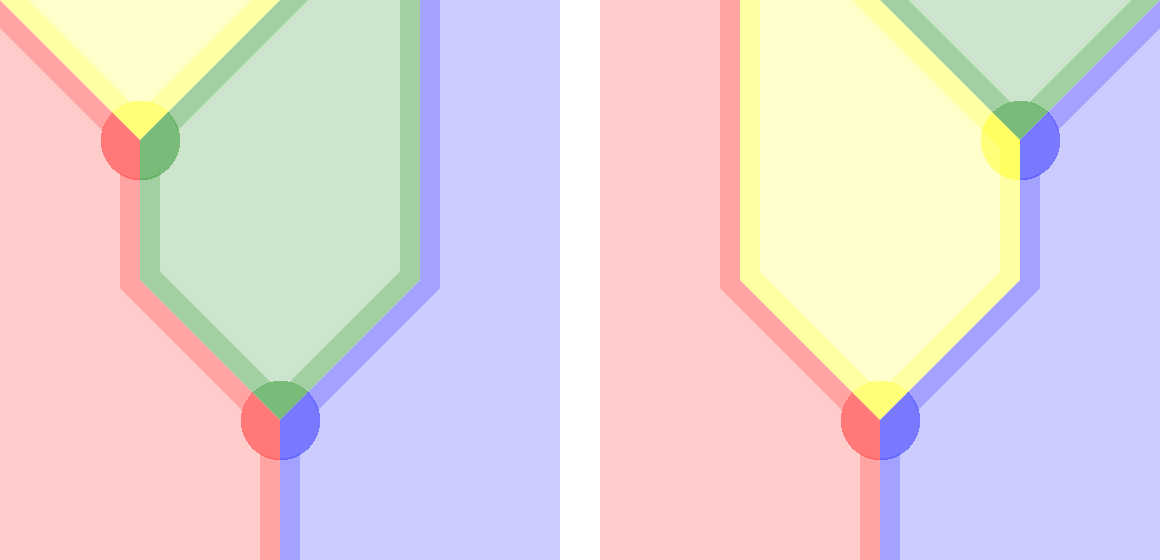}
\begin{picture}(0,0)
\put(-400,157){$c^{F}_{f,g}$}
\put(-433,185){$F(f)$}
\put(-375,185){$F(g)$}
\put(-298,105){$F(h)$}
\put(-407,105){$F(fg)$}
\put(-349,50){$c^{F}_{fg,h}$}
\put(-358,17){$F(fgh)$}
\put(-120,50){$c^{F}_{f,gh}$}
\put(-130,17){$F(fgh)$}
\put(-175,105){$F(f)$}
\put(-100,185){$F(g)$}
\put(-40,185){$F(h)$}
\put(-72,105){$F(gh)$}
\put(-65,157){$c^{F}_{g,h}$}
\put(-228,100){$=$}
\end{picture}
\end{figure}
i.e.
\be
\begin{matrix}[0.9]
c^{F}_{f,g}\id_{F(h)}\\
c^{F}_{fg,h}
\end{matrix}
=
\begin{matrix}[0.9]
\id_{F(f)}c^{F}_{g,h}\\
c^{F}_{f,gh}
\end{matrix}
\ee
must hold.
\end{enumerate}
If $c^{F}_{f,g}$ is the identity for all $f$ and $g$
in $\mC,$ then $F$ is said to be a 
\emph{\uline{strict functor}}.
\ed
\br
For each pair of composable 1-d defects $f$ and $g,$
the 0-d defect $c^{F}_{f,g}$ can be thought of
as filling in the triangle from the comments
after Definition \ref{defn:functor}
by enlarging the 1-d domains to 2-d domains
and enlarging the 0-d defects to 1-d defects.
Condition (d) resembles associativity. In fact, it
is an example of a cocycle condition and will be
discussed more in the following example
(in particular, this definition
allows one to define higher cocycles for non-abelian
groups). Condition (d) can also be re-written as
\begin{figure}[H]
\centering
\includegraphics[width=0.90\textwidth]{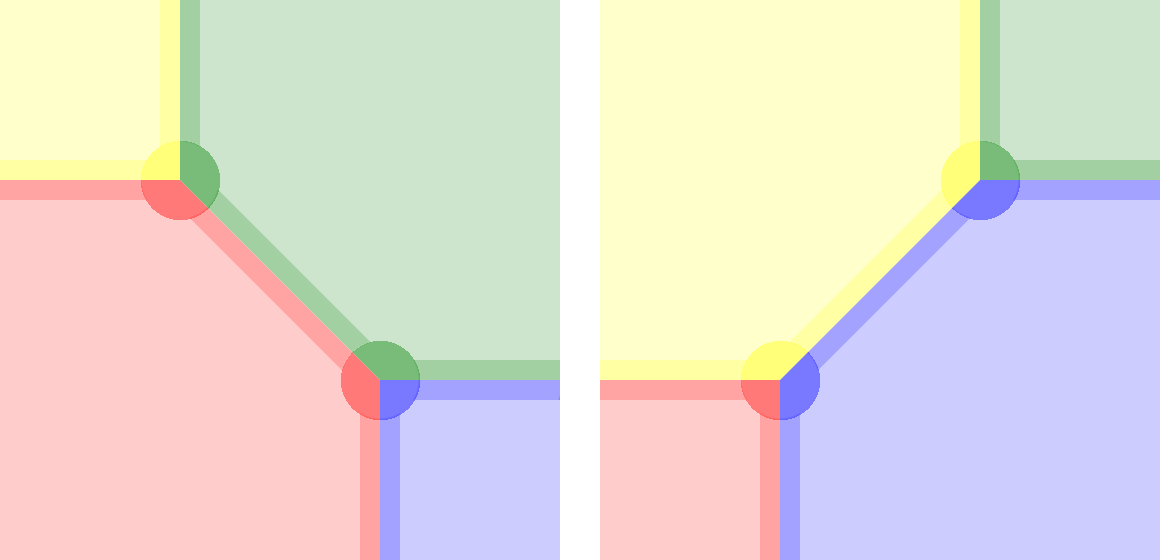}
\begin{picture}(0,0)
\put(-384,142){$c^{F}_{f,g}$}
\put(-424,141){$F(f)$}
\put(-388,185){$F(g)$}
\put(-271,66){$F(h)$}
\put(-350,100){$F(fg)$}
\put(-310,67){$c^{F}_{fg,h}$}
\put(-322,17){$F(fgh)$}
\put(-417,105){\xy0;/r.32pc/:
(-20,20)*+{}="3";
(20,-20)*+{}="1";
(20,20)*+{}="2";
(-20,-20)*+{}="4";
(14,5)*+{}="start";
(-2,-19)*+{}="end";
(-10,10)*+{\begin{matrix}\phantom{xy}\\\phantom{xy}\end{matrix}}="TL";
(10,-11)*+{\begin{matrix}\phantom{xy}\\\phantom{xy}\end{matrix}}="BR";
(0,0)*+{\begin{matrix}\phantom{xy}\\\phantom{xy}\end{matrix}}="center";
{\ar"1";"2"};
{\ar"2";"3"};
{\ar"3";"4"};
{\ar"1";"4"};
{\ar@{-}"2";"center"};
{\ar"center";"4"};
{\ar@{=}"3";"TL"};
{\ar@{=>}"TL";"center"};
{\ar@{=}"start";"BR"};
{\ar@{=>}"BR";"end"};
\endxy}
\put(-158,67){$c^{F}_{f,gh}$}
\put(-168,17){$F(fgh)$}
\put(-197,66){$F(f)$}
\put(-85,185){$F(g)$}
\put(-44,141){$F(h)$}
\put(-124,103){$F(gh)$}
\put(-80,142){$c^{F}_{g,h}$}
\put(-191,105){\xy0;/r.32pc/:
(-20,20)*+{}="3";
(20,-20)*+{}="1";
(20,20)*+{}="2";
(-20,-20)*+{}="4";
(-14,5)*+{}="start";
(2,-19)*+{}="end";
(10,10)*+{\begin{matrix}\phantom{xy}\\\phantom{xy}\end{matrix}}="TR";
(-10,-11)*+{\begin{matrix}\phantom{xy}\\\phantom{xy}\end{matrix}}="BL";
(0,0)*+{\begin{matrix}\phantom{xy}\\\phantom{xy}\end{matrix}}="center";
{\ar"1";"2"};
{\ar"2";"3"};
{\ar"3";"4"};
{\ar"1";"4"};
{\ar@{-}"1";"center"};
{\ar"center";"3"};
{\ar@{=}"2";"TR"};
{\ar@{=>}"TR";"center"};
{\ar@{=}"start";"BL"};
{\ar@{=>}"BL";"end"};
\endxy}
\put(-228,100){$=$}
\end{picture}
\end{figure}
\noindent
which illustrates more of a connection to Pachner
moves for triangulations of surfaces. 
However, this latter
presentation requires arrows to keep track
of incoming versus outgoing directions.
\er
Examples of weak functors abound. For example, 
projective representations are described by weak
functors that are not strict functors as will
be explained in the following example.
Weak functors can also be used to define the local cocycle data of 
higher bundles \cite{Wo11}. Since we will be working locally
for simplicity, we will make little use of weak functors, but have
included their discussion here for completeness and so that the 
standard definitions of higher bundles may be less mysterious 
\cite{Pa}, \cite{SW4}, \cite{Wo11}.
Strict functors will be used as a means
of \emph{defining} parallel transport along 
surfaces in gauge theory in Section 
\ref{sec:2dalgebra2dgt}. 
Natural transformations
will be used to \emph{define} gauge transformations
of such functors and their infinitesimal counterparts
will be \emph{derived} from these definitions.

\bx
\label{ex:projunrep}
Let $G$ be a group and $\B G$ its associated 
category (see Example \ref{ex:BG}). 
Every category, such as $\B G,$ 
can be given the structure of a 
2-category by adding only identity 0-d defects.
Namely, there is only a single 2-d domain,
the 1-d defects are elements of $G,$ and the 0-d
defects are all identities. This 2-category will also 
be denoted by $\B G.$ Let $\Hilb_{\isom}^{\proj}$
be the 2-category introduced in Example 
\ref{ex:Hilbip}. A weak normalized functor
$\rho:\B G\to\Hilbip$ encodes the data of a 
Hilbert space $\Hi,$ a function
$\rho:G\to U(\Hi),$ and a function 
$c^{\rho}:G\times G\to U(1)$ 
in such a way so that to every pair of elements
$g,h\in G$
\begin{figure}[H]
\centering
\includegraphics[width=0.235\textwidth]{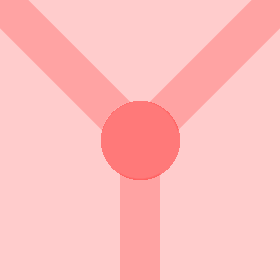}
\begin{picture}(0,0)
\put(-104,85){$\rho(g)$}
\put(-107,44){$\Hi$}
\put(-30,44){$\Hi$}
\put(-67,94){$\Hi$}
\put(-43,85){$\rho(h)$}
\put(-69,54){$c^{\rho}_{g,h}$}
\put(-76,18){$\rho(gh)$}
\end{picture}
\end{figure}
\noindent
i.e. 
\be
\rho(gh)=c^{\rho}_{g,h}\rho(g)\rho(h)
\ee
and also
\be
\rho(e)=\id_{\Hi}.
\ee
Furthermore, $c$ satisfies the condition
that to every triple $g,h,k\in G,$
\be
c^{\rho}_{gh,k}c^{\rho}_{g,h}
=c^{\rho}_{g,hk}c^{\rho}_{h,k}.
\ee
This provides the datum of a
(normalized) projective
unitary representation of $G$ on a Hilbert space
$\Hi$ (ignoring any continuity conditions). 
\ex

\bd
\label{defn:nattransf}
Let $F,G:\mC\to\mD$ be two weak functors between
two 2-categories. 
A \emph{\uline{natural transformation}}
$\s:F\Rightarrow G$ is an assignment sending
$k$-d domains/defects of $\mC$ to $(k-1)$-d defects
of $\mD$ for $k=1,2$ satisfying
the following conditions.
\begin{enumerate}[(a)]
\item
The assignment is such that
\begin{figure}[H]
\centering
\begin{subfigure}{0.30\textwidth}
\centering
    \includegraphics[width=0.70\textwidth]{red2d}
      \begin{picture}(0,0)
\put(-60,47){$R$}
\end{picture}
\end{subfigure} 
$\xmapsto{\quad\s\quad}$
\begin{subfigure}{0.60\textwidth}
\centering
    \includegraphics[width=0.70\textwidth]{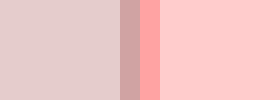}
      \begin{picture}(0,0)
\put(-63,35){$F(R)$}
\put(-178,35){$G(R)$}
\put(-119,35){$\s(R)$}
\end{picture}
\end{subfigure} 
\end{figure}
\noindent
and%
\footnote{The diagram on the right can be
thought of as filling in the square from the comments
after Definition \ref{defn:nattrans1cat} (rotate the square by
$45^{\circ}$ counterclockwise to see this more clearly).
}
\begin{figure}[H]
\centering
\begin{subfigure}{0.52\textwidth}
\centering
    \includegraphics[width=0.70\textwidth]{1ddefectfusion}
      \begin{picture}(0,0)
\put(-48,35){$A$}
\put(-148,35){$R$}
\put(-96,35){$g$}
\end{picture}
\end{subfigure} 
$\xmapsto{\quad\s\quad}$
\begin{subfigure}{0.33\textwidth}
\centering
    \includegraphics[width=0.70\textwidth]{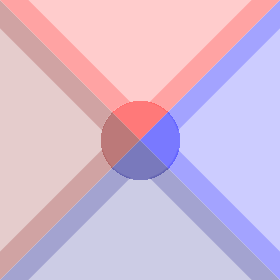}
      \begin{picture}(0,0)
\put(-71,53){$\s(g)$}
\put(-105,86){$\s(R)$}
\put(-39,86){$F(g)$}
\put(-105,20){$G(g)$}
\put(-38,20){$\s(A)$}
\end{picture}
\end{subfigure} 
\end{figure}
\item
To every pair of parallel composable 1-d defects
\begin{figure}[H]
\centering
\includegraphics[width=0.47\textwidth]{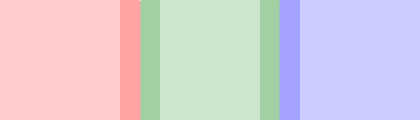}
\begin{picture}(0,0)
\put(-203,30){$R$}
\put(-160,30){$f$}
\put(-123,30){$V$}
\put(-83,30){$g$}
\put(-42,30){$A$}
\end{picture}
\end{figure}
the equality
\begin{figure}[H]
\centering
\includegraphics[width=0.90\textwidth]{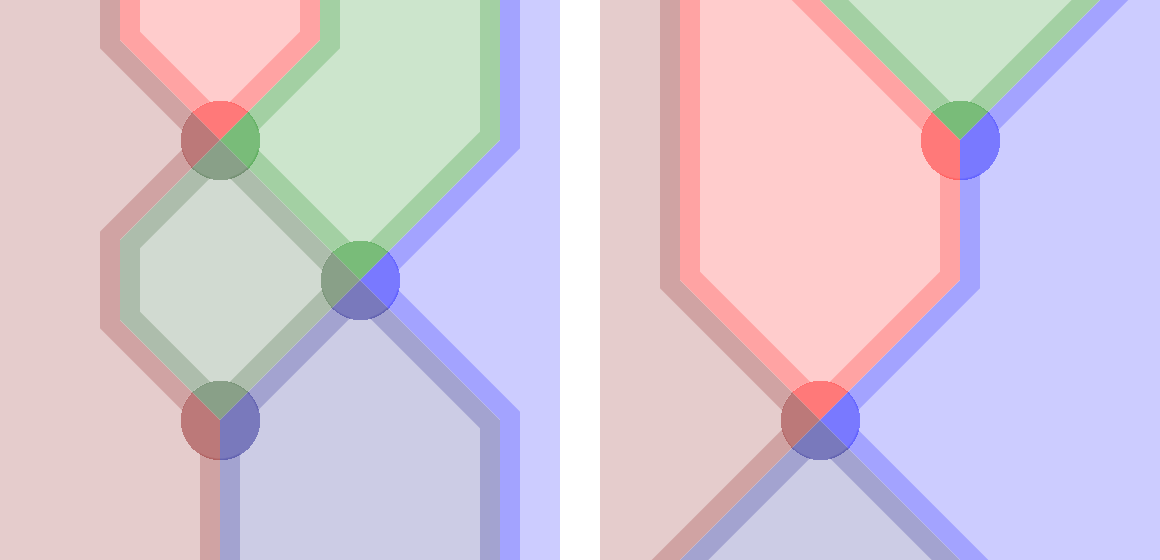}
\begin{picture}(0,0)
\put(-372,157){$\s(f)$}
\put(-268,157){$F(g)$}
\put(-411,195){$\s(R)$}
\put(-335,195){$F(f)$}
\put(-318,103){$\s(g)$}
\put(-345,130){$\s(V)$}
\put(-345,76){$G(g)$}
\put(-411,103){$G(f)$}
\put(-370,50){$c^{G}_{f,g}$}
\put(-267,50){$\s(A)$}
\put(-375,17){$G(fg)$}
\put(-147,50){$\s(fg)$}
\put(-181,17){$G(fg)$}
\put(-113,17){$\s(A)$}
\put(-195,101){$\s(R)$}
\put(-127,191){$F(f)$}
\put(-56,191){$F(g)$}
\put(-98,100){$F(fg)$}
\put(-87,157){$c^{F}_{f,g}$}
\put(-228,100){$=$}
\end{picture}
\end{figure}
\noindent
i.e. 
\be
\begin{matrix}[0.9]
\s(f)\id_{F(g)}\\
\id_{G(f)}\s(g)\\
c^{G}_{f,g}\s(A)
\end{matrix}
=
\begin{matrix}[0.9]
\s(R)c^{F}_{f,g}\\
\s(fg)
\end{matrix}
\ee
must hold.
\item
To every identity 1-d defect $\id_{R}$
the equality
%
\be
\s(\id_{R})=\id_{\s(R)}
\ee
must hold.
\item
To every 0-d defect
\begin{figure}[H]
\centering
\includegraphics[width=0.20\textwidth]{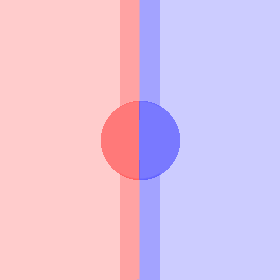}
\begin{picture}(0,0)
\put(-56,78){$f$}
\put(-56,46){$\l$}
\put(-56,16){$g$}
\end{picture}
\end{figure} 
\noindent
the equality
\begin{figure}[H]
\centering
\includegraphics[width=0.90\textwidth]{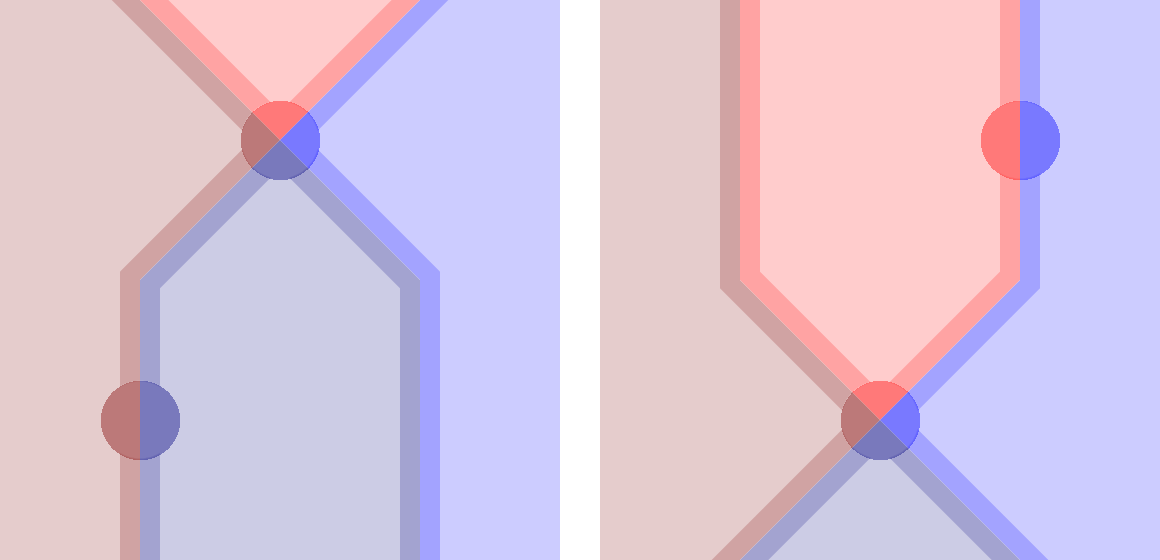}
\begin{picture}(0,0)
\put(-350,157){$\s(f)$}
\put(-391,195){$\s(R)$}
\put(-315,195){$F(f)$}
\put(-300,103){$\s(A)$}
\put(-403,103){$G(f)$}
\put(-403,50){$G(\l)$}
\put(-403,17){$G(g)$}
\put(-121,50){$\s(g)$}
\put(-157,17){$G(g)$}
\put(-90,17){$\s(A)$}
\put(-175,101){$\s(R)$}
\put(-69,191){$F(f)$}
\put(-73,100){$F(g)$}
\put(-70,157){$F(\l)$}
\put(-228,100){$=$}
\end{picture}
\end{figure}
\noindent
i.e. 
\be
\begin{matrix}[0.9]
\s(f)\\
G(\l)\id_{\s(A)}
\end{matrix}
=
\begin{matrix}[0.9]
\id_{\s(R)}F(\l)\\
\s(g)
\end{matrix}
\ee
must hold.
\end{enumerate}
\ed
Such string diagram pictures facilitate 
certain kinds of computations \cite{PoSh}
(for instance, compare the definition of natural
transformation in Figure 10 of said paper). 
Natural transformations between functors
can be thought of as symmetries. For example, just
as natural transformations of functors
between ordinary categories describe 
intertwiners for ordinary representations,
natural transformations of functors between
2-categories describe intertwiners of 
projective representations. 
\bx
Using the notation of Example \ref{ex:projunrep}, 
let $\rho,\pi:\B G\to\Hilbip$ be two projective
unitary representations on $\Hi$ and $\mK$ 
with cocycles $c^{\rho}$
and $c^{\pi},$ respectively. A natural transformation
$\s:\rho\Rightarrow\pi$ provides an isometry
$\s^{\Hi}_{\mK}:\Hi\to\mK$ 
and a function $\s:G\to U(1),$
whose value on $g$ is denoted by $\s_{g}$ and 
fits into
\begin{figure}[H]
\centering
\includegraphics[width=0.23\textwidth]{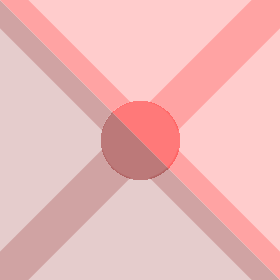}
\begin{picture}(0,0)
\put(-66,54){$\s_{g}$}
\put(-101,86){$\s^{\Hi}_{\mK}$}
\put(-39,86){$\rho(g)$}
\put(-105,20){$\pi(g)$}
\put(-34,20){$\s^{\Hi}_{\mK}$}
\put(-67,96){$\Hi$}
\put(-66,8){$\mK$}
\put(-106,53){$\mK$}
\put(-26,53){$\Hi$}
\end{picture}
,
\end{figure}
\noindent
which in particular says
\be
\pi(g)\s^{\Hi}_{\mK}=\s_{g}\s^{\Hi}_{\mK}\rho(g),
\ee
satisfying the condition 
\be
\s_{gh}c^{\rho}_{g,h}=c^{\pi}_{g,h}\s_{g}\s_{h}
\ee
for all $g,h\in G.$ This provides the data of an
intertwiner of projective unitary representations. 
\ex
It will be important to compose natural 
transformations. This will correspond to 
iterating gauge transformations successively. 
\bd
\label{defn:vertcompnattrans}
Let $E,F,G:\mC\to\mD$ be two weak functors between
two 2-categories and let $\s:F\Rightarrow G$ and
$\l:E\Rightarrow F$ be two natural transformations. 
The vertical composition of $\s$ with $\l,$
written as (read from top to bottom)
\be
\begin{matrix}[0.5]
\l\\
\s
\end{matrix}
,
\ee
is a natural transformation $E\Rightarrow G$
defined by the assignment
\begin{figure}[H]
\centering
\begin{subfigure}{0.30\textwidth}
\centering
    \includegraphics[width=0.40\textwidth]{red2d}
      \begin{picture}(0,0)
\put(-40,28){$R$}
\end{picture}
\end{subfigure} 
$\xrightarrow{\quad\quad}$
\begin{subfigure}{0.60\textwidth}
\centering
    \includegraphics[width=0.875\textwidth]{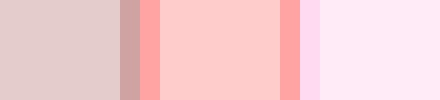}
\begin{picture}(0,0)
\put(-55,28){$E(R)$}
\put(-100,28){$\l(R)$}
\put(-147,28){$F(R)$}
\put(-192,28){$\s(R)$}
\put(-238,28){$G(R)$}
\end{picture}
\end{subfigure} 
\end{figure}
\noindent
on 2-d domains
and
\begin{figure}[H]
\centering
\begin{subfigure}{0.40\textwidth}
\centering
    \includegraphics[width=0.70\textwidth]{1ddefectfusion}
      \begin{picture}(0,0)
\put(-40,28){$A$}
\put(-115,28){$R$}
\put(-76,28){$g$}
\end{picture}
\end{subfigure} 
$\xrightarrow{\quad\quad}$
\begin{subfigure}{0.45\textwidth}
\centering
\includegraphics[width=0.90\textwidth]{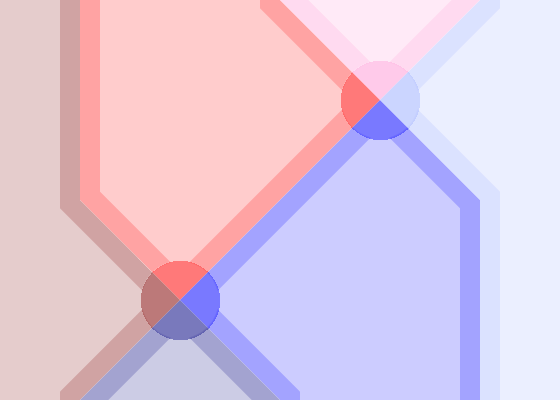}
\begin{picture}(0,0)
\put(-113,70){$F(g)$}
\put(-103,126){$\l(R)$}
\put(-78,103){$\l(g)$}
\put(-46,70){$\l(A)$}
\put(-55,126){$E(g)$}
\put(-185,70){$\s(R)$}
\put(-127,10){$\s(A)$}
\put(-149,33){$\s(g)$}
\put(-174,10){$G(g)$}
\end{picture}
\end{subfigure}
\end{figure}
\noindent
on 1-d domains.
\ed
Technically, one should check this indeed
defines a natural transformation. This is a good
exercise in two-dimensional algebra. 
There are actually similar symmetries between
natural transformations, called \emph{modifications},
which we define for completeness. 
\bd
\label{defn:modification}
Let $F,G:\mC\to\mD$ be two weak functors between
two 2-categories and 
$\s,\rho:F\Rightarrow G$ two natural transformations. 
A \emph{\uline{modification}} 
$m:\s\Rrightarrow\rho$ assigns
to every 2-d domain 
of $\mC$ a 0-d defect
in $\mD$ such that the following conditions hold.
\begin{enumerate}[(a)]
\item
The assignment is such that
\begin{figure}[H]
\centering
\begin{subfigure}{0.30\textwidth}
\centering
    \includegraphics[width=0.70\textwidth]{red2d}
      \begin{picture}(0,0)
\put(-60,47){$R$}
\end{picture}
\end{subfigure} 
$\xmapsto{\quad m\quad}$
\begin{subfigure}{0.40\textwidth}
\centering
    \includegraphics[width=0.70\textwidth]{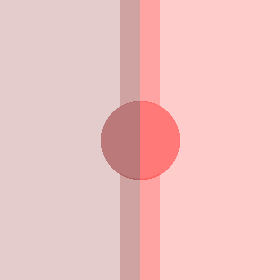}
      \begin{picture}(0,0)
\put(-43,65){$F(R)$}
\put(-130,65){$G(R)$}
\put(-84,110){$\s(R)$}
\put(-85,65){$m(R)$}
\put(-84,20){$\rho(R)$}
\end{picture}
\end{subfigure} 
\end{figure}
\noindent
\item
To every 1-d defect
\begin{figure}[H]
\centering
    \includegraphics[width=0.33\textwidth]{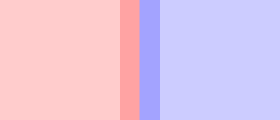}
      \begin{picture}(0,0)
\put(-45,33){$A$}
\put(-135,33){$R$}
\put(-89,33){$g$}
\end{picture}
\end{figure}
\noindent
the equality
\begin{figure}[H]
\centering
\includegraphics[width=0.90\textwidth]{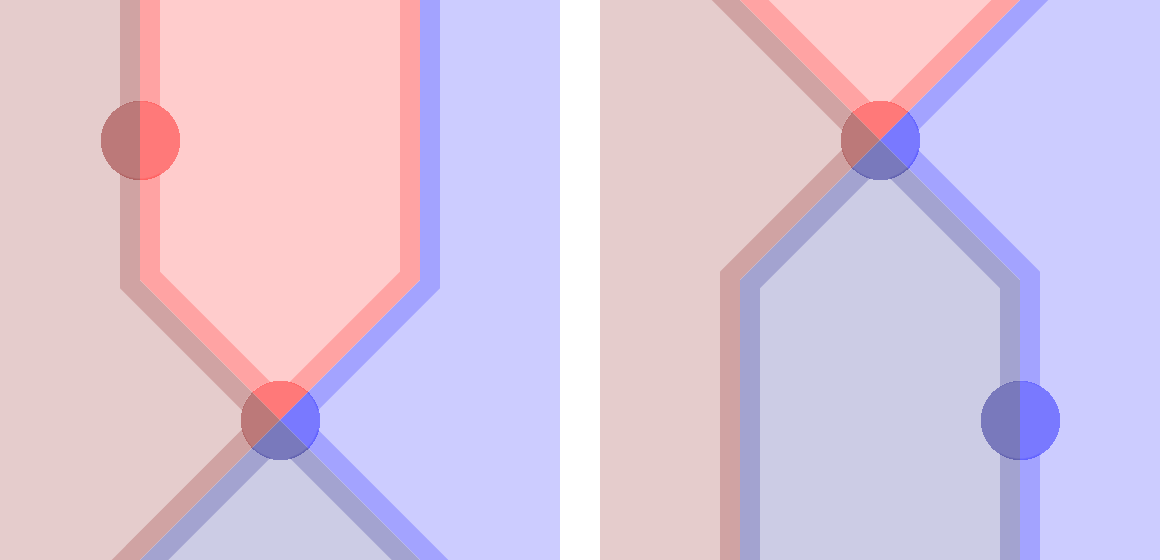}
\begin{picture}(0,0)
\put(-405,157){$m(R)$}
\put(-403,195){$\s(R)$}
\put(-300,103){$F(g)$}
\put(-403,103){$\rho(R)$}
\put(-349,50){$\rho(g)$}
\put(-385,17){$G(g)$}
\put(-315,17){$\rho(A)$}
\put(-121,157){$\s(g)$}
\put(-157,191){$\s(R)$}
\put(-70,17){$\rho(A)$}
\put(-175,101){$G(g)$}
\put(-87,191){$F(g)$}
\put(-70,100){$\s(A)$}
\put(-71,50){$m(A)$}
\put(-228,100){$=$}
\end{picture}
\end{figure}
\noindent
i.e.
\be
\begin{matrix}[0.9]
m(R)\id_{F(g)}\\
\rho(g)
\end{matrix}
=
\begin{matrix}[0.9]
\s(g)\\
\id_{G(g)}m(A)
\end{matrix}
\ee
must hold.
\end{enumerate}
\ed

\subsection{Two-dimensional group theory}
\label{sec:2dalgebra2group}
A convenient class of 2-categories are those for which
there is only a single 2-d domain and all defects are
invertible under all compositions. Such a 2-category
is called a \emph{\uline{2-group}}. 2-groups therefore
only have labels on 1-d and 0-d defects. They can be
described more concretely in terms of more familiar
objects, namely ordinary groups. 
\bd
\label{defn:crossedmodule2dalg}
A \emph{\uline{crossed module}} is a quadruple 
$\mathcal{G}:=(H, G , \t, \a)$ 
of two groups, $G$ and $H,$ group homomorphisms 
$\t : H \to G$ and $\a : G \to \mathrm{Aut} (H),$ 
satisfying the two conditions 
\be
\a_{\t(h)} (h') = h h' h^{-1} 
\ee
and
\be
\label{eq:cm2}
\t ( \a_{g} ( h) ) = g \t ( h ) g^{-1} 
\ee
for all $g\in G$ and $h,h'\in H.$ 
Here $\mathrm{Aut} (H)$ is the automorphism group of $H,$
i.e. invertible group homomorphisms from $H$ to itself. 
If the groups $G$ and $H$ are Lie groups and the maps 
$\t$ and $\a$ are smooth, then $(H, G , \t, \a)$ is called 
a \emph{\uline{Lie crossed module}}.
\ed
Examples of crossed modules abound.
\bx
Let $G$ be any group, $H:=G,$ $\t:=\mathrm{id}_{G},$
and let $\a$ be conjugation. 
\ex
\bx
Let $H$ be any group, $G:=\mathrm{Aut}(H),$
let $\t(h)$ be the automorphism defined by $\t(h)(h'):=hh'h^{-1}$
for all $h,h'\in H,$ 
and set $\a:=\mathrm{id}_{\mathrm{Aut}(H)}.$ 
\ex
\bx
Let $N$ be a normal subgroup of $G.$ Set $H:=N,$
$\t$ the inclusion, and $\a$ conjugation. 
\ex
\bx
Let $G$ be a Lie group, $\t:H\to G$ a covering space,
and $\a$ conjugation by a lift. For instance, 
$\exp\{2\pi i\ \cdot\ \}:\R\to S^1$ and the quotient map
$SU(n)\to SU(n)/Z(n)$ give examples. Here
$SU(n)$ is the set of $n\times n$ special unitary 
matrices and $Z(n)$ is its center, i.e. elements of
the form $e^{2\pi i k/n} \id_{n}$ with $k\in\Z.$
\ex
\bx
Let $G:=\{*\},$ the trivial group, $H$ any \emph{abelian} group,
$\t$ the trivial map, and $\a$ the trivial map.
\ex

\br
\label{rmk:abelianH}
It is \emph{not} possible for $H$ to be a non-abelian
group if $G$ is trivial. In fact, for an arbitrary
crossed module $(H,G,\t,\a),$ 
$\mathrm{ker}(\t)$ is always a central subgroup of $H.$
\er
We now use crossed modules to construct examples
of 2-categories, specifically 2-groups.
\bx
\label{ex:2groups}
Let $\mathcal{G}:=(H,G,\t,\a)$ be a crossed module.
From $\mathcal{G},$ one can construct a 2-category,
denoted by $\B\mathcal{G},$ consisting only of a 
single 2-d domain, the 1-d defects are labelled by
elements of $G$ and the 0-d defects are labelled
by elements of $H.$ However, such labels
must be of the form
\begin{figure}[H]
\centering
\includegraphics[width=0.20\textwidth]{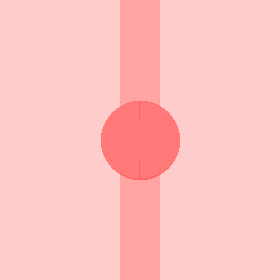}
\begin{picture}(0,0)
\put(-56,78){$g$}
\put(-56,46){$h$}
\put(-67,16){$\t(h)g$}
\end{picture}
\end{figure} 
\noindent
Composition of 1-d defects in parallel is the group
multiplication in $G$ just as in $\B G$
(see Example \ref{ex:BG}).  
Composition of 0-d defects in series is defined by
\vspace{-1mm}
\begin{figure}[H]
\centering
\hspace{5mm}
\begin{subfigure}{0.45\textwidth}
\centering
\includegraphics[width=0.45\textwidth]{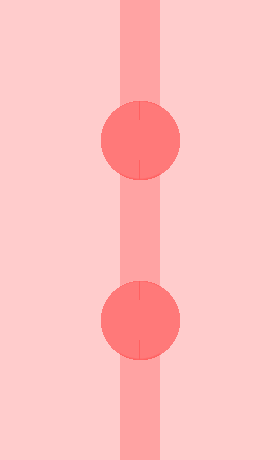}
\begin{picture}(0,0)
\put(-58,143){$g_{1}$}
\put(-57,110){$h$}
\put(-68,78){$\t(h)g_{1}$}
\put(-79,14){$\t(h')\t(h)g_{1}$}
\put(-57,45){$h'$}
\end{picture}
\end{subfigure}
$\to$
\begin{subfigure}{0.45\textwidth}
\centering
\includegraphics[width=0.45\textwidth]{0ddefect2grp}
\begin{picture}(0,0)
\put(-58,78){$g_{1}$}
\put(-61,46){$h'h$}
\put(-72,14){$\t(h'h)g_{1}$}
\end{picture}
\end{subfigure}
\end{figure} 
\noindent
Composition of 0-d defects in parallel is defined by 
\begin{figure}[H]
\centering
\begin{subfigure}{0.45\textwidth}
\centering
\includegraphics[width=0.825\textwidth]{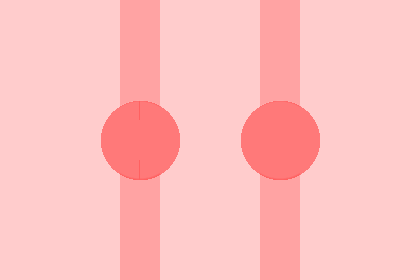}
\begin{picture}(0,0)
\put(-130,97){$g_{2}$}
\put(-142,18){$\t(h_{2})g_{2}$}
\put(-130,56){$h_{2}$}
\put(-69,97){$g_{1}$}
\put(-82,18){$\t(h_{1})g_{1}$}
\put(-69,56){$h_{1}$}
\end{picture}
\end{subfigure}
$\to$
\hspace{-10mm}
\begin{subfigure}{0.45\textwidth}
\centering
\includegraphics[width=0.55\textwidth]{0ddefect2grp}
\begin{picture}(0,0)
\put(-74,98){$g_{2}g_{1}$}
\put(-87,57){$h_{2}\a_{g_{2}}(h_{1})$}
\put(-100,18){$\t(h_{2})g_{2}\t(h_{1})g_{1}$}
\end{picture}
\end{subfigure}
\end{figure} 
\noindent
Notice that the outgoing 1-d defect is consistent with
our definitions because 
\be
\t\Big(h_{2}\a_{g_{2}}(h_{1})\Big)g_{2}g_{1}
=\t(h_{2})g_{2}\t(h_{1})g_{2}^{-1}g_{2}g_{1}
=\t(h_{2})g_{2}\t(h_{1})g_{1}
\ee
due to (\ref{eq:cm2}).

The identities are given as follows. The 1-d defect identity
associated to the single 2-d domain is the 1-d defect labelled
by $e,$ the identity of $G.$ 
The identity 0-d defect associated to a 1-d defect labelled by 
$g$ is labelled by slight abuse of notation $e,$ 
the identity of $H.$ 
It follows from these two
definitions that the identity 0-d defect associated to the
single 2-d domain is labelled by the identity on both the 1-d and
0-d defects. 
These three identities are depicted visually as
\begin{figure}[H]
\centering
\begin{subfigure}{0.32\textwidth}
\centering
\includegraphics[width=0.775\textwidth]{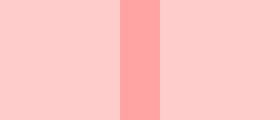}
\begin{picture}(0,0)
\put(-67,22){$e$}
\end{picture}
\end{subfigure}
\begin{subfigure}{0.32\textwidth}
\centering
\includegraphics[width=0.775\textwidth]{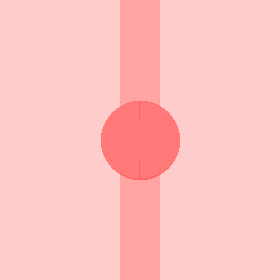}
\begin{picture}(0,0)
\put(-68,98){$g$}
\put(-68,58){$e$}
\put(-68,20){$g$}
\end{picture}
\end{subfigure}
\begin{subfigure}{0.32\textwidth}
\centering
\includegraphics[width=0.775\textwidth]{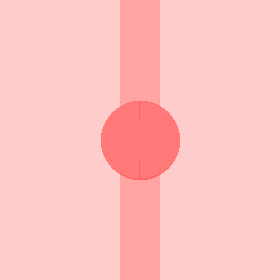}
\begin{picture}(0,0)
\put(-68,98){$e$}
\put(-68,58){$e$}
\put(-68,20){$e$}
\end{picture}
\end{subfigure}
\end{figure} 
\noindent
respectively.

The inverse of the 1-d defect labelled by $g$
for the parallel composition of 1-d defects is just 
the 1-d defect labelled by $g^{-1}.$ Inverses for
0-d defects are depicted 
for series composition by 
\begin{figure}[H]
\centering
\begin{subfigure}{0.30\textwidth}
\centering
\includegraphics[width=0.675\textwidth]{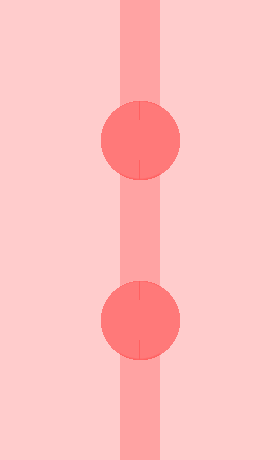}
\begin{picture}(0,0)
\put(-57,143){$g$}
\put(-57,110){$h$}
\put(-68,78){$\t(h)g$}
\put(-76,14){$\t(h^{-1}h)g$}
\put(-60,45){$h^{-1}$}
\end{picture}
\end{subfigure}
$=$
\begin{subfigure}{0.30\textwidth}
\centering
\includegraphics[width=0.675\textwidth]{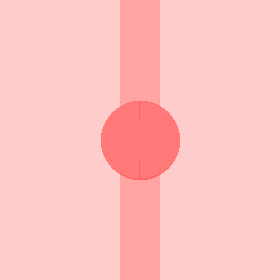}
\begin{picture}(0,0)
\put(-57,78){$g$}
\put(-57,46){$e$}
\put(-57,14){$g$}
\end{picture}
\end{subfigure}
=
\begin{subfigure}{0.30\textwidth}
\centering
\includegraphics[width=0.675\textwidth]{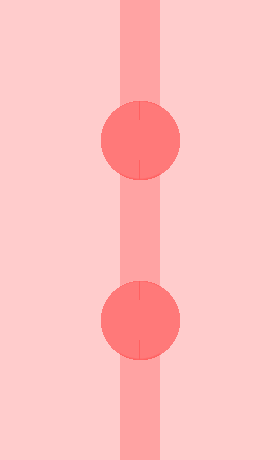}
\begin{picture}(0,0)
\put(-57,143){$g$}
\put(-60,110){$h^{-1}$}
\put(-72,78){$\t(h^{-1})g$}
\put(-76,14){$\t(hh^{-1})g$}
\put(-57,45){$h$}
\end{picture}
\end{subfigure}
\end{figure} 
\noindent
and parallel composition by 
\begin{figure}[H]
\centering
\begin{subfigure}{0.45\textwidth}
\centering
\includegraphics[width=0.825\textwidth]{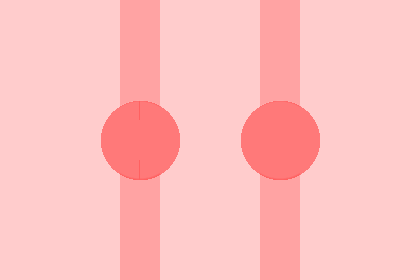}
\begin{picture}(0,0)
\put(-129,97){$g$}
\put(-140,18){$\t(h)g$}
\put(-129,56){$h$}
\put(-72,97){$g^{-1}$}
\put(-85,18){$g^{-1}\t(h^{-1})$}
\put(-86,56){$\a_{g^{-1}}(h^{-1})$}
\end{picture}
\end{subfigure}
$=$
\hspace{-10mm}
\begin{subfigure}{0.45\textwidth}
\centering
\includegraphics[width=0.55\textwidth]{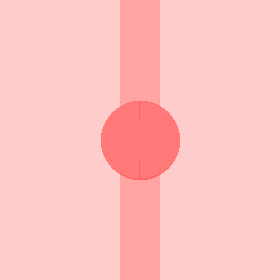}
\begin{picture}(0,0)
\put(-67,98){$e$}
\put(-67,57){$e$}
\put(-67,20){$e$}
\end{picture}
\end{subfigure}
\end{figure} 
\noindent
and similarly on the left. Notice that 0-d defects have
two inverses for the two compositions.
\ex

This last class of examples of 2-groups
from crossed modules will be used throughout
this paper. In fact, all 2-groups arise in this way.

\bt
For every 2-group, 
let $G$ be the set of 1-d defects and
let $H$ be the set of 0-d defects of the form 
\begin{figure}[H]
\centering
\includegraphics[width=0.20\textwidth]{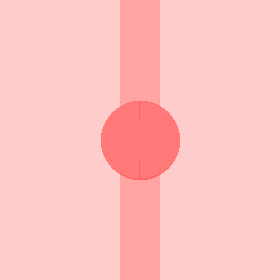}
\begin{picture}(0,0)
\put(-56,78){$e$}
\put(-56,46){$h$}
\put(-56,16){$g$}
\end{picture}
\end{figure} 
\noindent
(i.e. 0-d defects whose source 1-d defect is $e$).
Define $\t:H\to G$ by $\t(h):=g$ from 0-d defects
of the above form. Set $\a_{g}(h)$ to be the
resulting 0-d defect obtained from the composition
\begin{figure}[H]
\centering
\includegraphics[width=0.40\textwidth]{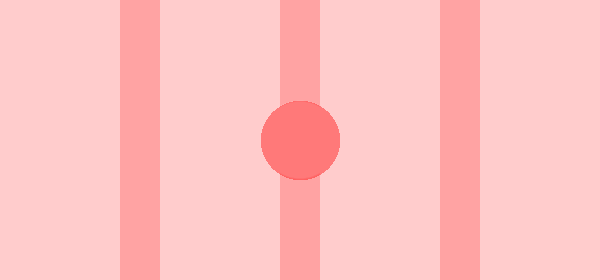}
\begin{picture}(0,0)
\put(-105,73){$e$}
\put(-106,43){$h$}
\put(-113,13){$\t(h)$}
\put(-156,43){$g$}
\put(-57,43){$g^{-1}$}
\end{picture}
.
\end{figure}
\noindent
The product in $G$ is obtained from the 
composition of 1-d defects in parallel
and the product in $H$ is obtained from the
composition of 0-d defects in series.
With this structure, $(G,H,\t,\a)$ is a crossed module. 
Furthermore, this correspondence between
crossed modules and 2-groups extends
to an equivalence of 2-categories \cite{BH}. 
\et

We now provide some
examples of 2-groups along with weak functors
between them to illustrate their meaning.

\bx
Let $G$ be a group and $\Hi$ a Hilbert space.
Let $U(\Hi)$ denote the unitary operators of $\Hi.$
Let $\mathcal{G}$ be the crossed module 
$(\{1\},G,!,!),$ where the $!$ stand for 
the trivial
map and trivial action, respectively. Let 
$\mathcal{U}(\Hi)$ be the crossed module
$(U(1),U(\Hi),\t,\a)$ with 
$\t(e^{i\theta}):=e^{i\theta}\mathrm{id}_{\Hi}$
and $\a$ the trivial action.
By definition, a weak functor 
$\rho:\mathcal{G}\to\mathcal{U}(\Hi)$ consists of 
a function $\rho:G\to U(\Hi)$ and a function
$c^{\rho}:G\times G\to U(1)$ 
of the form sending $(g,h)$ to 
\begin{figure}[H]
\centering
\includegraphics[width=0.235\textwidth]{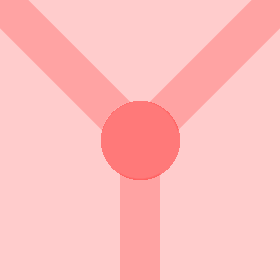}
\begin{picture}(0,0)
\put(-104,85){$\rho(g)$}
\put(-43,85){$\rho(h)$}
\put(-69,54){$c^{\rho}_{g,h}$}
\put(-76,18){$\rho(gh)$}
\end{picture}
\end{figure}
\noindent
which in particular says 
\be
c^{\rho}_{g,h}\rho(g)\rho(h)=\rho(gh),
\ee
satisfying
\be
c^{\rho}_{g,e}=1=c^{\rho}_{e,g}
\ee
for all $g\in G$ and 
\be
c^{\rho}_{gh,k}c^{\rho}_{g,h}
=c^{\rho}_{g,hk}c^{\rho}_{h,k}
\ee
for all $g,h,k\in G.$ This is
the definition of a (normalized) 
projective representation of $G$
on $\Hi$ and is really a special case of
Example \ref{ex:projunrep},
where the Hilbert space is
fixed from the start. The crossed module
$\mathcal{U}(\Hi)$ introduced here
is actually the automorphism crossed module
(in analogy to the automorphism group)
of the Hilbert space $\Hi$ viewed as
a 2-d domain in the 2-category $\Hilbip.$ 
\ex
The following fact will be used in
distinguishing two types of gauge transformations.
It allows one to decompose an arbitrary 
gauge transformation into a composition
of these two types. 
\bn
\label{prop:nattransftoa2group}
Let $\mC$ be a category viewed as a 2-category 
so that its 1-d domains become 2-d domains, 
its 0-d defects become 1-d defects, and its 0-d defects
are all identity 0-d defects. 
$\mathcal{G}:=(H,G,\t,\a)$ a crossed module
with associated 2-group $\B\mathcal{G},$ and 
$F,F':\mC\to\B\mathcal{G}$ two \emph{strict}
functors (so that $c^{F}$ and $c^{F'}$ are identities). 
A natural transformation $\s:F\Rightarrow F'$ 
consists of a function from 2-d defects of $\mC$ 
to $G,$ denoted by $g$ 
\begin{figure}[H]
\centering
\begin{subfigure}{0.30\textwidth}
\centering
    \includegraphics[width=0.475\textwidth]{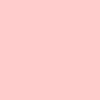}
      \begin{picture}(0,0)
\put(-42,33){$z$}
\end{picture}
\end{subfigure} 
$\xmapsto{\quad\s\quad}$
\begin{subfigure}{0.50\textwidth}
\centering
\includegraphics[width=0.66\textwidth]{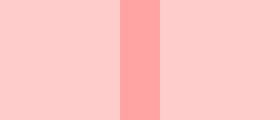}
\begin{picture}(0,0)
\put(-95,33){$g(z)$}
\end{picture}
\end{subfigure}
,
\end{figure}
\noindent
and a function from 1-d defects of $\mC$ to
$H,$ denoted by $h$ 
\begin{figure}[H]
\centering
\begin{subfigure}{0.45\textwidth}
\centering
\includegraphics[width=0.715\textwidth]{1ddefectRV}
\begin{picture}(0,0)
\put(-87,23){$\g$}
\put(-131,23){$z$}
\put(-47,23){$y$}
\end{picture}
\end{subfigure}
$\xmapsto{\quad\s\quad}$
\begin{subfigure}{0.45\textwidth}
\centering
\includegraphics[width=0.45\textwidth]{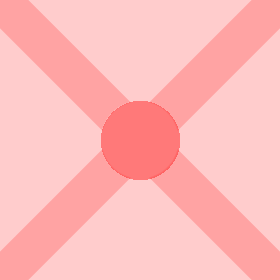}
\begin{picture}(0,0)
\put(-92,20){$F'(\g)$}
\put(-91,76){$g(z)$}
\put(-64,46){$h(\g)$}
\put(-35,20){$g(y)$}
\put(-36,76){$F(\g)$}
\end{picture}
\end{subfigure}
,
\end{figure}
\noindent
which says that
\be
\t\big(h(\g)\big)g(z)F(\g)=F'(\g)g(y),
\ee
satisfying the axioms in the definition of a
natural transformation. Thus, $\s$ can be
written as the pair $(g,h).$ Furthermore, there
exists a strict functor $F'':\mC\to\B\mathcal{G}$ 
such that the natural transformation $\s$ decomposes into a 
vertical composition (recall Definition \ref{defn:vertcompnattrans})
of the natural transformations $(g,e):F\Rightarrow F''$ and
$(e,h):F''\Rightarrow F,$ i.e.
\be
\s=\begin{matrix}[0.9](g,e)\\(e,h)\end{matrix}
\ee
namely, for any 1-d defect $z\xleftarrow{\g}y,$
\begin{figure}[H]
\centering
\begin{subfigure}{0.45\textwidth}
\centering
\includegraphics[width=0.45\textwidth]{nattransf2group}
\begin{picture}(0,0)
\put(-92,20){$F'(\g)$}
\put(-91,76){$g(z)$}
\put(-64,46){$h(\g)$}
\put(-35,20){$g(y)$}
\put(-36,76){$F(\g)$}
\end{picture}
\end{subfigure}
\hspace{-8mm}$=$\hspace{8mm}
\begin{subfigure}{0.45\textwidth}
\centering
\includegraphics[width=0.90\textwidth]{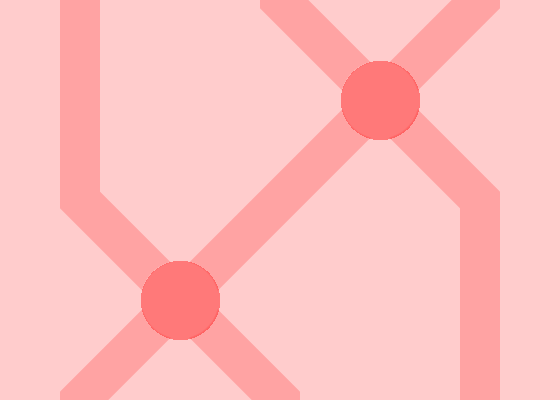}
\begin{picture}(0,0)
\put(-113,70){$F''(\g)$}
\put(-103,126){$g(z)$}
\put(-71,103){$e$}
\put(-46,70){$g(y)$}
\put(-55,126){$F(\g)$}
\put(-178,70){$e$}
\put(-118,11){$e$}
\put(-149,33){$h(\g)$}
\put(-175,11){$F'(\g)$}
\end{picture}
\end{subfigure}
\end{figure}
\en
\bprf
Define $F'':\mC\to\B\mathcal{G}$ by sending
a 1-d defect $z\xleftarrow{\g}y$ of $\mC$ to
\begin{figure}[H]
\centering
\begin{subfigure}{0.42\textwidth}
\centering
\includegraphics[width=0.535\textwidth]{id1ddefect}
\begin{picture}(0,0)
\put(-76,20){$F''(\g)$}
\end{picture}
\end{subfigure}
\hspace{-8mm}$:=$\hspace{6mm}
\begin{subfigure}{0.50\textwidth}
\centering
\includegraphics[width=0.95\textwidth]{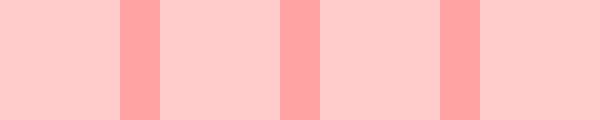}
\begin{picture}(0,0)
\put(-192,20){$g(z)$}
\put(-131,20){$F(g)$}
\put(-69,20){$g(y)^{-1}$}
\end{picture}
\end{subfigure} 
\end{figure}
\noindent
and sending a 0-d defect 
\begin{figure}[H]
\centering
\includegraphics[width=0.225\textwidth]{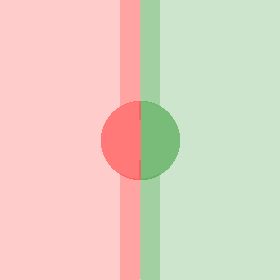}
\begin{picture}(0,0)
\put(-63,88){$\g$}
\put(-63,51){$\S$}
\put(-63,18){$\de$}
\put(-98,51){$z$}
\put(-26,51){$y$}
\end{picture}
\end{figure}
\noindent
of $\mC$ to
\begin{figure}[H]
\centering
\begin{subfigure}{0.45\textwidth}
\centering
\includegraphics[width=0.425\textwidth]{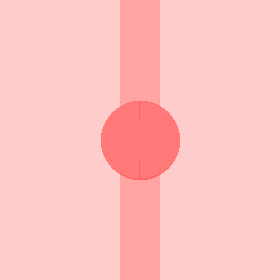}
\begin{picture}(0,0)
\put(-65,75){$F''(\g)$}
\put(-67,44){$F''(\S)$}
\put(-65,13){$F''(\de)$}
\end{picture}
\end{subfigure}
\hspace{-9mm}$:=$\hspace{9mm}
\begin{subfigure}{0.45\textwidth}
\centering
\includegraphics[width=0.91\textwidth]{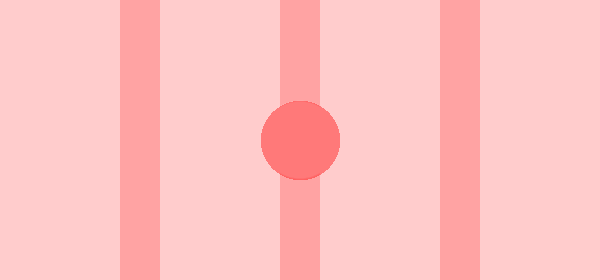}
\begin{picture}(0,0)
\put(-116,73){$F(\g)$}
\put(-117,43){$F(\S)$}
\put(-116,13){$F(\de)$}
\put(-168,43){$g(z)$}
\put(-62,43){$g(y)^{-1}$}
\end{picture}
\end{subfigure}
\end{figure} 
\noindent
Using these definitions, one should check
$F''$ is indeed a strict functor, 
both $(g,e):F\Rightarrow F''$
and $(e,h):F''\Rightarrow F'$ are natural
transformations, and $\s$ is the composition
of $(g,e)$ with $(e,h).$ 
\eprf

\section{Computing parallel transport}
\label{sec:2dalgebralcgt}
In classical electromagnetism,
or gauge theory in general, 
the equations of motion dictate the dynamics. In particular,
the field strength, and not the gauge potential, appear
in the equations of motion. The vector potential
becomes relevant when formulating the 
equations of motion as a variational principle which is
itself a reference point towards quantization \cite{Sc16}, \cite{Sc}. 
The exponentiated Action and parallel transports
of gauge theory are realized precisely in this intermediate
stage of local prequantum field theory which lies between
classical field theory and quantum field theory. We will
focus on special 1-d and 2-d field theories, i.e. particle
mechanics and string theory. The particle case is provided
as a review as well as to set the notation. We will use the
2-dimensional algebra of Sections 
\ref{sec:2dalgebra2cat}
and
\ref{sec:2dalgebra2group}
to explicitly compute parallel transport and its change
under gauge transformations. 
The novelty here, compared with the results of 
\cite{SW2} for instance, is the explicit calculations on a
cubic lattice and a direct derivation of the formula
for the parallel transport including convergence results. 
Although our main results are
Propositions \ref{prop:trivnconverges} and 
Theorem \ref{thm:fulltoreducedsurfacetransport}, 
the diagrammatic picture developed for
how these gauge fields interact with combinations of 
edges and plaquettes in a lattice might be fruitful 
for applications.

\subsection{One-dimensional algebra and parallel transport}
\label{sec:2dalgebra1dgt}
The solution to the initial value problem (IVP)
\be
\label{eq:ODEparalleltransport}
\frac{d\psi(t)}{dt}
=-A(t)\psi(t), \qquad \psi(0)\equiv \psi_{0}\in\mathbb{R}^{n}
\ee
at time $T$ with $A(t)$ a time-dependent $n\times n$ matrix is 
\be
\label{eq:pathorderedintegral}
\psi(T)=\psi_{0}+\sum_{k=1}^{\infty} \frac{(-1)^{k}}{k!}
\int_{0}^{T} dt_{k} \cdots 
\int_{0}^{T} dt_{1} \; \mathcal{T}\left[A(t_{k})\cdots A(t_{1})\right]
\psi_{0}
\ee
where $\mathcal{T}$ stands for time-ordering 
with earlier times appearing to the right, namely
\be
\mathcal{T}\big[A(t_{k})\cdots A(t_{1})\big]
:=A(t_{f(k)})\cdots A(t_{f(1)}),
\ee
where $f:\{1,\dots,k\}\to\{1,\dots,k\}$ is any bijection such that 
\be
t_{f(k)}\ge\cdots\ge t_{f(1)}.
\ee
The choice of sign convention (\ref{eq:ODEparalleltransport})
is to be consistent with references 
\cite{Pa}, \cite{BaMu}, and \cite{SW1}.%
\footnote{Be warned, however, as this sign will
lead to different conventions for other related
forms such as the curvature 2-form, the connection
2-form, and gauge transformation relations.
Certain authors use this other convention
\cite{Hu}, \cite{MS}.
Yet another convention is to include an imaginary factor
\cite{CT}.}
This IVP shows up in several contexts such as (a) solving Schr\"odinger's
equation with $A(t)=i H(t)$ for a time-dependent Hamiltonian and
$\psi$ a vector in the space on which $H$ acts and (b) calculating the
parallel transport along a curve in gauge theory, where $A$ is the local
vector potential, a matrix-valued (or Lie algebra-valued) differential form
on a smooth manifold $M.$ 
This integral goes under many names: 
Dyson series, Picard iteration, path/time-ordered exponential, Berry phase, 
etc.

As an approximation, the solution to this
differential equation can be obtained by breaking up a curve
into infinitesimal paths
\begin{figure}[H]
\centering
\includegraphics[width=0.375\textwidth]{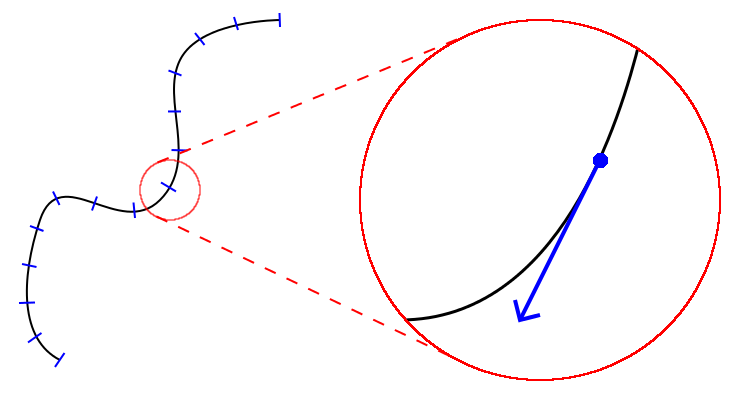}
\begin{picture}(0,0)
\put(-200,45){$x(t)$}
\put(-115,95){$x(t_{1})$}
\put(-170,5){$x(t_{n+1})$}
\put(-64,60){$x(t_{i})$}
\put(-47,31){$\cfrac{dx}{dt}\bigg|_{t_{i}}$}
\end{picture}
\end{figure} 
\noindent
and associating the group elements
\be
\label{eq:Agrp}
\exp\left\{-A_{\mu_{i}}\big(x(t_{i})\big)
\frac{dx^{\mu_{i}}}{dt}\Big|_{t_{i}}\D t_{i}\right\}
\ee
to these infinitesimal paths and multiplying those group 
elements in the order dictated by the path.
In this notation, we have used local coordinates $\{x^{\mu}\}$
and the Einstein summation convention for these local coordinates.
The $i$ subscript on $\mu$ 
is meant to distinguish the summations at different times $\{t_{i}\}.$
$\frac{dx^{\mu_{i}}}{dt}\big|_{t_{i}}$ stands for evaluating the 
derivative of the path at the time $t_{i}.$ 
Furthermore, $\D t_{i}$ should be thought of as the length
of the infinitesimal interval from $t_{i}$ to $t_{i+1},$ 
namely $\D t_{i}=t_{i+1}-t_{i},$ and will be used later as an 
approximation for calculating integrals. For simplicity, we may
take it to be $\D t_{i}=\frac{1}{n}$ if our parametrization
is defined on $[0,1]$ and if there are $n$ subintervals.
Furthermore, by locality, the group elements should be of this form
to lowest order in approximation.
Preserving the order dictated
by the path, the result of multiplying all these elements is
\be
\label{eq:grpproductpt}
\exp\left\{-A_{\mu_{n}}\big(x(t_{n})\big)
\frac{dx^{\mu_{n}}}{dt}\Big|_{t_{n}}\D t_{n}\right\}
\cdots
\exp\left\{-A_{\mu_{1}}\big(x(t_{1})\big)
\frac{dx^{\mu_{1}}}{dt}\Big|_{t_{1}}\D t_{1}\right\}.
\ee
Expanding out to lowest order (since the paths are infinitesimal) gives%
\footnote{$\mathds{1}$ denotes the identity matrix.}
\be
\label{eq:grpproductptexpansion}
\left(\mathds{1}-A_{\mu_{n}}\big(x(t_{n})\big)
\frac{dx^{\mu_{n}}}{dt}\Big|_{t_{n}}\D t_{n}\right)
\cdots
\left(\mathds{1}-A_{\mu_{1}}\big(x(t_{1})\big)
\frac{dx^{\mu_{1}}}{dt}\Big|_{t_{1}}\D t_{1}\right)
\ee
and reorganizing terms results in 
\be
\label{eq:grouppt}
\begin{split}
\mathds{1}&-\sum_{i=1}^{n}A_{\mu_{i}}\big(x(t_{i})\big)
\frac{dx^{\mu_{i}}}{dt}\Big|_{t_{i}}\D t_{i}
+\sum_{\substack{i,j\\i> j\ge 1}}^{n}
A_{\mu_{i}}\big(x(t_{i})\big)A_{\mu_{j}}\big(x(t_{j})\big)
\frac{dx^{\mu_{i}}}{dt}\Big|_{t_{i}}\frac{dx^{\mu_{j}}}{dt}\Big|_{t_{j}}
\D t_{i}\D t_{j}
\pm\cdots\\
&+(-1)^{n}A_{\mu_{n}}\big(x(t_{n})\big)\cdots A_{\mu_{1}}\big(x(t_{1})\big)
\frac{dx^{\mu_{n}}}{dt}\Big|_{t_{n}}\cdots\frac{dx^{\mu_{1}}}{dt}\Big|_{t_{1}}
\D t_{n}\cdots\D t_{1},
\end{split}
\ee
which is exactly the path-ordered integral 
appearing in (\ref{eq:pathorderedintegral})
after taking the $n\to\infty$ limit in which the 
$\D t_{i}$ are replaced by $dt_{i}.$ 
There are several things to check
to confirm this claim. 
First, to see that the limit as $n\to\infty$ of the partial products
coming from (\ref{eq:grpproductptexpansion}) converges,
we use the fact that this product converges if and only if%
\footnote{Technically, one should be a bit more precise
since the matrices change as a function of $n.$ This would
be correct if we replace $n$ with an arbitrary partition and look 
at subpartitions because any two partitions have a common
refinement. Another proof of convergence can be
done using Picard's method \cite{Ne69}.}
the sequence of partial sums
\be
\sum_{i=1}^{n}\left\lVert A_{\mu_{n}}\big(x(t_{n})\big)
\frac{dx^{\mu_{n}}}{dt}\Big|_{t_{n}}\right\rVert\D t_{n}
\ee
converges as $n\to\infty$ (cf Section 8.10 in \cite{We64}).  
Here, the norm can be taken to be the operator norm
for matrices in any representation. 
If one defines the real-valued function 
\be
[0,1]\ni t\mapsto \mathcal{A}(t):=\left\lVert A_{\mu}\big(x(t)\big)
\frac{dx^{\mu}}{dt}\right\rVert
\ee
on the domain of the path,
then the convergence of this sum is equivalent to the
existence of the Riemann integral of the function $\mathcal{A}$ over $[0,1]$
(one could have made these definitions for any partition of the interval
to relate it more precisely to the Riemann integral \cite{Ab15}).
Since the Lie algebra-valued differential form $A$ 
is smooth and since the path is smooth, 
$\mathcal{A}$ is smooth and therefore integrable so that 
\be
\lim_{n\to\infty}\left(\sum_{i=1}^{n}\left\lVert A_{\mu_{n}}\big(x(t_{n})\big)
\frac{dx^{\mu_{n}}}{dt}\Big|_{t_{n}}\right\rVert\D t_{n}\right)
=\int_{0}^{1}\mathcal{A}(t)dt.
\ee
Second, one should note that the 
sums in (\ref{eq:grouppt})
are automatically ordered so that they become 
integrals over simplices in the $n\to\infty$ limit. 
This follows from the equality 
\be
\int_{0}^{1}dt_{k}\int_{0}^{t_{k}}dt_{k-1} \cdots 
\int_{0}^{t_{2}} dt_{1} \; \mathcal{T}\left[A(t_{k})\cdots A(t_{1})\right]
=
\frac{1}{k!}
\int_{0}^{1} dt_{k} \cdots 
\int_{0}^{1} dt_{1} \; \mathcal{T}\left[A(t_{k})\cdots A(t_{1})\right]
\ee
giving an additional $\frac{1}{k!}$ for the volume of
the $k$-simplex. 
For example, the double sum term above with the $i>j\ge1$ 
becomes the double-integral term over the 2-simplex.
The lowest order terms resemble integrals while the latter terms
do not (for example, see the last term in (\ref{eq:grouppt})).
However, as $n\to\infty,$ the latter terms get ``pushed out'' to infinity
and (\ref{eq:pathorderedintegral}) is what remains. 
More precise derivations can be found in \cite{BaMu} and \cite{Br85}.
We picture the group element (\ref{eq:grouppt})
as all the number of ways
in which $A$ interacts with the particle preserving the order
of the path
\be
\begin{split}
\quad\;
\xy0;/r.25pc/:
(-30,0)*{}="1";
(30,0)*{}="2";
{\ar@{-}"1";"2"};
\endxy
&\qquad \mathds{1}
\\
-\;\;\int
\xy0;/r.25pc/:
(-30,0)*{}="1";
(30,0)*{}="2";
(10,0)*{\bullet};
(10,5)*{A_{\mu_{1}}(t_1)\frac{dx^{\mu_{1}}}{dt}\Big|_{t_{1}}};
{\ar@{-}"1";"2"};
\endxy
&\qquad
dt_{1}
\\
+\iint
\xy0;/r.25pc/:
(-30,0)*{}="1";
(30,0)*{}="2";
(-15,0)*{\bullet};
(-15,5)*{A_{\mu_{2}}(t_2)\frac{dx^{\mu_{2}}}{dt}\Big|_{t_{2}}};
(10,0)*{\bullet};
(10,5)*{A_{\mu_{1}}(t_1)\frac{dx^{\mu_{1}}}{dt}\Big|_{t_{1}}};
{\ar@{-}"1";"2"};
\endxy
&\qquad
dt_{1}dt_{2}
\\
\pm\cdots
\hspace{68mm}&
\end{split}
\ee
Thus, given a path $\g:[0,1]\to M,$ we denote the parallel
transport group element in (\ref{eq:grouppt}), after taking the
$n\to\infty$ limit, by 
$\triv(\g).$%
\footnote{The reason for the notation $\triv(\g)$ is
because we will always work in a local trivialization
of a bundle with connection. This choice is also
made to be consistent with earlier work \cite{Pa}
as well as the reference \cite{SW1}.
}
Three key properties of the parallel transport are that
(a) it is reparametrization invariant,
(b) if one had two paths connected at their endpoints as in 
\begin{figure}[H]
\centering
\includegraphics[width=0.5\textwidth]{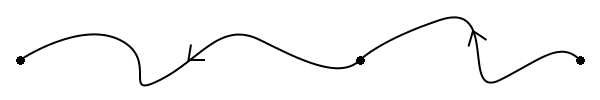}
\begin{picture}(0,0)
\put(-186,21){$\gamma$}
\put(-68,16){$\delta$}
\end{picture}
\end{figure} 
\noindent
then%
\be
\label{eq:trivgrouphomo}
\triv(\g\de)=\triv(\g)\triv(\de),
\ee
and finally (c) it is a smooth function from 
paths in $M$ to
the Lie group $G.$ 
This resembles the definition of a functor. To state the
relationship between parallel transport and functors more
precisely, we note that $\triv(\g)$ is invariant under more
than just reparametrizations of $\g.$ It is also invariant
under thin homotopy. The appropriate domain on which
$\triv$ is therefore defined is a (smooth) category $\mathcal{P}^{1}M$
known as the \emph{\uline{thin path groupoid}} of $M.$ 
A \emph{\uline{groupoid}} is a category all of whose
0-d defects are invertible. Briefly, the thin path groupoid
$\mathcal{P}^{1}M$ consists of points of $M$ and
certain equivalence classes of paths of $M.$ In terms of 1-d domains
and 0-d defects, we actually use the Poincar\'e dual so that
points in $M$ correspond to 1-d domains 
(which are now better thought of as objects)
and paths in $M$ correspond to 0-d defects
(which are now better thought of as morphisms). 
More details on the thin path groupoid
can be found in \cite{Pa} and \cite{SW1}
with a proof of thin homotopy invariance as well as smoothness of 
$\mathrm{triv}$ in the latter reference.
Fortunately, we will not need such technical details for
our calculations. 
All we should keep in mind is that 
$\triv:\mathcal{P}^{1}M\to\B G$ associates group elements
to paths
\begin{figure}[H]
\centering
\begin{subfigure}{0.45\textwidth}
\centering
    \includegraphics[width=0.75\textwidth]{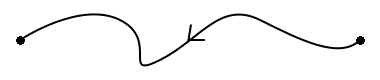}
      \begin{picture}(0,0)
\put(-70,17){$\gamma$}
\end{picture}
\end{subfigure}
$\to$
\begin{subfigure}{0.45\textwidth}
\centering
 \includegraphics[width=0.85\textwidth]{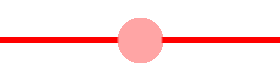}
      \begin{picture}(0,0)
\put(-109,24){$\scriptstyle\mathrm{triv}(\gamma)$}
\end{picture}
\end{subfigure}
\end{figure} 
\noindent
smoothly
and the path ordered integral arises from smoothness, breaking up
the path into infinitesimal pieces, and using the generalized
group homomorphism property. Namely, associated to such a path
$\g$ and a decomposition 
\be
\g=\g_{n}\cdots\g_{1}
\ee
let
\be
a_{i}:=\triv(\g_{i})\cong
\exp\left\{-A_{\mu_{i}}\big(x(t_{i})\big)
\frac{dx^{\mu_{i}}}{dt}\Big|_{t_{i}}\D t_{i}\right\}.
\ee
Then the parallel transport is the product 
\begin{figure}[H]
\centering
\includegraphics[width=0.45\textwidth]{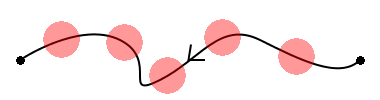}
\begin{picture}(0,0)
\put(-194,39){$a_{n}$}
\put(-160,41){\rotatebox{-35}{$a_{n-1}$}}
\put(-137,12){\rotatebox{27}{$\cdots$}}
\put(-101,40){$a_{2}$}
\put(-55,27){$a_{1}$}
\end{picture}
\end{figure} 
\noindent
given in (\ref{eq:grpproductpt}) and the time ordering is automatic. 
This is essentially
what we mean by one-dimensional algebra:
one-dimensional algebra is the theory of categories
and functors.

The symmetries associated with the parallel transport
are given by functions $M\to G.$ More precisely, let
$\triv,\triv':\mathcal{P}^{1}M\to\B G$ 
be two parallel transport functors
defined by vector potentials $A$ and $A',$ respectively.
A \emph{\uline{finite gauge transformation}} 
from $A$ to $A'$ is a smooth function $g:M\to G$
satisfying%
\footnote{As usual, we are thinking of $G$ as a
matrix group, though we do not need to be
for any statements made. It is only meant
to facilitate computations and simplify formulas.}
\be
\label{eq:usualgaugetransformation}
A'=gAg^{-1}-dgg^{-1}.
\ee
This condition for a gauge transformation is equivalent
(see \cite{SW1})
to the condition that for any path $\g$ from $y$ to $z,$
\be
\triv'(\g)g(y)=g(z)\triv(\g)
\ee
which in turn is equivalent to the statement that
$g:M\to G$ defines a smooth natural transformation
from $\triv$ to $\triv'$ (see Definition \ref{defn:nattrans1cat}).
A sketch of this equivalence can be seen by
discretizing a path $t\mapsto x(t)$ into $n$ pieces
and using the expression 
(\ref{eq:grpproductptexpansion})
for the approximation of the parallel transport. 
Applying a
gauge transformation to each piece gives
\be
\label{eq:trivprime}
\triv'(\g)\approxeq
\prod^{n}_{i=1}g\big(x(t_{i+1})\big)
\left(\mathds{1}-A_{\mu_{i}}\big(x(t_{i})\big)
\frac{dx^{\mu_{i}}}{dt}\Big|_{t_{i}}\D t_{i}\right)
g\big(x(t_{i})\big)^{-1}
\ee
where the product is in the specified order
as in (\ref{eq:grpproductptexpansion}). 
Taylor expanding out the latter group element
gives 
\be
g\big(x(t_{i+1})\big)\approxeq
g\big(x(t_{i})\big)
+\frac{\p g}{\p x^{\mu_{i}}}\frac{dx^{\mu_{i}}}{dt}
\Big|_{t_{i}}\D t_{i}
\ee
to first order in $\D t.$ 
Plugging this into (\ref{eq:trivprime}) gives
\be
\begin{split}
\triv&'(\g)\approxeq
\prod^{n}_{i=1}\left(g\big(x(t_{i})\big)
+\frac{\p g}{\p x^{\mu_{i}}}\frac{dx^{\mu_{i}}}{dt}
\Big|_{t_{i}}\D t_{i}\right)
\left(\mathds{1}-A_{\mu_{i}}\big(x(t_{i})\big)
\frac{dx^{\mu_{i}}}{dt}\Big|_{t_{i}}\D t_{i}\right)
g\big(x(t_{i})\big)^{-1}\\
&\approxeq
\prod^{n}_{i=1}
\left(\mathds{1}-g\big(x(t_{i})\big)A_{\mu_{i}}\big(x(t_{i})\big)
\frac{dx^{\mu_{i}}}{dt}\Big|_{t_{i}}
g\big(x(t_{i})\big)^{-1}\D t_{i}
+\frac{\p g}{\p x^{\mu_{i}}}\frac{dx^{\mu_{i}}}{dt}
\Big|_{t_{i}}
g\big(x(t_{i})\big)^{-1}\D t_{i}\right),
\end{split}
\ee
where we have dropped the term 
\be
\left(\frac{\p g}{\p x^{\mu_{i}}}\frac{dx^{\mu_{i}}}{dt}
\Big|_{t_{i}}\D t_{i}\right)
\left(-A_{\mu_{i}}\big(x(t_{i})\big)
\frac{dx^{\mu_{i}}}{dt}\Big|_{t_{i}}\D t_{i}\right)
\ee
since it is second order in $\D t_{i}.$ 
Finally, since
\be
\triv'(\g)\approxeq
\prod^{n}_{i=1}
\left(\mathds{1}-A'_{\mu_{i}}\big(x(t_{i})\big)
\frac{dx^{\mu_{i}}}{dt}\Big|_{t_{i}}\D t_{i}\right)
\ee
it is reasonable to identify corresponding terms giving
\be
A'_{\mu}=gA_{\mu}g^{-1}-\frac{\p g}{\p x^{\mu}}g^{-1},
\ee
which reproduces (\ref{eq:usualgaugetransformation}). 
This latter perspective of functors and
natural transformations will be used in the sequel to
\emph{define} parallel transport along two-dimensional
surfaces (with some data on orientations). 
This was first made precise in \cite{SW1} though
the formulation in terms of functors had been 
expressed earlier \cite{BS}. 
\br
Most of the calculations in this paper will follow
this sort of logic. 
Although similar techniques were used in \cite{GP}
and \cite{BS}, we
were largely motivated by the kinds of calculations
in \cite{CT} and hope that our treatment will be
more accessible to physicists. 
More rigorous results can be found in the 
references \cite{SW1}, \cite{SW2}, \cite{SW3}, \cite{SW4}. 
\er

\subsection{Two-dimensional algebra and surface transport}
\label{sec:2dalgebra2dgt}
Understanding higher form non-abelian gauge fields
has been a long-standing problem in physics, particularly
in string theory and M-theory 
(see for instance the end of \cite{Wit}).
Some progress is being made to answer some of these problems
with the use of higher gauge theory 
(see \cite{SS17} and the references therein). 
Although we do not aim to solve these problems,
we hope to indicate the important role played by
category theory in understanding certain aspects
of these theories. We will
show how 2-categories and the laws set
up in the previous sections naturally lead to the notion
of parallel transport along surfaces. 
This will also illustrate how explicit calculations can be done in 2-groups.
Parallel transport
will obey an important gluing condition analogous
to the gluing condition for paths. Gauge
transformations will be studied in the next section. 
Furthermore, we will produce an explicit
formula analogous to the Dyson series expansion
for paths. Although an integral formula is known in the literature 
\cite{SW2}, the derivation there is not entirely direct nor is it
obvious how the formulas are derived from, say, a
cubic lattice approximation. 
A sketch is included in \cite{BS} in Section 2.3.2 but further analysis
was done in path space, which we feel is more difficult---indeed,
the goal of that work was to relate gerbes with connection
on manifolds to connections on their corresponding path spaces. 
Furthermore, 
although experts are aware of how bigons are related to
more general surfaces, we explicitly perform our calculations
on ``reasonable'' surfaces, namely squares, for clearer visualization.
Our method is more in line with the types of calculations
done in lattice gauge theory \cite{Ma}. 

We feel it is important to express surface transport in
a more computationally explicit manner using a lattice
and \emph{derive}, from the ground up, a visualization
of the surface-ordered integral sketched
in Figure 15 in \cite{Pa}. This is done in 
Proposition \ref{prop:trivnconverges}, 
Theorem \ref{thm:fulltoreducedsurfacetransport}, 
and the surrounding text. 
Just as the group $G$-valued 
parallel transport along paths in a manifold
$M$ is described
by a functor $\triv:\mathcal{P}^{1}M\to\B G,$
crossed-module $\mathcal{G}$-valued 
parallel transport
along surfaces should be described
by a functor from some 2-category associated
with paths and surfaces in $M$ to the 2-group $\B\mathcal{G}.$
Ideally, such a 2-category should be a version of
the (extended) 2-dimensional cobordism 2-category over
the manifold $M$ to mimic the ideas of 
functorial field theories. However, this
has not yet been achieved in this form for 
non-abelian 2-groups. In fact, it
has only recently been achieved for the 1-dimensional
case by Berwick-Evans and Pavlov \cite{BE-P}. 
Earlier work on abelian gerbes
indicates this should be the case in general \cite{Pi}
though this has not been fully worked out. Part
of the reason is due to the fact that the representation
theory for higher groups is a rather young subject \cite{BBFW}. 

Fortunately, a related solution exists if one works with
a 2-category of paths and homotopies. 
This 2-category is denoted by $\mathcal{P}^{2}M.$
It is more natural to describe this category in
terms of the Poincar\'e dual of string diagrams. Namely,
objects of $\mathcal{P}^{2}M$ are points of $M,$
1-morphisms of $\mathcal{P}^{2}M$ are thin homotopy
classes of paths in $M,$ and 2-morphisms are
thin-homotopy classes of \emph{bigons} in $M.$ A
bigon is essentially a homotopy $\S$ between two
paths whose endpoints agree. 

\bd
\label{defn:bigon}
Let $\g$ and $\de$ be two paths from $x$ to $y$
parametrized by $t\in[0,1]$ such that
there exists an $\e>0$ with 
$\g(t)=\de(t)$ for all $t\in[0,\e]\cup[1-\e,1].$ A \emph{\uline{bigon}}
from $\g$ to $\de$ is a map $\S:[0,1]\times[0,1]\to M$ 
such that there exists an $\e>0$ with
\be
\label{eq:bigon}
\S(t,s)= 
\begin{cases}
x&\mbox{ for all }(t,s)\in[0,\e]\times[0,1]\\
y&\mbox{ for all }(t,s)\in[1-\e,1]\times[0,1] \\
\g(t)&\mbox{ for all }(t,s)\in[0,1]\times[0,\e]\\
\de(t)&\mbox{ for all }(t,s)\in[0,1]\times[1-\e,1]\\
\end{cases}  
\ee
\ed

It is helpful to visualize such a bigon as
\begin{figure}[H]
\centering
\begin{subfigure}{0.5\textwidth}
\centering
\includegraphics[width=0.40\textwidth]{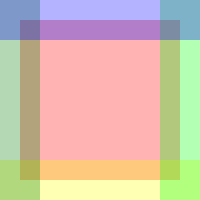}
\begin{picture}(0,0)
\put(-100,45){$y$}
\put(-55,90){$\g$}
\put(-55,3){$\de$}
\put(-55,45){$\S$}
\put(-12,45){$x$}
\end{picture}
\end{subfigure}
$\to$
\begin{subfigure}{0.45\textwidth}
\centering
\includegraphics[width=0.65\textwidth]{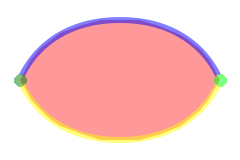}
\begin{picture}(0,0)
\put(-149,45){$y$}
\put(-80,92){$\g$}
\put(-80,81){$<$}
\put(-80,9){$<$}
\put(-80,0){$\de$}
\put(-83,45){$\xy0;/r.25pc/:
(0,0)*+{\displaystyle\S}="S";
(0,9)*{}="g";
(0,-9)*{}="d";
{\ar@{=}"g";"S"};
{\ar@{=>}"S";"d"};
\endxy$}
\put(-8,45){$x$}
\end{picture}
\end{subfigure}
\end{figure} 
\noindent

\bd
Two bigons $\S$ and $\G$ from paths $\g$ to $\de$ 
are \emph{\uline{thinly homotopic}} if there exists a
smooth map of a 3-dimensional cube into $M$ whose
top face is $\S,$ whose bottom face is $\G,$
and similarly for the other face for the paths $\g$ and $\de$
along with their endpoints
(all of these assume some constancy in a small neighborhood
of each face). 
Furthermore, and most importantly,
this map cannot sweep out any volume in $M,$ i.e.
its rank is strictly less than 3. 
\ed
More details can be found
in \cite{Pa} and \cite{SW2} though again such technicalities
will be avoided here. 
Thus, a strict smooth functor $\triv:\mathcal{P}^{2}M\to\B\mathcal{G}$
is a smooth assignment 
\begin{figure}[H]
\centering
\begin{subfigure}{0.45\textwidth}
\centering
\includegraphics[width=0.65\textwidth]{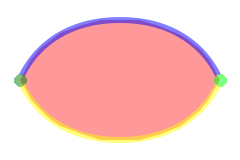}
\begin{picture}(0,0)
\put(-149,45){$y$}
\put(-80,92){$\g$}
\put(-80,81){$<$}
\put(-80,9){$<$}
\put(-80,0){$\de$}
\put(-83,45){$\xy0;/r.25pc/:
(0,0)*+{\displaystyle\S}="S";
(0,9)*{}="g";
(0,-9)*{}="d";
{\ar@{=}"g";"S"};
{\ar@{=>}"S";"d"};
\endxy$}
\put(-8,45){$x$}
\end{picture}
\end{subfigure}
$\to$
\begin{subfigure}{0.5\textwidth}
\centering
\includegraphics[width=0.45\textwidth]{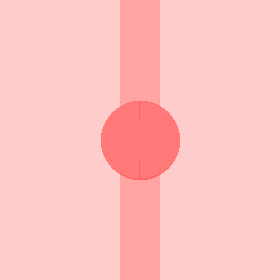}
\begin{picture}(0,0)
\put(-76,84){$\triv(\g)$}
\put(-77,52){$\triv(\S)$}
\put(-76,16){$\triv(\de)$}
\end{picture}
\end{subfigure}
\end{figure} 
\noindent
that, by the conventions of 2-groups in Section \ref{sec:2dalgebra2group}, says
\be
\label{eq:trivbigons}
\t\big(\triv(\S)\big)\triv(\g)=\triv(\de).
\ee
Furthermore, this assignment satisfies a 
homomorphism property in the following sense.
Bigons can be glued together in series and in parallel 
by a choice of parametrization. By the thin
homotopy assumption, the
value of the bigons is independent of such parametrizations.
It might seem undesirable to restrict ourselves to surfaces
of this form. However, this is no serious matter because
every compact surface can be expressed in this manner under
suitable identifications living on sets of measure zero. 
For example, a surface of genus two with three boundary
components with orientations shown 
(the orientation
of the surface itself is clockwise) 
\begin{figure}[H]
\centering
\begin{subfigure}{0.45\textwidth}
\centering
\includegraphics[width=0.75\textwidth]{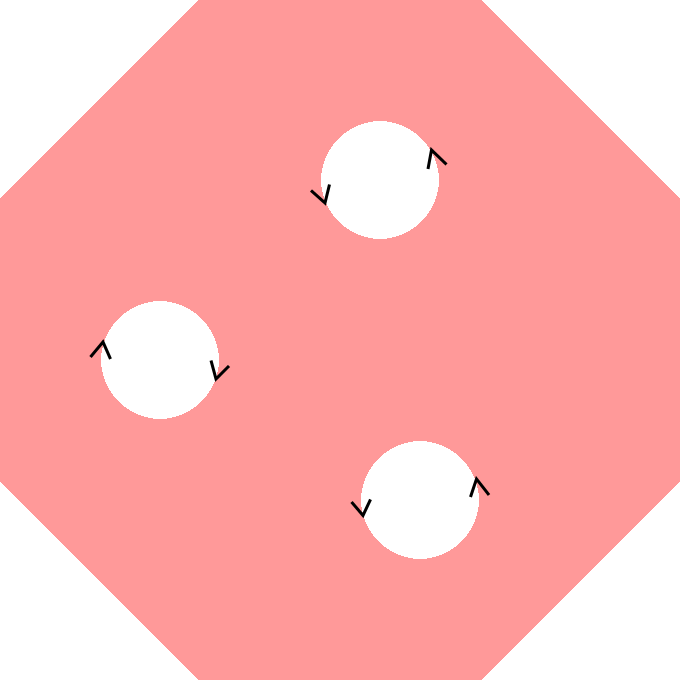}
\end{subfigure}
\qquad
\begin{subfigure}{0.45\textwidth}
\centering
\includegraphics[width=0.775\textwidth]{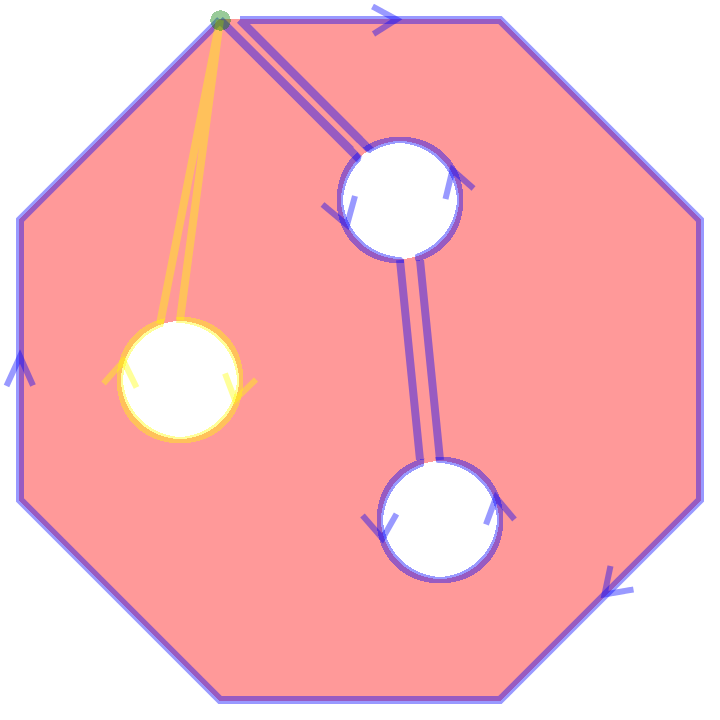}
\begin{picture}(0,0)
\put(-148,96){$\delta$}
\put(-4,114){$\gamma$}
\end{picture}
\end{subfigure}
\end{figure} 
\noindent
is depicted on the right as a bigon 
beginning at the path $\g$ (in blue)
and ending at the path $\de$ (in yellow) 
both of which are loops
beginning at the same basepoint which is the top left corner
of the octogon on the left. 
The identifications on the outer boundary
of the octagon are standard ways of representing a
genus two surface. Furthermore, one can always
triangulate or cubulate such a surface. 
If one chooses triangulations, then one merely
needs to know the parallel transport on triangles
\begin{figure}[H]
\centering
\begin{subfigure}{0.45\textwidth}
\centering
\includegraphics[width=0.65\textwidth]{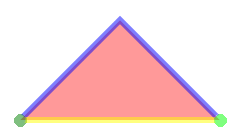}
\begin{picture}(0,0)
\put(-149,10){$z$}
\put(-120,45){$\b$}
\put(-40,45){$\g$}
\put(-80,80){$y$}
\put(-80,1){$\de$}
\put(-80,35){$\S$}
\put(-8,10){$x$}
\end{picture}
\end{subfigure}
$\to$
\begin{subfigure}{0.5\textwidth}
\centering
\includegraphics[width=0.45\textwidth]{weakgrouphomo}
\begin{picture}(0,0)
\put(-105,84){$\triv(\b)$}
\put(-47,84){$\triv(\g)$}
\put(-77,52){$\triv(\S)$}
\put(-76,16){$\triv(\de)$}
\end{picture}
\end{subfigure}
\end{figure} 
\noindent
and if one cubulates a surface, then one needs
to know it for squares
\begin{figure}[H]
\centering
\begin{subfigure}{0.45\textwidth}
\centering
\includegraphics[width=0.65\textwidth]{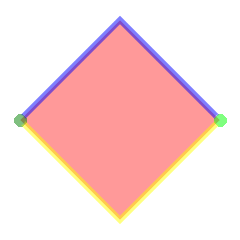}
\begin{picture}(0,0)
\put(-149,68){$z$}
\put(-110,108){$\b$}
\put(-47,108){$\g$}
\put(-80,138){$y$}
\put(-110,32){$\de$}
\put(-80,1){$w$}
\put(-47,32){$\e$}
\put(-80,68){$\S$}
\put(-8,68){$x$}
\end{picture}
\end{subfigure}
$\to$
\begin{subfigure}{0.5\textwidth}
\centering
\includegraphics[width=0.45\textwidth]{nattransf2group}
\begin{picture}(0,0)
\put(-107,84){$\triv(\b)$}
\put(-45,84){$\triv(\g)$}
\put(-77,52){$\triv(\S)$}
\put(-107,18){$\triv(\de)$}
\put(-45,18){$\triv(\e)$}
\end{picture}
\end{subfigure}
\end{figure} 
Thus, in order to find an explicit formula for the parallel
transport along surfaces with non-trivial topology, it suffices
to calculate the parallel transport along a square, say.
Squares are also more convenient to use for continuum
limiting procedures
as opposed to triangles \cite{Su}.
Functoriality for gluing squares together
implies
\begin{figure}[H]
\centering
\begin{subfigure}{0.45\textwidth}
\centering
\includegraphics[width=0.90\textwidth]{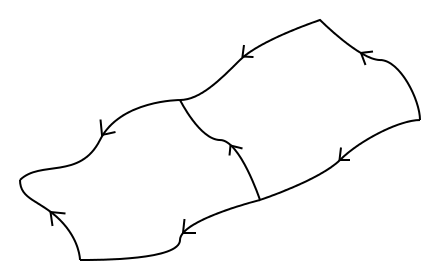}
\begin{picture}(0,0)
\put(-70,75){$\Sigma$}
\put(-90,109){$\gamma$}
\put(-32,103){$\delta$}
\put(-42,45){$\xi$}
\put(-108,50){$\zeta$}
\put(-140,38){$\Omega$}
\put(-156,75){$\beta$}
\put(-187,19){$\omega$}
\put(-117,10){$\rho$}
\end{picture}
\end{subfigure}
$\to$
\begin{subfigure}{0.45\textwidth}
\centering
\includegraphics[width=0.90\textwidth]{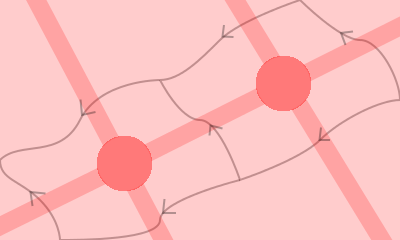}
\begin{picture}(0,0)
\put(-79,75){$\triv(\Sigma)$}
\put(-94,104){$\triv(\gamma)$}
\put(-36,96){$\triv(\delta)$}
\put(-54,43){$\triv(\xi)$}
\put(-113,54){$\triv(\zeta)$}
\put(-158,34){$\triv(\Omega)$}
\put(-170,73){$\triv(\beta)$}
\put(-192,16){$\triv(\omega)$}
\put(-141,10){$\triv(\rho)$}
\end{picture}
\end{subfigure}
\end{figure}
\noindent
and using the rules of two-dimensional algebra, 
this composition is 
\be
\label{eq:trbl}
\triv(\W)\a_{\triv(\b)}\big(\triv(\S)\big).
\ee
Similarly, for gluing along a different edge
\begin{figure}[H]
\centering
\begin{subfigure}{0.45\textwidth}
\centering
\includegraphics[width=0.90\textwidth]{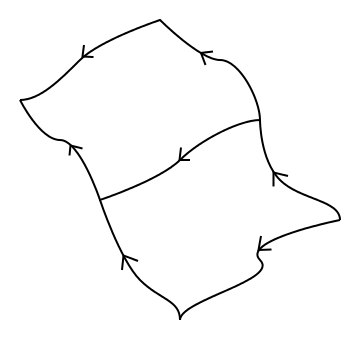}
\begin{picture}(0,0)
\put(-130,130){$\Sigma$}
\put(-163,167){$\gamma$}
\put(-84,160){$\delta$}
\put(-91,97){$\xi$}
\put(-172,92){$\zeta$}
\put(-146,35){$\chi$}
\put(-88,63){$\Pi$}
\put(-48,38){$\pi$}
\put(-38,87){$\epsilon$}
\end{picture}
\end{subfigure}
$\to$
\begin{subfigure}{0.45\textwidth}
\centering
\includegraphics[width=0.90\textwidth]{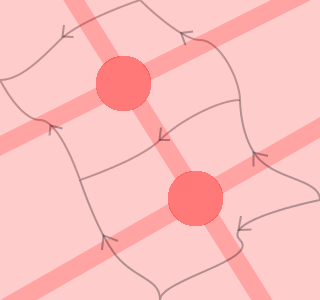}
\begin{picture}(0,0)
\put(-144,130){$\triv(\Sigma)$}
\put(-163,167){$\triv(\gamma)$}
\put(-84,160){$\triv(\delta)$}
\put(-114,94){$\triv(\xi)$}
\put(-184,108){$\triv(\zeta)$}
\put(-146,33){$\triv(\chi)$}
\put(-97,58){$\triv(\Pi)$}
\put(-67,22){$\triv(\pi)$}
\put(-48,87){$\triv(\epsilon)$}
\end{picture}
\end{subfigure}
\end{figure}
\noindent
the composition of the 0-d defects is
\be
\label{eq:tlbr}
\a_{\triv(\zeta)}\big(\triv(\Pi)\big)\triv(\Sigma).
\ee
These two ways of composing squares will form the basis for later
computations. 
If one also wishes to attach a square in a somewhat arbitrary
way such as 
\begin{figure}[H]
\centering
\includegraphics[width=0.25\textwidth]{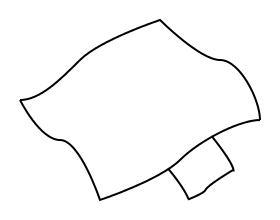}
\end{figure} 
\noindent
then this attachment must be oriented in such a way that
(a) the boundary orientation agrees with the orientation
of the first surface and
(b) the two surface orientations combine to form
a consistent orientation when glued together. 
So, for example, 
\begin{figure}[H]
\centering
\includegraphics[width=0.25\textwidth]{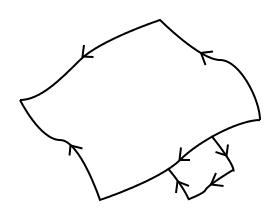}
\begin{picture}(0,0)
\put(-70,58){\rotatebox{-100}{$\Rightarrow$}}
\put(-43,27){\rotatebox{-60}{$\Rightarrow$}}
\end{picture}
\end{figure} 
\noindent
is an allowed glueing orientation
(more on orientations and their physical meaning 
is discussed in Section \ref{sec:2dalgebraorient}).
In this case, if we label
all the vertices, edges, and squares,
then the parallel transport along the glued surface is 
\begin{figure}[H]
\centering
\begin{subfigure}{0.45\textwidth}
\centering
\includegraphics[width=0.90\textwidth]{worldsheet_plus_tiny_with_arrows}
\begin{picture}(0,0)
\put(-137,130){$\g$}
\put(-50,119){$\delta$}
\put(-106,78){$\S$}
\put(-162,43){$\zeta$}
\put(-76,57){$\xi_{2}$}
\put(-63,33){$\Pi$}
\put(-42,72){$\xi_{1}$}
\put(-115,32){$\xi_{3}$}
\put(-38,48){$\pi$}
\put(-52,13){$\chi$}
\put(-88,19){$\psi$}
\end{picture}
\end{subfigure}
$\to$
\begin{subfigure}{0.45\textwidth}
\centering
\includegraphics[width=0.90\textwidth]{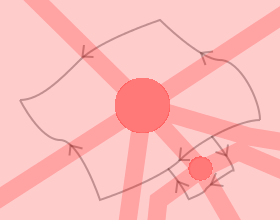}
\begin{picture}(0,0)
\put(-166,130){$\triv(\g)$}
\put(-50,121){$\triv(\delta)$}
\put(-118,78){$\triv(\S)$}
\put(-178,36){$\triv(\zeta)$}
\put(-89,54){$\scriptstyle\triv(\xi_{2})$}
\put(-72,34){$\scriptstyle\triv(\Pi)$}
\put(-40,58){$\scriptstyle\triv(\xi_{1})$}
\put(-120,32){$\scriptstyle\triv(\xi_{3})$}
\put(-38,42){$\scriptstyle\triv(\pi)$}
\put(-58,13){$\scriptstyle\triv(\chi)$}
\put(-98,13){$\scriptstyle\triv(\psi)$}
\end{picture}
\end{subfigure}
\end{figure}
\noindent
which reads
\be
\label{eq:addingepsilon}
\a_{\triv(\z\xi_{3})}\big(\triv(\Pi)\big)\triv(\S)
\ee
on the resulting 0-d defect. This result will play a crucial role in
Remark \ref{rmk:SW}. 

Using all of these results, we can take an arbitrary worldsheet
(with orientations giving it the structure of a bigon),
break it up into infinitesimal squares
\begin{figure}[H]
\centering
\includegraphics[width=0.75\textwidth]{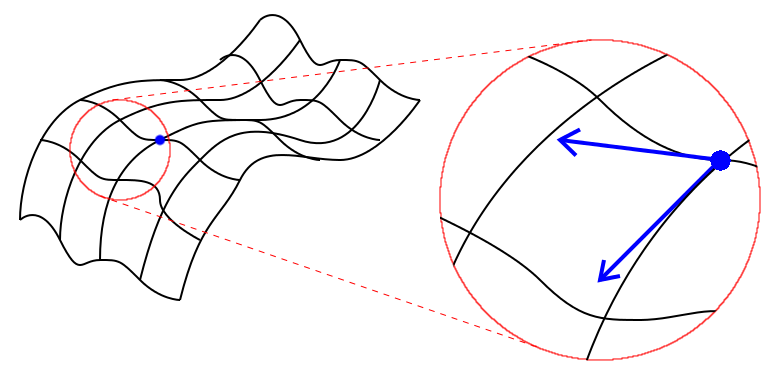}
\begin{picture}(0,0)
\put(-380,125){$x(s,t)$}
\put(-14,105){$x(s_{i},t_{j})$}
\put(-69,46){$\ds\frac{\p x}{\p t}\Big|_{(s_{i},t_{j})}$}
\put(-114,86){$\ds\frac{\p x}{\p s}\Big|_{(s_{i},t_{j})}$}
\end{picture}
\end{figure} 
\noindent
and approximate 
the parallel transport along an infinitesimal square 
\begin{figure}[H]
\centering
\begin{subfigure}{0.45\textwidth}
\centering
\includegraphics[width=0.65\textwidth]{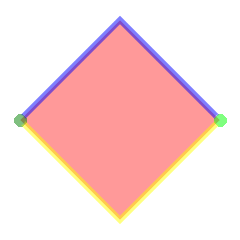}
\begin{picture}(0,0)
\put(-195,68){$(s_{i+1},t_{j+1})$}
\put(-95,138){$(s_{i+1},t_{j})$}
\put(-95,1){$(s_{i},t_{j+1})$}
\put(-8,68){$(s_{i},t_{j})$}
\end{picture}
\end{subfigure}
$\to$
\begin{subfigure}{0.5\textwidth}
\centering
\includegraphics[width=0.45\textwidth]{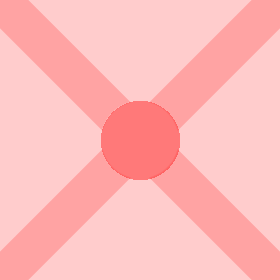}
\begin{picture}(0,0)
\put(-103,84){$a_{i+1,j}^{t}$}
\put(-36,84){$a_{ij}^{s}$}
\put(-65,52){$b_{ij}$}
\put(-103,21){$a_{i,j+1}^{s}$}
\put(-36,21){$a_{ij}^{t}$}
\end{picture}
\end{subfigure}
\end{figure} 
\noindent
where
\be
\label{eq:aijs}
a_{ij}^{s}:=\exp\left\{-A_{\mu_{i}}\big(x(s_{i},t_{j})\big)
\frac{\p x^{\mu_{i}}}{\p s}
\Big|_{(s_{i},t_{j})}\D s_{i}\right\}
\ee
and
\be
\label{eq:aijt}
a_{ij}^{t}:=\exp\left\{-A_{\nu_{j}}\big(x(s_{i},t_{j})\big)
\frac{\p x^{\nu_{j}}}{\p t}
\Big|_{(s_{i},t_{j})}\D t_{j}\right\}
\ee
denote the parallel transport along infinitesimal paths and%
\footnote{Our convention is to include all
combinatorial factors into our 
Einstein summation convention since these
are cumbersome to carry. 
With respect to the usual Einstein summation convention, such
an expression in the exponential (\ref{eq:bij})
would have a $\frac{1}{2}.$
This is due to the fact that $B$ is a 2-form. 
For a $k$-form, the factor we are leaving out is $\frac{1}{k!}.$ 
}
\be
\label{eq:bij}
b_{ij}:=\exp\left\{B_{\mu_{i}\nu_{j}}\big(x(s_{i},t_{j})\big)
\frac{\p x^{\mu_{i}}}{\p s}
\frac{\p x^{\nu_{j}}}{\p t}\Big|_{(s_{i},t_{j})}\D s_{i}\D t_{j}\right\}
\ee
denotes the parallel transport along infinitesimal squares.
Here 
\be
\D s_{i}=s_{i+1}-s_{i}
\aand
\D t_{j}=t_{j+1}-t_{j}
\ee
and for an $n\times n$ square grid these are both
$\D s_{i}=\frac{1}{n}=\D t_{j}.$ 
Note that
in order for this association to be
consistent with our 2-group conventions, it must be true that 
\be
a_{i,j+1}^{s}a_{ij}^{t}=\t(b_{ij})a_{i+1,j}^{t}a_{ij}^{s},
\ee
or equivalently
\be
\label{eq:taubiscurvature}
\t(b_{ij})=a_{i,j+1}^{s}a_{ij}^{t}
\left(a_{ij}^{s}\right)^{-1}\left(a_{i+1,j}^{t}\right)^{-1},
\ee
at least to lowest non-trivial order. The term on the
right-hand-side of (\ref{eq:taubiscurvature}) 
is precisely the parallel transport along
the infinitesimal square%
\footnote{For the purpose of this calculation, 
we have dropped the $\D s_{i}$ and $\D t_{j}$
from the notation to avoid clutter. This should cause
no confusion because these quantities are always
coupled with their corresponding derivatives
$\frac{\p}{\p s}$ and $\frac{\p}{\p t},$ respectively.
}
\be
\begin{split}
a_{i,j+1}^{s}a_{ij}^{t}&
\left(a_{ij}^{s}\right)^{-1}\left(a_{i+1,j}^{t}\right)^{-1}
\\
&\approxeq
\left(\mathds{1}-A_{\mu'_{i}}\big(x(s_{i},t_{j+1})\big)
\frac{\p x^{\mu'_{i}}}{\p s}\Big|_{(s_{i},t_{j+1})}\right)
\left(\mathds{1}-A_{\nu_{j}}\big(x(s_{i},t_{j})\big)
\frac{\p x^{\nu_{i}}}{\p t}\Big|_{(s_{i},t_{j})}\right)\\
&\quad\times
\left(\mathds{1}+A_{\mu_{i}}\big(x(s_{i},t_{j})\big)
\frac{\p x^{\mu_{i}}}{\p s}\Big|_{(s_{i},t_{j})}\right)
\left(\mathds{1}+A_{\nu'_{j}}\big(x(s_{i+1},t_{j})\big)
\frac{\p x^{\nu'_{j}}}{\p t}\Big|_{(s_{i+1},t_{j})}\right)\\
&\approxeq
\left(\mathds{1}-A_{\mu'_{i}}\frac{\p x^{\mu'_{i}}}{\p s}
-\frac{\p A_{\mu'_{i}}}{\p x^{\nu_{j+1}}}
\frac{\p x^{\nu_{j+1}}}{\p t}\frac{\p x^{\mu'_{i}}}{\p s}
-A_{\mu'_{i}}\frac{\p^2 x^{\mu'_{i}}}{\p s \p t}\right)
\left(\mathds{1}-A_{\nu_{j}}\frac{\p x^{\nu_{j}}}{\p t}\right)\\
&\quad\times
\left(\mathds{1}+A_{\mu_{i}}\frac{\p x^{\mu_{i}}}{\p s}\right)
\left(\mathds{1}+A_{\nu'_{j}}\frac{\p x^{\nu'_{j}}}{\p t}
+\frac{\p A_{\nu'_{j}}}{\p x^{\mu_{i+1}}}
\frac{\p x^{\mu_{i+1}}}{\p s}\frac{\p x^{\nu'_{j}}}{\p t}
+A_{\nu'_{j}}\frac{\p^2 x^{\nu'_{j}}}{\p t \p s}\right)
\Big|_{(s_{i},t_{j})}\\
&\approxeq
\Bigg(\mathds{1}-\frac{\p A_{\mu'_{i}}}{\p x^{\nu_{j+1}}}
\frac{\p x^{\nu_{j+1}}}{\p t}\frac{\p x^{\mu'_{i}}}{\p s}
+\frac{\p A_{\nu'_{j}}}{\p x^{\mu_{i+1}}}
\frac{\p x^{\mu_{i+1}}}{\p s}\frac{\p x^{\nu'_{j}}}{\p t}
+A_{\mu'_{i}}\frac{\p x^{\mu'_{i}}}{\p s}
A_{\nu_{j}}\frac{\p x^{\nu_{j}}}{\p t}
\\
&\quad
-A_{\nu_{j}}\frac{\p x^{\nu_{j}}}{\p t}
A_{\mu_{i}}\frac{\p x^{\mu_{i}}}{\p s}
-A_{\mu'_{i}}\frac{\p x^{\mu'_{i}}}{\p s}
A_{\nu'_{j}}\frac{\p x^{\nu'_{j}}}{\p t}
+A_{\mu_{i}}\frac{\p x^{\mu_{i}}}{\p s}
A_{\nu'_{j}}\frac{\p x^{\nu'_{j}}}{\p t}
\Bigg)\Big|_{(s_{i},t_{j})}
\\
&
=\mathds{1}+\left(
\frac{\p A_{\nu_{j}}}{\p x^{\mu_{i}}}
-\frac{\p A_{\mu_{i}}}{\p x^{\nu_{j}}}
+A_{\mu_{i}}A_{\nu_{j}}
-A_{\nu_{j}}A_{\mu_{i}}
\right)\frac{\p x^{\mu_{i}}}{\p s}
\frac{\p x^{\nu_{j}}}{\p t}
\Big|_{(s_{i},t_{j})}
\\
&=\mathds{1}+F_{\mu_{i}\nu_{j}}\frac{\p x^{\mu_{i}}}{\p s}
\frac{\p x^{\nu_{j}}}{\p t}\Big|_{(s_{i},t_{j})},
\end{split}
\ee
to lowest order,
which is a standard result, 
reproduced here to illustrate the methods that will be employed in more
involved calculations.
Here
\be
\label{eq:curvature2form}
F:=dA+A\wedge A
\ee
is the curvature of $A.$
Meanwhile, the left-hand-side of (\ref{eq:taubiscurvature}) is 
\be
\t(b_{ij})\approxeq
\mathds{1}+\un\t\left(B_{\mu_{i}\nu_{j}}\big(x(s_{i},t_{j})\big)
\frac{\p x^{\mu_{i}}}{\p s}
\frac{\p x^{\nu_{j}}}{\p t}\Big|_{(s_{i},t_{j})}\right)
\ee
to lowest order. Here $\un \t:\mathfrak{h}\to\mathfrak{g}$ 
is the derivative
of the map $\t:H\to G$ at the identity, i.e. on the 
Lie algebras (see Appendix \ref{app:dlcm} for more on
the infinitesimal version of $(H,G,\t,\a)$).
This therefore forces the condition
\be
\un\t(B)-F=0,
\ee
which is known in the literature as the \emph{vanishing of the fake
curvature}.
Finally, we can expand
out these exponentials of differential forms and multiply all
terms together analogously to what was done for a path.
An arbitrary worldsheet is described by a map (via some
reparametrization if necessary) from $[0,1]\times[0,1]$ 
to some target manifold and is naturally a bigon
with the orientation induced by having $(s,t)$ a right-handed
coordinate system.
Breaking up such a bigon into infinitesimal squares
(one can also use arbitrary partitions---see in Appendix \ref{app:spc}).
allows one to associate the above exponentials on the
Poincar\'e dual of the cubulation of the worldsheet.
\begin{figure}[H]
\centering
\begin{subfigure}{0.30\textwidth}
\centering
\includegraphics[width=1.0\textwidth]{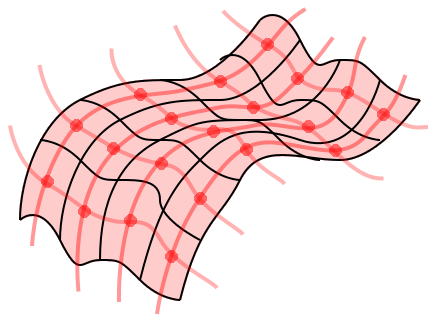}
\begin{picture}(0,0)
\put(-12,80){$\xy0;/r.25pc/: 
(0,0)*{}="1";
(-25,-23)*+{\small t}="t";
{\ar"1";"t"};
\endxy$}
\put(1,89){$\xy0;/r.25pc/: 
(0,0)*{}="1";
(-20,13)*+{\small s}="s";
{\ar"1";"s"};
\endxy$}
\end{picture}
\end{subfigure}
$\qquad\qquad\qquad$
\begin{subfigure}{0.30\textwidth}
\centering
\includegraphics[width=1.00\textwidth]{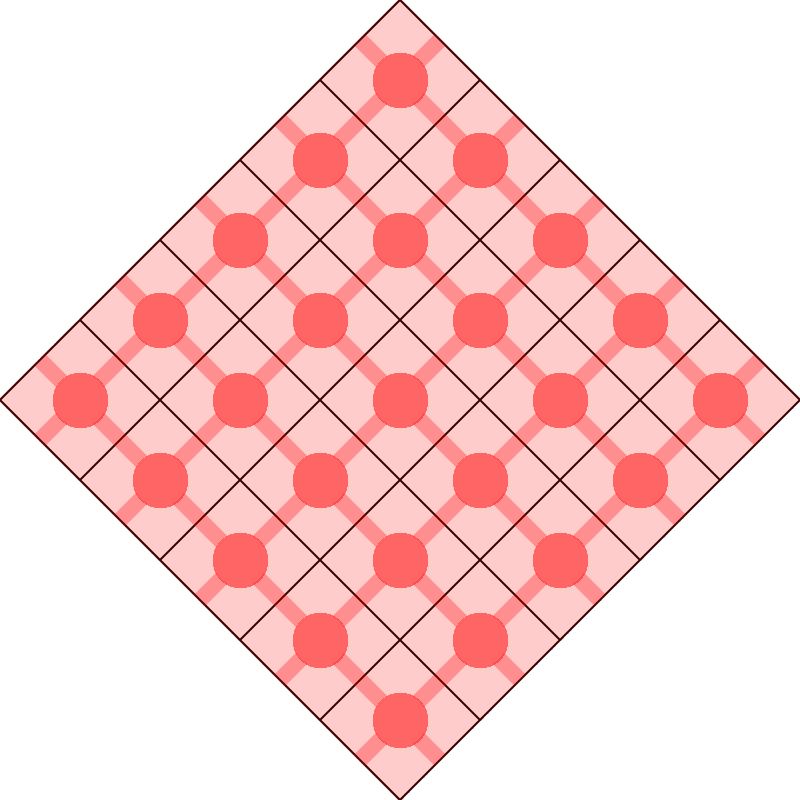}
\begin{picture}(0,0)
\put(10,82){$\xy0;/r.25pc/: 
(0,0)*{}="1";
(-20,-20)*+{\small t}="t";
{\ar"1";"t"};
\endxy$}
\put(10,91){$\xy0;/r.25pc/: 
(0,0)*{}="1";
(-20,20)*+{\small s}="s";
{\ar"1";"s"};
\endxy$}
\end{picture}
\end{subfigure}
\end{figure}  
\noindent
In the above figure, a cubulation of the domain $[0,1]\times[0,1]$ is
shown on the right together with its Poincar\'e dual.
This rotated $(s,t)$ coordinate system was chosen to agree
with our earlier convention on two-dimensional algebra
(cf. (\ref{eq:trbl}) and (\ref{eq:tlbr})).
To be a bit more clear, since a bigon is a map $[0,1]\times[0,1]\to M$ 
with conditions described by (\ref{eq:bigon}), we can visualize it as 
\begin{figure}[H]
\centering
\begin{subfigure}{0.5\textwidth}
\centering
\includegraphics[width=0.40\textwidth]{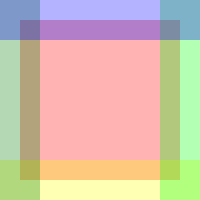}
\begin{picture}(0,0)
\put(-100,45){$y$}
\put(-55,90){$\g$}
\put(-55,3){$\de$}
\put(-55,45){$\S$}
\put(-12,45){$x$}
\end{picture}
\end{subfigure}
$\to$
\begin{subfigure}{0.45\textwidth}
\centering
\includegraphics[width=0.50\textwidth]{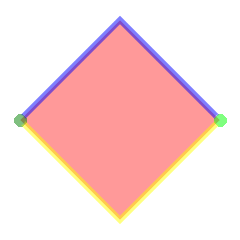}
\begin{picture}(0,0)
\put(-118,52){$y$}
\put(-63,107){$\g$}
\put(-63,-2){$\de$}
\put(-66,53){$\xy0;/r.25pc/:
(0,0)*+{\displaystyle\S}="S";
(0,9)*{}="g";
(0,-9)*{}="d";
{\ar@{=}"g";"S"};
{\ar@{=>}"S";"d"};
\endxy$}
\put(-7,52){$x$}
\end{picture}
\end{subfigure}
\end{figure} 
\noindent
We prefer to use these rectangular coordinates to more
easily express our results for cubic lattices. 
We will consider a $5\times5$ grid for concreteness.
The goal is to associate to each square in this grid the 2-group 
elements (\ref{eq:aijs}), (\ref{eq:aijt}), (\ref{eq:bij}), etc. and then to 
multiply all of these elements together using the rules for 2-group
multiplication set up in Section \ref{sec:2dalgebra2group}. 
In order to do this, 
we use the rules set up earlier on how to read such diagrams and
this requires
us to extend the 1-d defects of the Poincar\'e dual to the 
top and bottom of the page using identity 0-d defects drawn
on the $(s,t)$ domain of the worldsheet (the identities
are drawn in yellow to illustrate where they are and not because
the 2-d domain is different---there is only a single 2-d domain
in a 2-group---see Section \ref{sec:2dalgebra2group}).%
\footnote{One could have also added identities in 
many other consistent ways. The end results would
all be the same (to lowest order) due to the interchange law
(\ref{eq:interchangelaw}).
}
\begin{figure}[H]
\centering
    \includegraphics[width=0.652\textwidth]{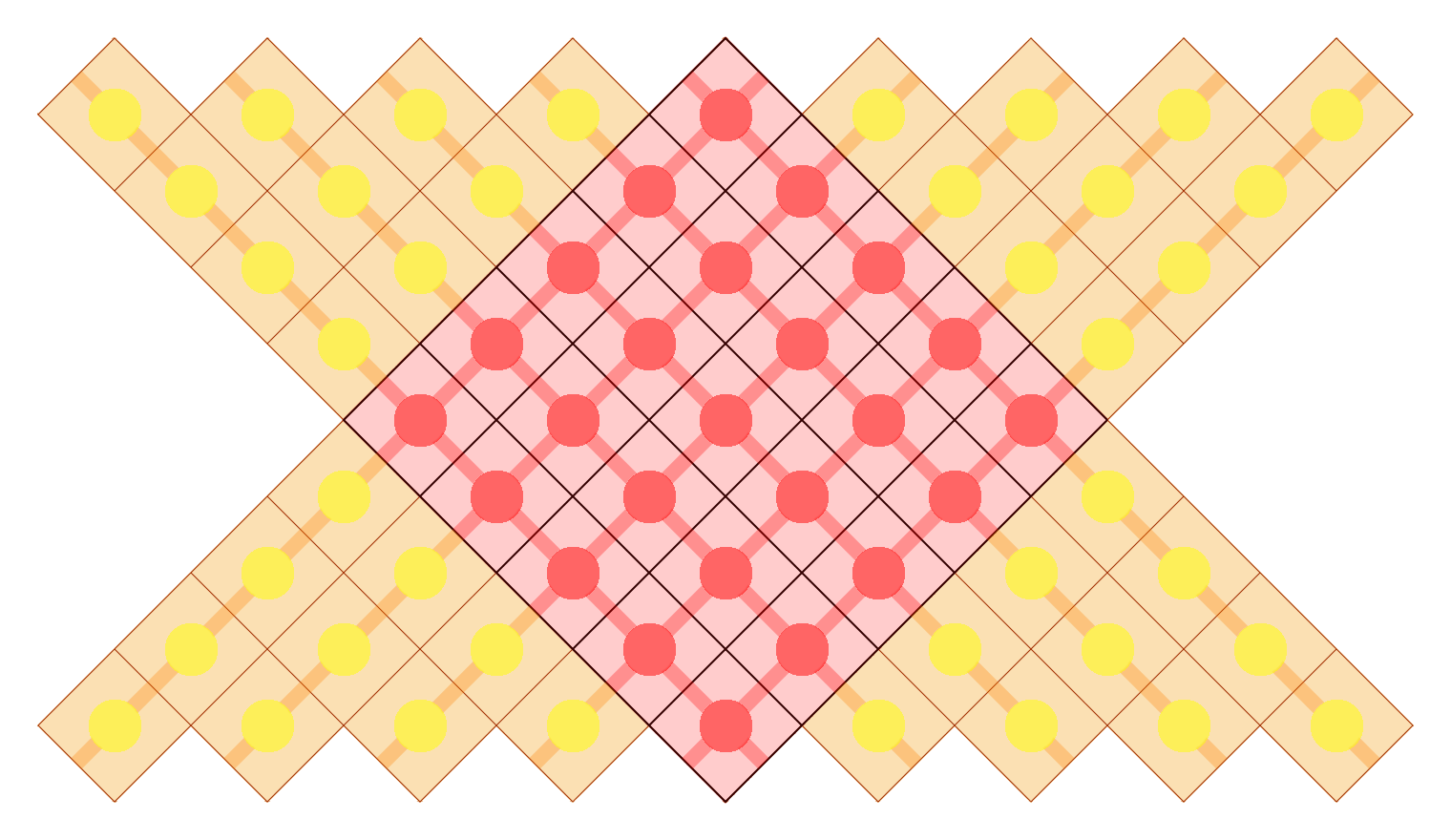}
    \begin{picture}(0,0)
    \put(-174,85){$\xy0;/r.25pc/: 
    	(0,0)*{}="1";
	(-32,-32)*+{\small t}="t";
	{\ar"1";"t"};
    	\endxy$}
    \put(-174,94){$\xy0;/r.25pc/: 
    	(0,0)*{}="1";
	(-32,32)*+{\small s}="s";
	{\ar"1";"s"};
    	\endxy$}
    \put(-82,90){$\bullet$}
    \put(-72,92){$(s_{1},t_{1})$}
    \put(-132,72){$\bullet$}
    \put(-165,74){$(s_{2},t_{3})$}
    \put(-166,7){$\bullet$}
    \put(-201,5){$(s_{1},t_{6})$}
    \put(-166,173){$\bullet$}
    \put(-200,178){$(s_{6},t_{1})$}
    \put(-250,90){$\bullet$}
    \put(-289,91){$(s_{6},t_{6})$}
    \put(-133,23){$e$}
    \put(-116,40){$e$}
    \put(-99,57){$e$}
    \put(-99,23){$e$}
    \put(-82,40){$e$}
    \put(-65,23){$e$}
    \put(-82,74){$e$}
    \put(-65,57){$e$}
    \put(-48,40){$e$}
    \put(-32,23){$e$}
    \put(-82,108){$e$}
    \put(-99,124){$e$}
    \put(-65,124){$e$}
    \put(-116,141){$e$}
    \put(-48,141){$e$}
    \put(-82,141){$e$}
    \put(-32,158){$e$}
    \put(-65,158){$e$}
    \put(-99,158){$e$}
    \put(-133,158){$e$}
    \put(-251,108){$e$}
    \put(-267,124){$e$}
    \put(-234,124){$e$}
    \put(-284,141){$e$}
    \put(-217,141){$e$}
    \put(-251,141){$e$}
    \put(-200,158){$e$}
    \put(-234,158){$e$}
    \put(-267,158){$e$}
    \put(-301,158){$e$}
    \put(-301,23){$e$}
    \put(-284,40){$e$}
    \put(-267,57){$e$}
    \put(-267,23){$e$}
    \put(-251,40){$e$}
    \put(-234,23){$e$}
    \put(-251,74){$e$}
    \put(-234,57){$e$}
    \put(-217,40){$e$}
    \put(-200,23){$e$}
    \end{picture}
\end{figure}  
To more easily relate this picture to earlier ones for 2-groups, it helps
to draw horizontal lines to distinguish the order
of multiplication
\begin{figure}[H]
\centering
\includegraphics[width=0.622\textwidth]{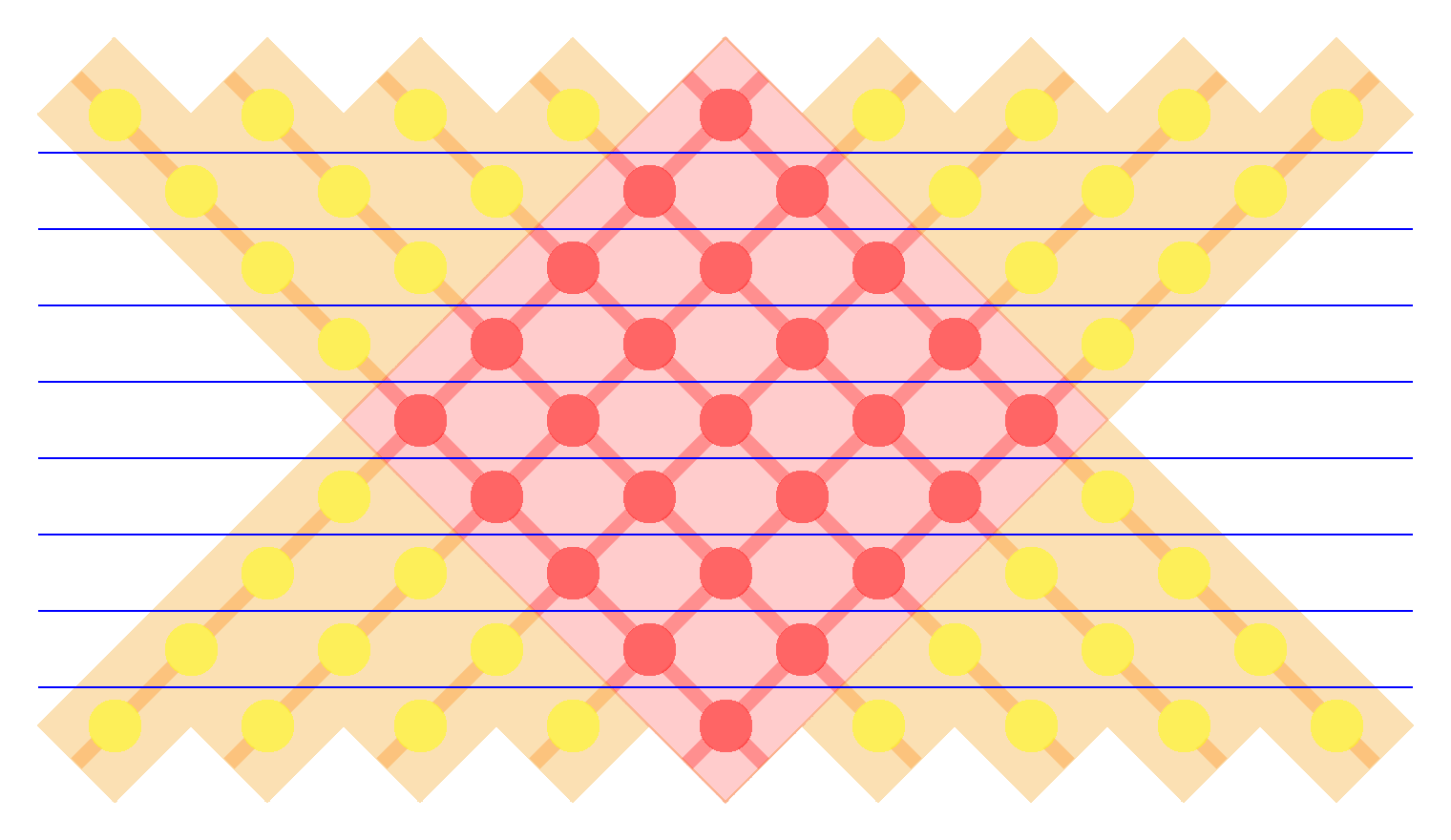}
\end{figure}  
\noindent
and then to tilt the angles of the identities
(only the top half is drawn)
\begin{figure}[H]
\centering
    \includegraphics[width=0.50\textwidth]{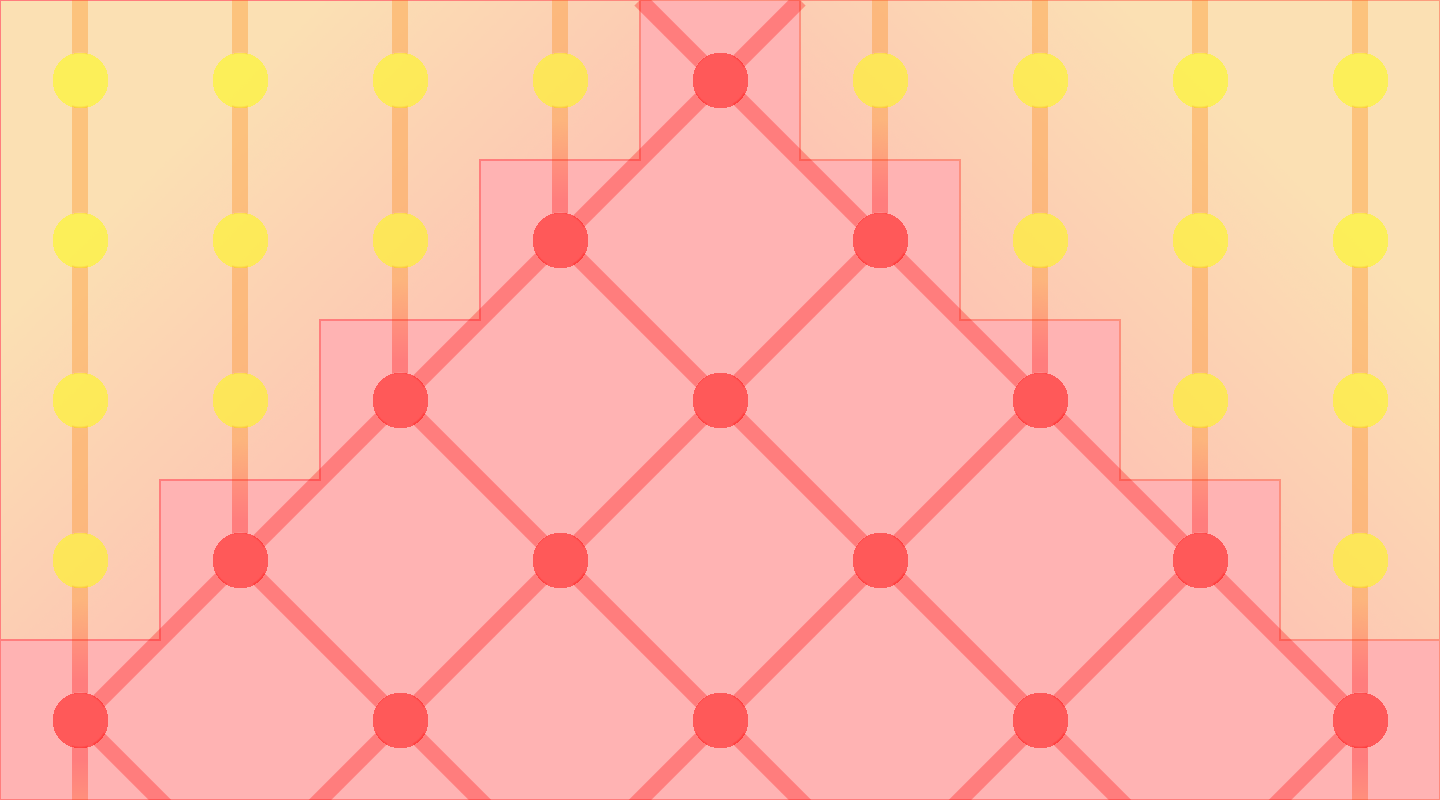}
\end{figure}  
\noindent
which now makes it easy to see we can first compose
each row in parallel and then compose the results in the 
remaining column in series. We explicitly label (some of) 
the 1-d and 0-d defects
\begin{figure}[H]
\centering
    \includegraphics[width=0.65\textwidth]{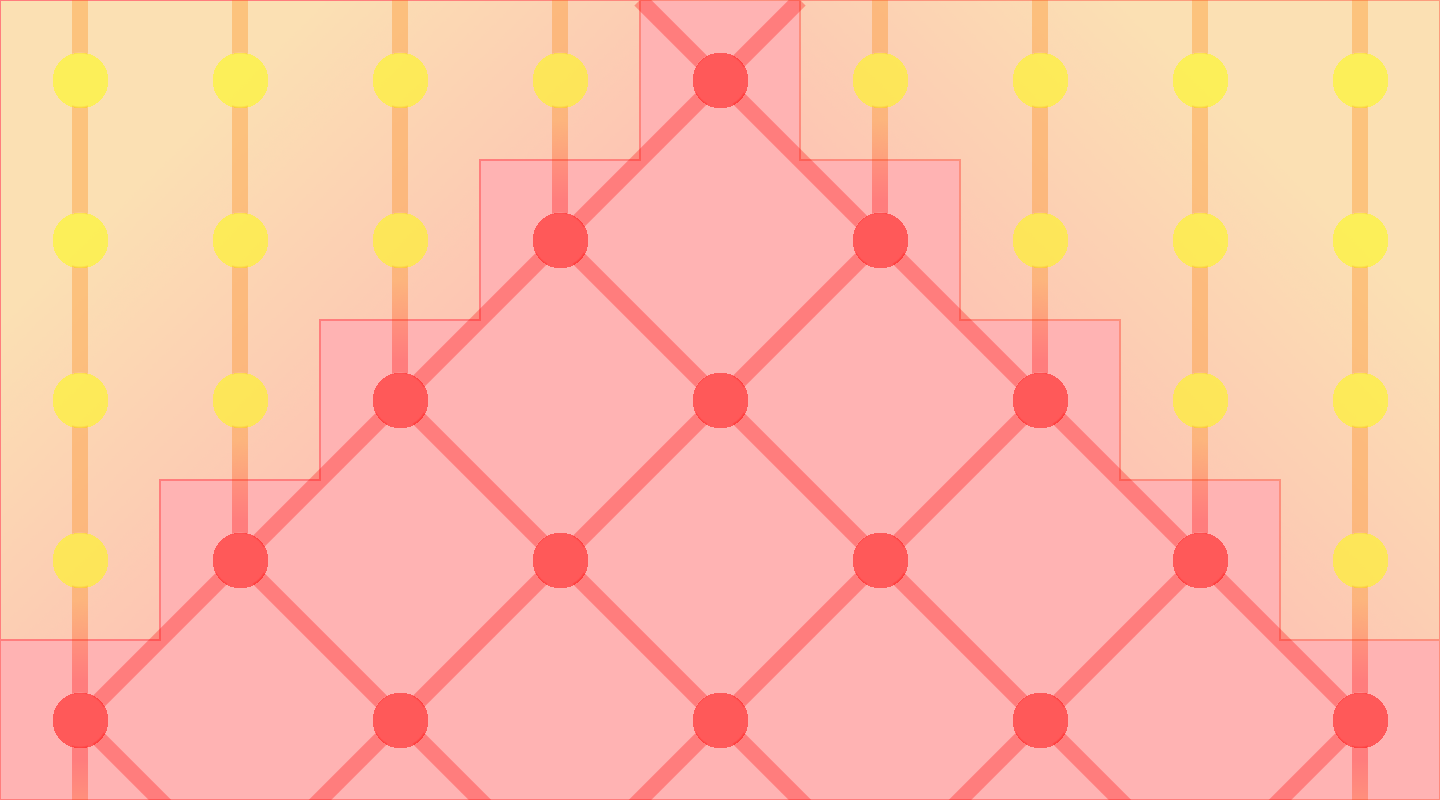}
    \begin{picture}(0,0)
     \put(-308,157){$e$}
     \put(-308,122){$e$}
     \put(-308,86){$e$}
     \put(-308,50){$e$}
     \put(-272,157){$e$}
     \put(-272,122){$e$}
     \put(-272,86){$e$}
     \put(-237,157){$e$}
     \put(-237,122){$e$}
     \put(-202,157){$e$}
     \put(-27,15){$b_{11}$}
     \put(-63,50){$b_{21}$}
     \put(-98,15){$b_{22}$}
     \put(-169,157){$b_{51}$}
     \put(-205,122){$b_{52}$}
     \put(-135,122){$b_{41}$}
     \put(-312,15){$b_{55}$}
     \put(-205,50){$b_{43}$}
     \put(-187,70){$a_{43}^{s}$}
     \put(-187,34){$a_{43}^{t}$}
     \put(-225,70){$a_{53}^{t}$}
     \put(-225,34){$a_{54}^{s}$}
     \put(-24,51){$e$}
     \put(-24,86){$e$}
     \put(-24,122){$e$}
     \put(-24,157){$e$}
     \put(-58,86){$e$}
     \put(-58,122){$e$}
     \put(-58,157){$e$}
     \put(-94,122){$e$}
     \put(-94,157){$e$}
     \put(-130,157){$e$}
     \put(-27,34){$a_{11}^{s}$}
     \put(-27,70){$a_{11}^{s}$}
     \put(-27,106){$a_{11}^{s}$}
     \put(-27,142){$a_{11}^{s}$}
     \put(-27,178){$a_{11}^{s}$}
     \put(-63,70){$a_{21}^{s}$}
     \put(-63,106){$a_{21}^{s}$}
     \put(-63,142){$a_{21}^{s}$}
     \put(-63,178){$a_{21}^{s}$}
     \put(-133,142){$a_{41}^{s}$}
     \put(-154,178){$a_{51}^{s}$}
     \put(-192,178){$a_{61}^{t}$}
     \put(-208,142){$a_{62}^{t}$}
     \put(-311,34){$a_{65}^{t}$}
     \put(-311,70){$a_{65}^{t}$}
     \put(-311,106){$a_{65}^{t}$}
     \put(-311,142){$a_{65}^{t}$}
     \put(-311,178){$a_{65}^{t}$}
     \put(-45,31){$a_{21}^{t}$}
     \put(-81,67){$a_{31}^{t}$}
     \put(-155,140){$a_{51}^{t}$}
     \put(-188,140){$a_{52}^{s}$}
     \put(-293,35){$a_{55}^{s}$}
     \put(-26,-02){$a_{11}^{t}$}
     \put(-46,-02){$a_{12}^{s}$}
     \put(-80,34){$a_{22}^{s}$}
     \put(-80,-02){$a_{22}^{t}$}
     \put(-116,-02){$a_{23}^{s}$}
     \put(-116,34){$a_{32}^{t}$}
     \put(-311,-02){$a_{56}^{s}$}
     \put(-294,-02){$a_{55}^{t}$}
     \end{picture}
\end{figure}  
\noindent
and then multiply each row in parallel. The first row
looks like
\begin{figure}[H]
\centering
    \includegraphics[width=0.95\textwidth]{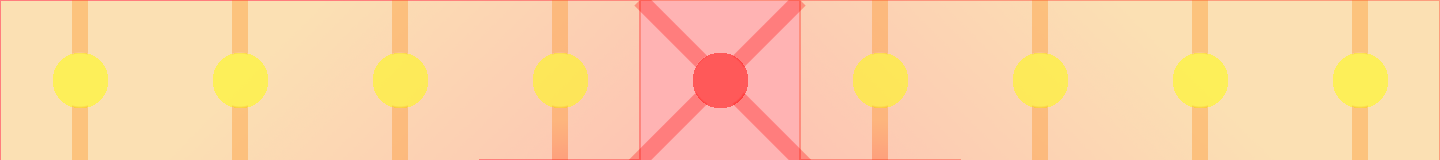}
    \begin{picture}(0,0)
    \put(-450,41){$a_{65}^{t}$}
    \put(-450,4){$a_{65}^{t}$}
    \put(-398,4){$a_{64}^{t}$}
    \put(-398,41){$a_{64}^{t}$}
    \put(-348,4){$a_{63}^{t}$}
    \put(-348,41){$a_{63}^{t}$}
    \put(-446,23){$e$}
    \put(-394,23){$e$}
    \put(-343,23){$e$}
    \put(-291,23){$e$}
    \put(-296,4){$a_{62}^{t}$}
    \put(-296,41){$a_{62}^{t}$}
    \put(-263,4){$a_{52}^{s}$}
    \put(-243,23){$b_{51}$}
    \put(-263,41){$a_{61}^{t}$}
    \put(-226,4){$a_{51}^{t}$}
    \put(-226,41){$a_{51}^{s}$}
    \put(-192,41){$a_{41}^{s}$}
    \put(-192,4){$a_{41}^{s}$}
    \put(-141,4){$a_{31}^{s}$}
    \put(-141,41){$a_{31}^{s}$}
    \put(-89,4){$a_{21}^{s}$}
    \put(-89,41){$a_{21}^{s}$}
    \put(-36,4){$a_{11}^{s}$}
    \put(-36,41){$a_{11}^{s}$}
    \put(-188,23){$e$}
    \put(-136,23){$e$}
    \put(-84,23){$e$}
    \put(-32,23){$e$}
    \end{picture}
    \end{figure}
    \noindent
The result on the 1-d defects is just the usual group
multiplication product while the result on the 0-d defects is
\be
\label{eq:squaretransportforn=5}
\a_{a_{65}^{t}a_{64}^{t}a_{63}^{t}a_{62}^{t}}\big(b_{51}\big).
\ee
The 0-d defects of the next several rows 
are all given by the following
\begin{figure}[H]
\centering
    \includegraphics[width=0.95\textwidth]{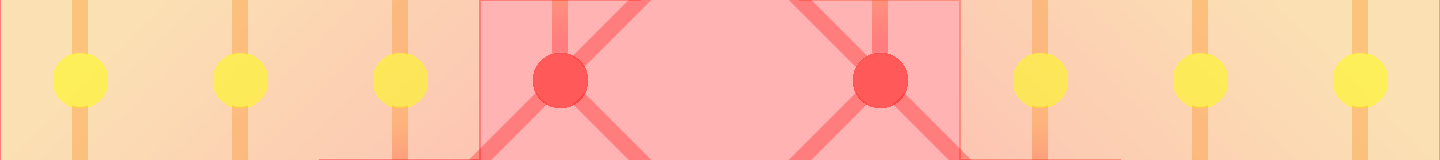}
    \begin{picture}(0,0)
    \put(-450,41){$a_{65}^{t}$}
    \put(-450,4){$a_{65}^{t}$}
    \put(-398,4){$a_{64}^{t}$}
    \put(-398,41){$a_{64}^{t}$}
    \put(-348,4){$a_{63}^{t}$}
    \put(-348,41){$a_{63}^{t}$}
    \put(-446,23){$e$}
    \put(-394,23){$e$}
    \put(-343,23){$e$}
    \put(-295,23){$b_{52}$}
    \put(-296,41){$a_{62}^{t}$}
    \put(-313,4){$a_{53}^{s}$}
    \put(-274,4){$a_{52}^{t}$}
    \put(-274,41){$a_{52}^{s}$}
    \put(-212,4){$a_{42}^{s}$}
    \put(-212,41){$a_{51}^{t}$}
    \put(-192,41){$a_{41}^{s}$}
    \put(-175,4){$a_{41}^{s}$}
    \put(-141,4){$a_{31}^{s}$}
    \put(-141,41){$a_{31}^{s}$}
    \put(-89,4){$a_{21}^{s}$}
    \put(-89,41){$a_{21}^{s}$}
    \put(-36,4){$a_{11}^{s}$}
    \put(-36,41){$a_{11}^{s}$}
    \put(-192,23){$b_{41}$}
    \put(-136,23){$e$}
    \put(-84,23){$e$}
    \put(-32,23){$e$}
    \end{picture}
    \end{figure}
    \noindent
\be
=\a_{a_{65}^{t}a_{64}^{t}a_{63}^{t}}\big(b_{52}\big)
    \a_{a_{65}^{t}a_{64}^{t}a_{63}^{t}a_{62}^{t}a_{52}^{s}}
    \big(b_{41}\big),
\ee
\begin{figure}[H]
\centering
    \includegraphics[width=0.95\textwidth]{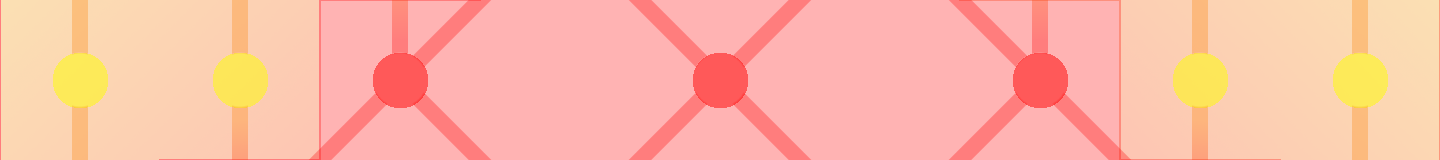}
    \begin{picture}(0,0)
    \put(-450,41){$a_{65}^{t}$}
    \put(-450,4){$a_{65}^{t}$}
    \put(-398,4){$a_{64}^{t}$}
    \put(-398,41){$a_{64}^{t}$}
    \put(-363,4){$a_{54}^{s}$}
    \put(-348,41){$a_{63}^{t}$}
    \put(-446,23){$e$}
    \put(-394,23){$e$}
    \put(-346,23){$b_{53}$}
    \put(-326,4){$a_{53}^{t}$}
    \put(-326,41){$a_{53}^{s}$}
    \put(-263,4){$a_{43}^{s}$}
    \put(-243,23){$b_{42}$}
    \put(-263,41){$a_{52}^{t}$}
    \put(-225,4){$a_{42}^{t}$}
    \put(-225,41){$a_{42}^{s}$}
    \put(-163,41){$a_{41}^{t}$}
    \put(-163,4){$a_{32}^{s}$}
    \put(-123,4){$a_{31}^{t}$}
    \put(-141,41){$a_{31}^{s}$}
    \put(-89,4){$a_{21}^{s}$}
    \put(-89,41){$a_{21}^{s}$}
    \put(-36,4){$a_{11}^{s}$}
    \put(-36,41){$a_{11}^{s}$}
    \put(-140,23){$b_{31}$}
    \put(-84,23){$e$}
    \put(-32,23){$e$}
    \end{picture}
    \end{figure}
    \noindent
\be
=\a_{a_{65}^{t}a_{64}^{t}}\big(b_{53}\big)
    \a_{a_{65}^{t}a_{64}^{t}a_{63}^{t}a_{53}^{s}}
    \big(b_{42}\big)
    \a_{a_{65}^{t}a_{64}^{t}a_{63}^{t}a_{53}^{s}a_{52}^{t}a_{42}^{s}}
    \big(b_{31}\big),
\ee
\begin{figure}[H]
\centering
    \includegraphics[width=0.95\textwidth]{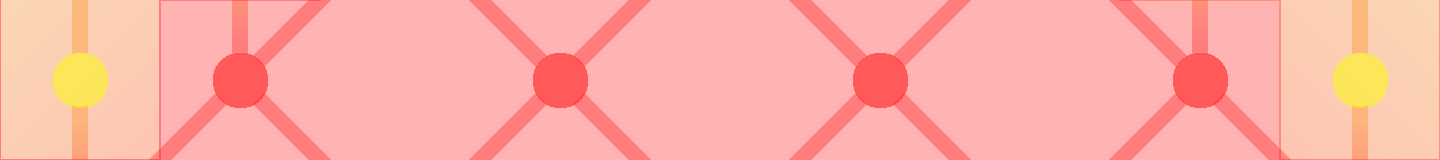}
    \begin{picture}(0,0)
    \put(-450,41){$a_{65}^{t}$}
    \put(-450,4){$a_{65}^{t}$}
    \put(-416,4){$a_{55}^{s}$}
    \put(-398,41){$a_{64}^{t}$}
    \put(-376,4){$a_{54}^{t}$}
    \put(-376,41){$a_{54}^{s}$}
    \put(-446,23){$e$}
    \put(-398,23){$b_{54}$}
    \put(-295,23){$b_{43}$}
    \put(-313,41){$a_{53}^{t}$}
    \put(-313,4){$a_{44}^{s}$}
    \put(-274,4){$a_{43}^{t}$}
    \put(-274,41){$a_{43}^{s}$}
    \put(-211,4){$a_{33}^{s}$}
    \put(-211,41){$a_{42}^{t}$}
    \put(-175,41){$a_{32}^{s}$}
    \put(-175,4){$a_{32}^{t}$}
    \put(-110,4){$a_{22}^{s}$}
    \put(-110,41){$a_{31}^{t}$}
    \put(-69,4){$a_{21}^{t}$}
    \put(-89,41){$a_{21}^{s}$}
    \put(-36,4){$a_{11}^{s}$}
    \put(-36,41){$a_{11}^{s}$}
    \put(-192,23){$b_{32}$}
    \put(-88,23){$b_{21}$}
    \put(-32,23){$e$}
    \end{picture}
    \end{figure}
    \noindent
\be
=\a_{a_{65}^{t}}\big(b_{54}\big)
    \a_{a_{65}^{t}a_{64}^{t}a_{54}^{s}}
    \big(b_{43}\big)
    \a_{a_{65}^{t}a_{64}^{t}a_{54}^{s}a_{53}^{t}a_{43}^{s}}
    \big(b_{32}\big)
    \a_{a_{65}^{t}a_{64}^{t}a_{54}^{s}a_{53}^{t}a_{43}^{s}a_{42}^{t}a_{32}^{s}}
    \big(b_{21}\big),
\ee
\begin{figure}[H]
\centering
    \includegraphics[width=0.95\textwidth]{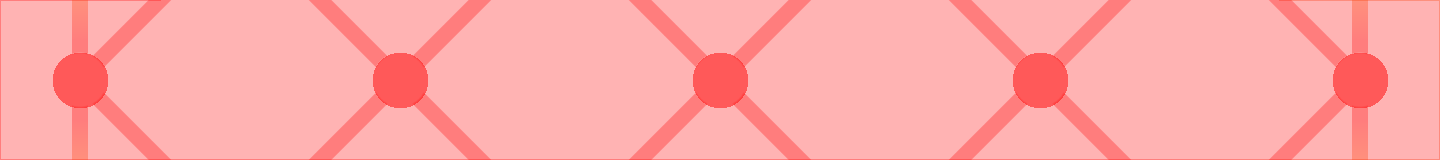}
    \begin{picture}(0,0)
    \put(-450,41){$a_{65}^{t}$}
    \put(-450,4){$a_{56}^{s}$}
    \put(-424,4){$a_{55}^{t}$}
    \put(-424,41){$a_{55}^{s}$}
    \put(-365,4){$a_{45}^{s}$}
    \put(-365,41){$a_{54}^{t}$}
    \put(-450,23){$b_{55}$}
    \put(-346,23){$b_{44}$}
    \put(-326,4){$a_{44}^{t}$}
    \put(-326,41){$a_{44}^{s}$}
    \put(-263,4){$a_{34}^{s}$}
    \put(-243,23){$b_{33}$}
    \put(-263,41){$a_{43}^{t}$}
    \put(-225,4){$a_{33}^{t}$}
    \put(-225,41){$a_{33}^{s}$}
    \put(-163,41){$a_{32}^{t}$}
    \put(-163,4){$a_{23}^{s}$}
    \put(-121,4){$a_{22}^{t}$}
    \put(-121,41){$a_{22}^{s}$}
    \put(-58,4){$a_{12}^{s}$}
    \put(-58,41){$a_{21}^{t}$}
    \put(-36,4){$a_{11}^{t}$}
    \put(-36,41){$a_{11}^{s}$}
    \put(-140,23){$b_{22}$}
    \put(-35,23){$b_{11}$}
    \end{picture}
    \end{figure}
    \noindent
\be
=b_{55}\a_{a_{65}^{t}a_{55}^{s}}\big(b_{44}\big)
    \a_{a_{65}^{t}a_{55}^{s}a_{54}^{t}a_{44}^{s}}
    \big(b_{33}\big)
    \a_{a_{65}^{t}a_{55}^{s}a_{54}^{t}a_{44}^{s}a_{43}^{t}a_{33}^{s}}
    \big(b_{22}\big)
    \a_{a_{65}^{t}a_{55}^{s}a_{54}^{t}a_{44}^{s}a_{43}^{t}a_{33}^{s}
    a_{32}^{t}a_{22}^{s}}
    \big(b_{11}\big),
\ee
\begin{figure}[H]
\centering
    \includegraphics[width=0.95\textwidth]{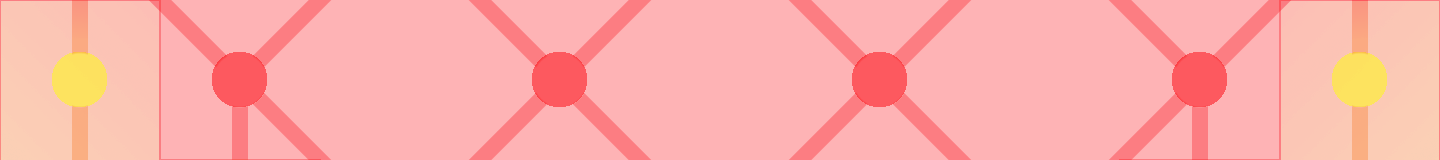}
    \begin{picture}(0,0)
    \put(-450,41){$a_{56}^{s}$}
    \put(-450,4){$a_{56}^{s}$}
    \put(-398,4){$a_{46}^{s}$}
    \put(-416,41){$a_{55}^{t}$}
    \put(-377,4){$a_{45}^{t}$}
    \put(-377,41){$a_{45}^{s}$}
    \put(-446,23){$e$}
    \put(-398,23){$b_{45}$}
    \put(-295,23){$b_{34}$}
    \put(-313,41){$a_{44}^{t}$}
    \put(-313,4){$a_{35}^{s}$}
    \put(-274,4){$a_{34}^{t}$}
    \put(-274,41){$a_{34}^{s}$}
    \put(-211,4){$a_{24}^{s}$}
    \put(-211,41){$a_{33}^{t}$}
    \put(-174,41){$a_{23}^{s}$}
    \put(-174,4){$a_{23}^{t}$}
    \put(-110,4){$a_{13}^{s}$}
    \put(-110,41){$a_{22}^{t}$}
    \put(-89,4){$a_{12}^{t}$}
    \put(-69,41){$a_{12}^{s}$}
    \put(-36,4){$a_{11}^{t}$}
    \put(-36,41){$a_{11}^{t}$}
    \put(-192,23){$b_{23}$}
    \put(-88,23){$b_{12}$}
    \put(-32,23){$e$}
    \end{picture}
    \end{figure}
    \noindent
\be
= \a_{a_{56}^{s}}\big(b_{45}\big)
    \a_{a_{56}^{s}a_{55}^{t}a_{45}^{s}}
    \big(b_{34}\big)
    \a_{a_{56}^{s}a_{55}^{t}a_{45}^{s}a_{44}^{t}a_{34}^{s}}
    \big(b_{23}\big)
    \a_{a_{56}^{s}a_{55}^{t}a_{45}^{s}a_{44}^{t}a_{34}^{s}a_{33}^{t}a_{23}^{s}}
    \big(b_{12}\big),
\ee
\begin{figure}[H]
\centering
    \includegraphics[width=0.95\textwidth]{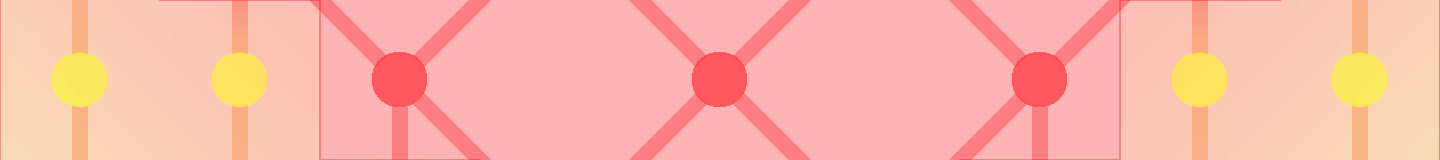}
    \begin{picture}(0,0)
    \put(-450,41){$a_{56}^{s}$}
    \put(-450,4){$a_{56}^{s}$}
    \put(-398,4){$a_{46}^{s}$}
    \put(-398,41){$a_{46}^{s}$}
    \put(-348,4){$a_{36}^{s}$}
    \put(-363,41){$a_{45}^{t}$}
    \put(-446,23){$e$}
    \put(-394,23){$e$}
    \put(-346,23){$b_{35}$}
    \put(-326,4){$a_{35}^{t}$}
    \put(-326,41){$a_{35}^{s}$}
    \put(-263,4){$a_{25}^{s}$}
    \put(-243,23){$b_{24}$}
    \put(-263,41){$a_{34}^{t}$}
    \put(-225,4){$a_{24}^{t}$}
    \put(-225,41){$a_{24}^{s}$}
    \put(-163,41){$a_{23}^{t}$}
    \put(-163,4){$a_{14}^{s}$}
    \put(-141,4){$a_{13}^{t}$}
    \put(-123,41){$a_{13}^{s}$}
    \put(-89,4){$a_{12}^{t}$}
    \put(-89,41){$a_{12}^{t}$}
    \put(-36,4){$a_{11}^{t}$}
    \put(-36,41){$a_{11}^{t}$}
    \put(-140,23){$b_{13}$}
    \put(-84,23){$e$}
    \put(-32,23){$e$}
    \end{picture}
    \end{figure}
    \noindent
\be
=\a_{a_{56}^{s}a_{46}^{s}}\big(b_{35}\big)
    \a_{a_{56}^{s}a_{46}^{s}a_{45}^{t}a_{35}^{s}}
    \big(b_{24}\big)
    \a_{a_{56}^{s}a_{46}^{s}a_{45}^{t}a_{35}^{s}a_{34}^{t}a_{24}^{s}}
    \big(b_{13}\big),
\ee
\begin{figure}[H]
\centering
    \includegraphics[width=0.95\textwidth]{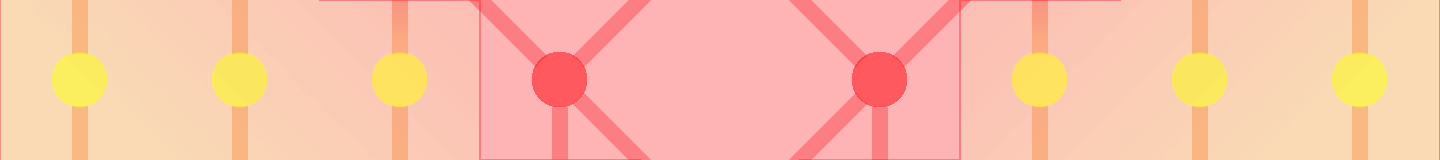}
    \begin{picture}(0,0)
    \put(-450,41){$a_{56}^{s}$}
    \put(-450,4){$a_{56}^{s}$}
    \put(-398,4){$a_{46}^{s}$}
    \put(-398,41){$a_{46}^{s}$}
    \put(-348,4){$a_{36}^{s}$}
    \put(-348,41){$a_{36}^{s}$}
    \put(-446,23){$e$}
    \put(-394,23){$e$}
    \put(-343,23){$e$}
    \put(-295,23){$b_{25}$}
    \put(-313,41){$a_{35}^{t}$}
    \put(-296,4){$a_{26}^{s}$}
    \put(-274,4){$a_{25}^{t}$}
    \put(-274,41){$a_{25}^{s}$}
    \put(-212,4){$a_{15}^{s}$}
    \put(-212,41){$a_{24}^{t}$}
    \put(-175,41){$a_{14}^{s}$}
    \put(-192,4){$a_{14}^{t}$}
    \put(-141,4){$a_{13}^{t}$}
    \put(-141,41){$a_{13}^{t}$}
    \put(-89,4){$a_{12}^{t}$}
    \put(-89,41){$a_{12}^{t}$}
    \put(-36,4){$a_{11}^{t}$}
    \put(-36,41){$a_{11}^{t}$}
    \put(-192,23){$b_{14}$}
    \put(-136,23){$e$}
    \put(-84,23){$e$}
    \put(-32,23){$e$}
    \end{picture}
    \end{figure}
    \noindent
\be
=\a_{a_{56}^{s}a_{46}^{s}a_{36}^{s}}\big(b_{25}\big)
    \a_{a_{56}^{s}a_{46}^{s}a_{36}^{s}a_{35}^{t}a_{25}^{s}}
    \big(b_{14}\big),
\ee
and finally 
\begin{figure}[H]
\centering
    \includegraphics[width=0.95\textwidth]{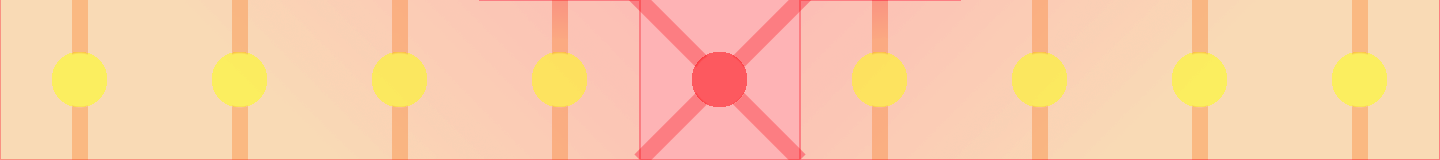}
    \begin{picture}(0,0)
    \put(-450,41){$a_{56}^{s}$}
    \put(-450,4){$a_{56}^{s}$}
    \put(-398,4){$a_{46}^{s}$}
    \put(-398,41){$a_{46}^{s}$}
    \put(-348,4){$a_{36}^{s}$}
    \put(-348,41){$a_{36}^{s}$}
    \put(-446,23){$e$}
    \put(-394,23){$e$}
    \put(-343,23){$e$}
    \put(-291,23){$e$}
    \put(-296,4){$a_{26}^{s}$}
    \put(-296,41){$a_{26}^{s}$}
    \put(-263,4){$a_{16}^{s}$}
    \put(-243,23){$b_{15}$}
    \put(-263,41){$a_{25}^{t}$}
    \put(-226,4){$a_{15}^{t}$}
    \put(-226,41){$a_{15}^{s}$}
    \put(-192,41){$a_{14}^{t}$}
    \put(-192,4){$a_{14}^{t}$}
    \put(-141,4){$a_{13}^{t}$}
    \put(-141,41){$a_{13}^{t}$}
    \put(-89,4){$a_{12}^{t}$}
    \put(-89,41){$a_{12}^{t}$}
    \put(-36,4){$a_{11}^{t}$}
    \put(-36,41){$a_{11}^{t}$}
    \put(-188,23){$e$}
    \put(-136,23){$e$}
    \put(-84,23){$e$}
    \put(-32,23){$e$}
    \end{picture}
    \end{figure}
    \noindent
\be
=\a_{a_{56}^{s}a_{46}^{s}a_{36}^{s}a_{26}^{s}}\big(b_{15}\big).
\ee
The result of composing all of these in series gives
\be
\label{eq:surfaceproductn=5}
\hspace{8mm}
\begin{matrix}[0.9]
\a_{a_{65}^{t}a_{64}^{t}a_{63}^{t}a_{62}^{t}}\big(b_{51}\big)
\\
\a_{a_{65}^{t}a_{64}^{t}a_{63}^{t}}\big(b_{52}\big)
    \a_{a_{65}^{t}a_{64}^{t}a_{63}^{t}a_{62}^{t}a_{52}^{s}}
    \big(b_{41}\big)
\\
\a_{a_{65}^{t}a_{64}^{t}}\big(b_{53}\big)
    \a_{a_{65}^{t}a_{64}^{t}a_{63}^{t}a_{53}^{s}}
    \big(b_{42}\big)
    \a_{a_{65}^{t}a_{64}^{t}a_{63}^{t}a_{53}^{s}a_{52}^{t}a_{42}^{s}}
    \big(b_{31}\big)
\\
\a_{a_{65}^{t}}\big(b_{54}\big)
    \a_{a_{65}^{t}a_{64}^{t}a_{54}^{s}}
    \big(b_{43}\big)
    \a_{a_{65}^{t}a_{64}^{t}a_{54}^{s}a_{53}^{t}a_{43}^{s}}
    \big(b_{32}\big)
    \a_{a_{65}^{t}a_{64}^{t}a_{54}^{s}a_{53}^{t}a_{43}^{s}a_{42}^{t}a_{32}^{s}}
    \big(b_{21}\big)
\\
b_{55}\a_{a_{65}^{t}a_{55}^{s}}\big(b_{44}\big)
    \a_{a_{65}^{t}a_{55}^{s}a_{54}^{t}a_{44}^{s}}
    \big(b_{33}\big)
    \a_{a_{65}^{t}a_{55}^{s}a_{54}^{t}a_{44}^{s}a_{43}^{t}a_{33}^{s}}
    \big(b_{22}\big)
    \a_{a_{65}^{t}a_{55}^{s}a_{54}^{t}a_{44}^{s}a_{43}^{t}a_{33}^{s}
    a_{32}^{t}a_{22}^{s}}
    \big(b_{11}\big)
\\
 \a_{a_{56}^{s}}\big(b_{45}\big)
    \a_{a_{56}^{s}a_{55}^{t}a_{45}^{s}}
    \big(b_{34}\big)
    \a_{a_{56}^{s}a_{55}^{t}a_{45}^{s}a_{44}^{t}a_{34}^{s}}
    \big(b_{23}\big)
    \a_{a_{56}^{s}a_{55}^{t}a_{45}^{s}a_{44}^{t}a_{34}^{s}a_{33}^{t}a_{23}^{s}}
    \big(b_{12}\big)
\\
\a_{a_{56}^{s}a_{46}^{s}}\big(b_{35}\big)
    \a_{a_{56}^{s}a_{46}^{s}a_{45}^{t}a_{35}^{s}}
    \big(b_{24}\big)
    \a_{a_{56}^{s}a_{46}^{s}a_{45}^{t}a_{35}^{s}a_{34}^{t}a_{24}^{s}}
    \big(b_{13}\big)
\\
\a_{a_{56}^{s}a_{46}^{s}a_{36}^{s}}\big(b_{25}\big)
    \a_{a_{56}^{s}a_{46}^{s}a_{36}^{s}a_{35}^{t}a_{25}^{s}}
    \big(b_{14}\big)
\\
\a_{a_{56}^{s}a_{46}^{s}a_{36}^{s}a_{26}^{s}}\big(b_{15}\big)
\end{matrix}
\ee
which, when expressed in terms of usual group multiplication
in $H$ becomes
\be
\label{eq:surfaceproductn=5inoneline}
\a_{a_{56}^{s}a_{46}^{s}a_{36}^{s}a_{26}^{s}}\big(b_{15}\big)
\a_{a_{56}^{s}a_{46}^{s}a_{36}^{s}}\big(b_{25}\big)
\a_{a_{56}^{s}a_{46}^{s}a_{36}^{s}a_{35}^{t}a_{25}^{s}}\big(b_{14}\big)
\cdots
\a_{a_{65}^{t}a_{64}^{t}a_{63}^{t}a_{62}^{t}a_{52}^{s}}\big(b_{41}\big)
\a_{a_{65}^{t}a_{64}^{t}a_{63}^{t}a_{62}^{t}}\big(b_{51}\big).
\ee
We can visualize this mess more easily 
by expanding out each
$b_{ij}$ to lowest order (since we already know that the $a$'s give
the one-dimensional parallel transport, we do not have to
expand them out) and examining the terms with zero $B_{ij}$'s,
terms with one $B_{ij},$ terms that involve the product
of two $B_{ij}$'s of different indices, and so on. 
For example, expanding out just the first two terms on the right 
in (\ref{eq:surfaceproductn=5inoneline}) gives
(a prime on the second $\nu_{1}$ index has been adjoined to remain consistent
with the Einstein summation convention)
\be
\begin{split}
&\a_{a_{65}^{t}a_{64}^{t}a_{63}^{t}a_{62}^{t}a_{52}^{s}}\big(b_{41}\big)
\a_{a_{65}^{t}a_{64}^{t}a_{63}^{t}a_{62}^{t}}\big(b_{51}\big)\\
&\approxeq\left(\mathds{1}+\un{\a_{a_{65}^{t}a_{64}^{t}a_{63}^{t}a_{62}^{t}a_{52}^{s}}}\left(B_{\mu_{4}\nu_{1}}\big(x(s_{4},t_{1})\big)
\frac{\p x^{\mu_{4}}}{\p s}
\frac{\p x^{\nu_{1}}}{\p t}\Big|_{(s_{4},t_{1})}\right)\D s_{4}\D t_{1}\right)\\
&\times\left(\mathds{1}+\un{\a_{a_{65}^{t}a_{64}^{t}a_{63}^{t}a_{62}^{t}}}\left(B_{\mu_{5}\nu'_{1}}\big(x(s_{5},t_{1})\big)
\frac{\p x^{\mu_{5}}}{\p s}
\frac{\p x^{\nu'_{1}}}{\p t}\Big|_{(s_{5},t_{1})}\right)\D s_{5}\D t_{1}\right).
\end{split}
\ee
Expanding all of these products out and separating the terms 
order by order (the order is now determined by the area elements) 
results in a single zeroth order term given by just the identity and
25 first order terms with a single $B$
(some of these terms are written underneath the pictures
to more clearly illustrate our convention)
\begin{figure}[H]
\centering
\begin{subfigure}{0.15\textwidth}
    \begin{picture}(0,0)
    \put(0,-75){$\scriptstyle\a_{a_{56}^{s}a_{46}^{s}a_{36}^{s}a_{26}^{s}}(B_{15})$}
    \end{picture}
    \includegraphics[width=1.00\textwidth]{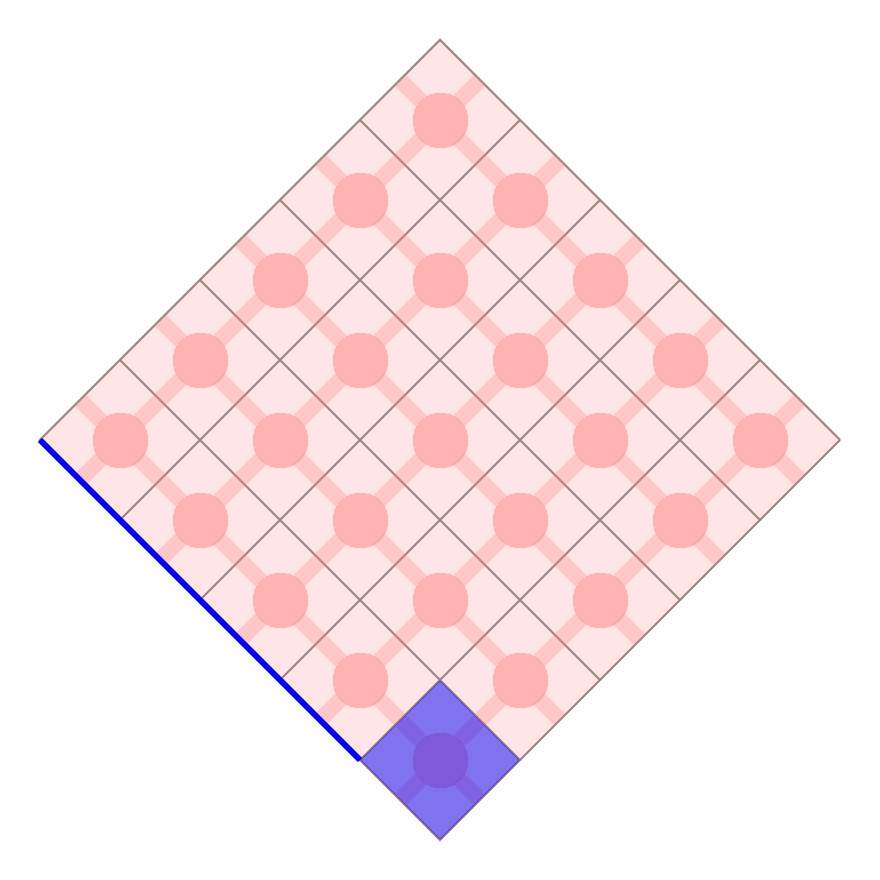}
\end{subfigure}
$+$
\begin{subfigure}{0.15\textwidth}
 \includegraphics[width=1.00\textwidth]{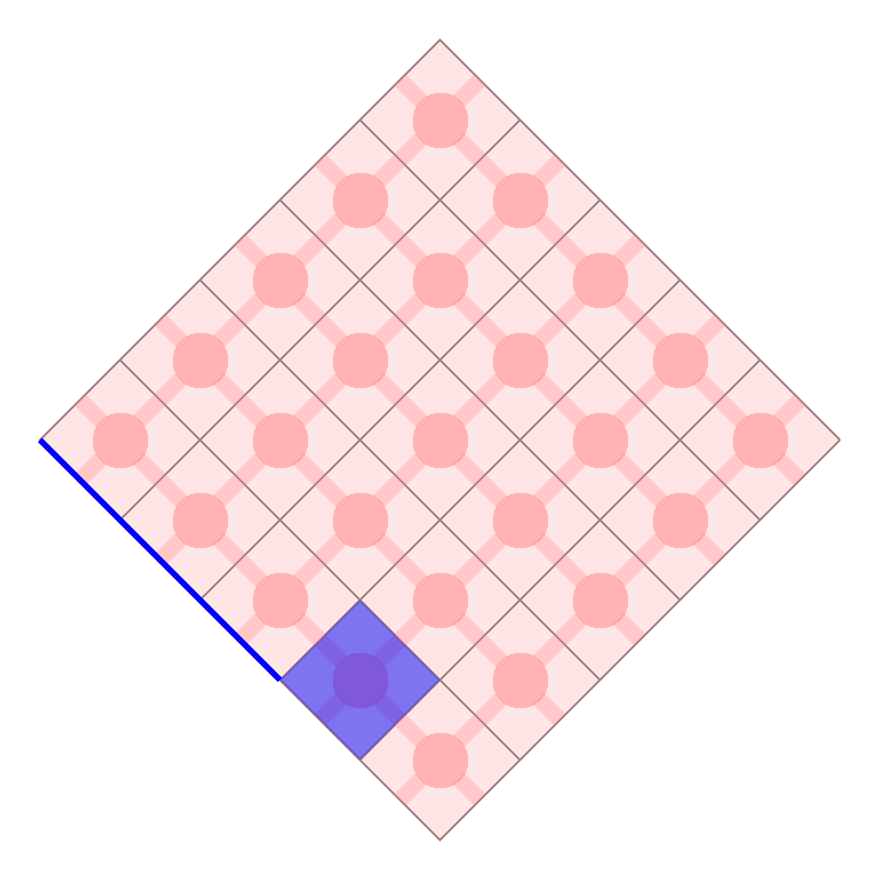}
\end{subfigure}
$+$
\begin{subfigure}{0.15\textwidth}
  \begin{picture}(0,0)
    \put(0,-75){$\scriptstyle\a_{a_{56}^{s}a_{46}^{s}a_{36}^{s}a_{35}^{t}a_{25}^{s}}
    (B_{14})$}
    \end{picture}
 \includegraphics[width=1.00\textwidth]{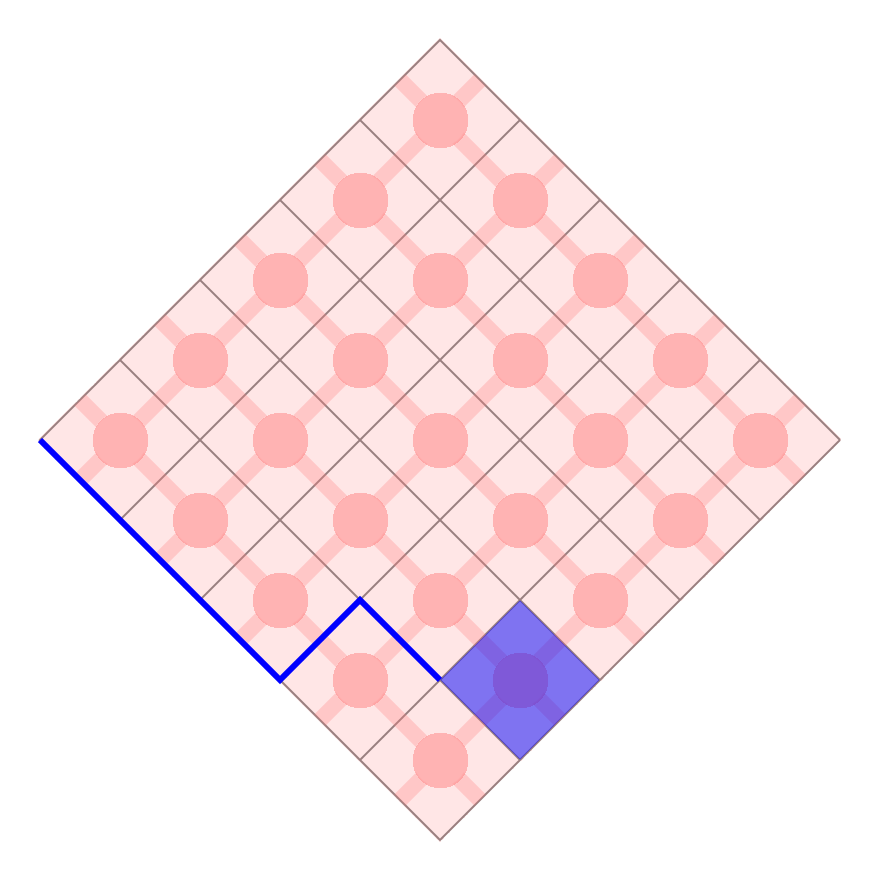}
\end{subfigure}
$+$
\begin{subfigure}{0.15\textwidth}
 \includegraphics[width=1.00\textwidth]{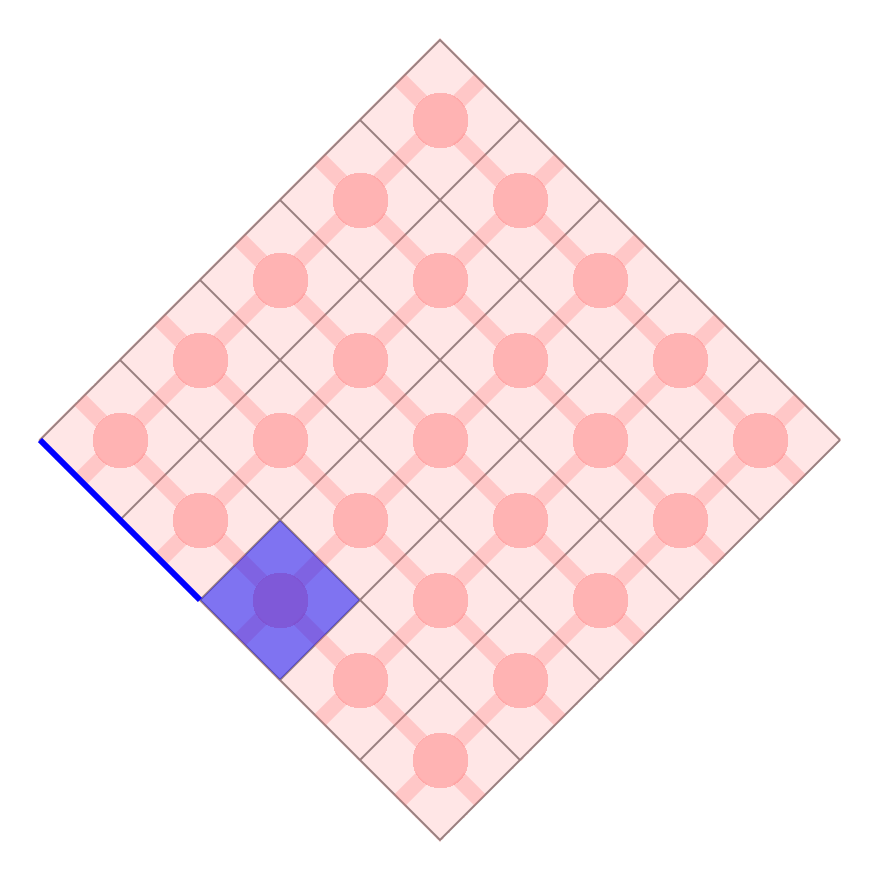}
\end{subfigure}
$+$
\begin{subfigure}{0.15\textwidth}
   \begin{picture}(0,0)
    \put(0,-75){$\scriptstyle\a_{a_{56}^{s}a_{46}^{s}a_{45}^{t}a_{35}^{s}}
    (B_{24})$}
    \end{picture}
 \includegraphics[width=1.00\textwidth]{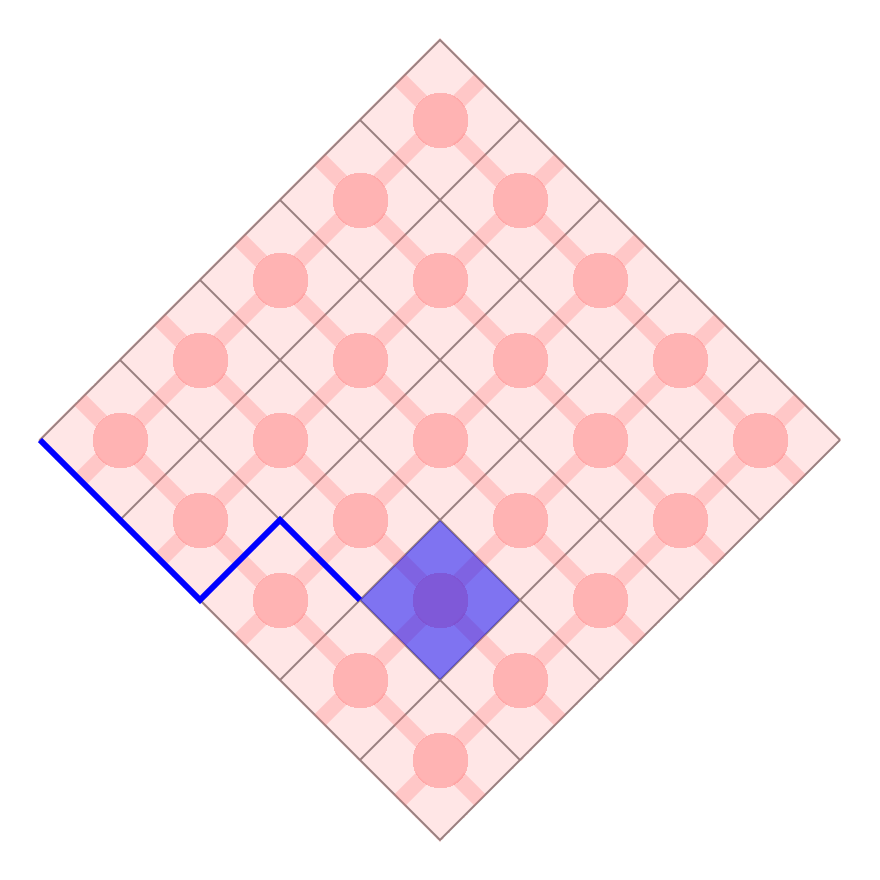}
\end{subfigure}
\end{figure} 
\vspace{-4mm}
\begin{figure}[H]
\centering
$+$
\begin{subfigure}{0.15\textwidth}
    \begin{picture}(0,0)
    \put(-10,-75){$\scriptstyle\a_{a_{56}^{s}a_{46}^{s}a_{45}^{t}a_{35}^{s}a_{34}^{t}a_{24}^{s}}
    (B_{13})$}
    \end{picture}
    \includegraphics[width=1.00\textwidth]{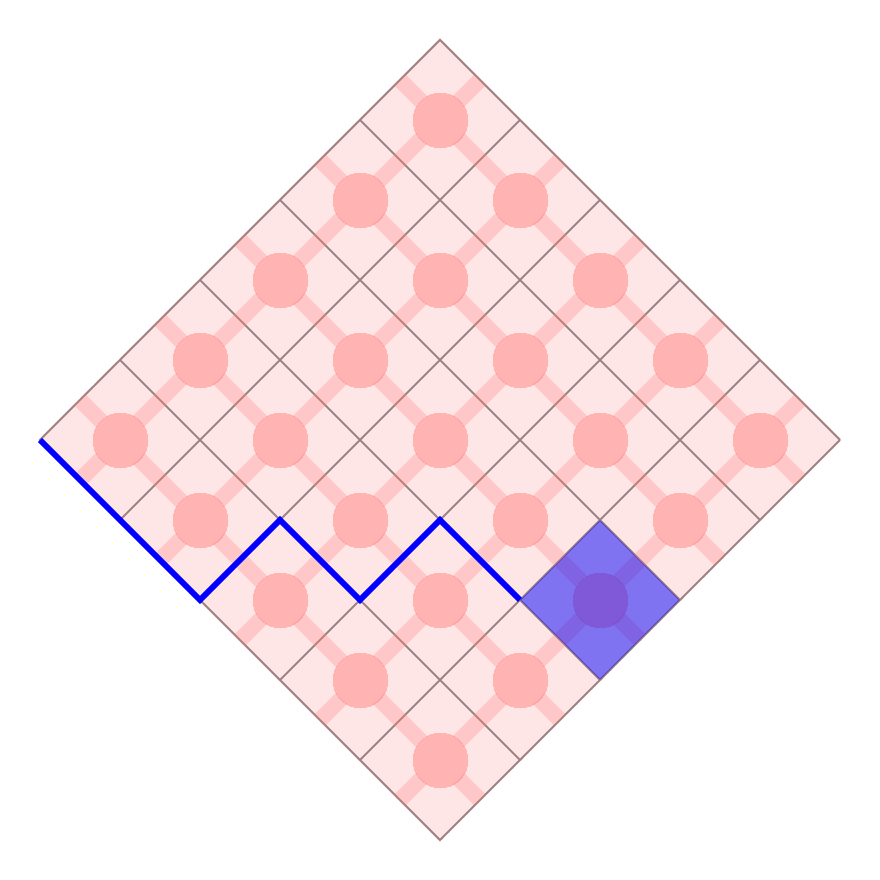}
\end{subfigure}
$+\cdots+$
\begin{subfigure}{0.15\textwidth}
 \includegraphics[width=1.00\textwidth]{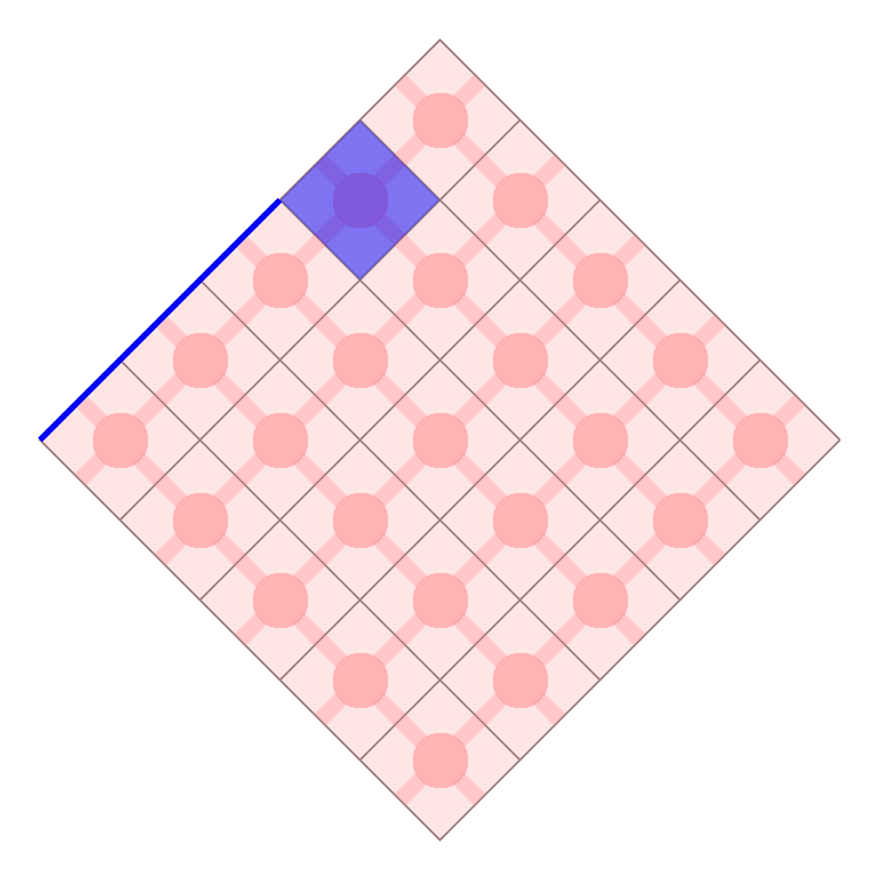}
\end{subfigure}
$+$
\begin{subfigure}{0.15\textwidth}
 \includegraphics[width=1.00\textwidth]{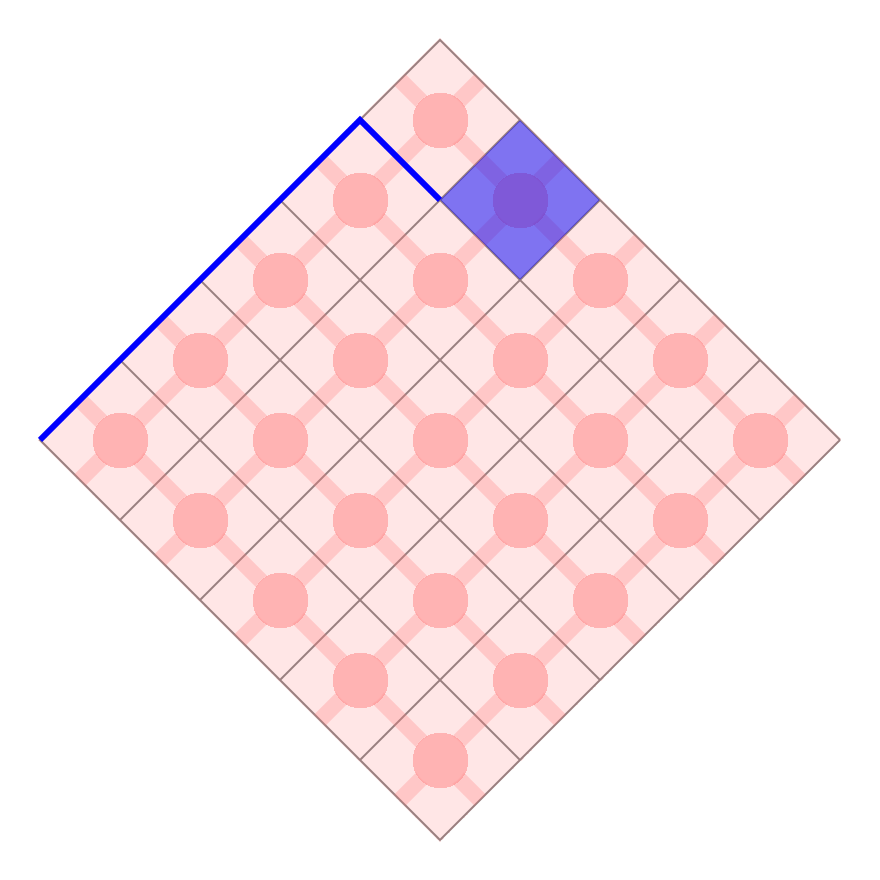}
\end{subfigure}
$+$
\begin{subfigure}{0.15\textwidth}
 \includegraphics[width=1.00\textwidth]{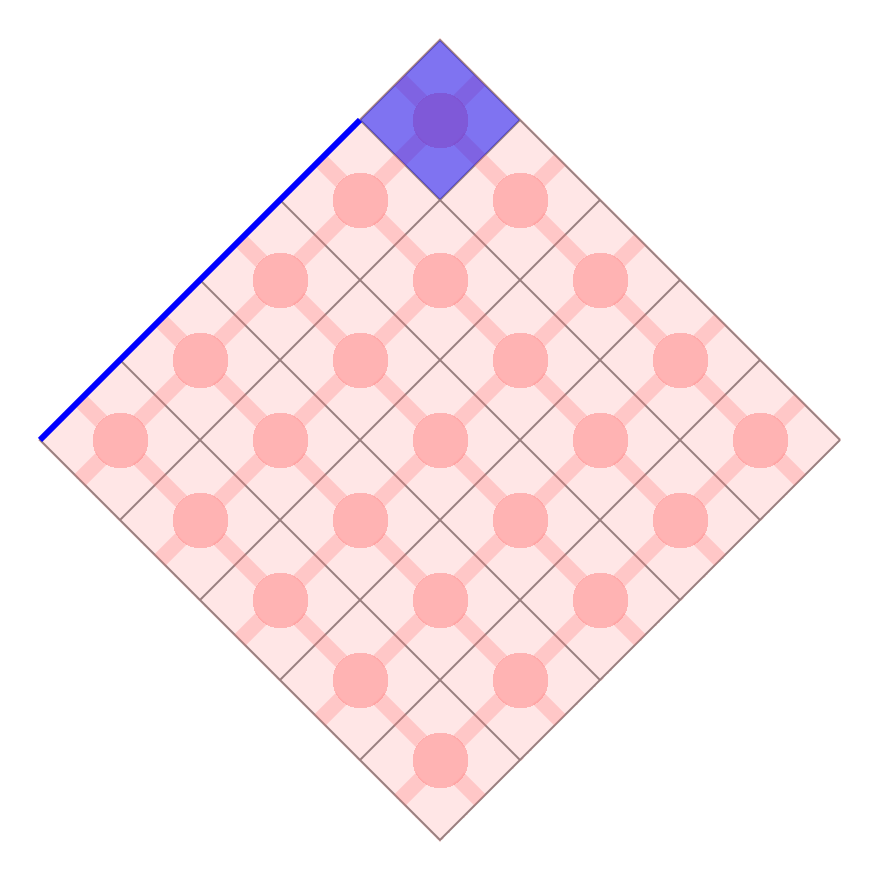}
\end{subfigure}
\end{figure} 
These pictures express the fact that 
we calculate the ordinary parallel transport along a 
specified path between the point $(s,t)=(s_{6},t_{6})$ and 
another point
$(s_{i+1},t_{j+1})$ (represented by a blue line) 
and conjugate each
$B$ field at $(s_{i},t_{j})$ (represented by a blue square) 
by that parallel transport using $\a.$ Then we sum over all points
at which $B$ has been specified.
There are 
$\sum_{k=1}^{24}k=\frac{24(25)}{2}=\binom{25}{2}=600$ 
second order terms, i.e. terms with two $B$'s:
\begin{figure}[H]
\centering
\hspace{5mm}
\begin{subfigure}{0.15\textwidth}
\centering
    \includegraphics[width=1.00\textwidth]{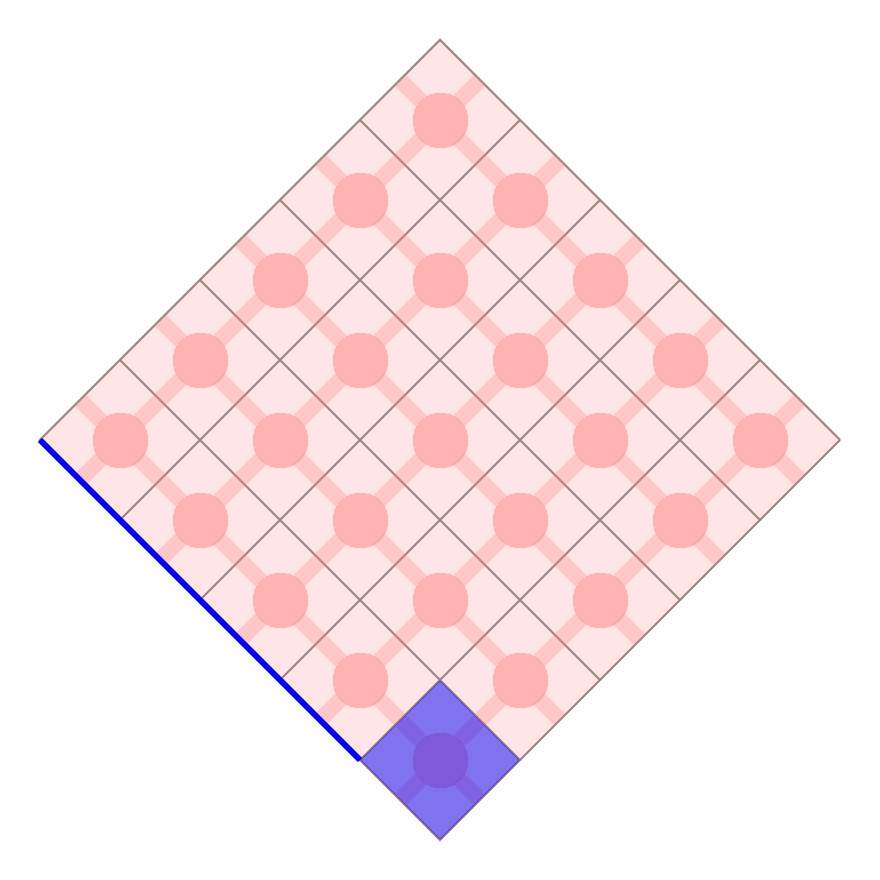}
\end{subfigure}
\hspace{-3mm}
$\cdot$
\hspace{-3mm}
\begin{subfigure}{0.15\textwidth}
\centering
 \includegraphics[width=1.00\textwidth]{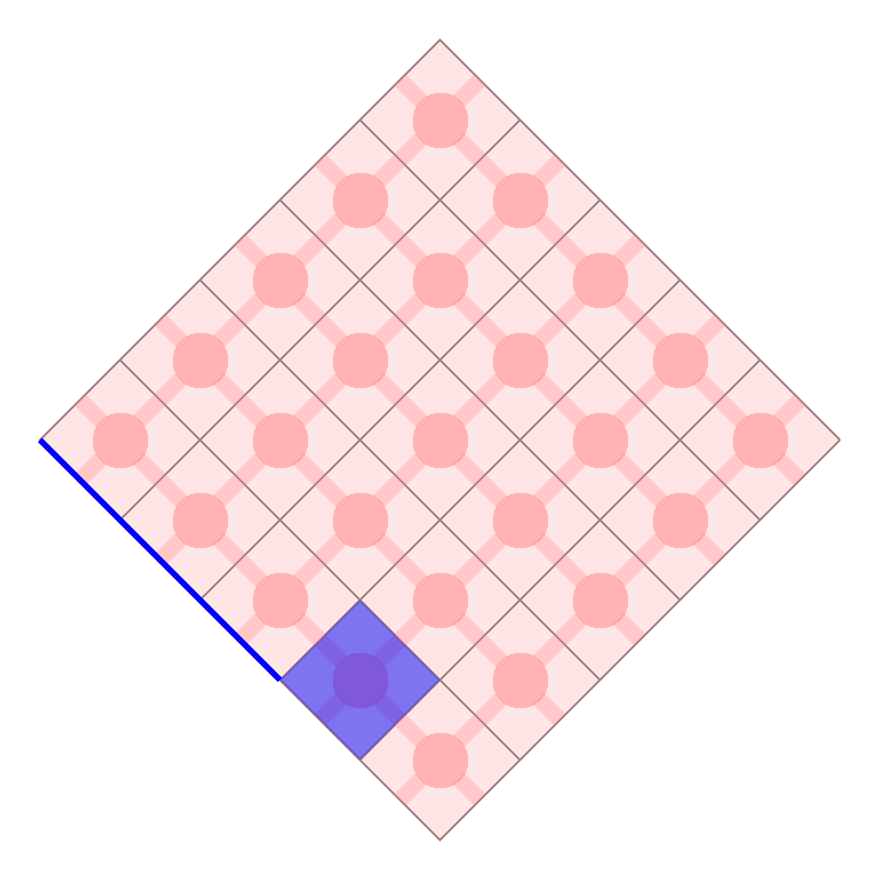}
\end{subfigure}
\hspace{-2mm}
$+$
\hspace{-2mm}
\begin{subfigure}{0.15\textwidth}
\centering
\includegraphics[width=1.00\textwidth]{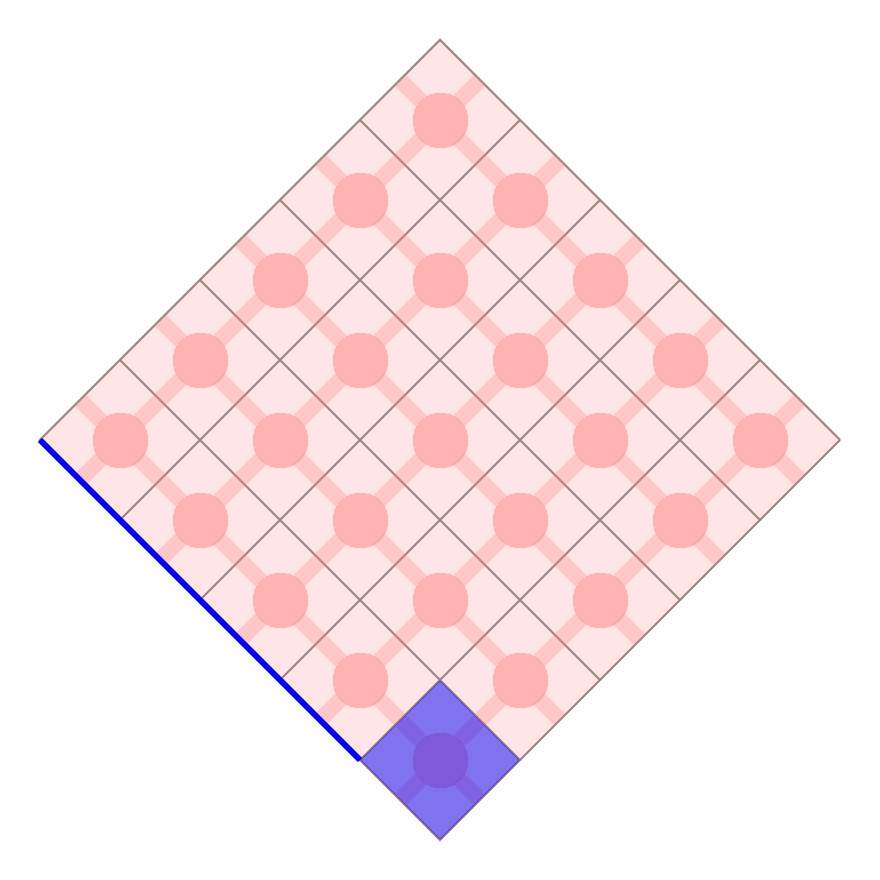}
\end{subfigure}
\hspace{-3mm}
$\cdot$
\hspace{-3mm}
\begin{subfigure}{0.15\textwidth}
\centering
 \includegraphics[width=1.00\textwidth]{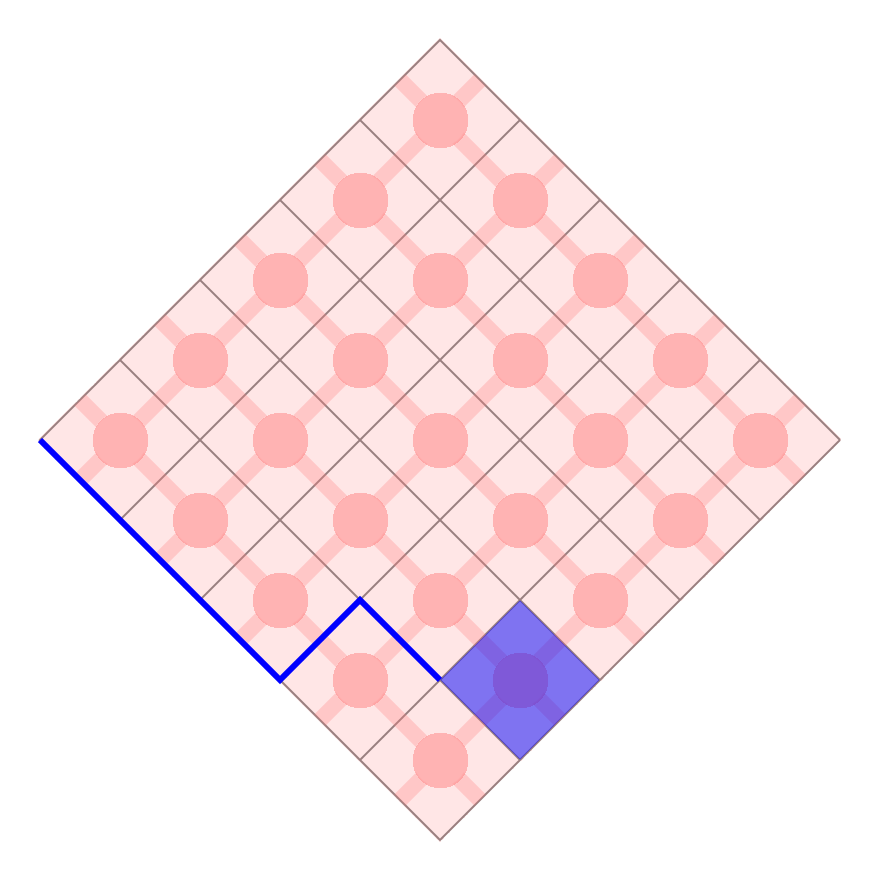}
\end{subfigure}
\hspace{-2mm}
$+$
\hspace{-2mm}
\begin{subfigure}{0.15\textwidth}
\centering
\includegraphics[width=1.00\textwidth]{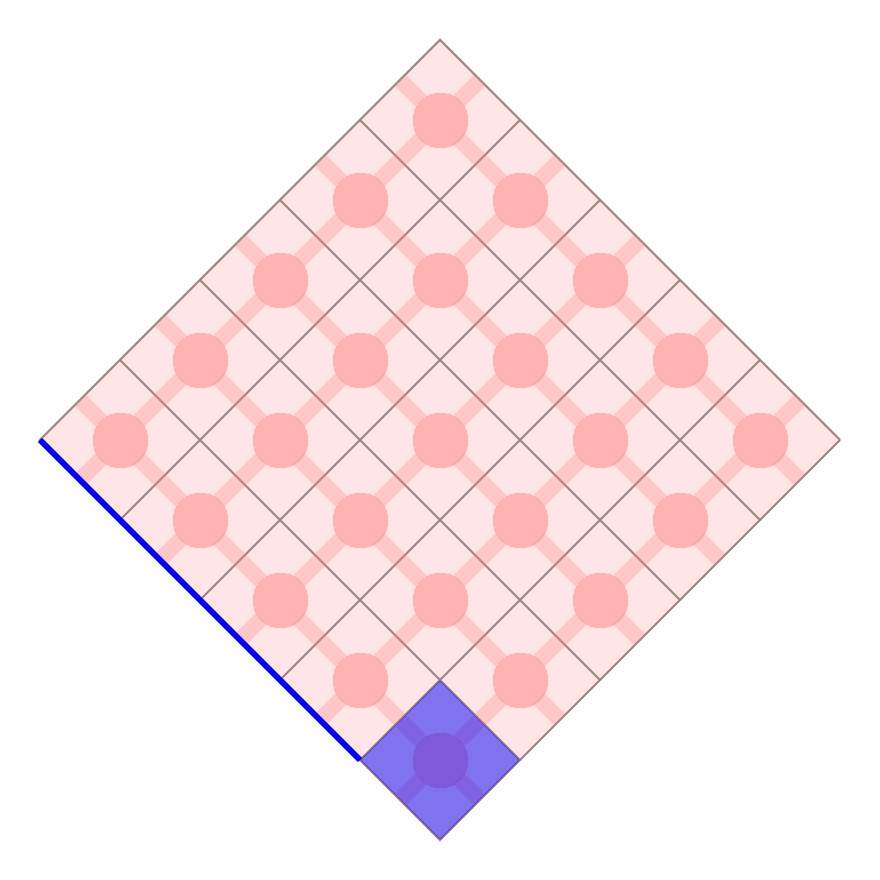}
\end{subfigure}
\hspace{-3mm}
$\cdot$
\hspace{-3mm}
\begin{subfigure}{0.15\textwidth}
\centering
\includegraphics[width=1.00\textwidth]{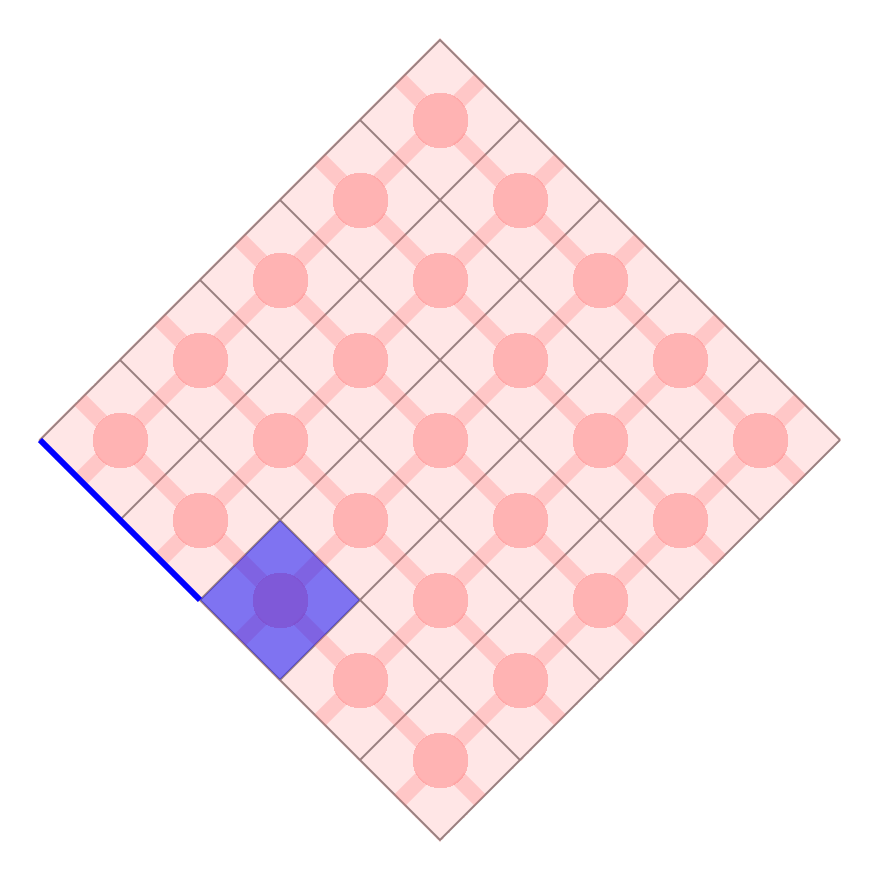}
\end{subfigure}
\end{figure} 
\vspace{-4mm}
\begin{figure}[H]
\centering
$+$
\hspace{-2mm}
\begin{subfigure}{0.15\textwidth}
\centering
    \includegraphics[width=1.00\textwidth]{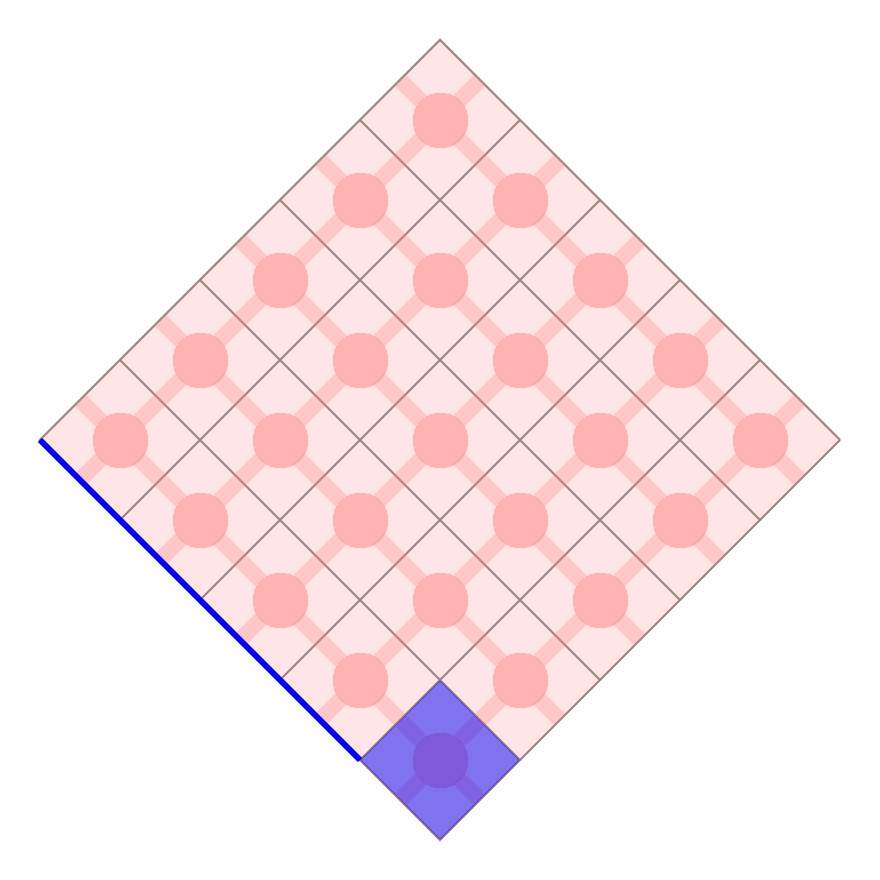}
\end{subfigure}
\hspace{-3mm}
$\cdot$
\hspace{-3mm}
\begin{subfigure}{0.15\textwidth}
\centering
 \includegraphics[width=1.00\textwidth]{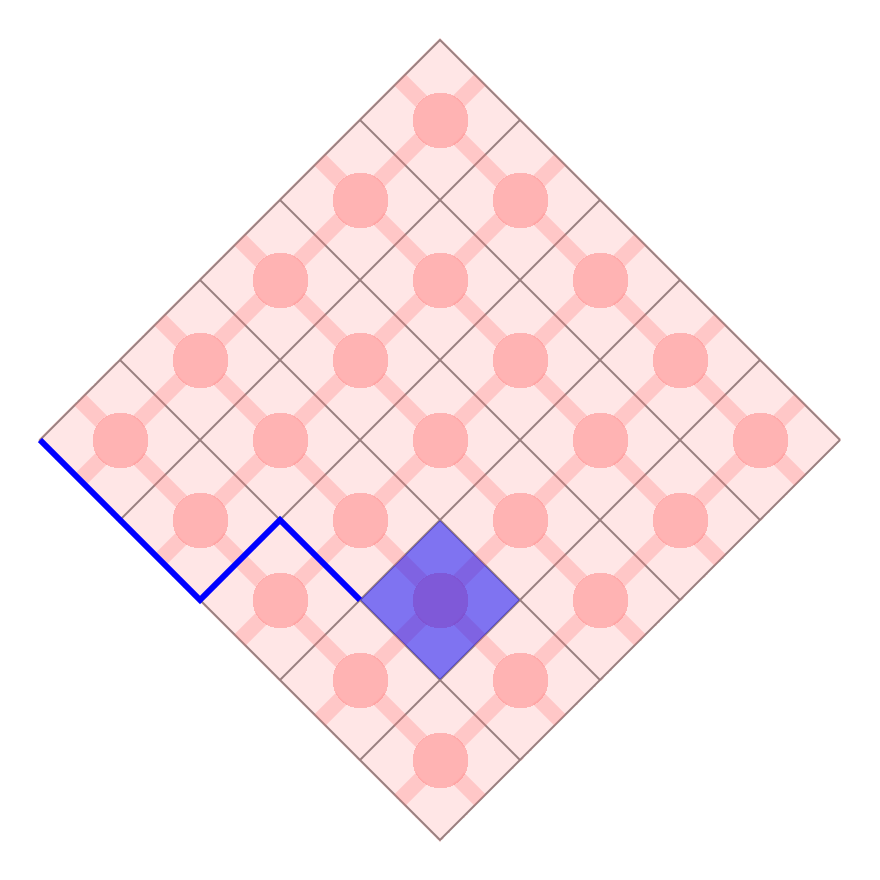}
\end{subfigure}
\hspace{-2mm}
$+$
\hspace{-2mm}
\begin{subfigure}{0.15\textwidth}
\centering
    \includegraphics[width=1.00\textwidth]{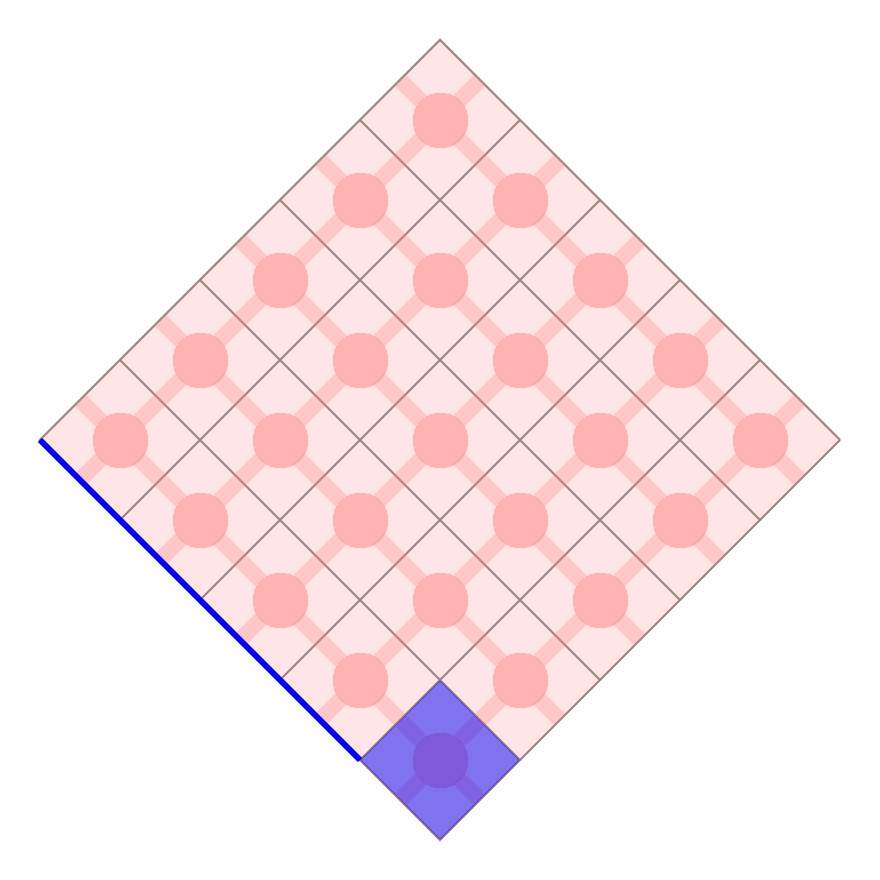}
\end{subfigure}
\hspace{-3mm}
$\cdot$
\hspace{-3mm}
\begin{subfigure}{0.15\textwidth}
\centering
    \includegraphics[width=1.00\textwidth]{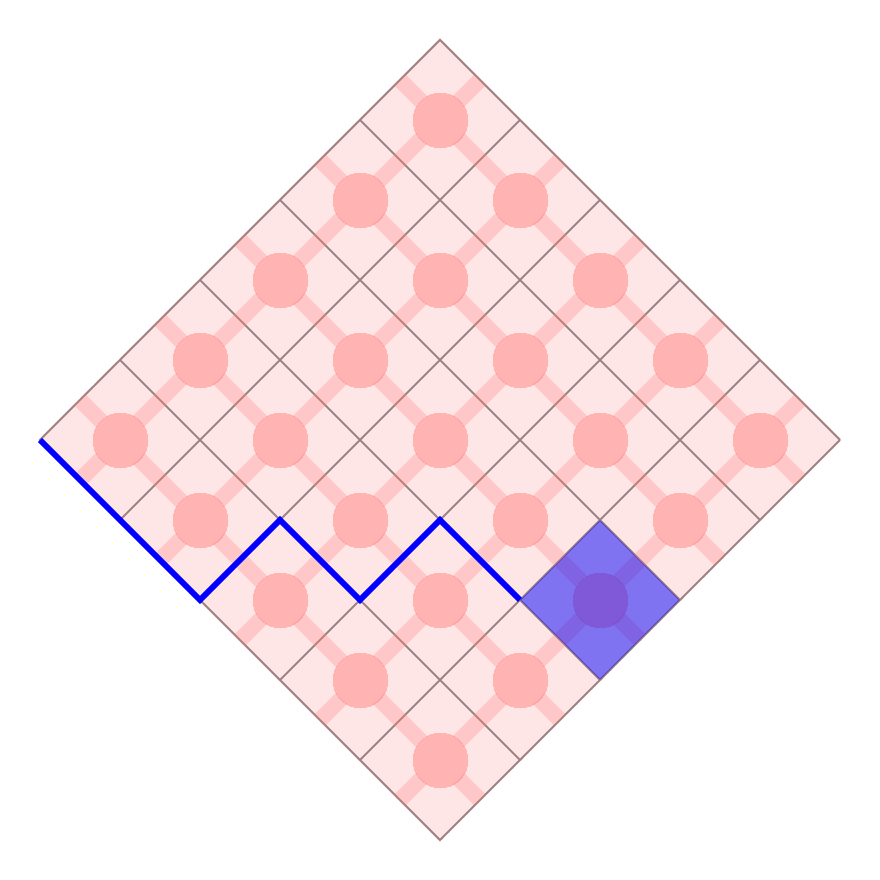}
\end{subfigure}
$+\cdots$
\end{figure} 
\vspace{-4mm}
\begin{figure}[H]
\centering
$+$
\hspace{-2mm}
\begin{subfigure}{0.15\textwidth}
\centering
    \includegraphics[width=1.00\textwidth]{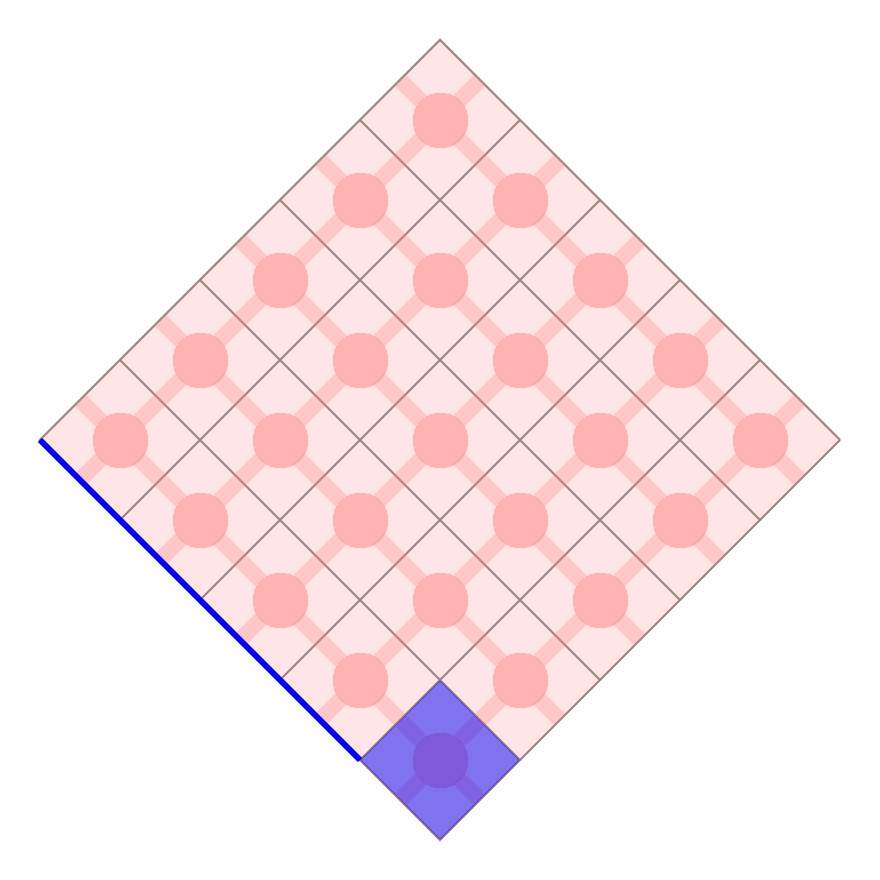}
\end{subfigure}
\hspace{-3mm}
$\cdot$
\hspace{-3mm}
\begin{subfigure}{0.15\textwidth}
\centering
 \includegraphics[width=1.00\textwidth]{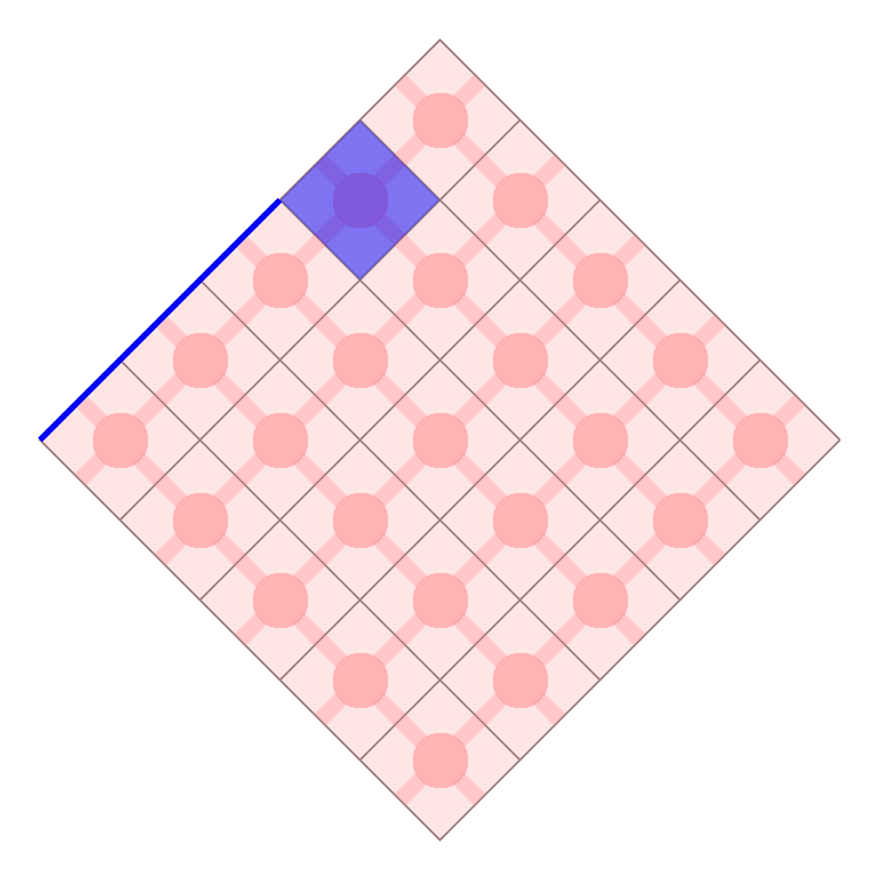}
\end{subfigure}
\hspace{-2mm}
$+$
\hspace{-2mm}
\begin{subfigure}{0.15\textwidth}
\centering
    \includegraphics[width=1.00\textwidth]{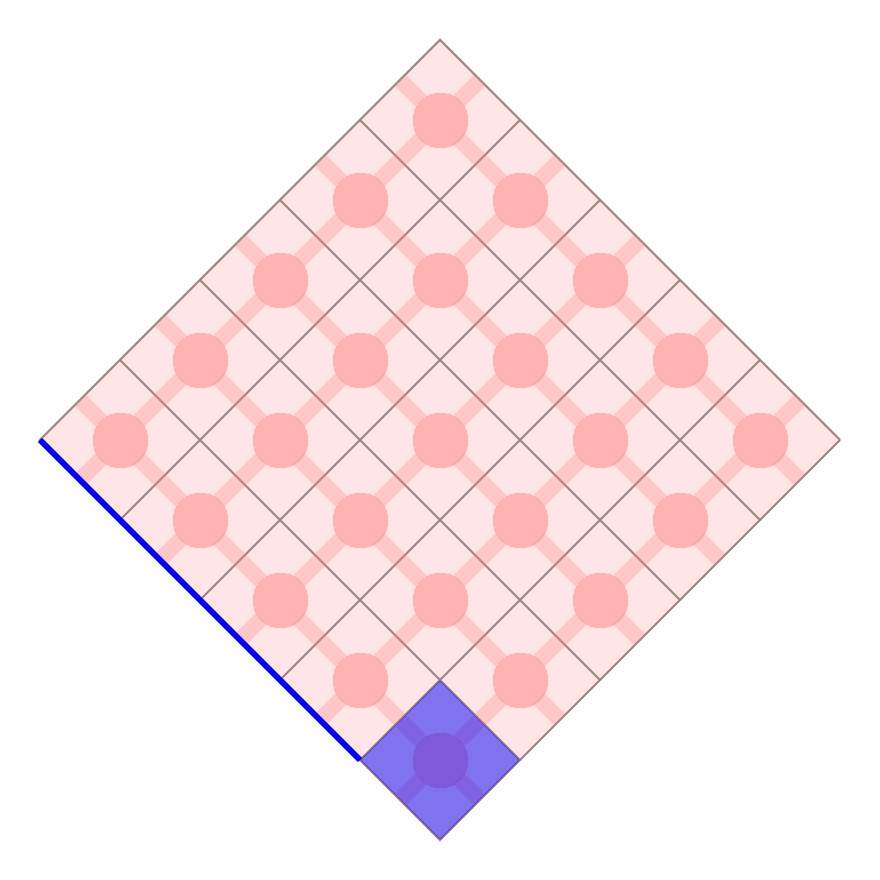}
\end{subfigure}
\hspace{-3mm}
$\cdot$
\hspace{-3mm}
\begin{subfigure}{0.15\textwidth}
\centering
 \includegraphics[width=1.00\textwidth]{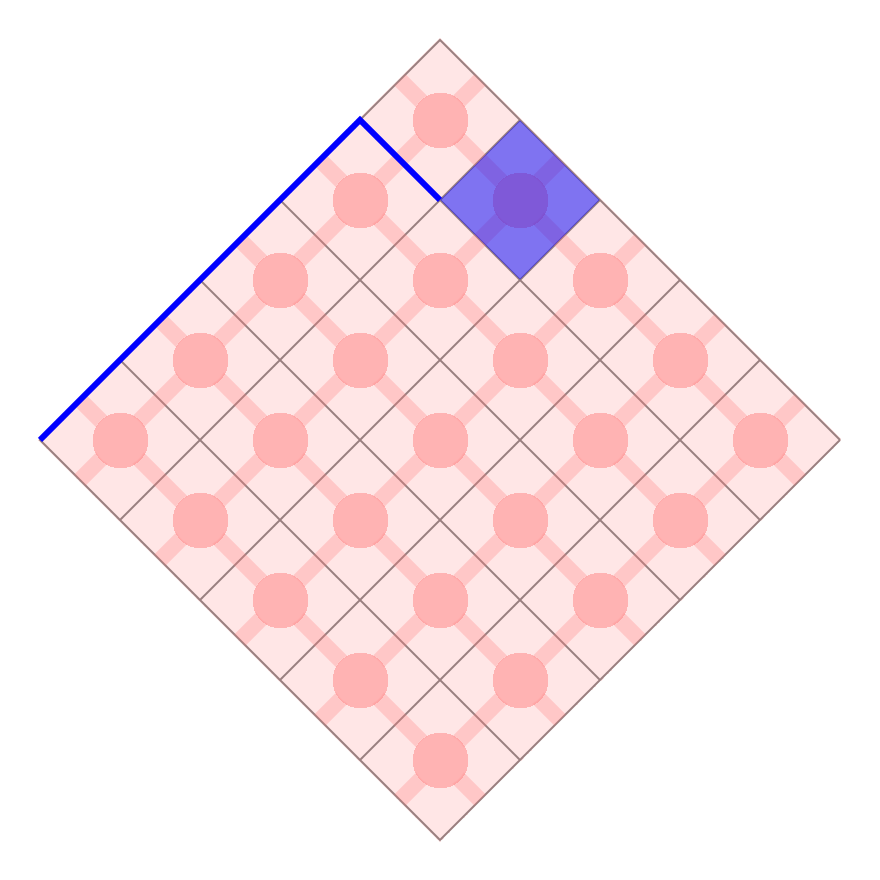}
\end{subfigure}
\hspace{-2mm}
$+$
\hspace{-2mm}
\begin{subfigure}{0.15\textwidth}
\centering
    \includegraphics[width=1.00\textwidth]{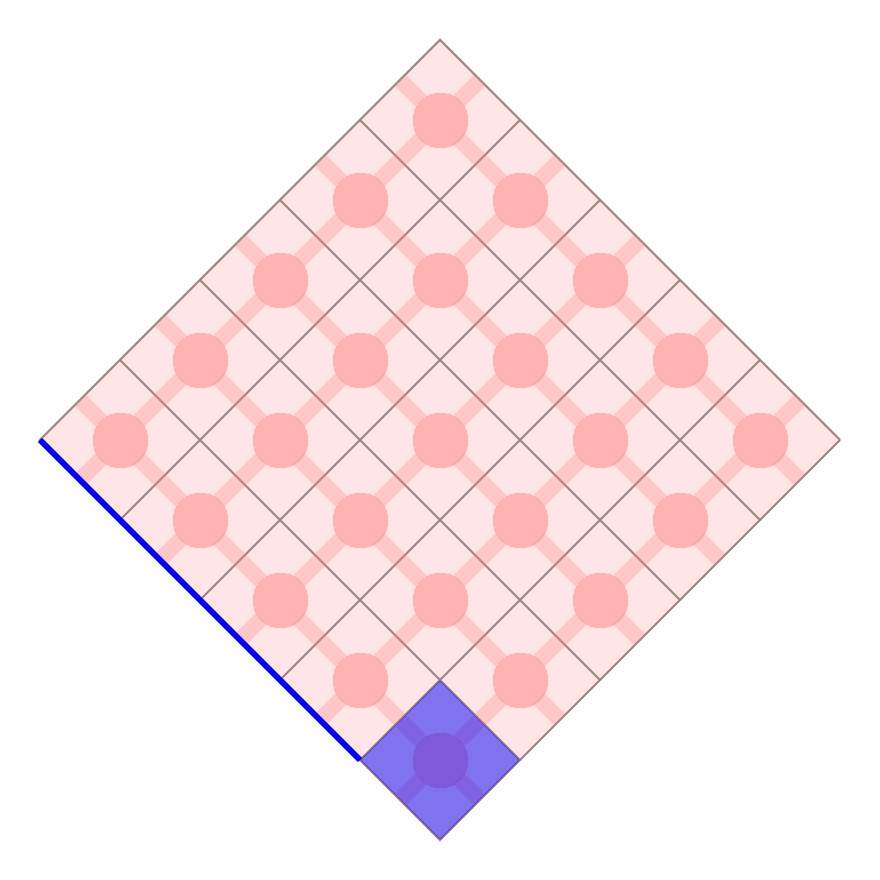}
\end{subfigure}
\hspace{-3mm}
$\cdot$
\hspace{-3mm}
\begin{subfigure}{0.15\textwidth}
\centering
 \includegraphics[width=1.00\textwidth]{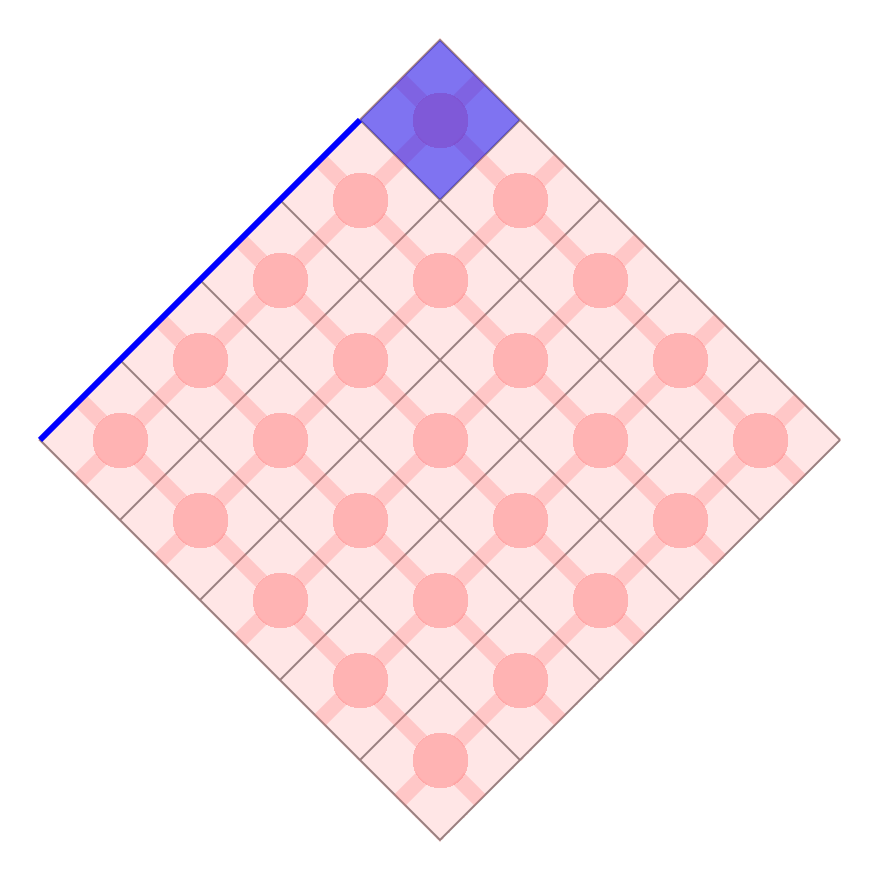}
\end{subfigure}
\end{figure} 
\vspace{-4mm}
\begin{figure}[H]
\centering
$+$
\hspace{-2mm}
\begin{subfigure}{0.15\textwidth}
\centering
    \includegraphics[width=1.00\textwidth]{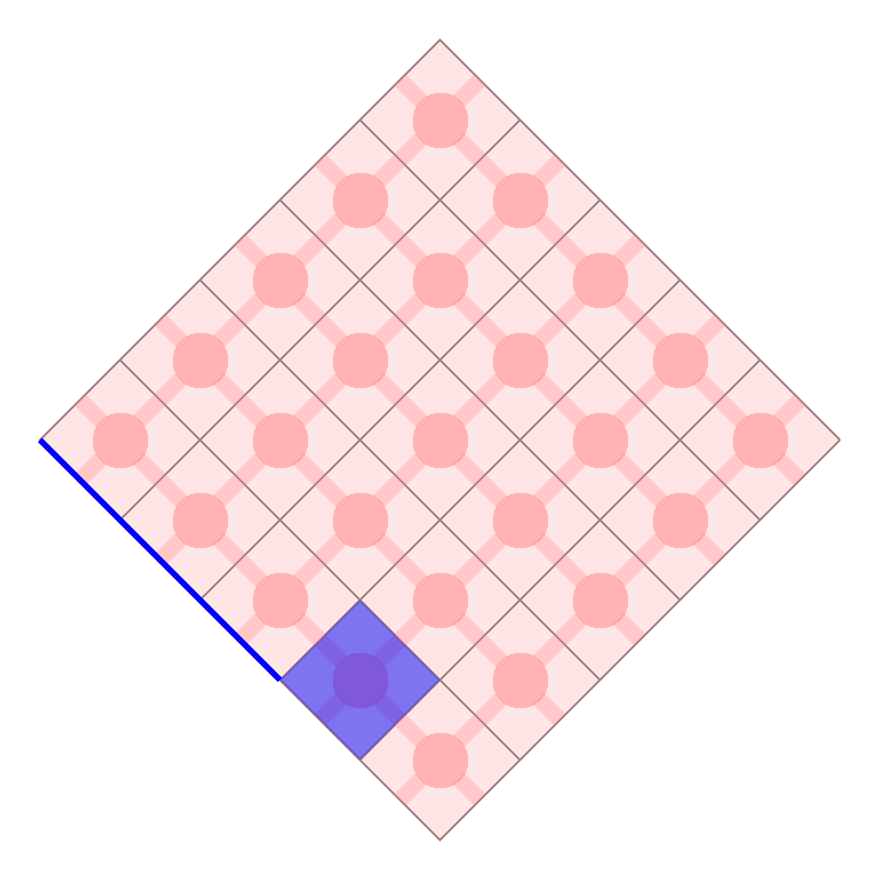}
\end{subfigure}
\hspace{-3mm}
$\cdot$
\hspace{-3mm}
\begin{subfigure}{0.15\textwidth}
\centering
 \includegraphics[width=1.00\textwidth]{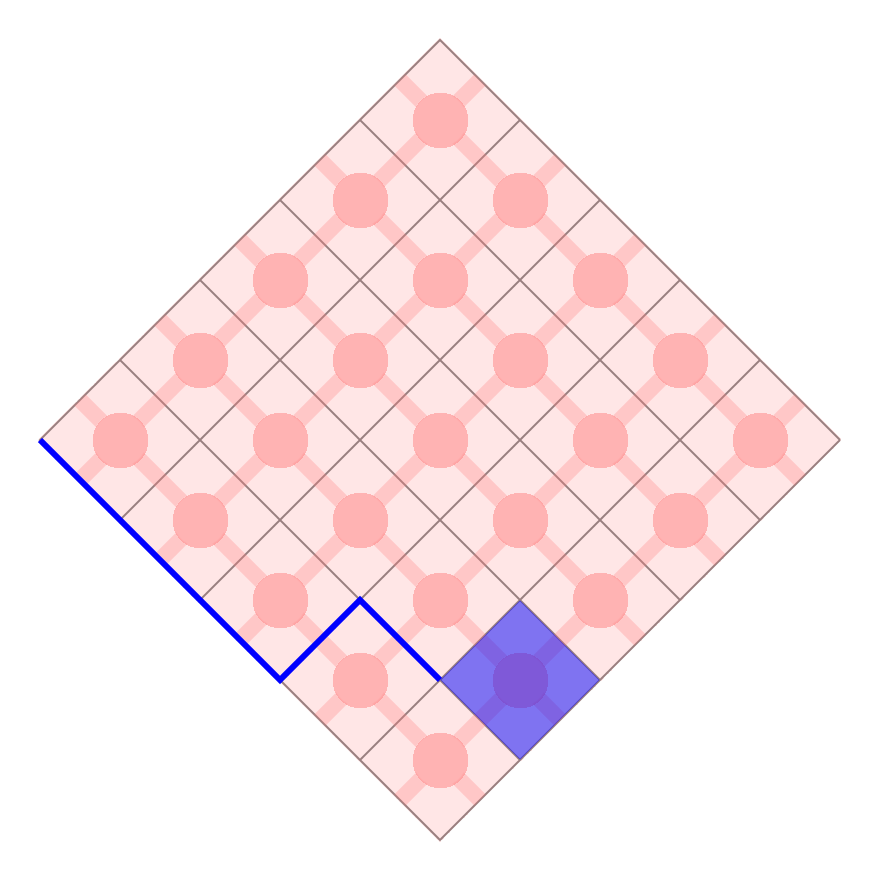}
\end{subfigure}
\hspace{-2mm}
$+$
\hspace{-2mm}
\begin{subfigure}{0.15\textwidth}
\centering
    \includegraphics[width=1.00\textwidth]{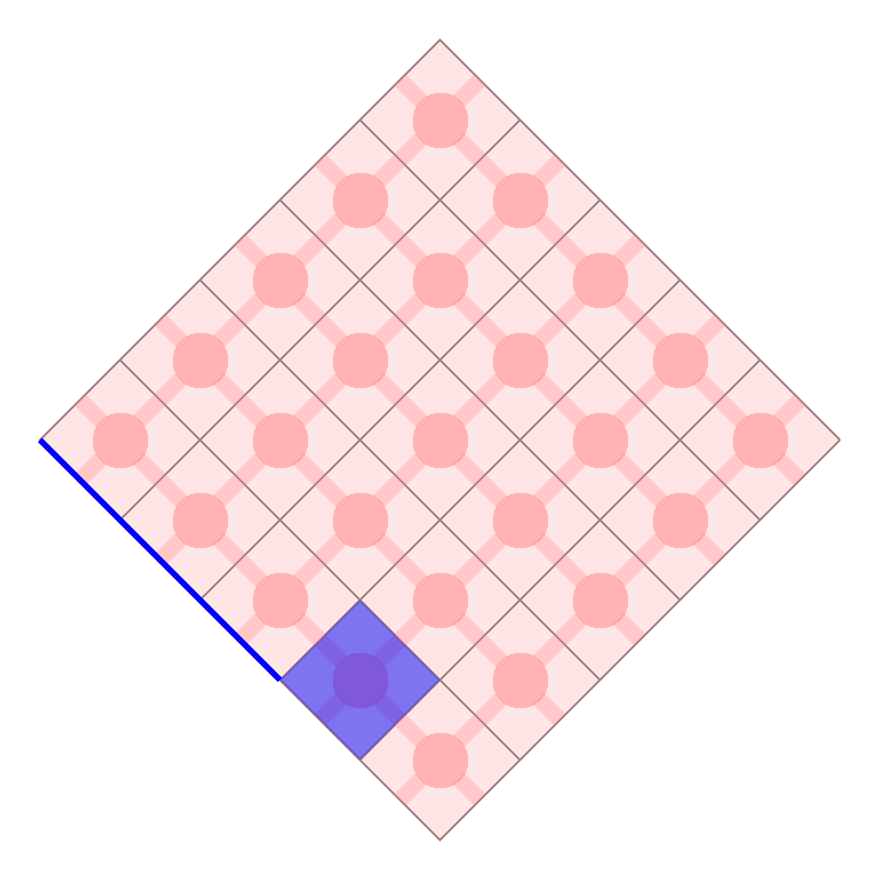}
\end{subfigure}
\hspace{-3mm}
$\cdot$
\hspace{-3mm}
\begin{subfigure}{0.15\textwidth}
\centering
 \includegraphics[width=1.00\textwidth]{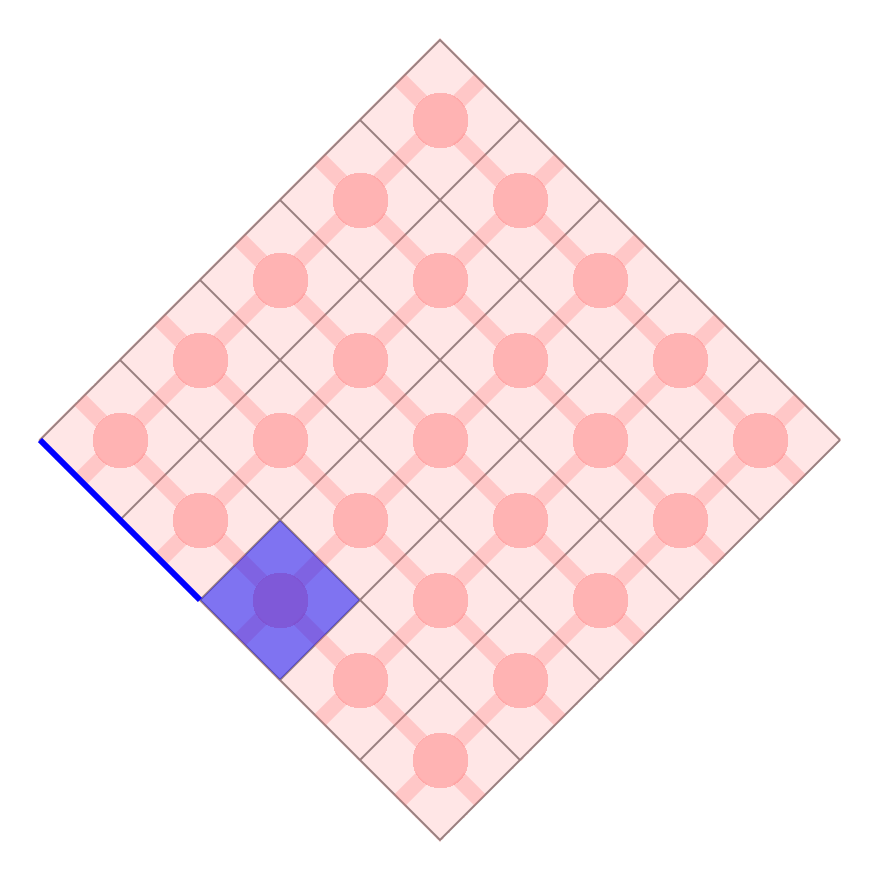}
\end{subfigure}
\hspace{-2mm}
$+$
\hspace{-2mm}
\begin{subfigure}{0.15\textwidth}
\centering
    \includegraphics[width=1.00\textwidth]{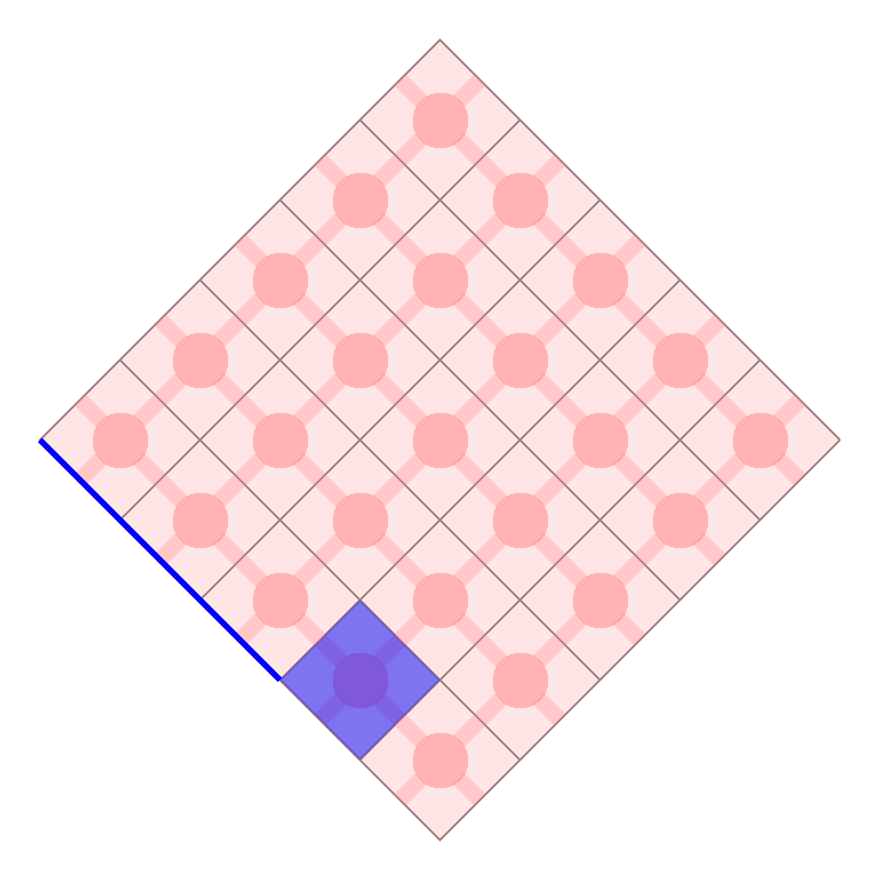}
\end{subfigure}
\hspace{-3mm}
$\cdot$
\hspace{-3mm}
\begin{subfigure}{0.15\textwidth}
\centering
 \includegraphics[width=1.00\textwidth]{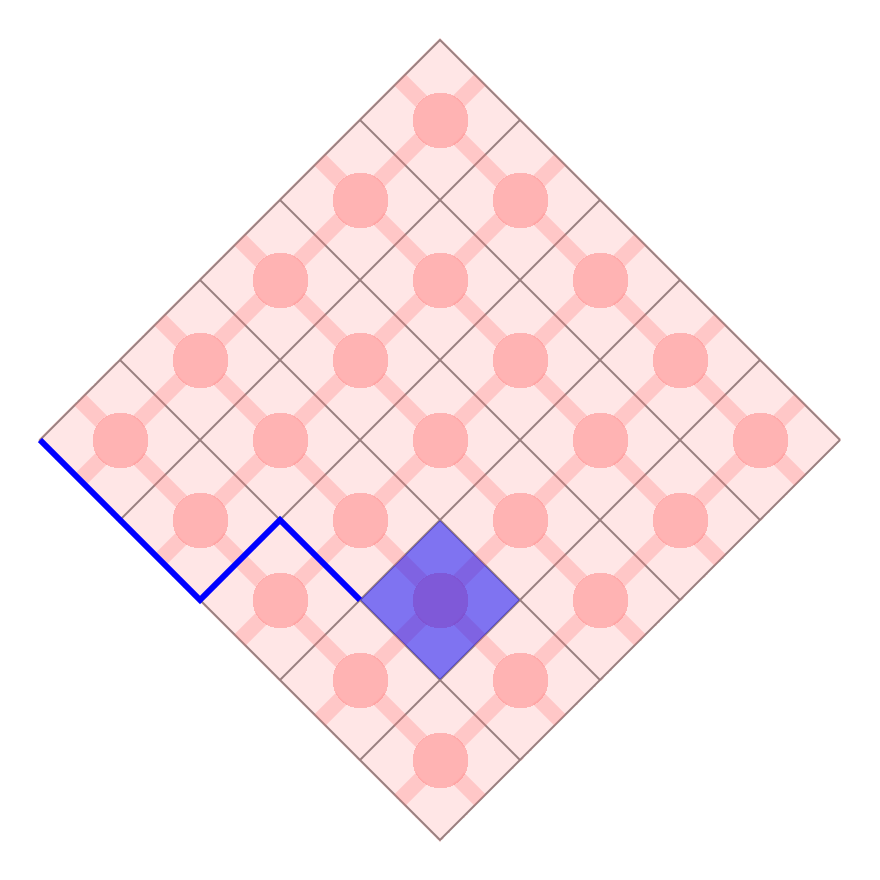}
\end{subfigure}
\end{figure} 
\vspace{-4mm}
\begin{figure}[H]
\centering
\hspace{-2mm}
$+$
\hspace{-3mm}
\begin{subfigure}{0.15\textwidth}
\centering
    \includegraphics[width=1.00\textwidth]{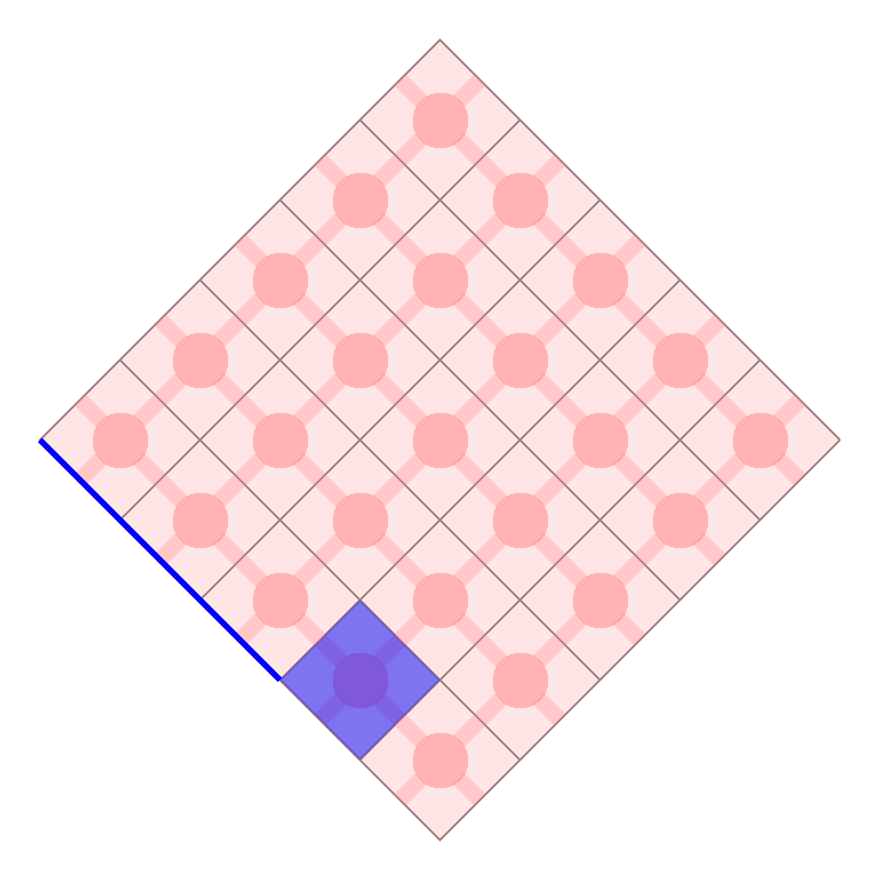}
\end{subfigure}
\hspace{-3mm}
$\cdot$
\hspace{-3mm}
\begin{subfigure}{0.15\textwidth}
\centering
 \includegraphics[width=1.00\textwidth]{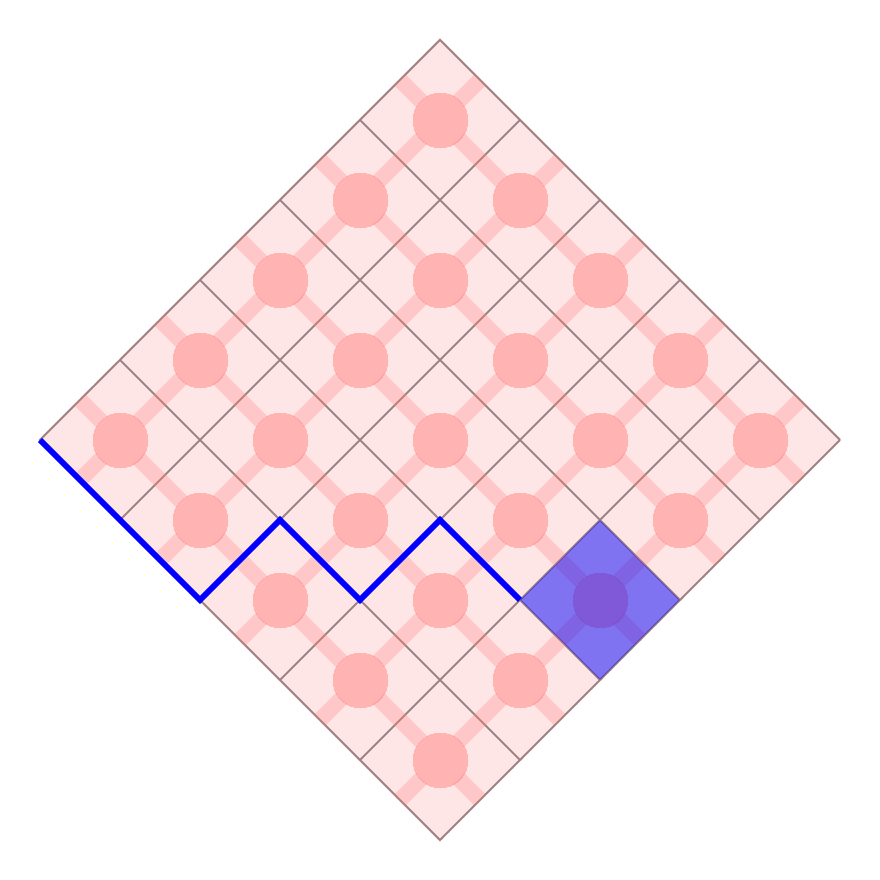}
\end{subfigure}
\hspace{-3mm}
$+\cdots+$
\hspace{-3mm}
\begin{subfigure}{0.15\textwidth}
\centering
    \includegraphics[width=1.00\textwidth]{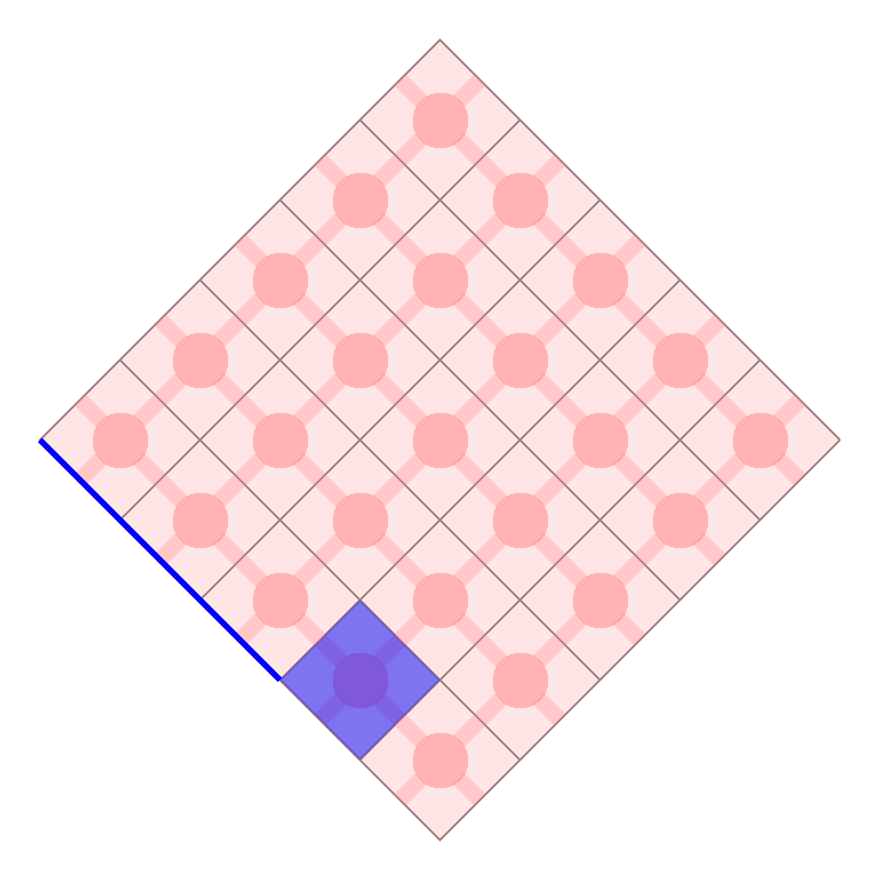}
\end{subfigure}
\hspace{-3mm}
$\cdot$
\hspace{-3mm}
\begin{subfigure}{0.15\textwidth}
\centering
 \includegraphics[width=1.00\textwidth]{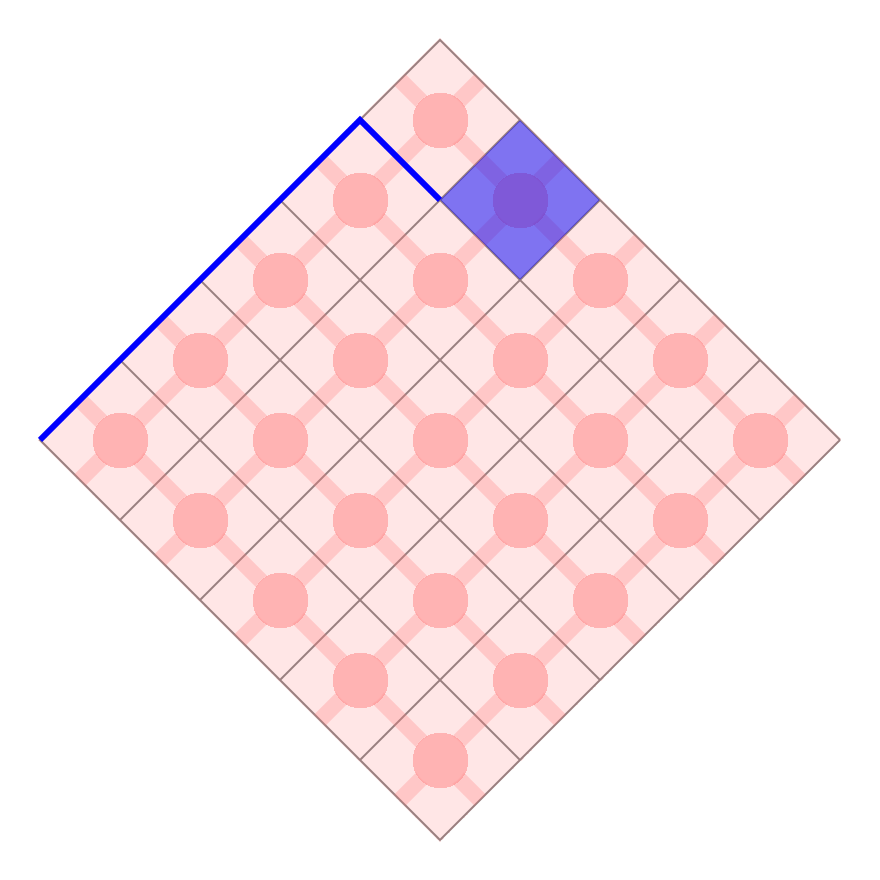}
\end{subfigure}
\hspace{-3mm}
$+$
\hspace{-3mm}
\begin{subfigure}{0.15\textwidth}
\centering
    \includegraphics[width=1.00\textwidth]{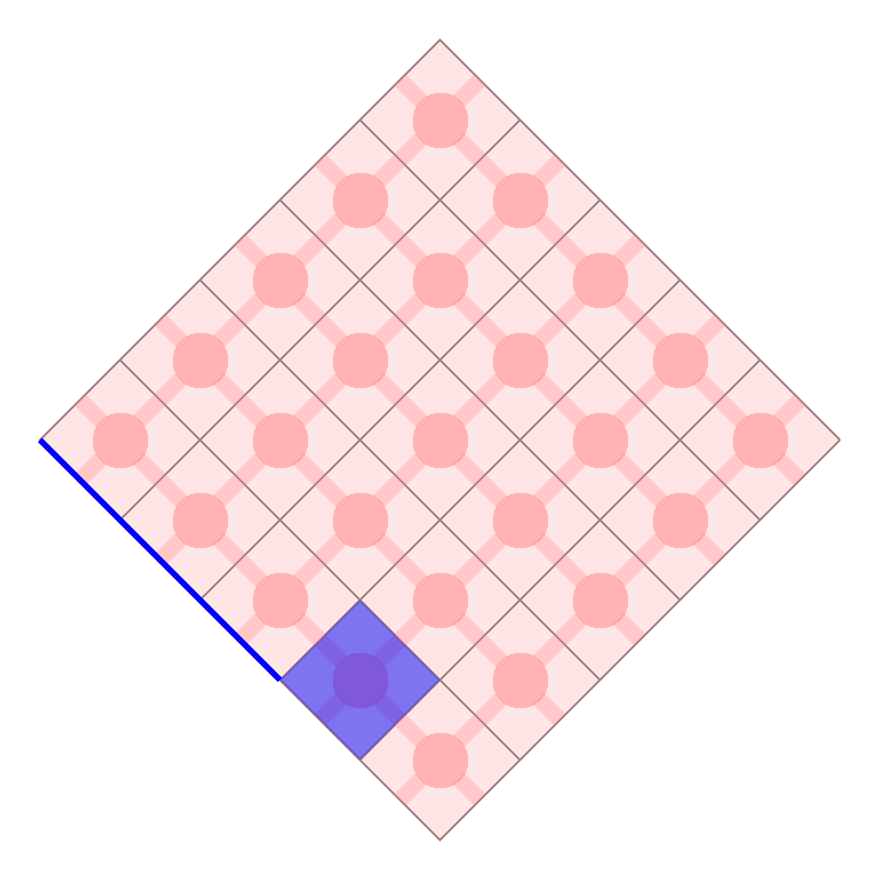}
\end{subfigure}
\hspace{-3mm}
$\cdot$
\hspace{-3mm}
\begin{subfigure}{0.15\textwidth}
\centering
 \includegraphics[width=1.00\textwidth]{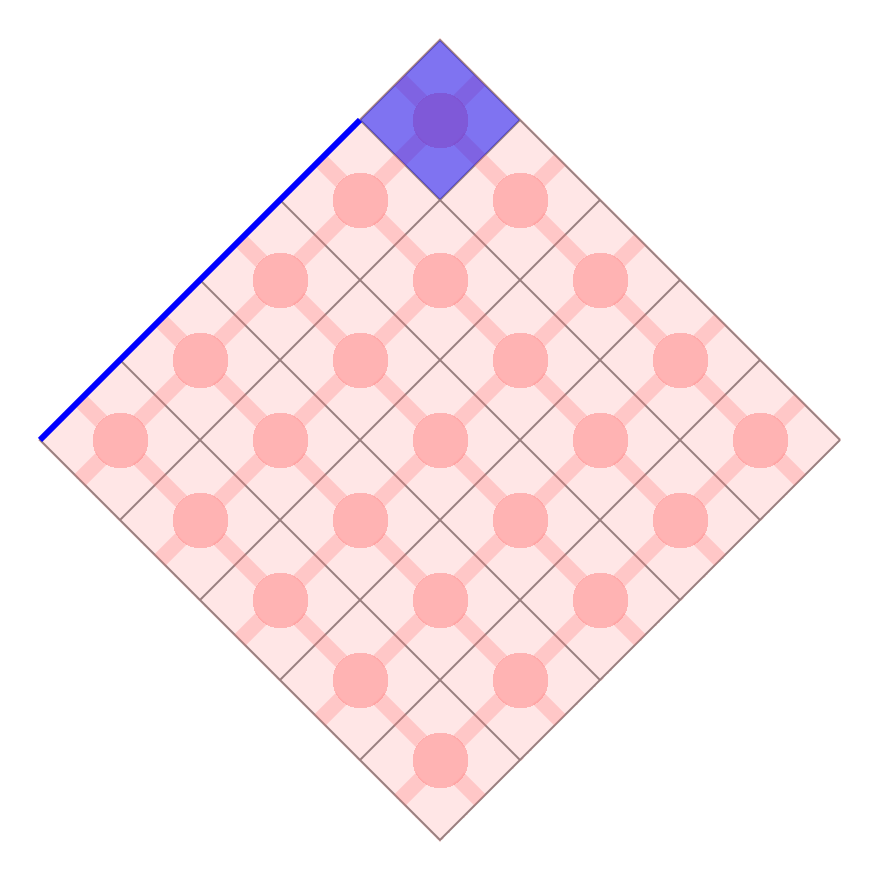}
\end{subfigure}
\hspace{-2mm}
\end{figure} 
\vspace{-4mm}
\begin{figure}[H]
\centering
\hspace{-2mm}
$+\cdots+$
\hspace{-2mm}
\begin{subfigure}{0.15\textwidth}
\centering
    \includegraphics[width=1.00\textwidth]{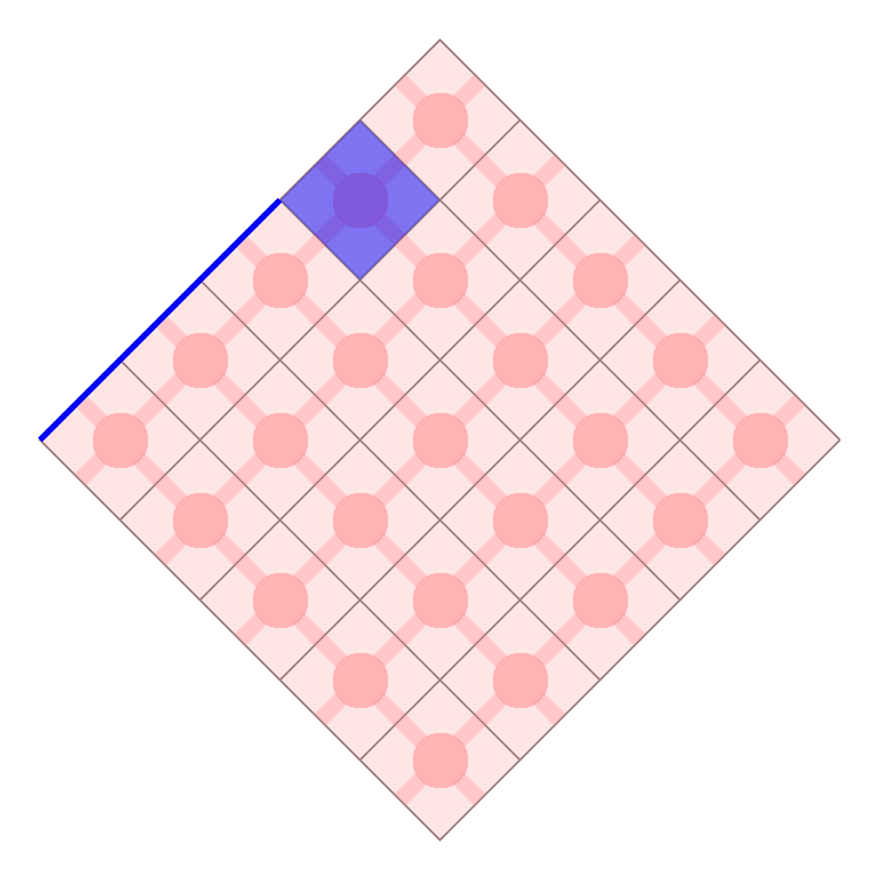}
\end{subfigure}
\hspace{-3mm}
$\cdot$
\hspace{-3mm}
\begin{subfigure}{0.15\textwidth}
\centering
 \includegraphics[width=1.00\textwidth]{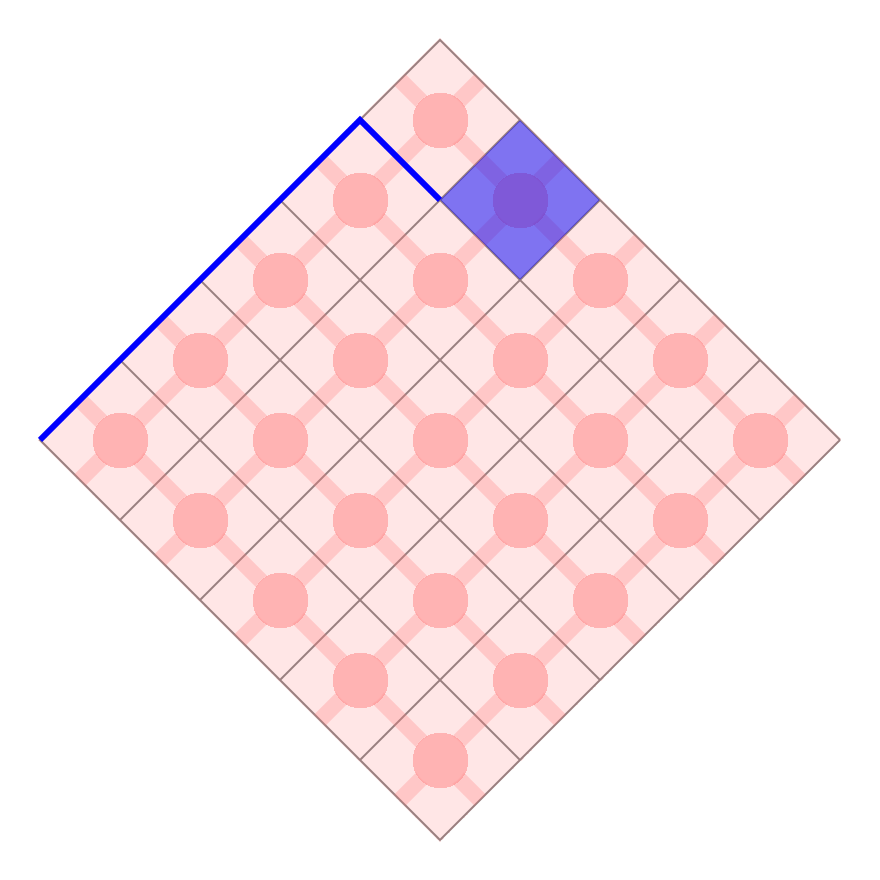}
\end{subfigure}
\hspace{-3mm}
$+$
\hspace{-3mm}
\begin{subfigure}{0.15\textwidth}
\centering
\includegraphics[width=1.00\textwidth]{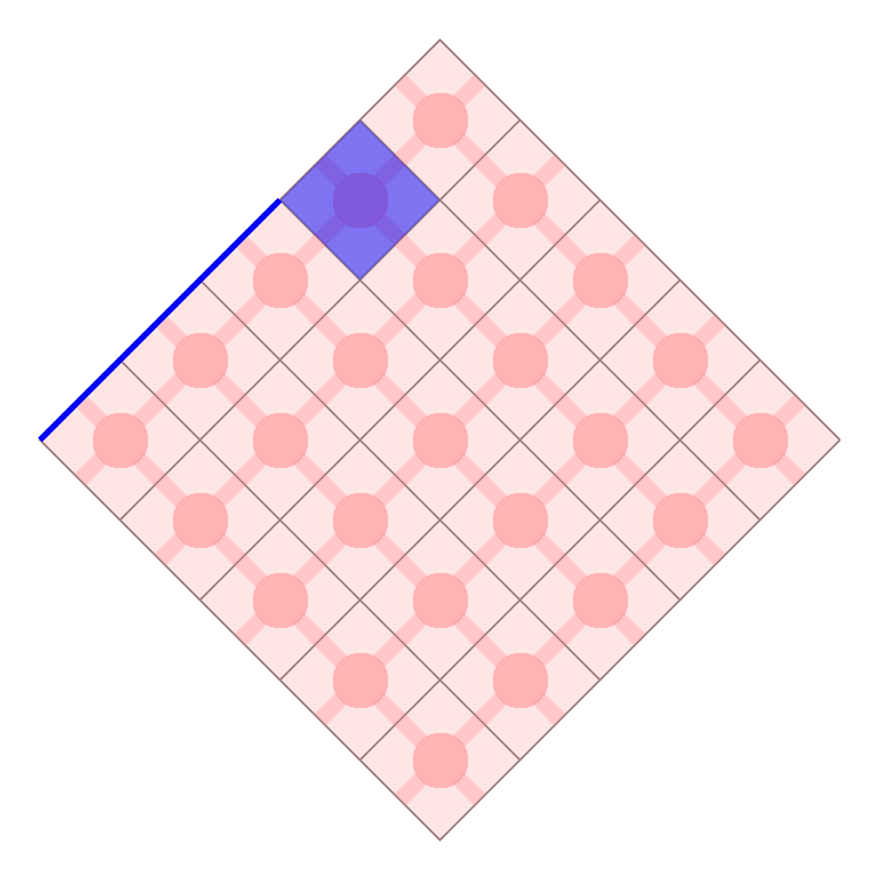}
\end{subfigure}
\hspace{-3mm}
$\cdot$
\hspace{-3mm}
\begin{subfigure}{0.15\textwidth}
\centering
 \includegraphics[width=1.00\textwidth]{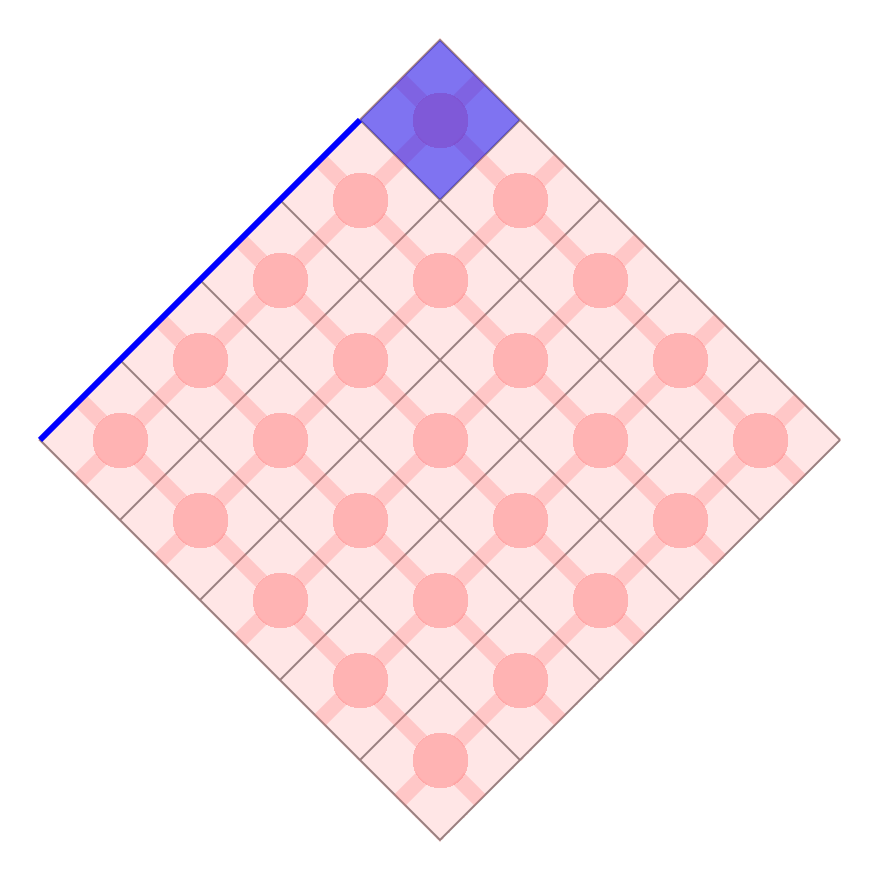}
\end{subfigure}
\hspace{-3mm}
$+$
\hspace{-3mm}
\begin{subfigure}{0.15\textwidth}
\centering
\includegraphics[width=1.00\textwidth]{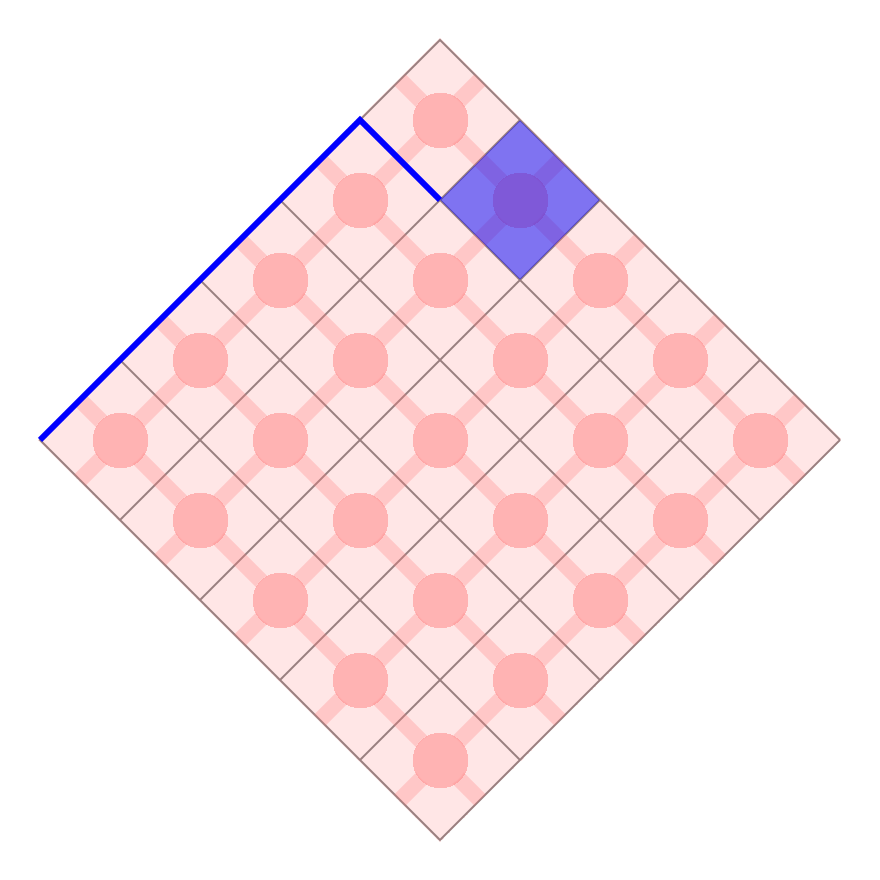}
\end{subfigure}
\hspace{-3mm}
$\cdot$
\hspace{-3mm}
\begin{subfigure}{0.15\textwidth}
\centering
 \includegraphics[width=1.00\textwidth]{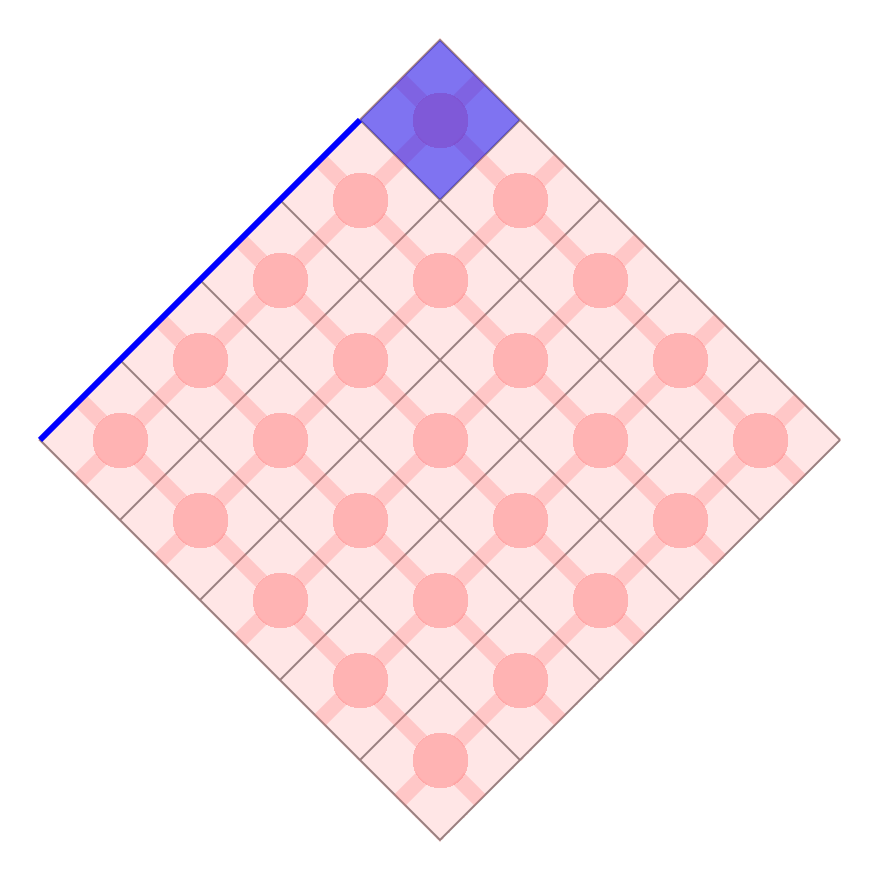}
\end{subfigure}
\hspace{-3mm}
\end{figure} 
\noindent
In this long expression,
there are $24$ terms
in the first 3 rows of pictures, 
$23$ in the two rows after that, 
up until we get $2+1$ shown in the last row.
This is consistent with the counting $\binom{25}{2}.$ 
Now, 
we should do this sum for all products of $B$'s 
ranging from terms with $0$ $B$'s to terms with $25$ $B$'s.
Just to be clear, for example, a term with 4 $B$'s
might look like
\begin{figure}[H]
\centering
\begin{subfigure}{0.15\textwidth}
\centering
\includegraphics[width=1.00\textwidth]{25}
\end{subfigure}
\hspace{-3mm}
$\cdot$
\hspace{-3mm}
\begin{subfigure}{0.15\textwidth}
\centering
 \includegraphics[width=1.00\textwidth]{14}
\end{subfigure}
\hspace{-3mm}
$\cdot$
\hspace{-3mm}
\begin{subfigure}{0.15\textwidth}
\centering
\includegraphics[width=1.00\textwidth]{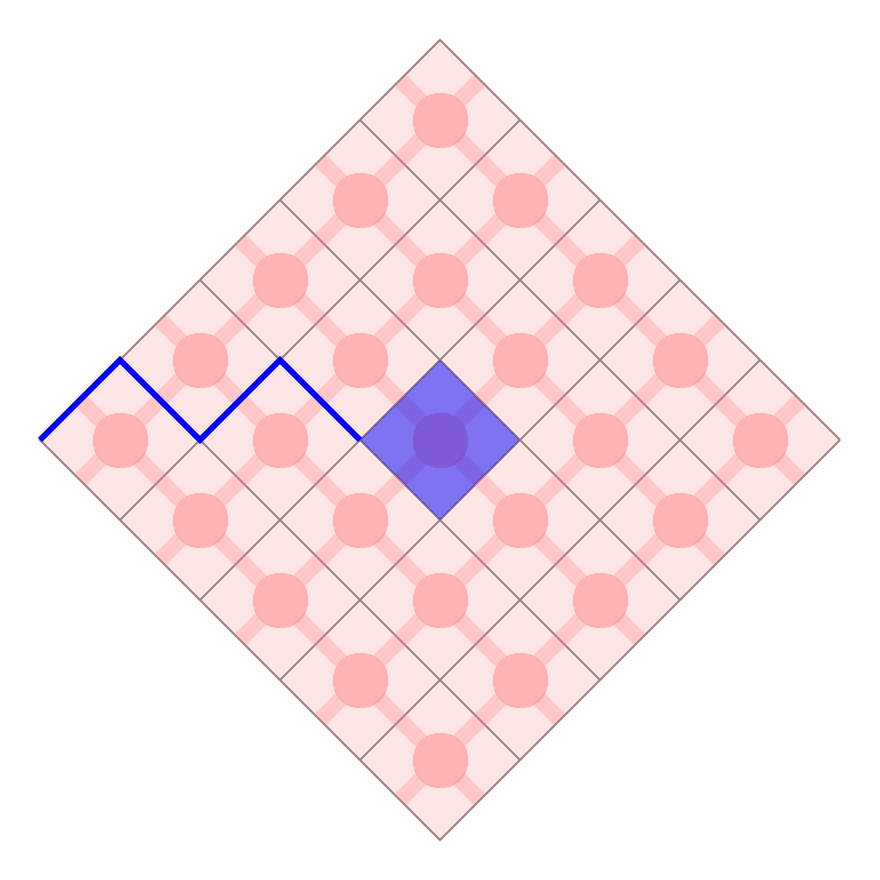}
\end{subfigure}
\hspace{-3mm}
$\cdot$
\hspace{-3mm}
\begin{subfigure}{0.15\textwidth}
\centering
\includegraphics[width=1.00\textwidth]{52}
\end{subfigure}
\hspace{-3mm}
\end{figure} 
\noindent
but a term such as 
\begin{figure}[H]
\centering
\begin{subfigure}{0.15\textwidth}
\centering
\includegraphics[width=1.00\textwidth]{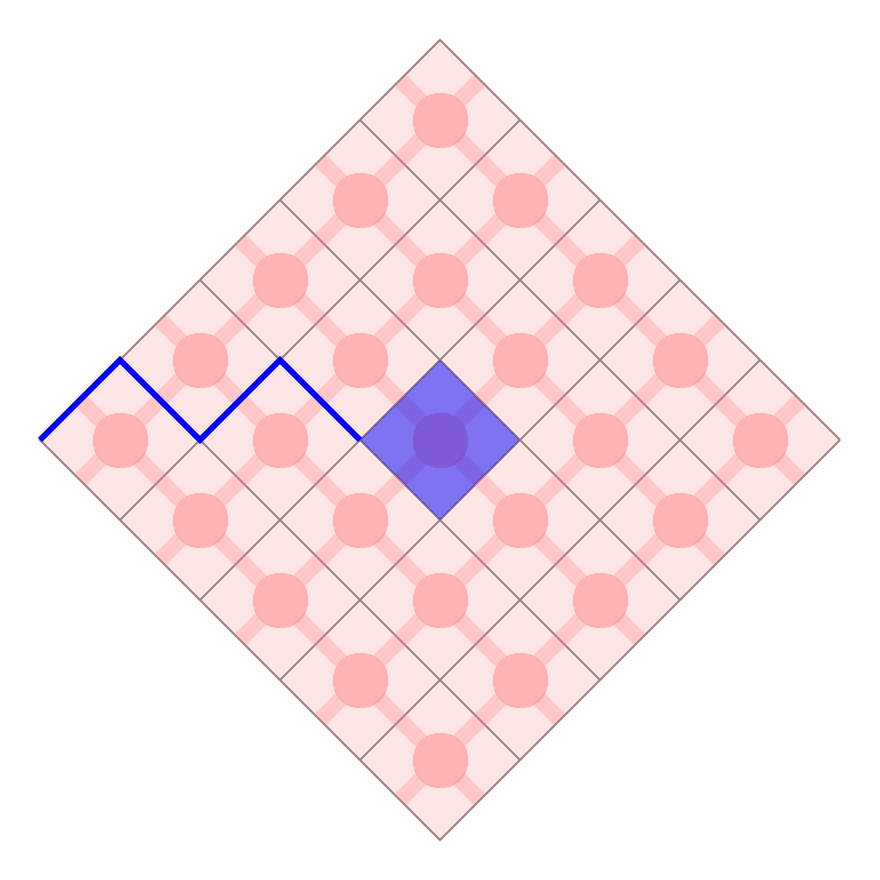}
\end{subfigure}
\hspace{-3mm}
$\cdot$
\hspace{-3mm}
\begin{subfigure}{0.15\textwidth}
\centering
 \includegraphics[width=1.00\textwidth]{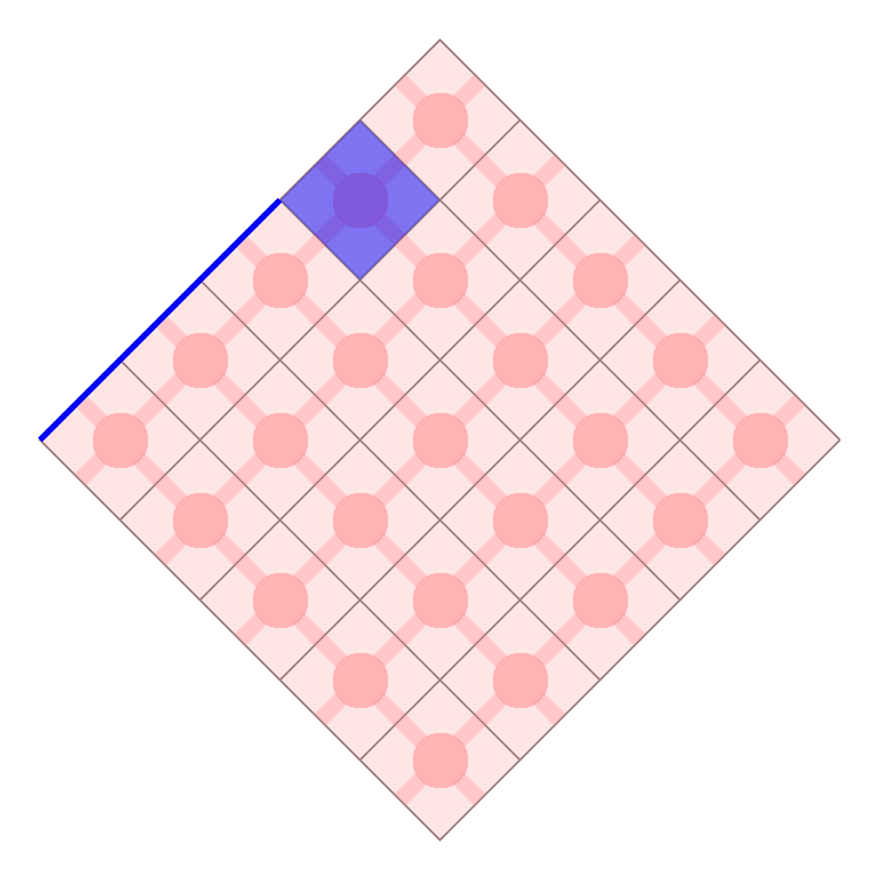}
\end{subfigure}
\hspace{-3mm}
$\cdot$
\hspace{-3mm}
\begin{subfigure}{0.15\textwidth}
\centering
\includegraphics[width=1.00\textwidth]{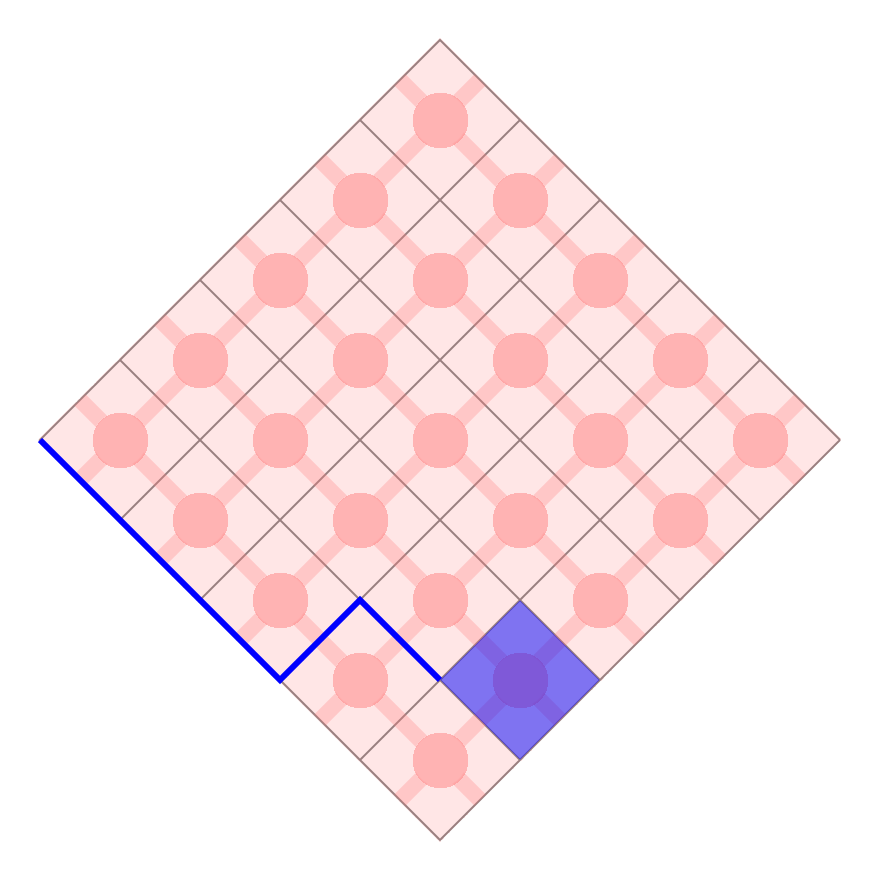}
\end{subfigure}
\hspace{-3mm}
$\cdot$
\hspace{-3mm}
\begin{subfigure}{0.15\textwidth}
\centering
\includegraphics[width=1.00\textwidth]{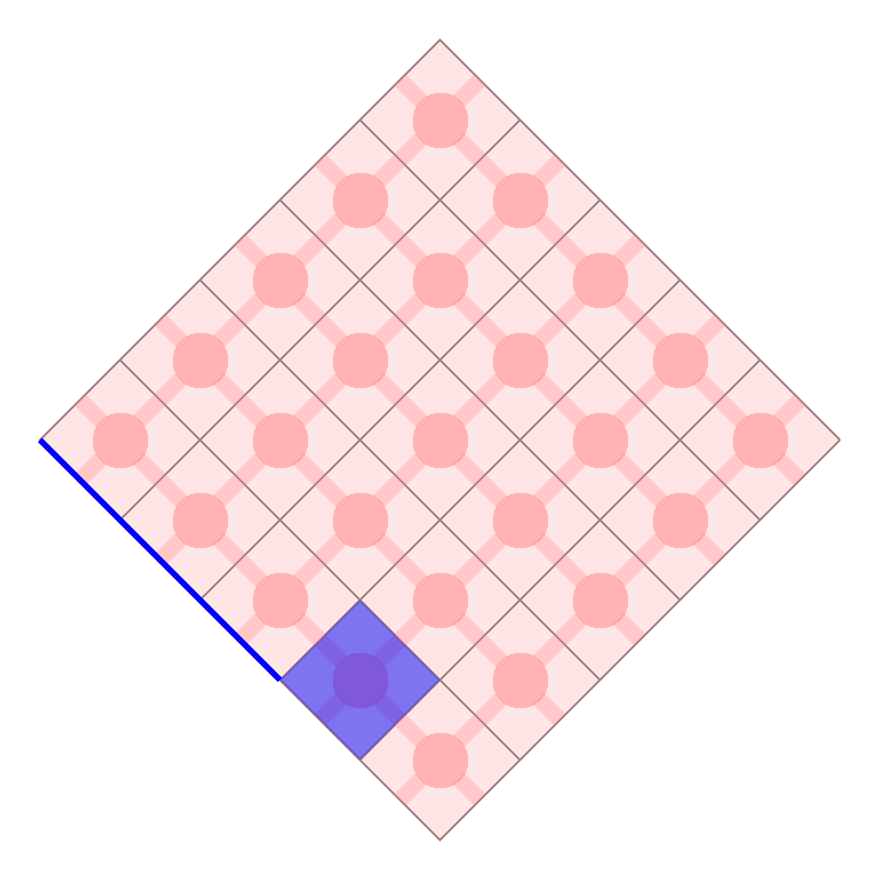}
\end{subfigure}
\hspace{-3mm}
\end{figure} 
\noindent
does \emph{not} appear in the expression  (\ref{eq:surfaceproductn=5inoneline})
 due to the automatic ordering. 
To be clear, this ordering is given as follows
\begin{figure}[H]
\centering
\includegraphics[width=0.30\textwidth]{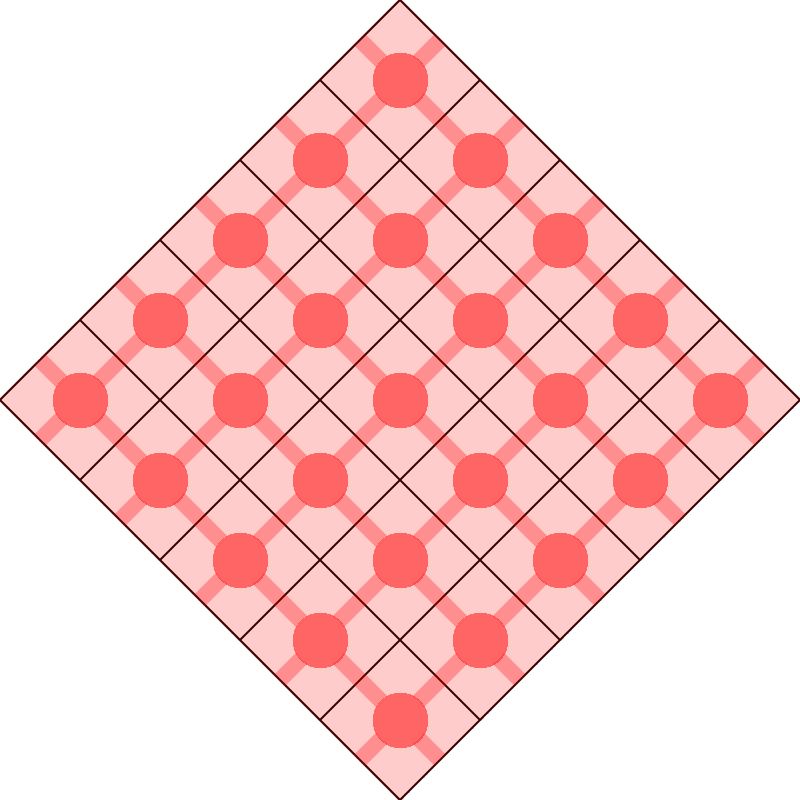}
\begin{picture}(0,0)
\put(-80,129){1}
\put(-66,114){2}
\put(-95,114){3}
\put(-51,99){4}
\put(-80,99){5}
\put(-109,99){6}
\put(-36,84){7}
\put(-66,84){8}
\put(-95,84){9}
\put(-128,84){10}
\put(-25,70){11}
\put(-54,70){12}
\put(-84,70){13}
\put(-113,70){14}
\put(-142,70){15}
\put(-40,54){16}
\put(-69,54){17}
\put(-98,54){18}
\put(-127,54){19}
\put(-54,40){20}
\put(-83,40){21}
\put(-113,40){22}
\put(-69,25){23}
\put(-98,25){24}
\put(-83,10){25}
\end{picture}
\end{figure}  
\noindent
where the earlier terms begin at $1$ and appear from right to left when
expressed algebraically using group multiplication. 
The total number of all terms in such an expansion is enormous 
and is given by 
\be
\label{eq:totalterms5by5}
\sum_{k=0}^{25}\binom{25}{k}=2^{25},
\ee
or more generally
\be
\sum_{k=0}^{n^{2}}\binom{n^{2}}{k}=2^{n^{2}}
\ee
if we have an $n\times n$ grid.
There are many things to check to make sense of the $n\to\infty$ limit. 
First, we need to argue why the product from  
(\ref{eq:surfaceproductn=5inoneline}) converges.
\bn
\label{prop:trivnconverges}
Let $\triv_{n}$ be the 
generalization of the expression given in 
(\ref{eq:squaretransportforn=5})
for an $n\times n$ grid decomposition. 
Then the sequence 
$\big\{\triv_{n}\big\}_{n\in\N}$
converges as $n\to\infty.$
\en
\bprf
To prove this, one can introduce $\triv_{P}$ for any 
partition $P$ of the unit square. Then one needs to show that
this quantity has a well-defined limit over all partitions 
ordered by refinement. 
This argument has been deferred to Appendix \ref{app:spc}. 
\eprf
Secondly, as a result of expanding out these products 
into sums of different orders, 
we should be sure that the sum of all such terms
coming from (\ref{eq:surfaceproductn=5inoneline})
for any given order converges
as the spacing goes to zero, i.e. as $n\to \infty.$ 
The terms with $k$ $B$'s have an additional factor of
$\frac{1}{n^{2k}}$ associated with the area elements
on which they are approximated.%
\footnote{We are ignoring the factors coming from 
the $a_{ij}^{s}$ and $a_{ij}^{t}$ terms because we can see from these
pictures that in the $n\to\infty$ limit these terms describe the 
parallel transport along a path as was discussed in Section 
\ref{sec:2dalgebra1dgt}. This is discussed in more detail in 
Appendix \ref{app:spc}. 
}
The ratio of the number
of \emph{all} such terms
for $k\le\left\lfloor\frac{n^2}{2}\right\rfloor$
to this factor is
\be
\label{eq:ratioktermstoallterms}
\cfrac{\binom{n^2}{k}}{n^{2k}}
=
\frac{n^2!}{k!(n^2-k)!n^{2k}}
=
\frac{1}{k!}\prod_{i=1}^{k}\left(1-\frac{i-1}{n^2}\right),
\ee
where $\left\lfloor\ \cdot \ \right\rfloor$
denotes the floor function. 
Note that the product term satisfies
\be
0\le\prod_{i=1}^{k}\left(1-\frac{i-1}{n^2}\right)\le1
\ee
because it is a product of numbers strictly
less than or equal to $1$ for all $i.$ Hence, 
\be
\cfrac{\binom{n^2}{k}}{n^{2k}}\le
\frac{1}{k!}.
\ee
For $k\ge\left\lfloor\frac{n^2}{2}\right\rfloor,$
this decays even more strongly because 
$\binom{n^2}{k}$ is symmetric at 
$\left\lfloor\frac{n^2}{2}\right\rfloor$ and hence
$\binom{n^2}{k}$
begins to decrease for larger values of $k$
while the $\frac{1}{n^{2k}}$ factor
remains and increases as $k$ gets larger. 
\bn
\label{prop:trivnkbounded}
For each $k\in\N$ and $n\ge k,$ let $\triv_{n,k}$ denote the $k$-th order terms 
obtained from expanding out $\triv_{n}$ to lowest order
(see Proposition \ref{prop:trivnconverges}). 
First, for each $n\in\N,$ there exists a positive real number $M_{n}>0$ 
such that 
\be
\lVert\triv_{n,k}\rVert\le\frac{M_{n}^k}{k!}
\ee
for all $k\in\{0,1,\dots,n^2\}.$
Second, for each $k\in\N,$ 
there exists an $N\in\N$ 
and a positive real number $M>0$ such that 
\be
\lVert\triv_{n,k}\rVert\le\frac{M^k}{k!}
\ee
for all $n\ge N.$ 
Finally, 
\be
\lim_{n\to\infty}\lVert\triv_{n}\rVert\le\exp\left\{\max_{(s,t)}\left\lVert\un{\a_{\triv(\g_{s,t})}}\big(B(s,t)\big)\right\rVert\right\}.
\ee
\en
\bprf
This argument has been deferred to Appendix \ref{app:spc}. 
\eprf
This result is analogous to the bound obtained for
ordinary parallel transport \cite{BaMu}.
Explicitly computing the $k$-th order sum as $n\to\infty$ is 
intractable due to the complicated ordering of terms present
(see the ordering on a $5\times5$ grid before 
equation (\ref{eq:totalterms5by5})). 
Fortunately, we can simplify the expression $\triv_{n}$
by rearranging and reorganizing all of these terms. 
For example, consider terms with two $B$'s. 
There are terms with two
$B$'s at different ``vertical heights'' such as
\begin{figure}[H]
\centering
\begin{subfigure}{0.25\textwidth}
\centering
    \includegraphics[width=1.00\textwidth]{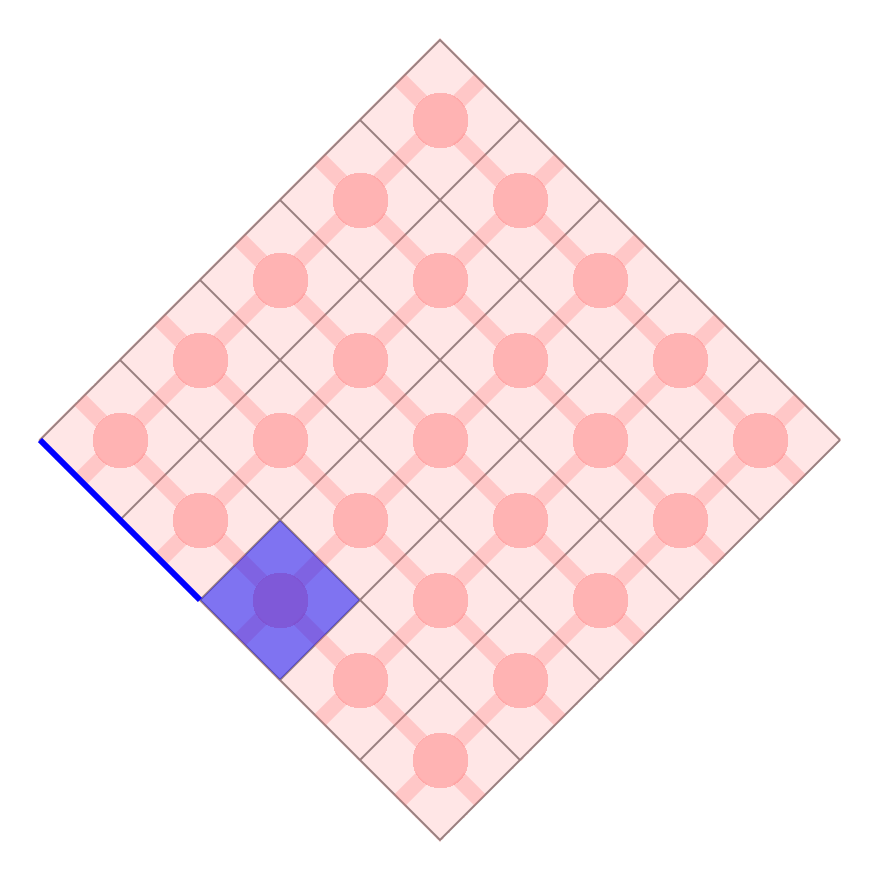}
\end{subfigure}
\hspace{-3mm}
$\cdot$
\hspace{-3mm}
\begin{subfigure}{0.25\textwidth}
\centering
 \includegraphics[width=1.00\textwidth]{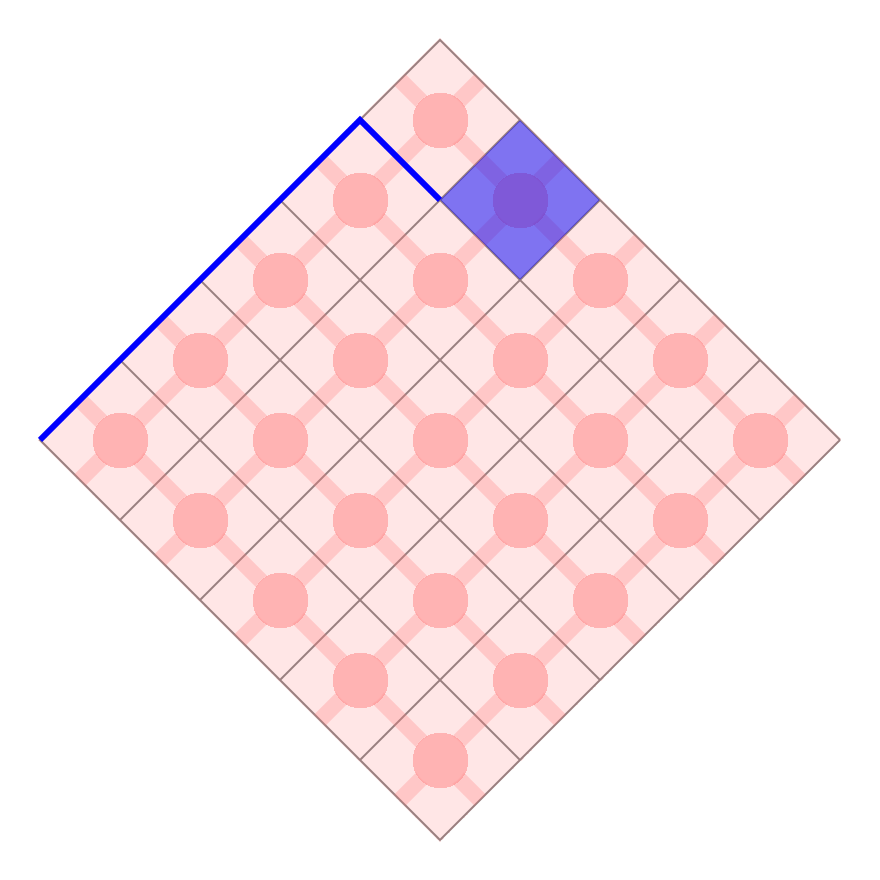}
\end{subfigure}
\end{figure}
\noindent
and terms with $B$'s at the same height such as 
\begin{figure}[H]
\centering
\begin{subfigure}{0.25\textwidth}
\centering
    \includegraphics[width=1.00\textwidth]{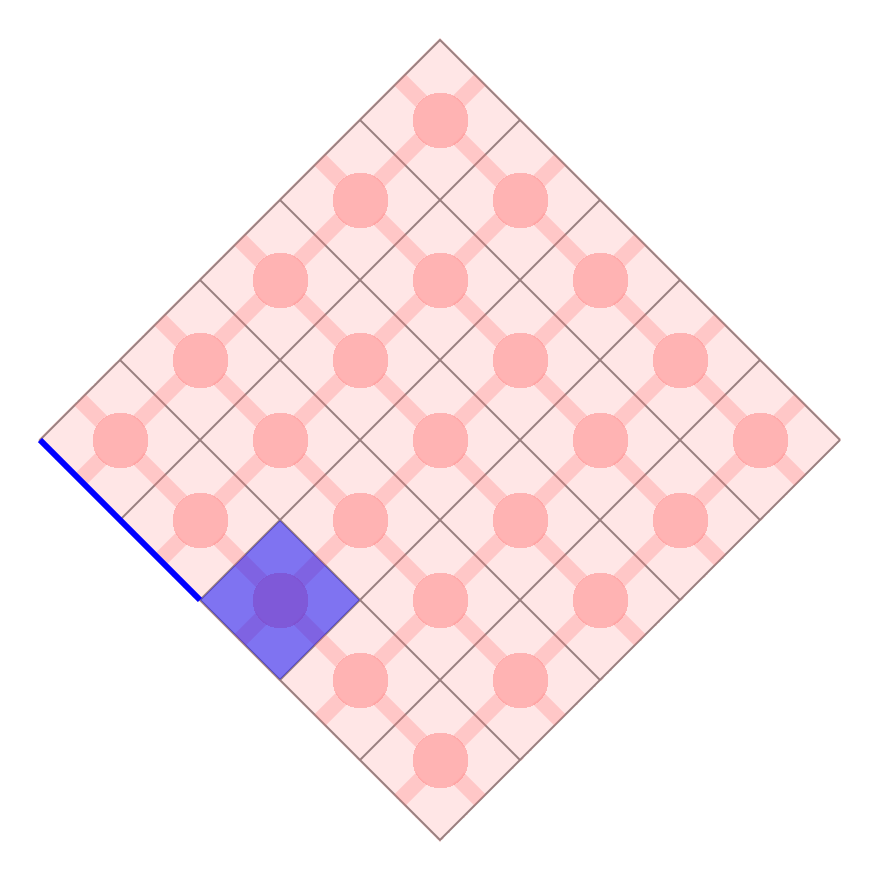}
\end{subfigure}
\hspace{-3mm}
$\cdot$
\hspace{-3mm}
\begin{subfigure}{0.25\textwidth}
\centering
 \includegraphics[width=1.00\textwidth]{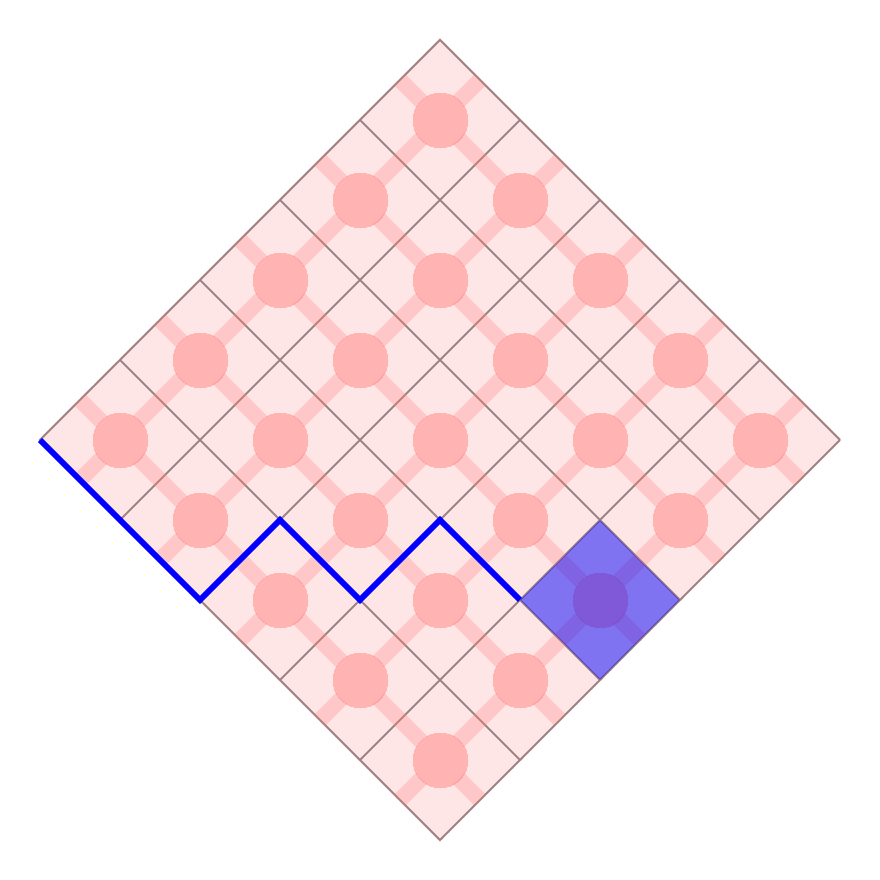}
\end{subfigure}
\end{figure}
As explained above (\ref{eq:totalterms5by5}), 
due to the automatic ordering, 
there do not exist terms with the order
flipped in the above two images.
Therefore, the number of terms with two $B$'s at the same height is
($n=5$ in our picture)
\be
\begin{split}
\sum_{m=2}^{n-1}\binom{m}{2}+\binom{n}{2}
+\sum_{m=2}^{n-1}\binom{m}{2}
&=2\sum_{m=2}^{n-1}\binom{m}{2}+\binom{n}{2}\\
&=2\binom{n}{3}+\binom{n}{2}\\
&=\frac{2n!}{3!(n-3)!}+\frac{n!}{2!(n-2)!}\\
&=\frac{n(n-1)(2n-1)}{3!},
\end{split}
\ee
where the second equality comes from 
a neat fact about Pascal's triangle
\be
\xy0;/r.25pc/:
(0,15)*+{1}="00";
(-5,10)*+{1}="10";
(5,10)*+{1}="11";
(-10,5)*+{1}="20";
(0,5)*+{2}="21";
(10,5)*+{1}="22";
(-15,0)*+{1}="30";
(-5,0)*+{3}="31";
(5,0)*+{3}="32";
(15,0)*+{1}="33";
(-20,-5)*+{1}="40";
(-10,-5)*+{4}="41";
(0,-5)*+{6}="42";
(10,-5)*+{4}="43";
(20,-5)*+{1}="44";
(-25,-10)*+{1}="50";
(-15,-10)*+{5}="51";
(-5,-10)*+{10}="52";
(5,-10)*+{10}="53";
(15,-10)*+{5}="54";
(25,-10)*+{1}="55";
(-30,-15)*+{1}="60";
(-20,-15)*+{6}="61";
(-10,-15)*+{15}="62";
(0,-15)*+{20}="63";
(10,-15)*+{15}="64";
(20,-15)*+{6}="65";
(30,-15)*+{1}="66";
{\ar@{}"22";"32"|-{+}};
{\ar@{}"32";"42"|-{+}};
{\ar@{}"42";"52"|-{+}};
{\ar@{=}"52";"63"};
\endxy
\ee
The ratio of terms with two $B$'s at the same height to the total
number of terms with two $B$'s is
\be
\cfrac{n(n-1)(2n-1)/3!}{\binom{n^2}{2}}
=\frac{2n-1}{3(n+1)n}.
\ee
Note that the limit of this quantity as $n\to\infty$ is
\be
\lim_{n\to\infty}\frac{2n-1}{3(n+1)n}
=0.
\ee
Hence, terms that involve a product of two $B$'s
that also appear at the same height become
negligible in the $n\to\infty$ limit. One might wonder
if this is true for any product of $B$'s. Clearly, this is
false when we have a product of $k$ $B$'s and
$k>2n-1$ since \emph{every} configuration has at
least one row in which $B$ occurs at least twice. 
However, it is true for $k$ sufficiently smaller than $n.$
This leads us to an interesting combinatorial problem
in its own right. 

The number of configurations of $k$ blocks in an $n\times n$ grid 
tilted $45^{\circ}$ such that
no two blocks appear at the same height is%
\footnote{We thank Zhibai Zhang and Scott O. Wilson
who both independently suggested the currently used approach
for this problem
and for discussions leading to this formula.}
\be
S_{n,k}:=
\sum_{2n-1\ge i_{k}>i_{k-1}>\cdots>i_{1}\ge1}l_{n}(i_{1})
\cdots l_{n}(i_{k}),
\ee
where
\be
l_{n}(i):=\begin{cases}
i&\mbox{ if } 1\le i\le n\\
2n-i&\mbox{ if } n<i\le 2n-1
\end{cases}
\ee
denotes the number of blocks of a given height $i.$ 
The ratio of this number to the total number of 
configurations of $k$ blocks is
\be
R_{n,k}:=\cfrac{(n^2-k)!k!S_{n,k}}{n^2!}.
\ee
\blem
\label{lem:twoboxeslimit}
For any $\e>0$ and $K\in\N,$
there exists an integer $N> K$
such that 
\be
1-R_{n,k}\le\e
\ee
for all $n\ge N$ and $k\le K,$
i.e. 
\be
\lim_{n\to\infty}R_{n,k}=1
\ee
for all $k\in\N.$ 
\elem

\begin{figure}[h]
\centering
\includegraphics[width=0.60\textwidth]{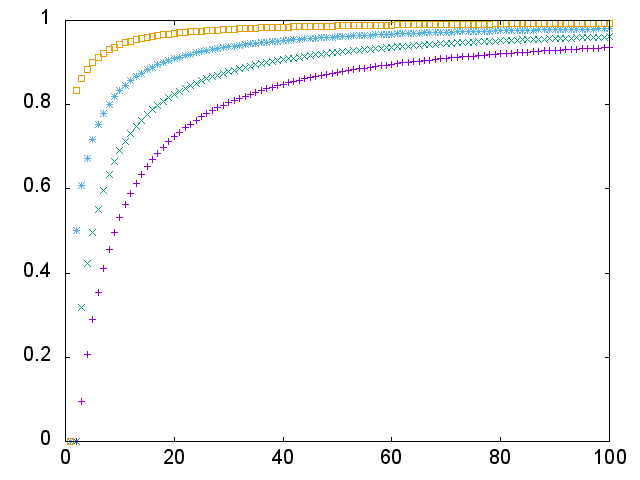}
\begin{picture}(0,0)
\put(-145,0){$n$}
\put(-150,217){$R_{n,k}$}
\put(-250,185){\rotatebox{21}{$\scriptsize k=2$}}
\put(-236,174){\rotatebox{26}{$\scriptsize k=3$}}
\put(-224,162){\rotatebox{31}{$\scriptsize k=4$}}
\put(-212,152){\rotatebox{33}{$\scriptsize k=5$}}
\end{picture}
\caption{A plot of $R_{n,k}$ for various values of $n$ and $k$
indicating $\lim_{n\to\infty}R_{n,k}=1.$}
\label{fig:ratio}
\end{figure}

The graph in Figure \ref{fig:ratio} should be convincing%
\footnote{We thank Steven Vayl
for teaching us some basics of C++ providing
the necessary tools to make this plot.
}
though of course it is not a substitute for a proof.

The proof of Lemma \ref{lem:twoboxeslimit} is quite
involved and is given in Appendix \ref{app:config}. 
Instead, we offer a rough estimate analysis
via averaging.
The average value of $l_{n}$ is
\be
\mathrm{avg}(l_{n}):=\cfrac{\sum_{i=1}^{2n-1}l_{n}(i)}{2n-1}
=\frac{n^2}{2n-1}.
\ee
Hence, to a good approximation for large $n$ and small $k,$
\be
\begin{split}
S_{n,k}&\approxeq
\sum_{2n-1\ge i_{k}>i_{k-1}>\cdots>i_{1}\ge1}
[\mathrm{avg}(l_{n})]^{k}\\
&=\left(\frac{n^2}{2n-1}\right)^{k}\binom{2n-1}{k}\\
&=\frac{n^{2k}(2n-1)(2n-2)\cdots(2n-k)}{k!(2n-1)^{k}},
\end{split}
\ee
where the second line comes from the fact that there are
$\binom{2n-1}{k}$ terms in the summation. 
Hence, to a good approximation 
\be
\begin{split}
R_{n,k}&\approxeq
\frac{n^{2k}(2n-1)(2n-2)\cdots(2n-k)}{
(2n-1)^{k}n^{2}(n^2-1)\cdots(n^2-(k-1))}\\
&=\frac{\left(1-\frac{1}{2n}\right)\cdots\left(1-\frac{k}{2n}\right)}{
\left(1-\frac{1}{2n}\right)^{k}\left(1-\frac{1}{n^2}\right)
\cdots\left(1-\frac{k-1}{n^2}\right)}
\end{split}
\ee
Since $k$ is fixed, the right-hand-side tends to $1$
as $n\to\infty.$ Again, the precise proof is given
in Appendix \ref{app:config}. 

\bt
\label{thm:fulltoreducedsurfacetransport}
Let $\triv_{n}$ be the 
generalization of the expression given in 
(\ref{eq:squaretransportforn=5})
for an $n\times n$ grid decomposition. 
Let $\triv^{\mathrm{red}}_{n}$ be the same
expression but with all terms in which
$B$ occurs at least twice at the same height removed. 
For any $\e>0,$ there exists an $N\in\N$ such that
\be
\left\lVert\triv_{n}-\triv_{n}^{\mathrm{red}}\right\rVert
\le\e
\ee
for all $n\ge N.$ 
\et
\bprf
Let $M$ be the maximum value of the norms of
all quantities of the form 
$\a_{a_{56}^{s}a_{46}^{s}a_{36}^{s}a_{35}^{t}a_{25}^{s}}(B_{14}).$
The difference $\triv_{n}-\triv_{n}^{\mathrm{red}}$
only consists of contributions from terms in which there
exist at least two $B$'s that occur at the same height.
Fix $\e>0.$ To begin, let $K$ be 
large enough so that 
\be
\sum_{k=K+1}^{\infty}\frac{M^k}{k!}
\le\frac{\e}{2},
\ee
which is possible since the series for the exponential converges.
Furthermore, by Lemma \ref{lem:twoboxeslimit},
for any $\e>0,$ there exists an $N$ large enough so that 
\be
1-R_{n,k}\le\frac{\e}{2e^{M}}\quad\forall\; k\le K,n\ge N.
\ee
Using these two results, 
the value of the norm of 
the difference $\triv_{n}-\triv_{n}^{\mathrm{red}}$
is bounded by 
\be
\begin{split}
\left\lVert\triv_{n}-\triv_{n}^{\mathrm{red}}\right\rVert
&\le\sum_{k=1}^{n^2}\left(\frac{M}{n^2}\right)^{k}\left[\binom{n^2}{k}-S_{n,k}\right]
=\sum_{k=1}^{n^2}\left(\frac{M}{n^2}\right)^{k}\binom{n^2}{k}[1-R_{n,k}]\\
&\le\sum_{k=1}^{n^2}\frac{M^k}{k!}[1-R_{n,k}]
=\sum_{k=1}^{K}\frac{M^k}{k!}[1-R_{n,k}]
+\sum_{k=K+1}^{n^2}\frac{M^k}{k!}[1-R_{n,k}]\\
&\le\sum_{k=1}^{K}\frac{M^k}{k!}\left(\frac{\e}{2e^{M}}\right)
+\sum_{k=K+1}^{n^2}\frac{M^k}{k!}\\
&\le\left(\frac{\e}{2e^{M}}\right)\sum_{k=1}^{\infty}\frac{M^k}{k!}
+\sum_{k=K+1}^{\infty}\frac{M^k}{k!}\\
&\le\frac{\e}{2}+\frac{\e}{2}
=\e.
\end{split}
\ee
\eprf

Thus, heuristically, as $n\to\infty,$ the number of terms for which
at least two $B$'s are at the same height is a set of measure zero
with respect to all possibilities and hence we can ignore them
in the calculation of the surface ordered parallel transport
after taking the $n\to\infty$ limit. 
This gives the following
picture for the surface-iterated integral.
Let $\g_{s,t}$ be the (thin) path 
\begin{figure}[H]
\centering
    \includegraphics[width=0.28\textwidth]{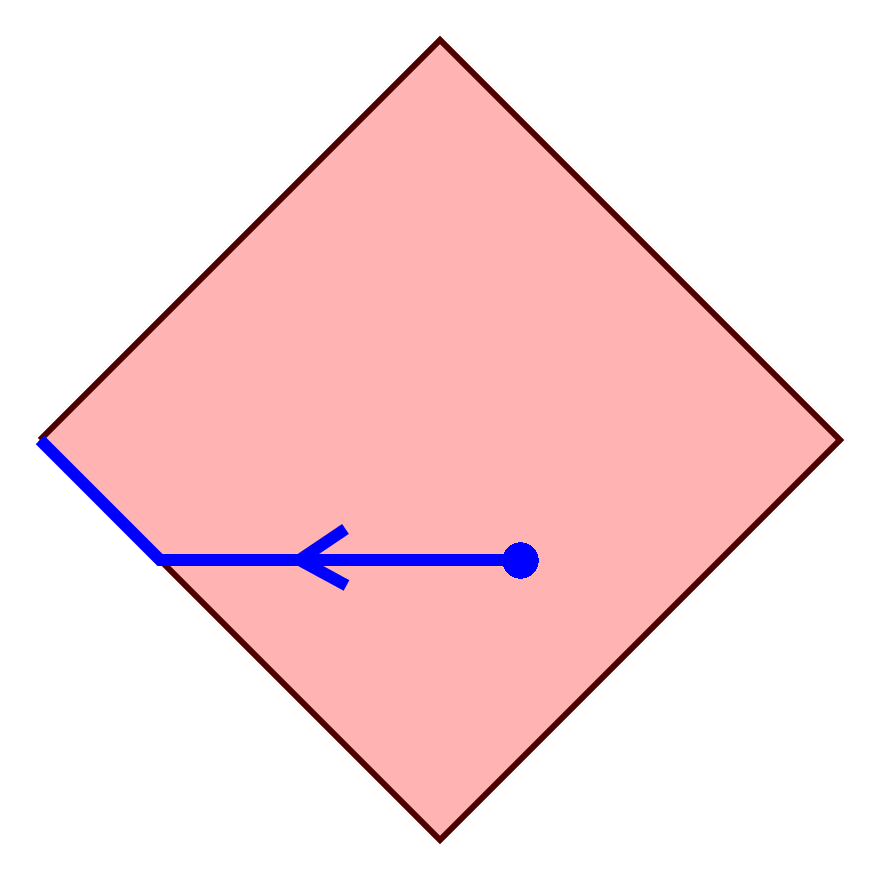}
     \begin{picture}(0,0)
         \put(-60,61){$\xy0;/r.25pc/: 
    	(0,0)*{}="1";
	(-18,-18)*+{\small t}="t";
	{\ar"1";"t"};
    	\endxy$}
    \put(-60,70){$\xy0;/r.25pc/: 
    	(0,0)*{}="1";
	(-18,18)*+{\small s}="s";
	{\ar"1";"s"};
    	\endxy$}
     \put(-56,48){$(s,t)$}
     \put(-185,42){$(s+1-t,1)$}
     \put(-100,60){$\g_{s,t}$}
     \end{picture}
\end{figure}  
The limit of the expression (\ref{eq:surfaceproductn=5}) as $n\to\infty$ 
is therefore given by a sum of iterated integrals
\begin{figure}[H]
\centering
\begin{subfigure}{0.25\textwidth}
\centering
    \includegraphics[width=1.00\textwidth]{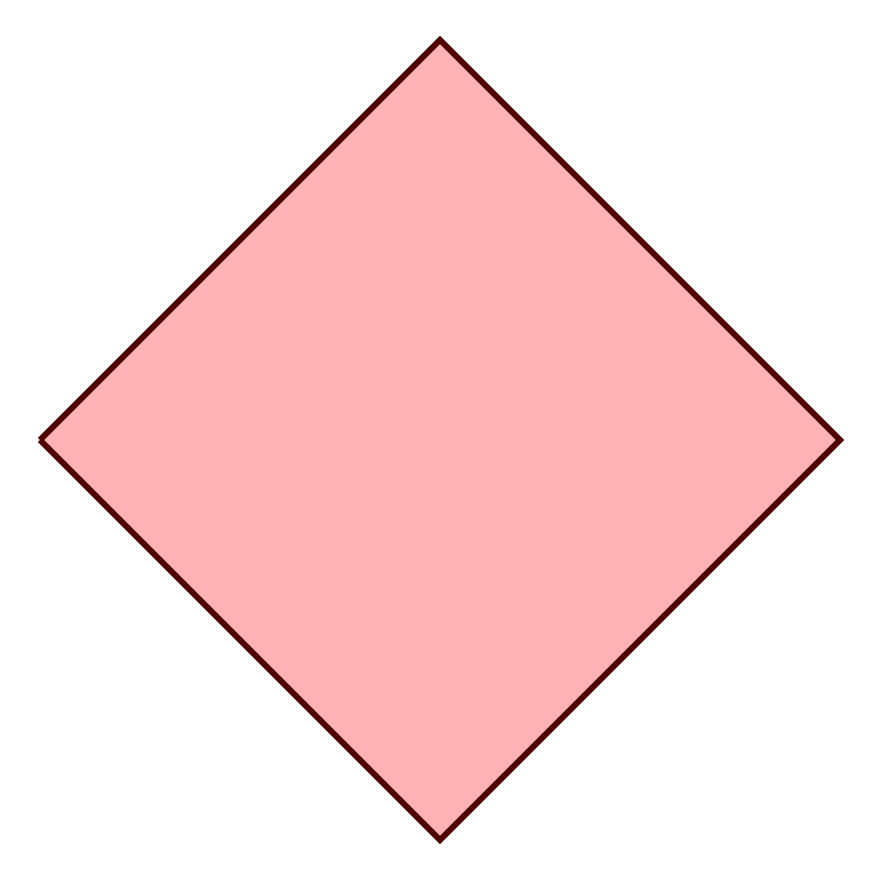}
\end{subfigure}
\hspace{-3mm}
$\displaystyle+\int$
\hspace{-3mm}
\begin{subfigure}{0.25\textwidth}
\centering
\includegraphics[width=1.00\textwidth]{gamma_st}
\end{subfigure}
\hspace{-3mm}
$\displaystyle+\int\int$
\hspace{-3mm}
\begin{subfigure}{0.25\textwidth}
\centering
\includegraphics[width=1.00\textwidth]{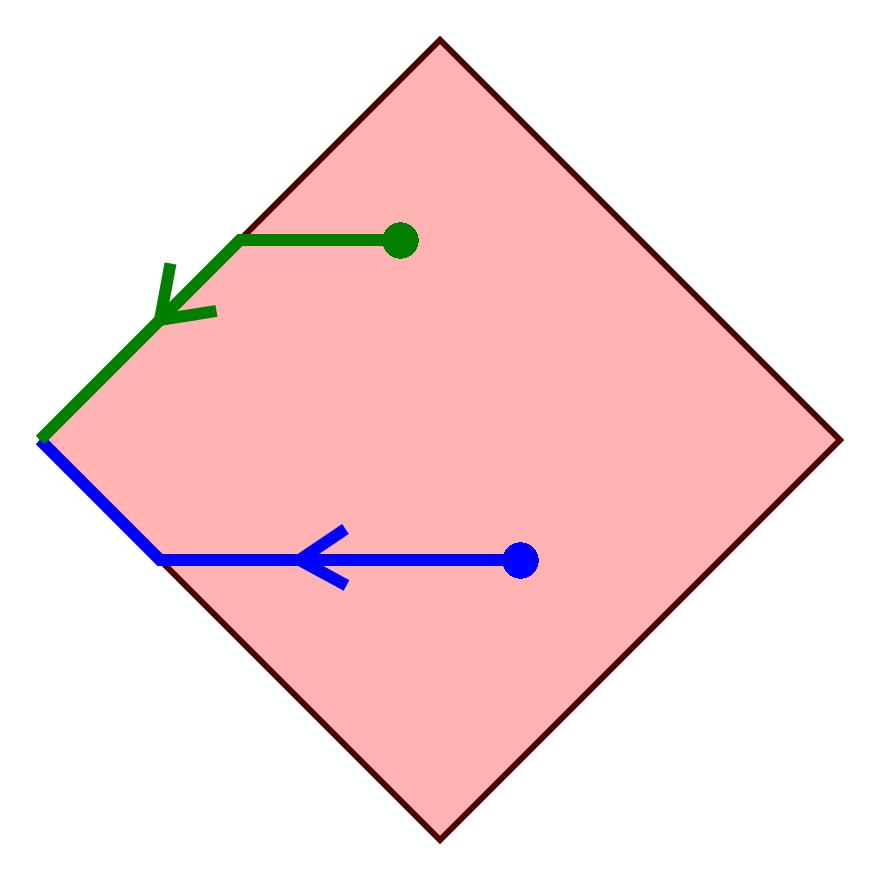}
\end{subfigure}
\hspace{-3mm}
$+\cdots$
\hspace{-3mm}
\end{figure} 
\noindent 
with path-ordering \emph{only} in the vertical direction. 
In more detail, 
the surface-ordered integral is depicted schematically 
as an infinite sum of terms expressed by placing $B$ at 
the endpoints of the drawn paths and conjugating it by parallel 
transport along the path connecting to it
using $A$ and $\a.$ Then we use an ordinary integral 
over the horizontal direction to get a 1-form
(similar to what is done in \cite{BS} and \cite{SW2}). Finally we 
use the usual path-ordered integral in the vertical direction. 
More explicitly, by changing coordinates to 
\be
u:=\frac{s+t}{\sqrt{2}}\aand v:=\frac{s-t}{\sqrt{2}},
\ee
one can express $\g_{s,t}$ in terms of $u$ and $v.$ 
We write this path as $\g_{u,v}.$ Using this, 
the surface parallel transport
is given by 
\be
\label{eq:surfaceiteratedintegral}
\begin{split}
1&+\int\left(\int\underline{\a_{\triv(\g_{u,v})}}
\big(B(u,v)\big)dv\right)du
\\
&+\int_{u_{2}\ge u_{1}}
\left(\int\underline{\a_{\triv(\g_{u_{2},v_{2}})}}
\big(B(u_{2},v_{2})\big)
\underline{\a_{\triv(\g_{u_{1},v_{1}})}}\big(B(u_{1},v_{1})\big)
dv_{2}dv_{1}\right)du_{2}du_{1}+\cdots
\\
&+
\int_{u_{n}\ge\cdots\ge u_{1}}
\left(\int\underline{\a_{\triv(\g_{u_{n},v_{n}})}}
\big(B(u_{n},v_{n})\big)\cdots
\underline{\a_{\triv(\g_{u_{1},v_{1}})}}\big(B(u_{1},v_{1})\big)
dv_{n}\cdots dv_{1}\right)du_{n}\cdots du_{1}
\\
&+\cdots,
\end{split}
\ee
where $B(u,v)$ stands for
\be
B(u,v):=B\left(\frac{\p\S}{\p s},\frac{\p\S}{\p t}\right)
\Big|_{\left(s=\frac{u+v}{\sqrt{2}},t=\frac{u-v}{\sqrt{2}}\right)}.
\ee
The sum in (\ref{eq:surfaceiteratedintegral}) is absolutely convergent
by Propositions \ref{prop:trivnconverges} and 
\ref{prop:trivnkbounded} and Theorem \ref{thm:fulltoreducedsurfacetransport}. 
In other words, 
\be
\lim_{n\to\infty}\triv_{n}=\text{(\ref{eq:surfaceiteratedintegral})}. 
\ee
\br
\label{rmk:SW}
Although our formula for the surface ordered product for parallel transport
has a similar form to the one given by Schreiber and Waldorf in
their equation (2.27) of \cite{SW2}, 
they do not look equal. Here, we provide an argument that shows our
two formulas are equal. This fact will follow from the defining
properties of a crossed module. 
In terms of bigons, the following figure depicts the differences
between the conventions of defining the surface ordered product
\begin{figure}[H]
\centering
    \includegraphics[width=0.80\textwidth]{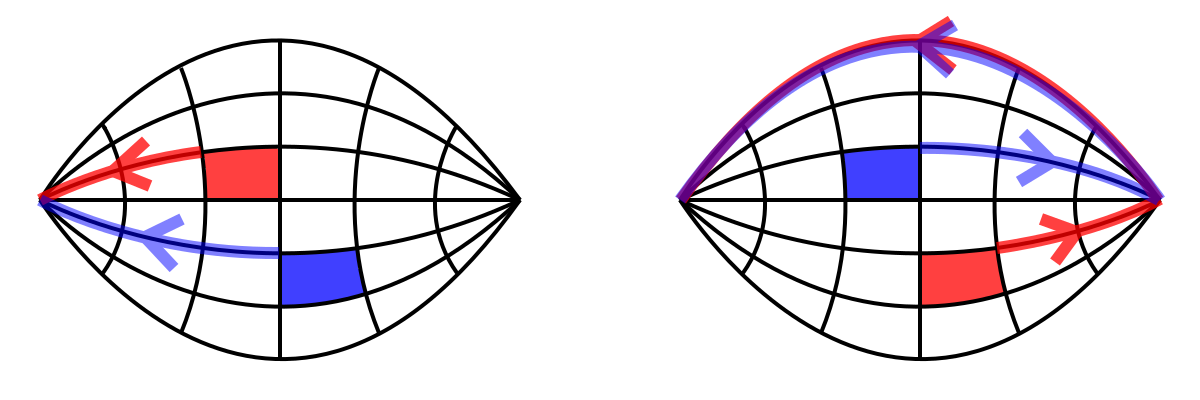}
    \begin{picture}(0,0)
    \put(-396,63){$y$} 
    \put(-220,63){$x$}
    \put(-308,124){$\g$}
    \put(-308,0){$\de$}
    \put(-347,81){$\b$}
    \put(-330,38){$\b'$}
    \put(-322,70){$B$}
    \put(-296,36){$B'$}
    \put(-187,63){$y$}
    \put(-11,63){$x$}
    \put(-99,124){$\g$}
    \put(-99,0){$\de$}
    \put(-113,70){$B$}
    \put(-87,36){$B'$}
    \put(-84,85){$\tilde{\b}$}
    \put(-61,40){$\tilde{\b}'$}
    \end{picture}
\end{figure}
\noindent
for a bigon $\S:\g\Rightarrow\de$ between paths $\g,\de:x\to y.$ 
The left bigon depicts our 
automatic ordering that was derived directly from 2-group multiplication
with the red quantity appearing to the right of the blue quantity, i.e. 
\be
{\color{blue}\un{\a_{\triv(\b')}}(B')}{\color{red}\un{\a_{\triv(\b)}}(B)}.
\ee
The right bigon depicts the ordering convention chosen by 
\cite{SW2}, which can be seen by noting that in their formula, some
inverses and minus signs appear that we have avoided. 
These amount to computing the parallel transport on the leftover form,
which they call $\mathcal{A}_{\S},$ along the \emph{reverse} direction. 
Also notice that the path they use to act on the $B$ field via the $\a$
action from the crossed module actually goes around 
quite differently than ours. 
For them, the order is swapped and the expression is written as
\be
{\color{blue}\un{\a_{\triv(\g\tilde{\b})}}(B)}{\color{red}\un{\a_{\triv(\g\tilde{\b}')}}(B')}.
\ee
Term-by-term, these expressions are in fact \emph{different}. 
However, when they are combined and all terms are taken to 
account in computing the full parallel transport, 
the resulting 2-group elements describing the parallel transport
are the \emph{same} so that our formulas agree. 
To see this, we will compare what happens in both of our conventions
if we add on an infinitesimal square to an already computed
surface transport (cf. (\ref{eq:addingepsilon})). 
Fix $(s,t)\in[0,1]\times[0,1]$ and consider a bigon 
whose parallel transport has been computed up to $s$ and an
additional bigon is to be added at this point
\be
\xy0;/r.35pc/:
(25,0)*+{x}="x";
(-25,0)*+{y}="y";
(3,-8)*+{\bullet}="stb";
(3,-16)*+{\bullet}="stbd";
(-5,-8)*+{\bullet}="steb";
(-5,-16)*+{\bullet}="stebd";
(5,-5)*+{{}_{(s,t)}}="st";
(-5,-5)*+{{}_{(s,t+\e)}}="ste";
{\ar@/_5.0pc/"x";"y"_{\displaystyle \g}};
{\ar@/^0.9pc/"x";"stb"_{\displaystyle \g_{s,t}^{SW}}};
{\ar@/^0.9pc/"steb";"y"_{\displaystyle \g_{s,t}^{P}}};
{\ar@/^0.20pc/"stb";"steb"_{\g_{1}}};
{\ar"stb";"stbd"};
{\ar"steb";"stebd"};
{\ar@/^0.20pc/"stbd";"stebd"};
{\ar@{=>}"steb";"stbd"|-{\D\S}};
{\ar@{=>}(0,12);(0,-6)_{\displaystyle \S_{s}}};
\endxy
\ee
Set
\be
\begin{split}
h:=\triv(\S_{s}),\quad
g_{P}:=\triv(\g^{P}_{s,t}),\quad
g_{SW}:=\triv(\g^{SW}_{s,t}),\\
g:=\triv(\g),\quad
h_{\D}:=\triv(\D\S),\quad
g_{1}:=\triv(\g_{1}).
\end{split}
\ee
Our convention is that the resulting parallel transport of adding this extra piece
is given by $\a_{g_{P}}(h_{\D})h,$ whereas Schreiber and Waldorf's 
convention (due to the inverses appearing in their formulas) would give
$h\a_{gg_{SW}^{-1}}(h_{\D}).$ To better compare these, notice that 
$\t(h)g=g_{P}g_{1}g_{SW},$ which is equivalent to 
$gg_{SW}^{-1}=\t(h^{-1})g_{P}g_{1}.$ Hence, 
\be
h\a_{gg_{SW}^{-1}}(h_{\D})
=h\a_{\t(h^{-1})g_{P}g_{1}}(h_{\D})
=\a_{g_{P}}(h_{\D})h
\ee
by the properties of crossed module multiplication. 
Now, by continuity of $\triv,$ these actions, and the group
multiplications and since $\ds\lim_{\e\to0}g_{1}=e,$ the identity element
in the group $G,$ we have
\be
\lim_{\e\to0}\big(\a_{g_{P}}(h_{\D})h\big)
=h\a_{gg_{SW}^{-1}}(h_{\D})
\ee
so that our formulas for the parallel transport along surfaces agree. 
\er

\subsection{Gauge transformations for surface transport}
\label{sec:2dalgebragtst}
In Section \ref{sec:2dalgebra1dgt}, we described gauge transformations
as natural transformations of parallel transport functors
for paths. In this section, we will use this as the \emph{definition}
of a gauge transformation and derive the corresponding
formulas for differential forms. 
As before, let $\mathcal{G}:=(H,G,\t,\a)$
be a crossed module, $\B\mathcal{G}$ its associated
2-group, and $M$ a smooth manifold. 
A \emph{\uline{(first order) gauge transformation}} 
from a parallel transport
functor $\triv:\mathcal{P}^{2}M\to\B\mathcal{G}$ to another
$\triv':\mathcal{P}^{2}(M)\to\B\mathcal{G}$ is a natural
transformation $\triv\Rightarrow\triv'.$ 
By Definition \ref{defn:nattransf} and Proposition
\ref{prop:nattransftoa2group}, such a natural
transformation consists of a pair
of functions $g:M\to G$ and $h:P^{1}M\to H$ 
satisfying the conditions described in that Proposition. 
Namely, to every thin path $z\xleftarrow{\g}y,$
\be
\label{eq:1stordergaugetransformationpaths}
\t\big(h(\g)\big)g(z)\triv(\g)=\triv'(\g)g(y),
\ee
to ever pair of composable thin paths
$z\xleftarrow{\g}y\xleftarrow{\de}x,$
\be
\label{eq:nattranscomposable}
h(\g\de)=\a_{\triv'(\g)}\big(h(\de)\big)h(\g),
\ee
to every point $x\in M,$
\be
h(\id_{x})=e,
\ee
and finally to any worldsheet
\begin{figure}[H]
\centering
\includegraphics[width=0.295\textwidth]{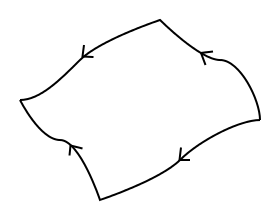}
\begin{picture}(0,0)
\put(-75,60){$\Sigma$}
\put(-115,90){$\gamma$}
\put(-32,88){$\delta$}
\put(-48,26){$\xi$}
\put(-120,30){$\zeta$}
\put(-10,45){$x$}
\put(-62,108){$y$}
\put(-145,60){$z$}
\put(-102,3){$w$}
\put(-86,75){\rotatebox{-105}{$\Longrightarrow$}}
\end{picture}
\end{figure}
\noindent
viewed as a bigon from $\g\de$ to $\z\xi,$ 
the equality%
\footnote{This follows from condition (d) in
Definition \ref{defn:nattransf}.
}
\begin{figure}[H]
\centering
\begin{subfigure}{0.47\textwidth}
\centering
    \includegraphics[width=0.60\textwidth]{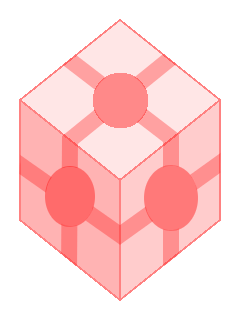}
      \begin{picture}(0,0)
\put(-53,67){$h(\xi)$}
\put(-26,84){$g(x)$}
\put(-83,50){$g(w)$}
\put(-111,68){$h(\zeta)$}
\put(-133,82){$g(z)$}
\put(-90,125){$\mathrm{triv}(\Sigma)$}
\put(-117,50){\rotatebox{-39}{$\mathrm{triv}'(\zeta)$}}
\put(-113,110){\rotatebox{-39}{$\mathrm{triv}(\zeta)$}}
\put(-65,154){\rotatebox{-39}{$\mathrm{triv}(\delta)$}}
\put(-113,132){\rotatebox{41}{$\mathrm{triv}(\gamma)$}}
\put(-59,91){\rotatebox{40}{$\mathrm{triv}(\xi)$}}
\put(-58,28){\rotatebox{40}{$\mathrm{triv}'(\xi)$}}
\end{picture}
\end{subfigure}
=
\begin{subfigure}{0.47\textwidth}
\centering
    \includegraphics[width=0.60\textwidth]{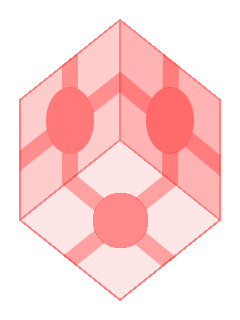}
      \begin{picture}(0,0)
\put(-116,67){\rotatebox{40}{$\mathrm{triv}'(\gamma)$}}
\put(-91,54){$\mathrm{triv}'(\Sigma)$}
\put(-135,91){$g(z)$}
\put(-27,89){$g(x)$}
\put(-82,131){$g(y)$}
\put(-112,111){$h(\gamma)$}
\put(-54,111){$h(\delta)$}
\put(-64,25){\rotatebox{40}{$\mathrm{triv}'(\xi)$}}
\put(-116,129){\rotatebox{40}{$\mathrm{triv}(\gamma)$}}
\put(-60,88){\rotatebox{-40}{$\mathrm{triv}'(\delta)$}}
\put(-57,147){\rotatebox{-40}{$\mathrm{triv}(\delta)$}}
\put(-110,45){\rotatebox{-40}{$\mathrm{triv}'(\zeta)$}}
\end{picture}
\end{subfigure}
\end{figure}
\noindent
holds. Reading this diagram is a bit tricky without
the arrows (recall Remark \ref{rmk:2catstandard}). 
More explicitly, this equality says
\begin{figure}[H]
\centering
    \includegraphics[width=0.90\textwidth]{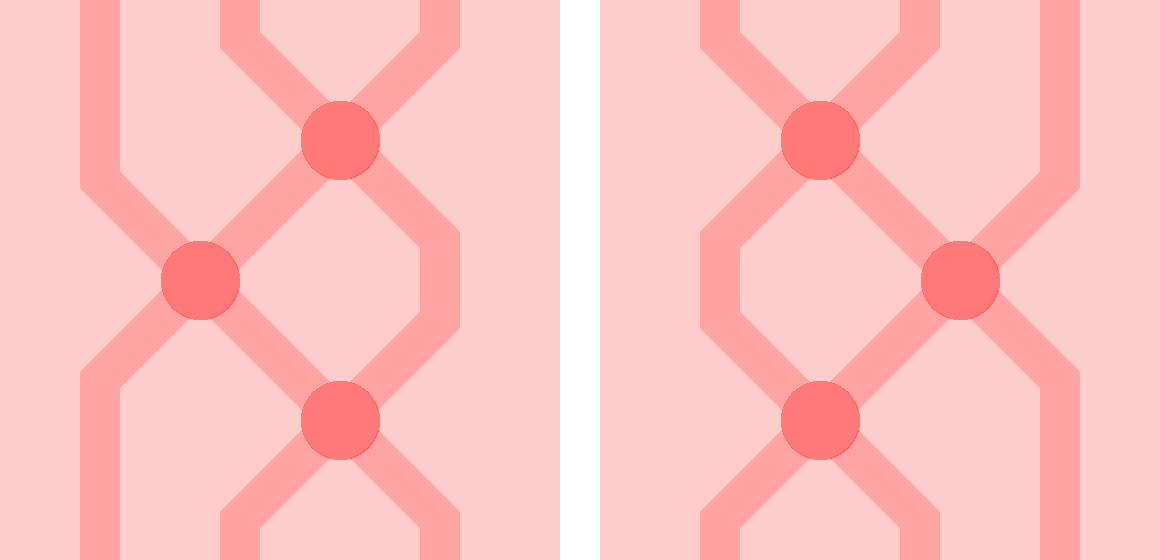}
\begin{picture}(0,0)
\put(-333,157){$\triv(\S)$}
\put(-416,175){$g(z)$}
\put(-366,195){$\triv(\g)$}
\put(-295,195){$\triv(\de)$}
\put(-379,103){$h(\zeta)$}
\put(-360,130){$\triv(\z)$}
\put(-355,76){$g(w)$}
\put(-292,103){$\triv(\xi)$}
\put(-325,50){$h(\xi)$}
\put(-366,16){$\triv'(\xi)$}
\put(-422,70){$\triv'(\z)$}
\put(-288,16){$g(x)$}
\put(-152,50){$\triv'(\S)$}
\put(-184,16){$\triv'(\z)$}
\put(-116,16){$\triv'(\xi)$}
\put(-192,101){$\triv'(\g)$}
\put(-183,194){$g(z)$}
\put(-118,191){$\triv(\g)$}
\put(-62,175){$\triv(\de)$}
\put(-117,130){$g(y)$}
\put(-90,103){$h(\de)$}
\put(-144,157){$h(\g)$}
\put(-122,76){$\triv'(\de)$}
\put(-55,68){$g(x)$}
\put(-228,100){$=$}
\end{picture}
\end{figure}
\noindent
i.e.
\be
\label{eq:1stordergaugetransformationforsurface}
\a_{\triv'(\zeta)}\big(h(\xi)\big)h(\zeta)\a_{g(z)}\big(\triv(\S)\big)
=
\triv'(\S)\a_{\triv'(\gamma)}\big(h(\delta)\big)h(\gamma)
\ee
or equivalently by our earlier condition 
(\ref{eq:nattranscomposable})
\be
h(\zeta\xi)\a_{g(z)}\big(\triv(\Sigma)\big)
=
\triv'(\Sigma)h(\gamma\delta).
\ee
By Proposition \ref{prop:nattransftoa2group}, such a
natural transformation can be decomposed into
\be
(g,h)=\begin{matrix}[0.9](g,e)\\(e,h)\end{matrix}.
\ee
A gauge transformation of the 
type $(g,e)$ is typically called a (first order) 
\emph{\uline{thin gauge transformation}} and one of the type
$(e,h)$ is called a (first order) \emph{\uline{fat gauge transformation}}
\cite{MM}. Thus, Proposition 
\ref{prop:nattransftoa2group} implies that an arbitrary
gauge transformation of the first kind can be decomposed
into a thin and fat gauge transformation. 
Using this, we can calculate infinitesimal versions of the 
functions $g:M\to G$ and $h:P^{1}M\to H$ for small
paths, i.e. for a point $x\in M$ and a tangent vector at $x.$ 
This was already done for $g:M\to G$ at the end of
Section \ref{sec:2dalgebra1dgt} with result 
(\ref{eq:usualgaugetransformation}). 
For $h:P^{1}M\to H,$ let $t\mapsto x(t)$ parametrize an
infinitesimal path $\g,$ then to lowest order in $\D t$
\be
\label{eq:fatgtvfform}
h(\g)=\exp\left\{\varphi_{\mu}\big(x(t)\big)
\frac{dx^{\mu}}{dt}\Big|_{t}\D t\right\}
\ee
for some 1-form $\varphi\in\W^{1}(M;\mathfrak{h}).$ 
Thus, plugging these expressions into 
(\ref{eq:1stordergaugetransformationpaths}),
a fat gauge transformation from $(A,B)$
to $(A',B')$ infinitesimally gives
\be
\t\left(\exp\left\{\varphi_{\mu}\big(x(t)\big)
\frac{dx^{\mu}}{dt}\Big|_{t}\D t\right\}\right)
\exp\left\{-A_{\nu}\big(x(t)\big)
\frac{dx^{\nu}}{dt}\Big|_{t}\D t\right\}
=
\exp\left\{-A'_{\nu}\big(x(t)\big)
\frac{dx^{\nu}}{dt}\Big|_{t}\D t\right\}
\ee
because $g$ has been set to be the identity. 
Expanding out to lowest order in $\D t$ gives
\be
\mathds{1}+\un\t\left(\varphi_{\mu}\big(x(t)\big)
\frac{dx^{\mu}}{dt}\Big|_{t}\right)\D t
-A_{\nu}\big(x(t)\big)
\frac{dx^{\nu}}{dt}\Big|_{t}\D t
=
\mathds{1}-A'_{\nu}\big(x(t)\big)
\frac{dx^{\nu}}{dt}\Big|_{t}\D t
\ee
giving the relationship
\be
A'=A-\un\t(\varphi)
\ee
for a fat gauge transformation. We already
calculated what happens for a thin gauge transformation
in Section \ref{sec:2dalgebra1dgt}. Using Proposition
\ref{prop:nattransftoa2group},
combining (\ref{eq:usualgaugetransformation}) with this
gives 
\be
\label{eq:firstordergtA}
A'=gAg^{-1}-dg g^{-1}-\un\t(\varphi)
\ee
for an arbitrary gauge transformation. 
The $B$ field under an arbitrary gauge transformation
changes according to 
(\ref{eq:1stordergaugetransformationforsurface}).
By substituting the necessary forms, this expression
on the left-hand-side of 
(\ref{eq:1stordergaugetransformationforsurface})
becomes (to avoid clutter, we have not explicitly
written $\D s$ and $\D t$)
\be
\begin{split}
&\a_{\exp\left\{-A'_{\nu}\big(x(s,t+\e)\big)
\frac{\p x^{\nu}}{\p s}\Big|_{(s,t+\e)}\right\}}
\left(\exp\left\{\varphi_{\mu}\big(x(s,t)\big)
\frac{\p x^{\mu}}{\p t}\Big|_{(s,t)}\right\}\right)\\
&\quad\times
\exp\left\{\varphi_{\rho}\big(x(s,t+\e)\big)
\frac{\p x^{\rho}}{\p s}\Big|_{(s,t+\e)}\right\}
\a_{g(x(s+\e,t+\e))}\left(\exp\left\{B_{\s\t}\big(x(s,t)\big)
\frac{\p x^{\s}}{\p s}
\frac{\p x^{\t}}{\p t}\Big|_{(s,t)}\right\}\right)\\
&=
\a_{\mathds{1}-A'_{\nu}\frac{\p x^{\nu}}{\p s}
-\frac{\p A'_{\nu}}{\p x^{\l}}\frac{\p x^{\l}}{\p t}\frac{\p x^{\nu}}{\p s}
-A'_{\nu}\frac{\p^2 x^{\nu}}{\p s\p t}
}
\left(\mathds{1}+\varphi_{\mu}\frac{\p x^{\mu}}{\p t}\right)\\
&\quad\times
\left(\mathds{1}+\varphi_{\rho}\frac{\p x^{\rho}}{\p s}
+\frac{\p\varphi_{\rho}}{\p x^{\b}}\frac{\p x^{\b}}{\p t}\frac{\p x^{\rho}}{\p s}
+\varphi_{\rho}\frac{\p^2 x^{\rho}}{\p s\p t}\right)
\a_{g+\frac{\p g}{\p s}+\frac{\p g}{\p t}+\frac{\p^2 g}{\p s\p t}}
\left(\mathds{1}+
B_{\s\t}\frac{\p x^{\s}}{\p s}\frac{\p x^{\t}}{\p t}\right)\\
&=\mathds{1}+\varphi_{\mu}\varphi_{\rho}\frac{\p x^{\mu}}{\p t}
\frac{\p x^{\rho}}{\p s}
+\frac{\p\varphi_{\rho}}{\p x^{\b}}\frac{\p x^{\b}}{\p t}\frac{\p x^{\rho}}{\p s}
+\varphi_{\rho}\frac{\p^2 x^{\rho}}{\p s\p t}
+\underline{\a_{g}}(B_{\s\t})\frac{\p x^{\s}}{\p s}\frac{\p x^{\t}}{\p t}
-\underline{\a}_{A'_{\nu}}(\varphi_{\mu})\frac{\p x^{\nu}}{\p s}\frac{\p x^{\mu}}{\p t}\\
&=\mathds{1}+\varphi_{\l}\frac{\p^2 x^{\l}}{\p s\p t}
+\left(\varphi_{\nu}\varphi_{\mu}
+\frac{\p\varphi_{\mu}}{\p x^{\nu}}
+\underline{\a_{g}}(B_{\mu\nu})
-\underline{\a}_{A'_{\mu}}(\varphi_{\nu})
\right)\frac{\p x^{\mu}}{\p s}\frac{\p x^{\nu}}{\p t},
\end{split}
\ee
where it is understood that all terms now are evaluated
at $(s,t),$ to lowest order. Meanwhile, the right-hand-side of 
(\ref{eq:1stordergaugetransformationforsurface})
is 
\be
\begin{split}
&\hspace{-5mm}\exp\left\{B'_{\s\t}\big(x(s,t)\big)
\frac{\p x^{\s}}{\p s}
\frac{\p x^{\t}}{\p t}\Big|_{(s,t)}\right\}
\a_{\exp\left\{-A'_{\nu}\big(x(s+\e,t)\big)
\frac{\p x^{\nu}}{\p t}\Big|_{(s+\e,t)}\right\}}
\left(\exp\left\{\varphi_{\mu}\big(x(s,t)\big)
\frac{\p x^{\mu}}{\p s}\Big|_{(s,t)}\right\}\right)\\
&\quad\times
\exp\left\{\varphi_{\l}\big(x(s+\e,t)\big)
\frac{\p x^{\l}}{\p t}\Big|_{(s+\e,t)}\right\}\\
&=\left(\mathds{1}+B'_{\s\t}\frac{\p x^{\s}}{\p s}
\frac{\p x^{\t}}{\p t}\right)
\a_{\mathds{1}-A'_{\nu}\frac{\p x^{\nu}}{\p t}
-\frac{\p A'_{\nu}}{\p x^{\b}}\frac{\p x^{\b}}{\p s}\frac{\p x^{\nu}}{\p t}
-A'_{\nu}\frac{\p^{2} x^{\nu}}{\p t\p s}}
\left(\mathds{1}+\varphi_{\mu}\frac{\p x^{\mu}}{\p s}\right)\\
&\quad\times
\left(\mathds{1}+\varphi_{\l}\frac{\p x^{\l}}{\p t}
+\frac{\p\varphi_{\l}}{\p x^{\a}}\frac{\p x^{\a}}{\p s}\frac{\p x^{\l}}{\p t}
+\varphi_{\l}\frac{\p^2 x^{\l}}{\p t\p s}\right)\\
&=\mathds{1}+B'_{\s\t}\frac{\p x^{\s}}{\p s}\frac{\p x^{\t}}{\p t}
+\varphi_{\mu}\varphi_{\l}\frac{\p x^{\mu}}{\p s}\frac{\p x^{\l}}{\p t}
-\un\a_{A'_{\nu}}(\varphi_{\mu})\frac{\p x^{\nu}}{\p t}\frac{\p x^{\mu}}{\p s}
+\frac{\p\varphi_{\l}}{\p x^{\a}}\frac{\p x^{\a}}{\p s}\frac{\p x^{\l}}{\p t}
+\varphi_{\l}\frac{\p^2 x^{\l}}{\p t\p s}\\
&=\mathds{1}+\varphi_{\l}\frac{\p^2 x^{\l}}{\p t\p s}
+\left(\varphi_{\mu}\varphi_{\nu}
+B'_{\mu\nu}
-\un\a_{A'_{\nu}}(\varphi_{\mu})
+\frac{\p\varphi_{\nu}}{\p x^{\mu}}
\right)\frac{\p x^{\mu}}{\p s}\frac{\p x^{\nu}}{\p t}
\end{split}
\ee
again to lowest order. 
Equating these two expressions gives 
\be
B'_{\mu\nu}=\underline{\a_{g}}(B_{\mu\nu})
-\vf_{\mu}\vf_{\nu}-\vf_{\nu}\vf_{\mu}
-\p_{\mu}\vf_{\nu}-\p_{\nu}\vf_{\mu}
-\un\a_{A'_{\mu}}(\vf_{\nu})+\un\a_{A'_{\nu}}(\vf_{\mu})
\ee
in components or 
\be
B'=\underline{\a_{g}}(B)-\vf\wedge\vf-d\vf-\un\a_{A'}(\vf)
\ee
as an equation in terms of differential forms. 
We write such a gauge transformation as
\be
\label{eq:1stordergt}
(A,B)\xrightarrow{(g,\varphi)}(A',B').
\ee
This and (\ref{eq:firstordergtA}) agrees with 
Proposition 2.10 of \cite{SW2}. We can express this
purely in terms of $A,$ $B,$ $g,$ and $\vf$
as
\be
\label{eq:BundergtintermsofABgvf}
\begin{split}
B'&=\underline{\a_{g}}(B)-\vf\wedge\vf-d\vf-
\un\a_{gAg^{-1}-dg g^{-1}-\un\t(\varphi)}(\vf)\\
&=\underline{\a_{g}}(B)-\vf\wedge\vf-d\vf-
\un\a_{gAg^{-1}-dg g^{-1}}(\vf)+[\vf,\vf]\\
&=\underline{\a_{g}}(B)+\vf\wedge\vf-d\vf-
\un\a_{gAg^{-1}-dg g^{-1}}(\vf).
\end{split}
\ee
This will be useful later.

Now suppose 
$(g,h),(g',h'):\triv\Rightarrow\triv'$ are two
first order gauge transformations. A
\emph{\uline{second order gauge transformation}}
$a:(g,h)\Rrightarrow(g',h')$ 
is a modification from $(g,h)$ to $(g',h').$ 
By Definition \ref{defn:modification}, this
consists of a function $a:M\to H$ fitting into
\begin{figure}[H]
\centering
\includegraphics[width=0.21\textwidth]{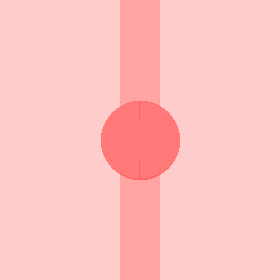}
\begin{picture}(0,0)
\put(-65,79){$g(x)$}
\put(-65,48){$a(x)$}
\put(-67,15){$g'(x)$}
\end{picture}
,
\end{figure}
\noindent
which in particular says
\be
\label{eq:2ndordergtsourcetarget}
\t(a)g=g',
\ee
satisfying the condition that to any path
$y\xleftarrow{\g}x,$ 
\be
h'(\g)a(y)=\a_{\triv'(\g)}\big(a(x)\big)h(\g).
\ee
Expanding out this expression on infinitesimal paths gives
\be
\left(\mathds{1}+\varphi'_{\mu}\frac{dx^{\mu}}{dt}\right)
\left(a+\frac{\p a}{\p x^{\nu}}\frac{d x^{\nu}}{d t}\right)
=\a_{\mathds{1}-A'_{\mu}\frac{dx^{\mu}}{dt}}(a)
\left(\mathds{1}+\varphi_{\nu}\frac{dx^{\nu}}{dt}\right),
\ee
which to lowest order says 
\be
\mathds{1}+\varphi'_{\mu}a\frac{dx^{\mu}}{dt}
+\frac{\p a}{\p x^{\nu}}\frac{d x^{\nu}}{d t}
=\mathds{1}+a\varphi_{\nu}\frac{dx^{\nu}}{dt}
-\un{\a_{a}}(A'_{\mu})\frac{dx^{\mu}}{dt}.
\ee
Note that if $h\in H,$ the function $\a_{h}:G\to H$ is defined
by $G\ni g\mapsto\a_{g}(h)$ so that $\un{\a_{h}}:\mathfrak{g}\to\mathfrak{h}$
is the derivative.  
This result gives the condition (after multiplying by $a^{-1}$ on the right) 
\be
\varphi'_{\mu}=a\varphi_{\mu}a^{-1}
-\un{\a_{a}}(A'_{\mu})a^{-1}
-(\p_{\mu}a) a^{-1}
\ee
on components and
\be
\varphi'=a\varphi a^{-1}-da a^{-1}-\un{\a_{a}}(A')a^{-1}
\ee
as $\mathfrak{h}$-valued differential forms.
This and (\ref{eq:2ndordergtsourcetarget}) 
exactly agree with Proposition 2.11 of \cite{SW2}. 
Physically, a second order gauge transformation 
is a gauge transformation for the 1-form $\vf,$ which itself
is a type of field even though it appears as a (fat) gauge transformation
for the 1-form and 2-form gauge potentials. 

\subsection{Orientations and inverses}
\label{sec:2dalgebraorient}
It is well-known that given a path $y\xleftarrow{\g}x,$ 
the parallel transport along the
reversed oriented path $\g^{-1}$ is the inverse
\be
\label{eq:1dparalleltransportinverse}
\triv(\g^{-1})=\triv(\g)^{-1},
\ee
where $\triv:\mathcal{P}^{1}M\to\B G$ is the (local) parallel
transport functor.
This can be viewed as a consequence of thin homotopy invariance
and functoriality of parallel transport. Namely, 
although the paths $\g\g^{-1}$ and $\g^{-1}\g$ are \emph{not}
the constant paths (the notation $\g^{-1}$ is therefore
a bit abusive), they are \emph{thinly homotopic} to 
constant paths and hence give the same value on $\triv.$ 
Thus, 
\be
\xy0;/r.30pc/:
(0,10)*+{\displaystyle\triv(\g^{-1}\g)}="1";
(-10,0)*+{\displaystyle\triv(\id_{x})}="4";
(10,0)*+{\quad\displaystyle\triv(\g^{-1})\triv(\g)}="2";
(0,-10)*+{\displaystyle1}="3";
{\ar@{=}@/^0.75pc/"1";"2"};
{\ar@{=}@/_0.75pc/"4";"3"};
{\ar@{=}@/^0.75pc/"4";"1"};
\endxy
\ee
verifying (\ref{eq:1dparalleltransportinverse}). 
In this section, we will explore analogous results for reversing
different kinds of orientations on bigons. We therefore
include arrows for clarity.
\begin{figure}[H]
\centering
\begin{subfigure}{0.32\textwidth}
\centering
\includegraphics[width=0.914\textwidth]{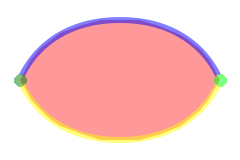}
\begin{picture}(0,0)
\put(-149,45){$y$}
\put(-80,92){$\g$}
\put(-80,81){$<$}
\put(-80,9){$<$}
\put(-80,-1){$\de$}
\put(-83,45){$\xy0;/r.25pc/:
(0,0)*+{\displaystyle\S}="S";
(0,9)*{}="g";
(0,-9)*{}="d";
{\ar@{=}"g";"S"};
{\ar@{=>}"S";"d"};
\endxy$}
\put(-8,45){$x$}
\end{picture}
\end{subfigure}
\begin{subfigure}{0.32\textwidth}
\centering
\includegraphics[width=0.914\textwidth]{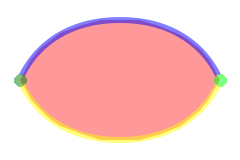}
\begin{picture}(0,0)
\put(-149,45){$y$}
\put(-80,92){$\g$}
\put(-80,81){$<$}
\put(-80,9){$<$}
\put(-80,-1){$\de$}
\put(-83,45){$\xy0;/r.25pc/:
(0,0)*+{\displaystyle\overline{\S}}="S";
(0,9)*{}="g";
(0,-9)*{}="d";
{\ar@{=}"d";"S"};
{\ar@{=>}"S";"g"};
\endxy$}
\put(-8,45){$x$}
\end{picture}
\end{subfigure}
\begin{subfigure}{0.32\textwidth}
\centering
\includegraphics[width=0.914\textwidth]{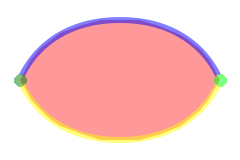}
\begin{picture}(0,0)
\put(-149,45){$y$}
\put(-80,92){$\g^{-1}$}
\put(-80,81){$>$}
\put(-80,9){$>$}
\put(-80,-1){$\de^{-1}$}
\put(-88,45){$\xy0;/r.25pc/:
(0,0)*+{\displaystyle\S^{-1}}="S";
(0,9)*{}="g";
(0,-9)*{}="d";
{\ar@{=}"g";"S"};
{\ar@{=>}"S";"d"};
\endxy$}
\put(-8,45){$x$}
\end{picture}
\end{subfigure}
\end{figure}
\noindent
Technically, there is one more possibility given by 
$\un{\S}:\de^{-1}\Rightarrow\g^{-1}.$ However,
this possibility is a combination of the above two, namely
$\un{\S}=\ov{\S^{-1}}.$ 
The meaning of these different possibilities is given physically as follows.
A given string may be given the additional datum of an orientation. 
Furthermore, as it moves in time, it has an additional directionality. 
These two directional orientations 
are precisely encoded in the definition of a bigon/2-morphism.
These different orientation reversals are given by time reversal 
for $\ov\S$ and spatial orientation reversal for $\S^{-1}.$ 
The case $\un\S$ corresponds to both reversals 
(the order of reversal does not matter since the operations commute).
Note that the different orientations on a bigon
can be expressed as an orientation of edges on the boundary
and an orientation of the surface. The above bigons
correspond to the following surfaces with associated
orientations
\begin{figure}[H]
\centering
\begin{subfigure}{0.32\textwidth}
\includegraphics[width=0.914\textwidth]{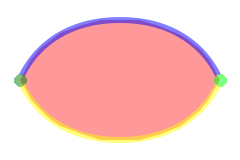}
\begin{picture}(0,0)
\put(-149,45){$y$}
\put(-80,92){$\g$}
\put(-80,81){$<$}
\put(-80,9){$<$}
\put(-80,-1){$\de$}
\put(-60,60){$\circlearrowleft$}
\put(-80,45){$\S$}
\put(-8,45){$x$}
\end{picture}
\end{subfigure}
\begin{subfigure}{0.32\textwidth}
\centering
\includegraphics[width=0.914\textwidth]{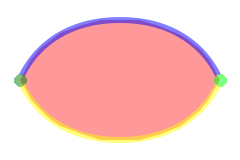}
\begin{picture}(0,0)
\put(-149,45){$y$}
\put(-80,92){$\g$}
\put(-80,81){$<$}
\put(-80,9){$<$}
\put(-80,-1){$\de$}
\put(-60,60){$\circlearrowright$}
\put(-80,45){$\ov\S$}
\put(-8,45){$x$}
\end{picture}
\end{subfigure}
\begin{subfigure}{0.32\textwidth}
\centering
\includegraphics[width=0.914\textwidth]{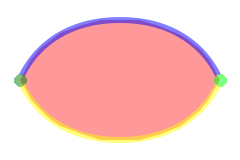}
\begin{picture}(0,0)
\put(-149,45){$y$}
\put(-80,92){$\g^{-1}$}
\put(-80,81){$>$}
\put(-80,9){$>$}
\put(-80,-1){$\de^{-1}$}
\put(-60,60){$\circlearrowright$}
\put(-85,45){$\S^{-1}$}
\put(-8,45){$x$}
\end{picture}
\end{subfigure}
\end{figure}
\noindent
respectively. 

A necessary and sufficient condition for such orientations 
on surfaces and edges to give rise to
a bigon is the following. Given a map of a polygon $\S$
into $M,$ the boundary consists of the edges of the
polygon. The union of the oriented edges consistent with 
the orientation of the polygon must be connected. Similarly,
the union of the orientated edges with negative orientation
with respect to the induced one from the polygon must
also be connected. Then, the source of the bigon is the
union of the consistent edges and the target is the
union of the oppositely oriented edges.
An example together with a non-example are
\begin{figure}[H]
\centering
\begin{subfigure}{0.35\textwidth}
\centering
\raisebox{0.5em}{\includegraphics[width=0.8\textwidth]{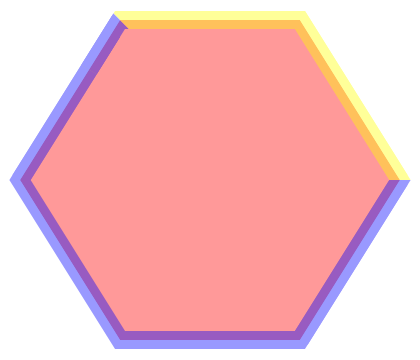}}
\begin{picture}(0,7)
\put(-122,40){\rotatebox{-60}{$>$}}
\put(-77,114){$\gg$}
\put(-32,93){\rotatebox{-60}{$\gg$}}
\put(-77,10){$>$}
\put(-123,86){\rotatebox{61}{$<$}}
\put(-74,63){$\circlearrowleft$}
\put(-32,44){\rotatebox{-119}{$<$}}
\end{picture}
\end{subfigure}
\&
\begin{subfigure}{0.35\textwidth}
\centering
\includegraphics[width=0.8\textwidth]{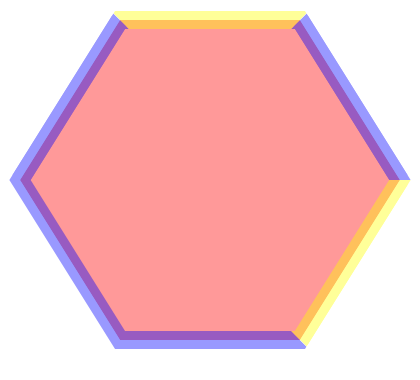}
\begin{picture}(0,0)
\put(-122,40){\rotatebox{-60}{$>$}}
\put(-77,114){$\gg$}
\put(-32,93){\rotatebox{-60}{$<$}}
\put(-77,10){$>$}
\put(-123,86){\rotatebox{61}{$<$}}
\put(-74,63){$\circlearrowleft$}
\put(-32,44){\rotatebox{-119}{$\gg$}}
\end{picture}
\end{subfigure}
,
\end{figure}
\noindent
respectively (blue, with arrows written using $>,$ 
corresponds to an orientation agreeing
with the induced one from the surface while yellow, 
with arrows written using $\gg,$
disagrees with that orientation). 

Going back to the three bigons and their orientations at
the beginning of this section, we notice that several
of these bigons can be composed with one another. 
For instance, 
\begin{figure}[H]
\centering
\includegraphics[width=0.65\textwidth]{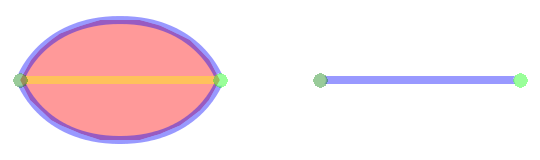}
\begin{picture}(0,0)
\put(-323,45){$y$}
\put(-256,10){$<$}
\put(-256,1){$\g$}
\put(-256,45){$<$}
\put(-256,24){$\ov\S$}
\put(-271,35){$\de$}
\put(-256,80){$<$}
\put(-256,90){$\g$}
\put(-256,62){$\S$}
\put(-186,45){$x$}
\put(-166,45){$=$}
\put(-144,45){$y$}
\put(-77,58){$\g$}
\put(-77,45){$<$}
\put(-9,45){$x$}
\end{picture}
\end{figure}
\noindent
and
\begin{figure}[H]
\centering
\includegraphics[width=0.65\textwidth]{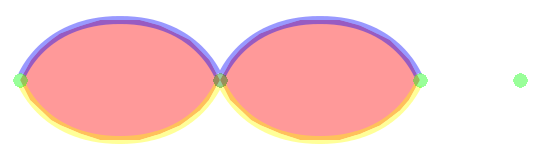}
\begin{picture}(0,0)
\put(-323,45){$x$}
\put(-256,10){$<$}
\put(-256,1){$\de$}
\put(-256,80){$<$}
\put(-256,90){$\g$}
\put(-256,45){$\S$}
\put(-196,33){$y$}
\put(-139,10){$<$}
\put(-139,1){$\de^{-1}$}
\put(-139,80){$<$}
\put(-139,90){$\g^{-1}$}
\put(-139,45){$\S^{-1}$}
\put(-44,45){$=$}
\put(-68,45){$x$}
\put(-9,45){$x$}
\end{picture}
\end{figure}
\noindent
after applying thin homotopies. Therefore, 
these bigons provide inverses in series and
in parallel, respectively, of $\S.$ This implies,
together with functoriality of $\triv$ and
the inverses discussed in Example \ref{ex:2groups},
\be
\triv\big(\ov\S\big)=\triv(\S)^{-1}
\aand
\triv\big(\S^{-1}\big)=\a_{\triv(\g)^{-1}}\big(\triv(\S)^{-1}\big)
\ee
and therefore describes how parallel transport
along surfaces changes under reversals in surface
orientations and boundary orientations, respectively. 
For completeness, 
for the last possible orientation $\un\S:\de^{-1}\Rightarrow\g^{-1},$
we have
\be
\triv\left(\un\S\right)=\triv\left(\ov{\S^{-1}}\right)
=\left(\a_{\triv(\g)^{-1}}\big(\triv(\S)\big)^{-1}\right)^{-1}
=\a_{\triv(\g)^{-1}}\big(\triv(\S)\big). 
\ee

\subsection{The 3-curvature}
\label{sec:2dalgebra3curv}
In the following, we make some further calculations.
Just as the curvature $F$ of a 1-form connection $A$ 
can be obtained by calculating the parallel transport
along an infinitesimal loop, the 2-curvature
of a 2-form connection $(A,B)$ can be obtained
by calculating the surface transport along an 
infinitesimal sphere, which on a lattice corresponds
to a cube. We will perform this calculation explicitly
and study some properties of the resulting 3-form curvature.
Similar analysis was done on a tetrahedron in
\cite{GP}.

Let $(r,s,t)\mapsto x(r,s,t)$ be an infinitesimal cube and
consider the following domain for that cube along with 
the infinitesimal path that goes first along the $r$ direction,
then in the $s$ direction, and finally in the $t$ direct. 
Our convention is that $(r,s,t)$ is a right-handed
coordinate frame, i.e. $dr\wedge ds\wedge dt$ is the
volume form. 
\begin{figure}[H]
\centering
\includegraphics[width=0.35\textwidth]{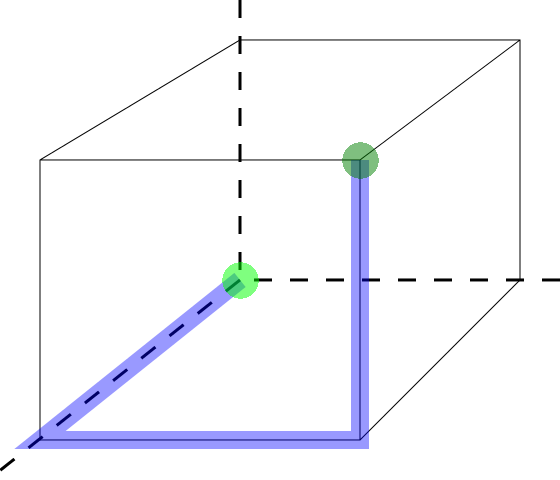}
\begin{picture}(0,0)
\put(-184,-2){$r$}
\put(-1,59){$s$}
\put(-104,152){$t$}
\put(-145,62){$(0,0,0)$}
\put(-58,87){$(\D r,\D s,\D t)$}
\put(-136,34){\rotatebox{45}{\Large$<$}}
\put(-118,8){\Large$>$}
\put(-70,47){\rotatebox{90}{\Large$>$}}
\end{picture}
\end{figure}
Such a cube can be expressed as a bigon by the following
composition of plaquette bigons 
that begin and end at the same path starting at the top left
and moving clockwise. 
\begin{figure}[H]
\centering
\begin{subfigure}{0.3\textwidth}
\centering
\includegraphics[width=0.9\textwidth]{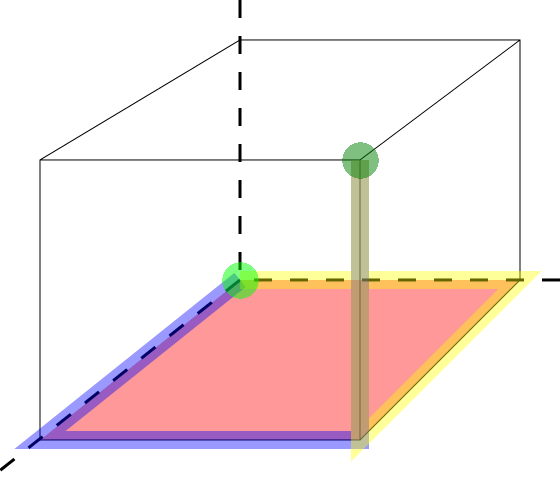}
\begin{picture}(0,0)
\put(-145,-3){$r$}
\put(0,45){$s$}
\put(-81,120){$t$}
\put(-75,24){\rotatebox{25}{$\Rightarrow$}}
\end{picture}
\end{subfigure}
$\rightsquigarrow$
\begin{subfigure}{0.3\textwidth}
\centering
\includegraphics[width=0.9\textwidth]{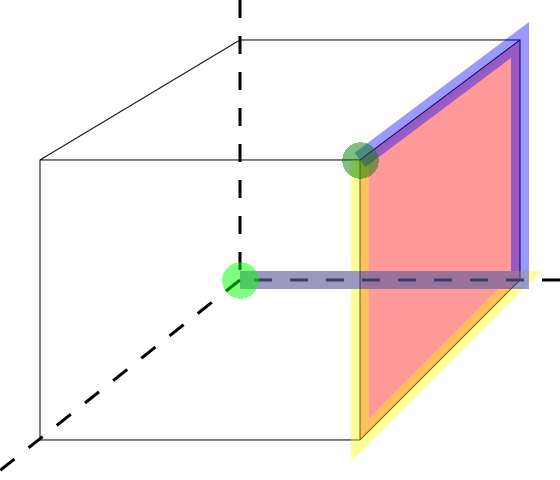}
\begin{picture}(0,0)
\put(-145,-3){$r$}
\put(0,45){$s$}
\put(-81,120){$t$}
\put(-36,54){\rotatebox{65}{$\Rightarrow$}}
\end{picture}
\end{subfigure}
$\rightsquigarrow$
\begin{subfigure}{0.3\textwidth}
\centering
\includegraphics[width=0.9\textwidth]{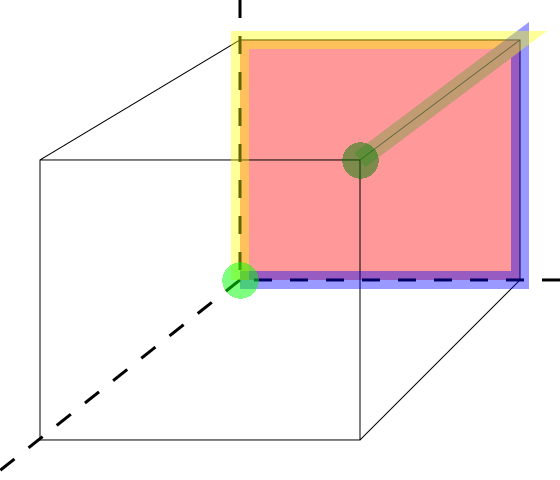}
\begin{picture}(0,0)
\put(-145,-3){$r$}
\put(0,45){$s$}
\put(-81,120){$t$}
\put(-44,69){\rotatebox{135}{$\Rightarrow$}}
\end{picture}
\end{subfigure}
\end{figure} 
\noindent
\begin{figure}[H]
\centering
\begin{subfigure}{0.3\textwidth}
\centering
\includegraphics[width=0.9\textwidth]{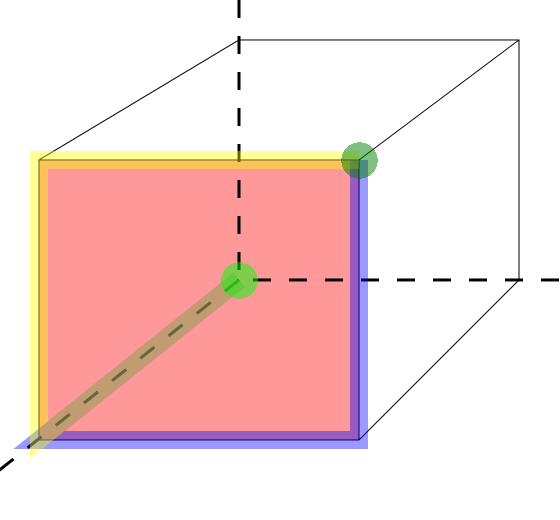}
\begin{picture}(0,0)
\put(-145,-3){$r$}
\put(0,45){$s$}
\put(-81,120){$t$}
\put(-95,52){\rotatebox{317}{$\Rightarrow$}}
\end{picture}
\end{subfigure}
\rotatebox{180}{$\rightsquigarrow$}
\begin{subfigure}{0.3\textwidth}
\centering
\includegraphics[width=0.9\textwidth]{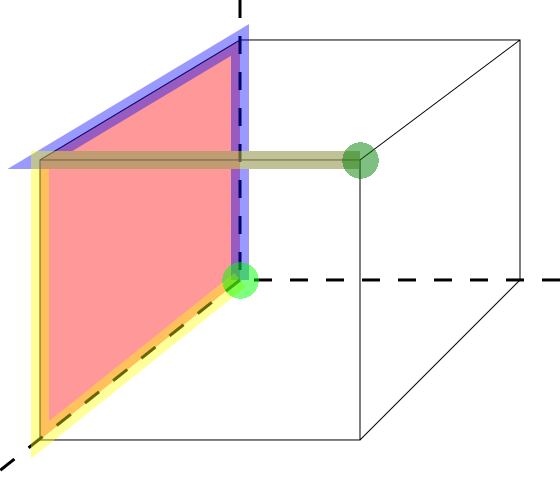}
\begin{picture}(0,0)
\put(-145,-3){$r$}
\put(0,45){$s$}
\put(-81,120){$t$}
\put(-108,64){\rotatebox{243}{$\Rightarrow$}}
\end{picture}
\end{subfigure}
\rotatebox{180}{$\rightsquigarrow$}
\begin{subfigure}{0.3\textwidth}
\centering
\includegraphics[width=0.9\textwidth]{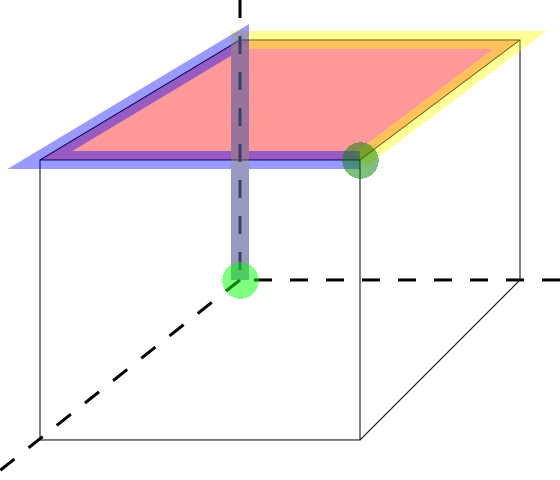}
\begin{picture}(0,0)
\put(-65,135){\rotatebox{-90}{$\rightsquigarrow$}}
\put(-145,-3){$r$}
\put(0,45){$s$}
\put(-81,120){$t$}
\put(-73,95){\rotatebox{200}{$\Rightarrow$}}
\end{picture}
\end{subfigure}
\end{figure} 
\noindent
The corresponding 2-group elements are given as
follows. We begin with the first surface 
\begin{figure}[H]
\centering
\begin{subfigure}{0.3\textwidth}
\centering
\includegraphics[width=0.9\textwidth]{cubecurvature1}
\begin{picture}(0,0)
\put(-145,-3){$r$}
\put(0,45){$s$}
\put(-81,120){$t$}
\put(-75,24){\rotatebox{25}{$\Rightarrow$}}
\end{picture}
\end{subfigure}
$\leftrightarrow$ 
\begin{subfigure}{0.6\textwidth}
\centering
\includegraphics[width=0.9\textwidth]{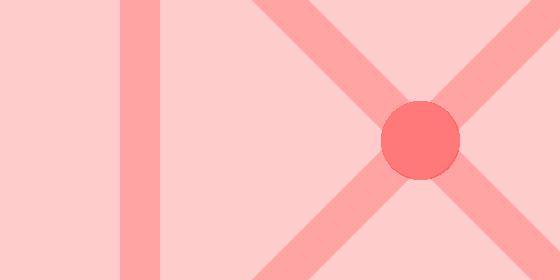}
\begin{picture}(0,0)
\put(-235,62){$e^{-A_{\mu}\p_{t}x^{\mu}|_{(\e,\e,0)}}$}
\put(-140,100){$e^{-A_{\nu}\p_{s}x^{\nu}|_{(\e,0,0)}}$}
\put(-60,100){$e^{-A_{\l}\p_{r}x^{\l}|_{(0,0,0)}}$}
\put(-105,62){$e^{B_{\l\nu}\p_{r}x^{\l}\p_{s}x^{\nu}|_{(0,0,0)}}$}
\put(-140,20){$e^{-A_{\r}\p_{r}x^{\r}|_{(0,\e,0)}}$}
\put(-60,20){$e^{-A_{\s}\p_{s}x^{\s}|_{(0,0,0)}}$}
\end{picture}
\end{subfigure}
\end{figure} 
\noindent
where we use the shorthand notation 
\be
\p_{r} x:=\frac{\p x}{\p r},\qquad
\p_{s} x:=\frac{\p x}{\p s},\aand
\p_{t} x:=\frac{\p x}{\p t}
\ee
as well as
\be
e^{-A_{\mu}\p_{t}x^{\mu}|_{(\e,\e,0)}}
:=
\exp\left\{-A_{\mu}\big(x(\e,\e,0)\big)
\frac{\p x^{\mu}}{\p t}\Big|_{(\e,\e,0)}\right\}
\ee
and similarly for the other terms. 
We also write $\e$ instead of $\D r, \D s,$ or $\D t$
and use the derivatives to remind ourselves of the direction.
We have also assumed for simplicity that our coordinates are
centered at the origin and the lattice spacing is $\e$ in each 
direction. Working out this diagram infinitesimally on
the 0-d defect gives
\be
\a_{e^{-A_{\mu}\p_{t}x^{\mu}|_{(\e,\e,0)}}}
\left(e^{B_{\l\nu}\p_{r}x^{\l}\p_{s}x^{\nu}|_{(0,0,0)}}\right)
=\mathds{1}+B_{\l\nu}\p_{r}x^{\l}\p_{s}x^{\nu}
-\un\a_{A_{\mu}}(B_{\l\nu})\p_{t}x^{\mu}\p_{r}x^{\l}\p_{s}x^{\nu}
\ee
to lowest order. As usual, rather than writing out the
$\D r,\D s,\D t,$ we use the number and type of
derivatives appearing to keep track of the order. 
The other terms are given by the following
\begin{figure}[H]
\centering
\begin{subfigure}{0.3\textwidth}
\centering
\includegraphics[width=0.9\textwidth]{cubecurvature2}
\begin{picture}(0,0)
\put(-145,-3){$r$}
\put(0,45){$s$}
\put(-81,120){$t$}
\put(-36,54){\rotatebox{65}{$\Rightarrow$}}
\end{picture}
\end{subfigure}
$\leftrightarrow$ 
\begin{subfigure}{0.6\textwidth}
\centering
\includegraphics[width=0.9\textwidth]{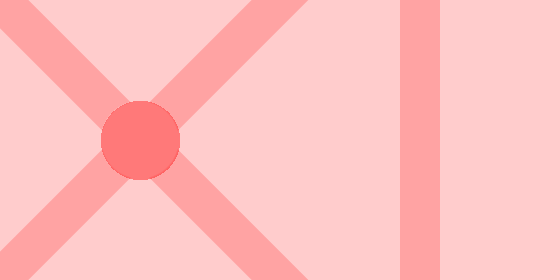}
\begin{picture}(0,0)
\put(-270,100){$e^{-A_{\mu}\p_{t}x^{\mu}|_{(\e,\e,0)}}$}
\put(-185,100){$e^{-A_{\r}\p_{r}x^{\r}|_{(0,\e,0)}}$}
\put(-245,62){$e^{B_{\r\pi}\p_{r}x^{\r}\p_{t}x^{\pi}|_{(0,\e,0)}}$}
\put(-100,62){$e^{-A_{\s}\p_{s}x^{\s}|_{(0,0,0)}}$}
\put(-270,20){$e^{-A_{\a}\p_{r}x^{\a}|_{(0,\e,\e)}}$}
\put(-185,20){$e^{-A_{\t}\p_{t}x^{\t}|_{(0,\e,0)}}$}
\end{picture}
\end{subfigure}
\end{figure} 
\be
\begin{split}
e^{B_{\r\pi}\p_{r}x^{\r}\p_{t}x^{\pi}|_{(0,\e,0)}}
&=\mathds{1}+B_{\r\pi}\p_{r}x^{\r}\p_{t}x^{\pi}
+\p_{c}B_{\r\pi}\p_{s}x^{c}\p_{r}x^{\r}\p_{t}x^{\pi}\\
&\quad+B_{\r\pi}\p_{s}\p_{r}x^{\r}\p_{t}x^{\pi}
+B_{\r\pi}\p_{r}x^{\r}\p_{s}\p_{t}x^{\pi}
\end{split}
\ee
\begin{figure}[H]
\centering
\begin{subfigure}{0.3\textwidth}
\centering
\includegraphics[width=0.9\textwidth]{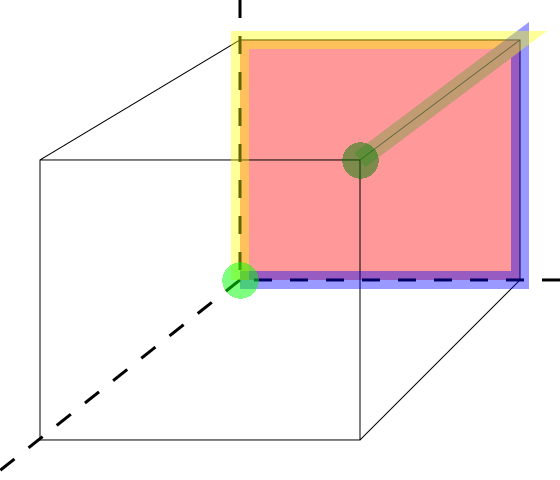}
\begin{picture}(0,0)
\put(-145,-3){$r$}
\put(0,45){$s$}
\put(-81,120){$t$}
\put(-44,69){\rotatebox{135}{$\Rightarrow$}}
\end{picture}
\end{subfigure}
$\leftrightarrow$ 
\begin{subfigure}{0.6\textwidth}
\centering
\includegraphics[width=0.9\textwidth]{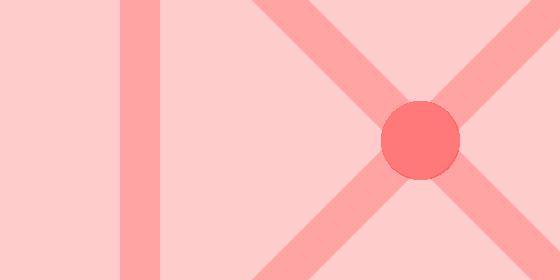}
\begin{picture}(0,0)
\put(-235,62){$e^{-A_{\a}\p_{r}x^{\a}|_{(0,\e,\e)}}$}
\put(-140,100){$e^{-A_{\t}\p_{t}x^{\t}|_{(0,\e,0)}}$}
\put(-60,100){$e^{-A_{\s}\p_{s}x^{\s}|_{(0,0,0)}}$}
\put(-105,62){$e^{B_{\s\t}\p_{s}x^{\s}\p_{t}x^{\t}|_{(0,0,0)}}$}
\put(-140,20){$e^{-A_{\b}\p_{s}x^{\b}|_{(0,0,\e)}}$}
\put(-60,20){$e^{-A_{\g}\p_{t}x^{\g}|_{(0,0,0)}}$}
\end{picture}
\end{subfigure}
\end{figure}
\be
\a_{e^{-A_{\a}\p_{r}x^{\a}|_{(0,0,\e)}}}
\left(e^{B_{\s\t}\p_{s}x^{\s}\p_{t}x^{\t}|_{(0,0,0)}}\right)
=
\mathds{1}+B_{\s\t}\p_{s}x^{\s}\p_{t}x^{\t}
-\un\a_{A_{\a}}(B_{\s\t})\p_{r}x^{\a}\p_{s}x^{\s}\p_{t}x^{\t}
\ee
\begin{figure}[H]
\centering
\begin{subfigure}{0.3\textwidth}
\centering
\includegraphics[width=0.9\textwidth]{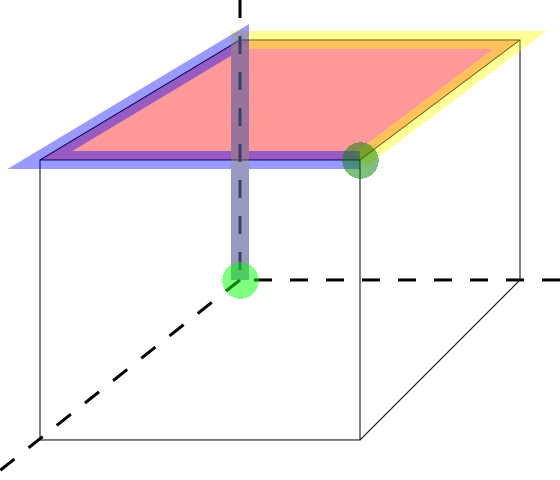}
\begin{picture}(0,0)
\put(-145,-3){$r$}
\put(0,45){$s$}
\put(-81,120){$t$}
\put(-73,95){\rotatebox{200}{$\Rightarrow$}}
\end{picture}
\end{subfigure}
$\leftrightarrow$ 
\begin{subfigure}{0.6\textwidth}
\centering
\includegraphics[width=0.9\textwidth]{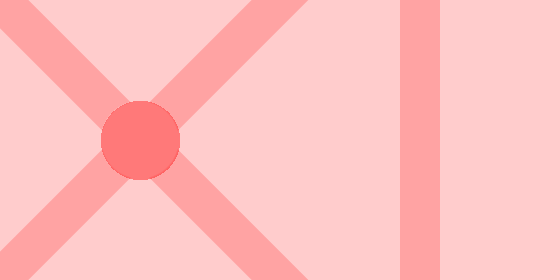}
\begin{picture}(0,0)
\put(-270,100){$e^{-A_{\a}\p_{r}x^{\a}|_{(0,\e,\e)}}$}
\put(-185,100){$e^{-A_{\b}\p_{s}x^{\b}|_{(0,0,\e)}}$}
\put(-245,62){$e^{B_{\b\k}\p_{s}x^{\b}\p_{r}x^{\k}|_{(0,0,\e)}}$}
\put(-100,62){$e^{-A_{\g}\p_{t}x^{\g}|_{(0,0,0)}}$}
\put(-270,20){$e^{-A_{\q}\p_{s}x^{\q}|_{(\e,0,\e)}}$}
\put(-185,20){$e^{-A_{\h}\p_{r}x^{\h}|_{(0,0,\e)}}$}
\end{picture}
\end{subfigure}
\end{figure} 
\be
\begin{split}
e^{B_{\b\k}\p_{s}x^{\b}\p_{r}x^{\k}|_{(0,0,\e)}}
&=\mathds{1}+B_{\b\k}\p_{s}x^{\b}\p_{r}x^{\k}
+\p_{b}B_{\b\k}\p_{t}x^{b}\p_{s}x^{\b}\p_{r}x^{\k}\\
&\quad+B_{\b\k}\p_{t}\p_{s}x^{\b}\p_{r}x^{\k}
+B_{\b\k}\p_{s}x^{\b}\p_{t}\p_{r}x^{\k}
\end{split}
\ee
\begin{figure}[H]
\centering
\begin{subfigure}{0.3\textwidth}
\centering
\includegraphics[width=0.9\textwidth]{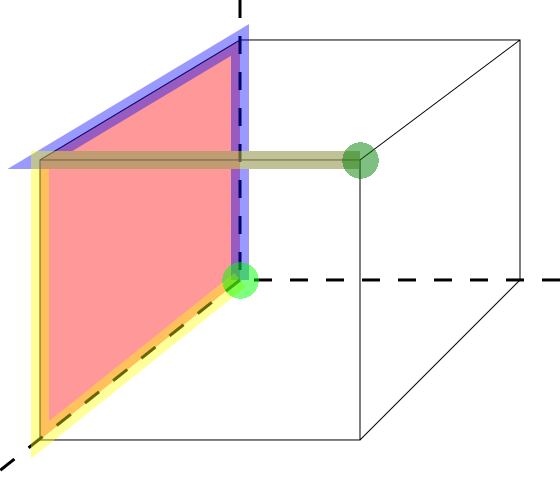}
\begin{picture}(0,0)
\put(-145,-3){$r$}
\put(0,45){$s$}
\put(-81,120){$t$}
\put(-108,64){\rotatebox{243}{$\Rightarrow$}}
\end{picture}
\end{subfigure}
$\leftrightarrow$ 
\begin{subfigure}{0.6\textwidth}
\centering
\includegraphics[width=0.9\textwidth]{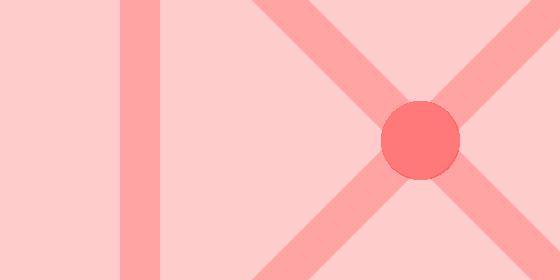}
\begin{picture}(0,0)
\put(-235,62){$e^{-A_{\q}\p_{s}x^{\q}|_{(\e,0,\e)}}$}
\put(-140,100){$e^{-A_{\h}\p_{r}x^{\h}|_{(0,0,\e)}}$}
\put(-60,100){$e^{-A_{\g}\p_{t}x^{\g}|_{(0,0,0)}}$}
\put(-105,62){$e^{B_{\g\h}\p_{t}x^{\g}\p_{r}x^{\h}|_{(0,0,0)}}$}
\put(-140,20){$e^{-A_{\w}\p_{t}x^{\w}|_{(\e,0,0)}}$}
\put(-60,20){$e^{-A_{\l}\p_{r}x^{\l}|_{(0,0,0)}}$}
\end{picture}
\end{subfigure}
\end{figure}
\be
\a_{e^{-A_{\q}\p_{s}x^{\q}|_{(\e,0,\e)}}}
\left(e^{B_{\g\h}\p_{t}x^{\g}\p_{r}x^{\h}|_{(0,0,0)}}\right)
=\mathds{1}+B_{\g\h}\p_{t}x^{\g}\p_{r}x^{\h}
-\un\a_{A_{\q}}(B_{\g\h})\p_{s}x^{\q}\p_{t}x^{\g}\p_{r}x^{\h}
\ee
\begin{figure}[H]
\centering
\begin{subfigure}{0.3\textwidth}
\centering
\includegraphics[width=0.9\textwidth]{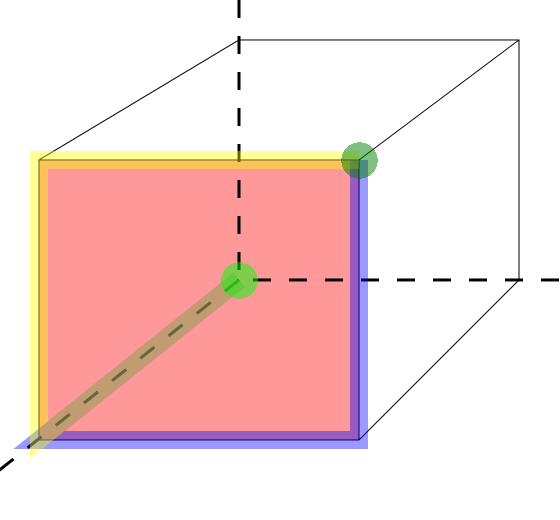}
\begin{picture}(0,0)
\put(-145,-3){$r$}
\put(0,45){$s$}
\put(-81,120){$t$}
\put(-95,52){\rotatebox{317}{$\Rightarrow$}}
\end{picture}
\end{subfigure}
$\leftrightarrow$ 
\begin{subfigure}{0.6\textwidth}
\centering
\includegraphics[width=0.9\textwidth]{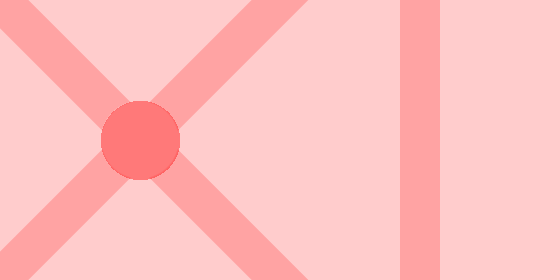}
\begin{picture}(0,0)
\put(-270,100){$e^{-A_{\q}\p_{s}x^{\q}|_{(\e,0,\e)}}$}
\put(-185,100){$e^{-A_{\w}\p_{t}x^{\w}|_{(\e,0,0)}}$}
\put(-245,62){$e^{B_{\w\psi}\p_{t}x^{\w}\p_{s}x^{\psi}|_{(\e,0,0)}}$}
\put(-100,62){$e^{-A_{\l}\p_{r}x^{\l}|_{(0,0,0)}}$}
\put(-270,20){$e^{-A_{\mu}\p_{t}x^{\mu}|_{(\e,\e,0)}}$}
\put(-185,20){$e^{-A_{\nu}\p_{s}x^{\nu}|_{(\e,0,0)}}$}
\end{picture}
\end{subfigure}
\end{figure} 
\be
\begin{split}
e^{B_{\w\psi}\p_{t}x^{\w}\p_{s}x^{\psi}|_{(\e,0,0)}}
&=\mathds{1}+B_{\w\psi}\p_{t}x^{\w}\p_{s}x^{\psi}
+\p_{a}B_{\w\psi}\p_{r}x^{a}\p_{t}x^{\w}\p_{s}x^{\psi}\\
&\quad+B_{\w\psi}\p_{r}\p_{t}x^{\w}\p_{s}x^{\psi}
+B_{\w\psi}\p_{t}x^{\w}\p_{r}\p_{s}x^{\psi}
\end{split}
\ee
The composition of all of these elements
is given by the following diagram
(with the light shaded blue squares depicting
the faces on the cube). 
\begin{figure}[H]
\centering
\includegraphics[width=0.4\textwidth]{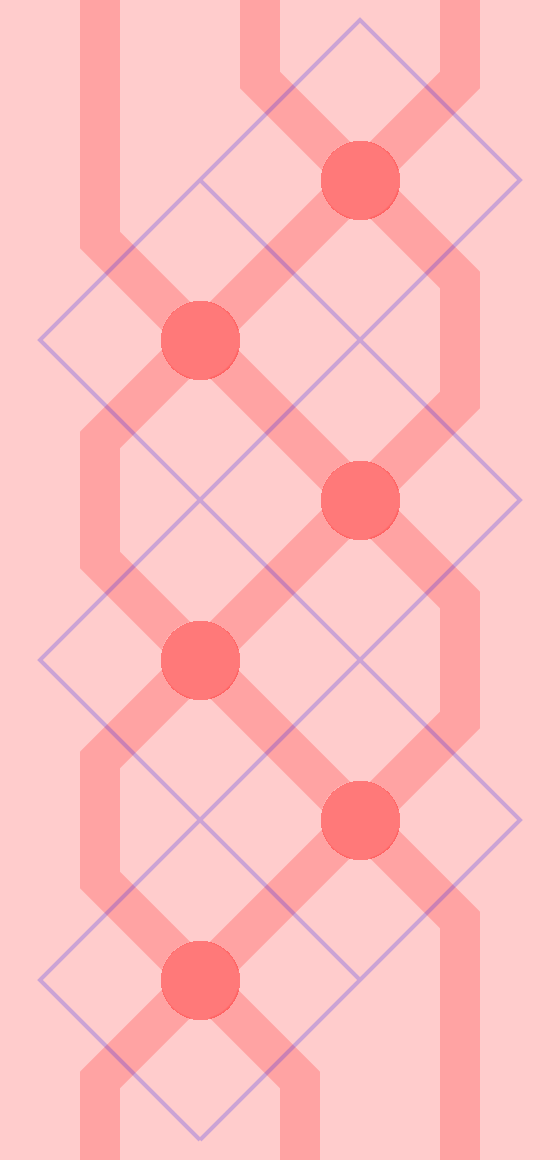}
\begin{picture}(0,0)
\put(-193,338){$e^{-A_{\mu}\p_{t}x^{\mu}|_{(\e,\e,0)}}$}
\put(-115,338){$e^{B_{\l\nu}\p_{r}x^{\l}\p_{s}x^{\nu}|_{(0,0,0)}}$}
\put(-175,282){$e^{B_{\r\pi}\p_{r}x^{\r}\p_{t}x^{\pi}|_{(0,\e,0)}}$}
\put(-193,226){$e^{-A_{\a}\p_{r}x^{\a}|_{(0,0,\e)}}$}
\put(-115,226){$e^{B_{\s\t}\p_{s}x^{\s}\p_{t}x^{\t}|_{(0,0,0)}}$}
\put(-175,170){$e^{B_{\b\k}\p_{s}x^{\b}\p_{r}x^{\k}|_{(0,0,\e)}}$}
\put(-193,114){$e^{-A_{\q}\p_{s}x^{\q}|_{(\e,0,\e)}}$}
\put(-115,114){$e^{B_{\g\h}\p_{t}x^{\g}\p_{r}x^{\h}|_{(0,0,0)}}$}
\put(-175,58){$e^{B_{\w\psi}\p_{t}x^{\w}\p_{s}x^{\psi}|_{(\e,0,0)}}$}
\end{picture}
\end{figure} 
\noindent
And the result of multiplying these out gives
\be
\begin{split}
e^{B_{\w\psi}\p_{t}x^{\w}\p_{s}x^{\psi}|_{(\e,0,0)}}
\a_{e^{-A_{\q}\p_{s}x^{\q}|_{(\e,0,\e)}}}
\left(e^{B_{\g\h}\p_{t}x^{\g}\p_{r}x^{\h}|_{(0,0,0)}}\right)
e^{B_{\b\k}\p_{s}x^{\b}\p_{r}x^{\k}|_{(0,0,\e)}}\\
\times
\a_{e^{-A_{\a}\p_{r}x^{\a}|_{(0,0,\e)}}}
\left(e^{B_{\s\t}\p_{s}x^{\s}\p_{t}x^{\t}|_{(0,0,0)}}\right)
e^{B_{\r\pi}\p_{r}x^{\r}\p_{t}x^{\pi}|_{(0,\e,0)}}
\a_{e^{-A_{\mu}\p_{t}x^{\mu}|_{(\e,\e,0)}}}
\left(e^{B_{\l\nu}\p_{r}x^{\l}\p_{s}x^{\nu}|_{(0,0,0)}}\right).
\end{split}
\ee
This is yet another manifestation of two-dimensional
algebra. 
The result of multiplying all these terms is given as follows,
order by order. 
The zeroth order term is $\mathds{1}.$ There are no first
order terms. The second order terms are given by
\be
\begin{split}
B_{\w\psi}\p_{t}x^{\w}\p_{s}x^{\psi}
&+B_{\g\h}\p_{t}x^{\g}\p_{r}x^{\h}
+B_{\b\k}\p_{s}x^{\b}\p_{r}x^{\k}
+B_{\s\t}\p_{s}x^{\s}\p_{t}x^{\t}
+B_{\r\pi}\p_{r}x^{\r}\p_{t}x^{\pi}
+B_{\l\nu}\p_{r}x^{\l}\p_{s}x^{\nu}
\\
&=(B_{\s\t}+B_{\t\s})\p_{s}x^{\s}\p_{t}x^{\t}
+(B_{\l\nu}+B_{\nu\l})\p_{r}x^{\l}\p_{s}x^{\nu}
+(B_{\rho\pi}+B_{\pi\rho})\p_{r}x^{\r}\p_{t}x^{\pi}\\
&=0
\end{split}
\ee
by anti-symmetry of $B_{\mu\nu}$ in the 
$\mu$ and $\nu$ indices. Thus, the only non-zero
terms are the zeroth and third order terms (up to third order).
One type of the third order terms are given by 
\be
\begin{split}
\Big(&B_{\w\psi}\p_{r}\p_{t}x^{\w}\p_{s}x^{\psi}
+B_{\w\psi}\p_{t}x^{\w}\p_{r}\p_{s}x^{\psi}
+B_{\b\k}\p_{t}\p_{s}x^{\b}\p_{r}x^{\k}\Big)\\
+\Big(&
B_{\b\k}\p_{s}x^{\b}\p_{t}\p_{r}x^{\k}
+B_{\r\pi}\p_{s}\p_{r}x^{\r}\p_{t}x^{\pi}
+B_{\r\pi}\p_{r}x^{\r}\p_{s}\p_{t}x^{\pi}
\Big)\\
&=(B_{\b\k}+B_{\k\b})\p_{s}x^{\b}\p_{t}\p_{r}x^{\k}
+(B_{\w\psi}+B_{\psi\w})\p_{t}x^{\w}\p_{r}\p_{s}x^{\psi}
+(B_{\r\pi}+B_{\pi\r})\p_{r}x^{\r}\p_{s}\p_{t}x^{\pi}
\end{split}
\ee
and vanish again by anti-symmetry of $B_{\mu\nu}$
and commutativity of partial derivatives.  
The final result is therefore 
\be
\begin{split}
\mathds{1}&+\p_{a}B_{\w\psi}\p_{r}x^{a}\p_{t}x^{\w}\p_{s}x^{\psi}
-\un\a_{A_{\q}}(B_{\g\h})\p_{s}x^{\q}\p_{t}x^{\g}\p_{r}x^{\h}
+\p_{b}B_{\b\k}\p_{t}x^{b}\p_{s}x^{\b}\p_{r}x^{\k}\\
&-\un\a_{A_{\a}}(B_{\s\t})\p_{r}x^{\a}\p_{s}x^{\s}\p_{t}x^{\t}
+\p_{c}B_{\r\pi}\p_{s}x^{c}\p_{r}x^{\r}\p_{t}x^{\pi}\
-\un\a_{A_{\mu}}(B_{\l\nu})\p_{t}x^{\mu}\p_{r}x^{\l}\p_{s}x^{\nu}\\
&=\mathds{1}-\Big(\p_{\mu}B_{\nu\l}+\p_{\l}B_{\mu\nu}+\p_{\nu}B_{\l\mu}
+\un\a_{A_{\mu}}(B_{\nu\l})+\un\a_{A_{\nu}}(B_{\l\mu})+\un\a_{A_{\l}}(B_{\mu\nu})
\Big)\p_{r}x^{\mu}\p_{s}x^{\nu}\p_{t}x^{\l}.
\end{split}
\ee
In analogy to the curvature 2-form associated to a
1-form potential $A$ obtained by calculating the
holonomy along an infinitesimal square, 
we define this third order term to be the 
\emph{\uline{3-form curvature}} associated
to the pair $(A,B)$ and denote it by $H.$ In terms of components, 
it is given by
\be
H_{\mu\nu\l}:=\p_{\mu}B_{\nu\l}+\p_{\l}B_{\mu\nu}+\p_{\nu}B_{\l\mu}
+\un\a_{A_{\mu}}(B_{\nu\l})+\un\a_{A_{\nu}}(B_{\l\mu})+\un\a_{A_{\l}}(B_{\mu\nu})
\ee
and using differential form notation
\be
\label{eq:3formcurvH}
H:=dB+\un\a_{A}(B).
\ee
This definition and result agrees with (3.28) of \cite{GP}
and Lemma A.11 in \cite{SW2}.
As was also pointed out in \cite{GP}, 
\be
\un\t(H)=\un\t(dB)+\un\t\big(\un\a_{A}(B)\big)
=d\un\t(B)+[A,\un\t(B)]=dF+[A,F]=0
\ee
by the Bianchi identity. Since $\ker\un\t$ is a central
Lie subalgebra of $\mathfrak{h},$ this means $H$ is a
3-form with values in an abelian Lie algebra
(see Remark \ref{rmk:abelianH}).
Under a first order gauge transformation $(A,B)\xrightarrow{(g,\vf)}(A',B')$
as in (\ref{eq:1stordergt}) and using
(\ref{eq:BundergtintermsofABgvf}), 
the 3-form curvature changes to
\be
\label{eq:3curvgaugetransform}
\begin{split}
H'&=dB'+\underline{\a}_{A'}(B')\\
&=d\Big(\underline{\a_{g}}(B)+\vf\wedge\vf-d\vf
-\un\a_{gAg^{-1}-dg g^{-1}}(\vf)\Big)\\
&\quad+\un\a_{gAg^{-1}-dg g^{-1}
-\un\t(\varphi)}\Big(\underline{\a_{g}}(B)+\vf\wedge\vf-d\vf-
\un\a_{gAg^{-1}-dg g^{-1}}(\vf)\Big)\\
&=\underbrace{d\big(\underline{\a_{g}}(B)\big)}_{(\ref{eq:3curvdagB})}
+\uwave{d\vf\wedge\vf-\vf\wedge d\vf}
-\underbrace{d\big(\un\a_{gAg^{-1}-dg g^{-1}}(\vf)\big)}_{(\ref{eq:3curvdaAprimephi})}\\
&\quad
+
\underbrace{\un\a_{gAg^{-1}-dg g^{-1}}\big(\underline{\a_{g}}(B)\big)}_{(\ref{eq:3curvagaAB})}
-\un\a_{dg g^{-1}}\big(\underline{\a_{g}}(B)\big)
+\underbrace{\un\a_{gAg^{-1}-dg g^{-1}}\big(\vf\wedge\vf\big)}_{(\ref{eq:3curveaAprimephiphi})}\\
&\quad-\underbrace{\un\a_{gAg^{-1}-dg g^{-1}}\big(\un\a_{gAg^{-1}-dg g^{-1}}(\vf)\big)}_{(\ref{eq:3curvaAprimeaAprimephi})}
-\underbrace{\un\a_{gAg^{-1}-dg g^{-1}}(d\vf)}_{(\ref{eq:3curvdaAprimephi})}\\
&\quad-[\vf,\underline{\a_{g}}(B)]
-\underbrace{[\vf,\vf\wedge\vf]}_{0}
+\uwave{[\vf,d\vf]}
+[\vf,\un\a_{gAg^{-1}-dg g^{-1}}(\vf)],
\end{split}
\ee
where the underlined terms cancel and the other terms
with underbraces will be calculated and compared momentarily. 
At this point, it is useful to simplify some of these terms
by applying $\un\t$ and calculating the results
in terms of commutators and such. For example, 
\be
\label{eq:3curvdagB}
\begin{split}
\un\t\Big(d\underline{\a_{g}}(B)\big)\Big)
&=d\Big(\un\t\big(\underline{\a_{g}}(B)\big)\Big)\\
&=d\big(g\un\t(B)g^{-1}\big)\\
&=dg\un\t(B)g^{-1}+g\un\t(dB)g^{-1}+g\un\t(B)dg^{-1}\\
&=dgg^{-1}g\un\t(B)g^{-1}+\un\t\big(\underline{\a_{g}}(dB)\big)
-g\un\t(B)g^{-1}dg g^{-1}\\
&=\un\t\Big(\underline{\a_{g}}(dB)
+\un\a_{dg g^{-1}}\big(\underline{\a_{g}}(B)\big)\Big).
\end{split}
\ee
We can safely equate the terms inside the $\un\t.$
This is a helpful trick and we will use it to calculate
all other terms. For instance, one can show using this trick that
\be
\label{eq:3curvagaAB}
\un\a_{gAg^{-1}}\big(\underline{\a_{g}}(B)\big)
=\underline{\a_{g}}\big(\un\a_{A}(B)\big).
\ee
Although this is a trick and not completely rigorous, 
it works for all of the calculations we will do. 
These formulas can all be found in Appendix A
of \cite{Wal1} and can be derived more rigorously. 
Since $\un\a$ is a derivation, 
\be
\label{eq:3curveaAprimephiphi}
\un\a_{gAg^{-1}-dg g^{-1}}(\vf\wedge\vf)
=\un\a_{gAg^{-1}-dg g^{-1}}(\vf)\wedge\vf
-\vf\wedge\un\a_{gAg^{-1}-dg g^{-1}}(\vf),
\ee
which cancels with the term
$[\vf,\un\a_{gAg^{-1}-dg g^{-1}}(\vf)]$
in (\ref{eq:3curvgaugetransform}).  
Furthermore, note that
\be
\begin{split}
\un\t\Big(\un\a_{X}\big(\un\a_{X}(Y)\big)\Big)
&=\Big[X,\un\t\big(\un\a_{X}(Y)\big)\Big]\\
&=\Big[X,[X,\un\t(Y)\big]\Big]\\
&=\big[X,X\wedge\un\t(Y)+\un\t(Y)\wedge X\big]\\
&=X\wedge X\wedge\un\t(Y)-X\wedge\un\t(Y)\wedge X
+X\un\t(Y)\wedge X-\un\t(Y)\wedge X\wedge X\\
&=\un\t\big(\un\a_{X\wedge X}(Y)\big)
\end{split}
\ee
for any $\mathfrak{g}$-valued 1-form $X$ and for any
$\mathfrak{h}$-valued 1-form $Y.$ Although $\un\t$ has been applied
to derive these equalities, the expressions inside $\un\t$ 
are still equal. This equality implies
\be
\label{eq:3curvaAprimeaAprimephi}
\begin{split}
\un\a_{gAg^{-1}-dg g^{-1}}\big(\un\a_{gAg^{-1}-dg g^{-1}}(\vf)\big)
&=\un\a_{gAAg^{-1}}(\vf)-\un\a_{dgAg^{-1}}(\vf)\\
&\quad-\un\a_{gAg^{-1}dgg^{-1}}(\vf)
+\un\a_{dgg^{-1}dgg^{-1}}(\vf).
\end{split}
\ee
One of the more cumbersome set of terms is
\be
\label{eq:3curvdaAprimephi}
\begin{split}
\un\t\Big(&d\big(\un\a_{gAg^{-1}-dg g^{-1}}(\vf)\big)
+\un\a_{gAg^{-1}-dg g^{-1}}(d\vf)\Big)
=d[gAg^{-1},\un\t(\vf)]-d[dgg^{-1},\un\t(\vf)]\\
&\qquad\qquad\qquad\qquad\qquad\qquad\qquad\qquad
+[gAg^{-1},\un\t(d\vf)]-[dgg^{-1},\un\t(d\vf)]\\
&=dgAg^{-1}\un\t(\vf)+gdAg^{-1}\un\t(\vf)
-gAdg^{-1}\un\t(\vf)-\uuline{gAg^{-1}\un\t(d\vf)}\\
&+\dashuline{\un\t(d\vf)gAg^{-1}}-\un\t(\vf)dgAg^{-1}
-\un\t(\vf)gdAg^{-1}+\un\t(\vf)gAdg^{-1}\\
&+dgdg^{-1}\un\t(\vf)+\dotuline{dgg^{-1}\un\t(d\vf)}
-\uwave{\un\t(d\vf)dgg^{-1}}-\un\t(\vf)dgdg^{-1}\\
&+\uuline{gAg^{-1}\un\t(d\vf)}-\dashuline{\un\t(d\vf)gAg^{-1}}
-\dotuline{dgg^{-1}\un\t(d\vf)}+\uwave{\un\t(d\vf)dgg^{-1}}\\
&=\un\t\Big(
\un\a_{dgAg^{-1}}(\vf)
+\un\a_{gdAg^{-1}}(\vf)
-\un\a_{gAdg^{-1}}(\vf)
+\un\a_{dgdg^{-1}}(\vf)
\Big).
\end{split}
\ee
Combining this with the 
result preceding it gives just a single term
$\un\a_{gFg^{-1}}(\vf).$
Putting all of this together, we obtain
\be
\begin{split}
H'
&=\underline{\a_{g}}\big(dB+\un\a_{A}(B)\big)
-[\vf,\underline{\a_{g}}(B)]
-\un\a_{gFg^{-1}}(\vf)\\
&=\underline{\a_{g}}(H)
-[\vf,\underline{\a_{g}}(B)]
-\un\a_{gFg^{-1}}(\vf).
\end{split}
\ee
Finally, by the properties of crossed modules and by the 
vanishing of the fake curvature, 
\be
\un\a_{gFg^{-1}}(\vf)
=\un\a_{g\un\t(B)g^{-1}}(\vf)
=\un\a_{\un\t(\un{\a_{g}}(B))}(\vf)
=\big[\un{\a_{g}}(B),\vf\big]
\ee
so that the above formula reduces further to just simply 
\be
H'=\underline{\a_{g}}(H).
\ee
In particular, $H$ is invariant under fat gauge transformations. 
This result agrees with what was discovered in \cite{GP}. 


\section{Conclusion}
\label{sec:2dalgebraconc}
We have illustrated that 2-category theory can be implemented
and used in such a way as to calculate parallel
transport along two-dimensional surfaces, 
such as worldsheets of strings, explicitly 
for gauge groups that are not necessarily abelian 
via an approximation technique that can
be implemented numerically. 
We have done this using string diagram techniques to
facilitate 2-categorical techniques and bring 
higher category theory to a wider audience. 
Although Girelli and Pfeiffer
have calculated infinitesimal gauge transformations
and curvature forms via similar techniques \cite{GP}
and Schreiber and Waldorf provided a 
formula for the parallel transport along a surface
\cite{SW2}, our infinitesimal methods give a much
more explicit and direct construction of the iterated surface integral
from elementary building blocks filling in some of the 
arguments sketched by Baez and Schreiber in \cite{BS}, 
particularly in Section 2.3.2
(Section 5.1 of a draft of this paper even
contains a nice picture that unfortunately did not
make it to the final version of their paper but is present
in Section 11.4.1 of Schreiber's thesis \cite{Scthesis}).
Schreiber and Waldorf's integral in \cite{SW2} 
was obtained from consistency conditions
and then they proved that it satisfies the necessary
functorial properties expected of surface holonomy.
In relation to other work, 
such surface-ordered integrals have been used
recently in constructing a Hochschild complex 
for surface transport \cite{Mi15}. 
The novelty of our result is that we derived the
formula for surface parallel transport 
from scratch using a discretization of our surface. 
To our knowledge, this is the first appearance of such
an explicit construction together with analytical results on
convergence (Proposition \ref{prop:trivnconverges}) 
and a simplification providing a manageable 
surface-ordered integral by reducing the surface ordering
to a single direction as opposed to two (this result is embodied in
Theorem \ref{thm:fulltoreducedsurfacetransport}). 
By implementing string diagrams,
we have also provided a more friendly visualization.
Furthermore, we have avoided using path spaces explicitly and have
simplified many arguments. 

We hope that
we have illustrated how two-dimensional
algebra can be used for explicit calculations. 
If developed further, these ideas might be used to explain
physical phenomena that utilize algebraic manipulations in more than
one dimension more naturally. 
Such higher-dimensional algebra appears in many
situations. For example, elements and molecules combine in
a variety of ways forming complicated compounds, amino acids, and 
proteins. 
These are objects that use three dimensions to
configure themselves. Therefore a natural and faithful 
representation of them would involve a sort of 3-dimensional algebra.
Another example occurs in painting. Given a painting, it is
much simpler for us to ``read'' a 2-dimensional painting than to view all the
pixels making it up in a straight 1-dimensional list. Both perspectives
contain the same information-theoretic data, but the 2-d form is
naturally and immediately recognized. 
As another even more speculative example, it is known that the entropy
of a black hole is proportional to the surface area of its horizon.
This may lead one to believe that the microstates of the theory
can be expressed as living on a lower-dimensional world. 
This in turn then suggests the possibility that a lower-dimensional
algebra might be useful in describing some of the properties
of these microstates. 
Although these ideas are entirely speculative, our point is that one can imagine
that the one-dimensional algebra we have forced upon ourselves
is only the tip of an iceberg of algebraic structures. 
Higher category theory opens us to these other possibilities. 

There are still many open questions in this relatively young field. 
One is how to construct useful Actions in physics that model 
phenomena with non-abelian higher form gauge fields
and also the interactions with matter fields. 
Some recent progress in this direction has been made by S\"amann
and others---see \cite{SS17} and the references therein. 
Work on the pure gauge field side was initiated
in the work of Pfeiffer \cite{Pf} using a 2-categorical approach. 
To proceed, it seems that a better suited representation
theory for 2-categories will be useful \cite{BBFW}. Furthermore, 
characters for 2-groups \cite{GK}, \cite{GU} 
and traces \cite{PoSh}, \cite{HPT} may need to be studied further
to better understand what gauge invariant combinations are possible.
Although the number of higher gauge theory examples are increasing
\cite{BW}, \cite{CLS}, \cite{FSS}, \cite{MM}, \cite{PaSa}, \cite{Pa}, \cite{SS17}, 
some work is still required to solidify the role of higher gauge theory
in lattice gauge theory and other areas of physics. 
Other lattice gauge theory approaches existed earlier
\cite{Or80}, \cite{Or83}, \cite{Or84} with a renewed interest
in \cite{LR}
but it is not clear to us how these approaches to higher lattice
gauge theory are related to the rest of the literature.

In the realm of string theory and M-theory,
beginning with early work of
Witten, Myers, and others \cite{Wit96}, \cite{My}, 
a more precise
construction of the non-abelian gauge theories on a stack
of $D$-branes \cite{Zw} and its low energy effective Action
is still lacking.
These effective Actions are swarmed with higher form
non-abelian gauge fields, but the precise mathematical
formulation is still lacking though it is likely that non-abelian
differential cohomology \cite{Sc} is relevant suggested by
recent work on M5-branes in which it plays an essential
role \cite{FSS}.
Some arguments used to describe such effective Actions
are not always entirely straightforward and involve
consistency conditions (such as T-duality \cite{My} and
scattering amplitude calculations \cite{DST}) rather than direct
derivations. 
It is therefore possible that a more thorough investigation
may involve understanding nonperturbative effects, one
of which is dictated by transport. On the other hand, due to the
non-commutative nature of the normal coordinates to these
branes \cite{Mo}, this may involve a modification of
such transport to the setting of non-commutative geometry. 
These and many other ideas have also been briefly discussed in 
\cite{Scthesis}, and several such open questions can be found there.





\begin{appendices}
\section{Differential Lie crossed modules}
\label{app:dlcm}
Here we briefly review the infinitesimal version of
a Lie crossed module $(H,G,\t,\a),$ which we write
as $(\mathfrak{h},\mathfrak{g},\un\t,\un\a),$
including the many relations that these maps satisfy
that we use throughout our calculations.
We also make some comments on how this is used
for differential forms with values in $\mathfrak{g}$ and
$\mathfrak{h}.$ This information can also be found in many
articles on the subject of higher gauge theory
such as \cite{BS}, \cite{GP}, and especially Waldorf's concise
one page formula sheet in Appendix A of \cite{Wal1}. 
Martins and Mikovi\'c also have an exceptionally
clear and thorough exposition in Section 2.1 of \cite{MM}.

$\un\t:\mathfrak{h}\to\mathfrak{g}$ is the derivative
of $\t:H\to G$ at the identity and is a Lie algebra homomorphism
since $\t$ is a Lie group homomorphism. 
Notice that $\a$ can be equivalently described as a function
$\a:G\times H\to H$ that is a group homomorphism
in each component separately. As a result, for any
fixed $g\in G,$ $\a_{g}:H\to H$ is a Lie group homomorphism
and hence has a derivative at the identity denoted
by $\underline{\a_{g}}:\mathfrak{h}\to\mathfrak{h}.$ 
This map, besides being a Lie algebra
homomorphism, satisfies the additional property that
\be
\label{eq:differentialcrossedmoduleidentity1}
\un\t\big(\underline{\a_{g}}(Y)\big)=g\un\t(Y)g^{-1}
\ee
for all $Y\in\mathfrak{h}$ and $g\in G.$  
Similarly, although
$\a:G\times H\to H$ is not a group homomorphism, it is
smooth and its derivative 
$\un\a:\mathfrak{g}\times\mathfrak{h}\to\mathfrak{h}$
is a well-defined linear map. It is a derivation once the
$\mathfrak{g}$ coordinate is fixed, i.e.
\be
\un\a_{X}\big([Y,Z])=\big[\un\a_{X}(Y),Z\big]+\big[Y,\un\a_{X}(Z)\big]
\ee
for all $X\in\mathfrak{g}$ and $Y,Z\in\mathfrak{h}.$
$\un\a$ also satisfies 
\be
\un\a_{[X,X']}(Y)=\un\a_{X}\big(\un\a_{X'}(Y)\big)
-\un\a_{X'}\big(\un\a_{X}(Y)\big)
\ee
for all $X,X'\in\mathfrak{g}$ and $Y\in\mathfrak{h}.$ 
Finally, 
\be
\un\t\big(\un\a_{X}(Y)\big)=\big[X,\un\t(Y)\big]
\ee
and
\be
\un\a_{\t(Y)}(Z)=[Y,Z]
\ee
for all $X\in\mathfrak{g}$ and $Y,Z\in\mathfrak{h}.$ 

Once combined with differential forms, the maps
$\un\a$ and $\un\t$ are extended in the appropriate way
(see Part II Chapter 3 in the section on the Bianchi 
Identity in \cite{BaMu} for details on differential forms
with values in Lie algebras). 
For instance, $\un\a$ is a graded derivation in its
second coordinate. 
To clarify the notation used throughout, 
consider differential forms 
$A\in\W^{1}(M;\mathfrak{g}),$ $F\in\W^{2}(M;\mathfrak{g}),$
$\varphi\in\W^{1}(M;\mathfrak{h}),$ $B\in\W^{2}(M;\mathfrak{h}).$
When we write expressions such as $\un\a_{A}(\varphi)$ or 
$\un\a_{A}(B)$ we mean the following. First, let 
$\{t^{a}\}_{a\in\{1,\dots,\dim\mathfrak{g}\}}$ be a basis for $\mathfrak{g}$ and
$\{s^{b}\}_{b\in\{1,\dots,\dim\mathfrak{h}\}}$ be a basis for $\mathfrak{h}.$ Then
\be
A=A_{a}t^{a},\qquad
F=F_{a}t^{a},\qquad
\varphi=\vf_{b}s^{b},
\aand
B=B_{b}s^{b},
\ee
where a summation over repeated indices is assumed and where 
$A_{a},\vf_{b}\in\W^{1}(M)$ and $F_{a},B_{b}\in\W^{2}(M)$ for all
indices. 
Then by definition,
\be
\un\a_{A}(\vf)\equiv\un\a_{A_{a}t^{a}}\big(\vf_{b}s^{b}\big)
:=(A_{a}\wedge\vf_{b})\un\a_{t^{a}}\big(s^{b}\big)
\ee
and similarly for any other forms.
Because we use Lie algebra valued forms,
the bracket is graded, so for instance
\be
[\vf,\vf]:=(\vf_{b}\wedge\vf_{b'})\big[s^{b},s^{b'}\big]
=\vf\wedge\vf+\vf\wedge\vf
\ee
but
\be
[\vf,B]:=(\vf_{b}\wedge B_{b'})\big[s^{b},s^{b'}\big]
=\vf\wedge B-B\wedge\vf
\ee
since $\vf$ is a 1-form and $B$ is a 2-form. 
The last two equalities follow if we think of our Lie algebras 
as coming from matrix Lie algebras, which we often do.
The general formula is
\be
[\w,\eta]=\w\wedge\eta-(-1)^{|\w||\eta|}\eta\wedge\w,
\ee
where
$|\w|$ and $|\eta|$ are the degrees of the forms 
$\w\in\W^{|\w|}(M;\mathfrak{h})$ 
and $\eta\in\W^{|\eta|}(M;\mathfrak{h}).$ 
Other properties are derived as needed in calculations
in the body of the article.

\section{Surface product convergence}
\label{app:spc}
This appendix serves to prove the convergence of the surface-ordered
product (\ref{eq:surfaceproductn=5inoneline}) as $n\to\infty$
and to also prove upper bounds on the $k$-th order terms when
expanded out. For this,
we will first relax our conditions and work with arbitrary partitions of 
the unit square. We will follow the conventions of \cite{Mu91} and
use the results there without further reference. 
The surface-ordered product is well-defined for each partition
$P$ and will be denoted by $\triv_{P}.$ We will also use the notation 
$\g_{s,t}$ to denote the path defined after 
Theorem \ref{thm:fulltoreducedsurfacetransport}. 
It will be helpful to define the function
\be
[0,1]\times[0,1]\ni(s,t)\mapsto \mathcal{B}(s,t):=
\left\lVert\un{\a_{\triv(\g_{s,t})}}\left(B_{\mu\nu}\big(x(s,t)\big)\frac{\p x^{\mu}}{\p t}\frac{\p x^{\nu}}{\p s}\right)\right\rVert. 
\ee
\bprf[Proof of Proposition \ref{prop:trivnconverges}]
By smoothness of $A,B,$ and parallel transport along paths, 
$\mathcal{B}$ is a smooth function. 
First note that for any point $(s_{i},t_{j})\in[0,1]$ on the intersection
points of the grid formed by some partition $P$ 
\be
\lim_{Q\ge P}\un{\a_{\g_{s_{i},t_{j}}^{Q}}}\left(B_{\mu_{i}\nu_{j}}\big(x(s_{i},t_{j})\big)\frac{\p x^{\mu_{i}}}{\p t}\frac{\p x^{\nu_{j}}}{\p s}\Big |_{(s_{i},t_{j})}\right)
=\un{\a_{\triv(\g_{s_{i},t_{j}})}}\left(B_{\mu\nu}\big(x(s_{i},t_{j})\big)\frac{\p x^{\mu}}{\p t}\frac{\p x^{\nu}}{\p s}\Big |_{(s_{i},t_{j})}\right)
\ee
where $\g_{s_{i},t_{j}}^{Q}$ is the path obtained from expanding out 
the 2-group multiplication with respect to the partition $Q.$ 
The ordering on partitions is given by refinement and the above limit is
taken over all refinements of $P.$ This limit is valid due to the smoothness
of all expressions. Visually, this limit is also reasonable. For example, 
consider the following examples of refinements 
\begin{figure}[H]
\centering
\hspace{-9mm}
\begin{subfigure}{0.23\textwidth}
\centering
\includegraphics[width=0.92\textwidth,angle=-45,origin=c]{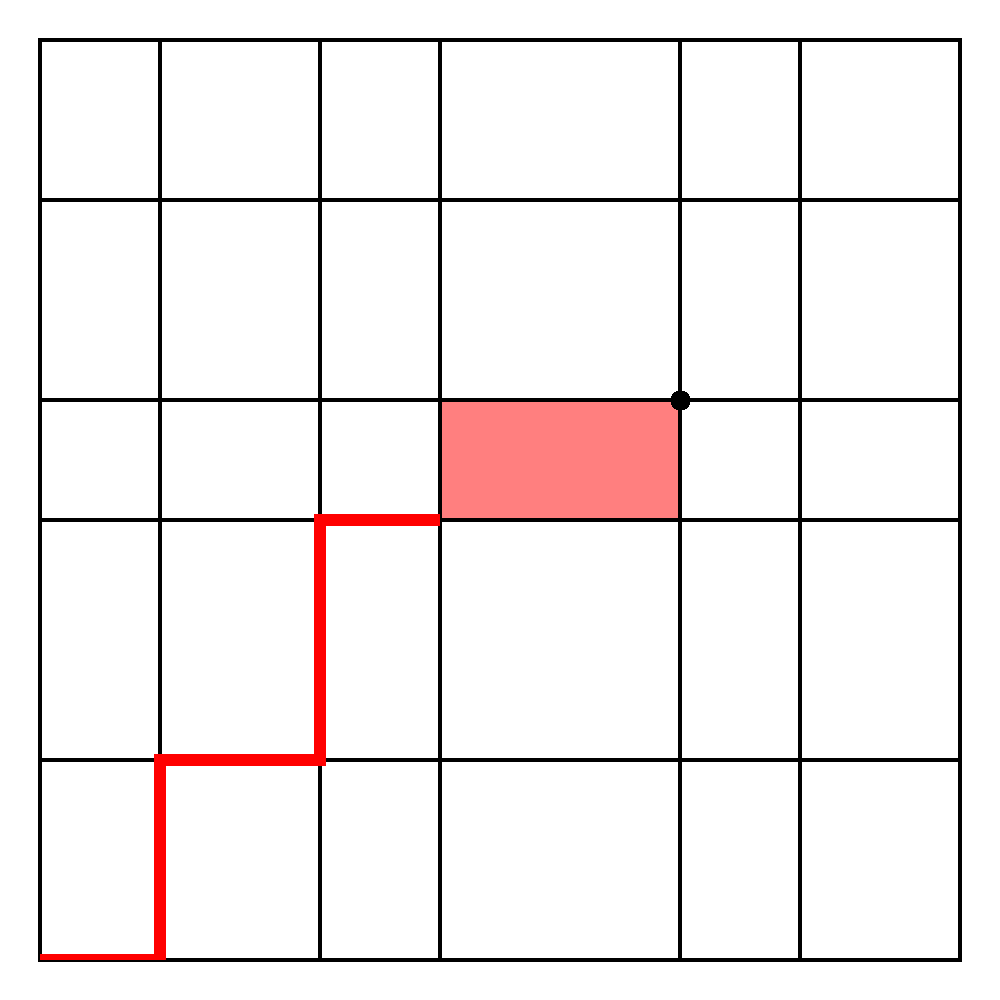}
\begin{picture}(0,0)
\put(10,62){$\xy0;/r.28pc/: 
(0,0)*{}="1";
(-20,-20)*+{\small t}="t";
{\ar"1";"t"};
\endxy$}
\put(10,80){$\xy0;/r.28pc/: 
(0,0)*{}="1";
(-20,20)*+{\small s}="s";
{\ar"1";"s"};
\endxy$}
\end{picture}
\end{subfigure}
$\qquad\quad\to$
\begin{subfigure}{0.23\textwidth}
\centering
\includegraphics[width=0.92\textwidth,angle=-45,origin=c]{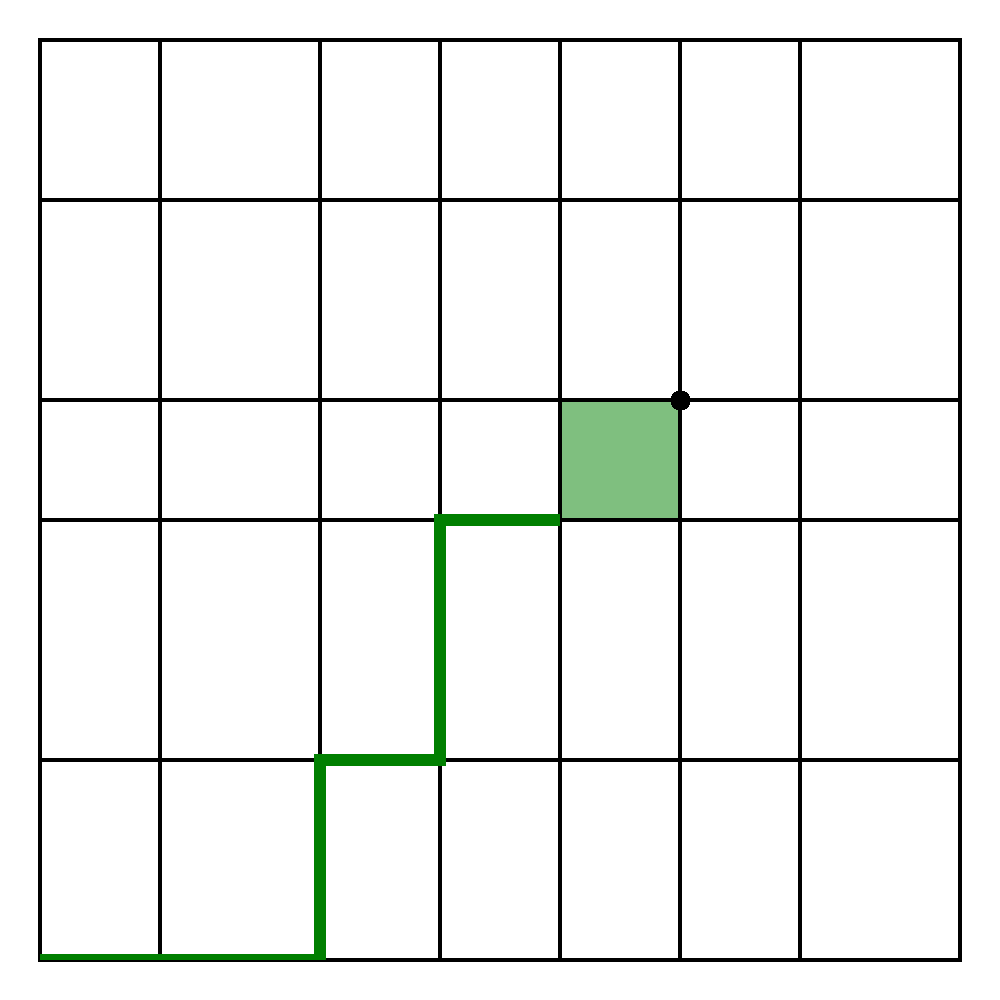}
\begin{picture}(0,0)
\put(10,62){$\xy0;/r.28pc/: 
(0,0)*{}="1";
(-20,-20)*+{\small t}="t";
{\ar"1";"t"};
\endxy$}
\put(10,80){$\xy0;/r.28pc/: 
(0,0)*{}="1";
(-20,20)*+{\small s}="s";
{\ar"1";"s"};
\endxy$}
\end{picture}
\end{subfigure}
$\qquad\quad\to$
\begin{subfigure}{0.23\textwidth}
\centering
\includegraphics[width=0.92\textwidth,angle=-45,origin=c]{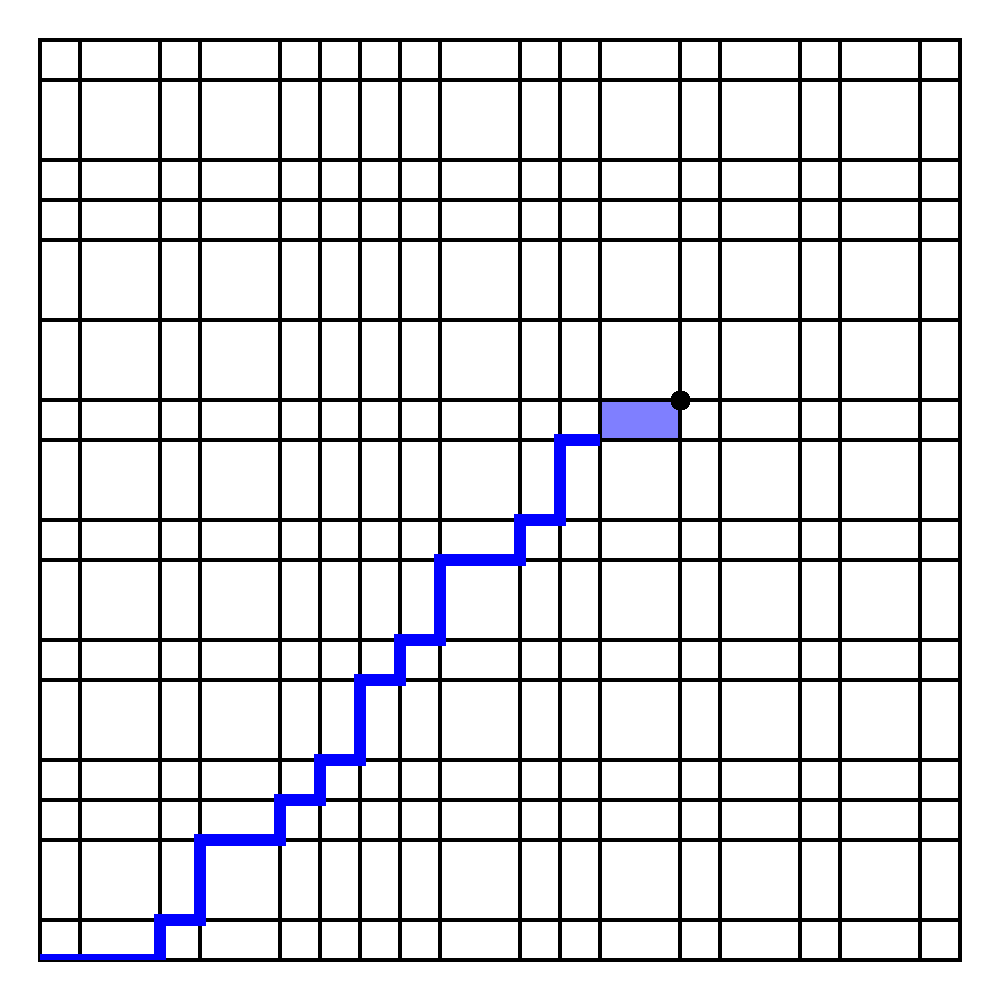}
\begin{picture}(0,0)
\put(10,62){$\xy0;/r.28pc/: 
(0,0)*{}="1";
(-20,-20)*+{\small t}="t";
{\ar"1";"t"};
\endxy$}
\put(10,80){$\xy0;/r.28pc/: 
(0,0)*{}="1";
(-20,20)*+{\small s}="s";
{\ar"1";"s"};
\endxy$}
\end{picture}
\end{subfigure}
\end{figure}
\noindent
Taking the limit over partitions 
\begin{figure}[H]
\centering
\includegraphics[width=0.30\textwidth,angle=-45,origin=c]{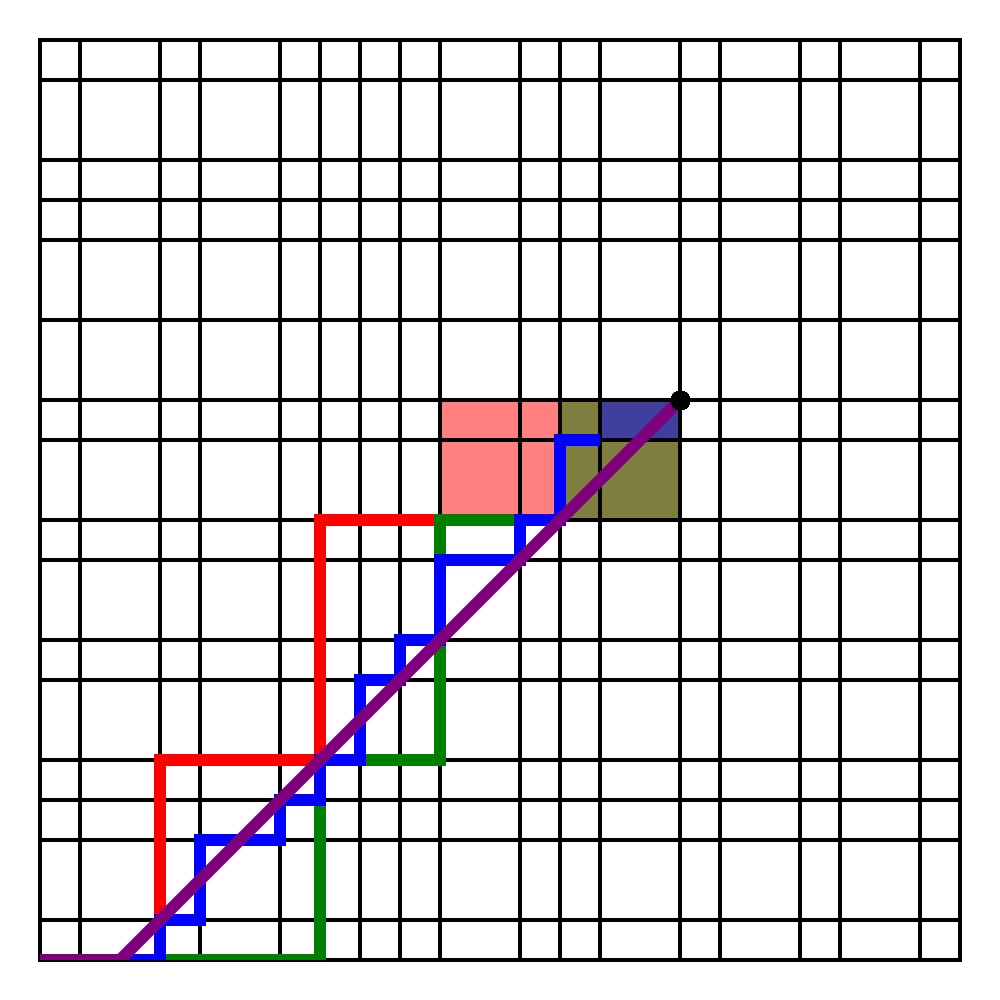}
\begin{picture}(0,0)
\put(-110,62){$\xy0;/r.35pc/: 
(0,0)*{}="1";
(-23,-23)*+{\small t}="t";
{\ar"1";"t"};
\endxy$}
\put(-110,80){$\xy0;/r.35pc/: 
(0,0)*{}="1";
(-23,23)*+{\small s}="s";
{\ar"1";"s"};
\endxy$}
\end{picture}
\end{figure}
\noindent
shows that the paths converge to $\g_{s,t}$ and the value of 
the alpha action along the path at $B$ depends only on the point $(s,t).$ 
Hence, the partial \emph{products} of terms of the form 
\be
\mathds{1}+\un{\a_{\triv(\g_{s_i,t_j})}}\left(B_{\mu_{i}\nu_{j}}\big(x(s_{i},t_{j})\big)\frac{\p x^{\mu_{i}}}{\p t}\frac{\p x^{\nu_{j}}}{\p s}\Big |_{(s_{i},t_{j})}\right)\D s_{i}\D t_{j},
\ee
which define $\triv_{P}$ to lowest order, converge if and only if the
\emph{sum} of terms of the form 
\be
\left\lVert\un{\a_{\triv(\g_{s_i,t_j})}}\left(B_{\mu_{i}\nu_{j}}\big(x(s_{i},t_{j})\big)\frac{\p x^{\mu_{i}}}{\p t}\frac{\p x^{\nu_{j}}}{\p s}\Big |_{(s_{i},t_{j})}\right)\right\rVert\D s_{i}\D t_{j}
\ee 
converges (cf Section 8.10 in \cite{We64}). 
However, the limit of the sum of these terms over all 
partitions is exactly the definition of the Riemann integral of the (smooth) 
function $\mathcal{B}$ over the unit cube
\be
\lim_{P}\sum_{(i,j)\in P}
\left\lVert\un{\a_{\triv(\g_{s_i,t_j})}}\left(B_{\mu_{i}\nu_{j}}\big(x(s_{i},t_{j})\big)\frac{\p x^{\mu_{i}}}{\p t}\frac{\p x^{\nu_{j}}}{\p s}\Big |_{(s_{i},t_{j})}\right)\right\rVert\D s_{i}\D t_{j}
=\iint\displaylimits_{[0,1]\times[0,1]}\mathcal{B}(s,t)\;dsdt.
\ee
Since the integral exists, the sum converges, and hence the product
converges. 
\eprf

\bprf[Proof of Proposition \ref{prop:trivnkbounded}]
For the first claim, set $M_{n}$ to be the maximum value of the
norm of the expressions of the form 
$\a_{a_{56}^{s}a_{46}^{s}a_{36}^{s}a_{35}^{t}a_{25}^{s}}(B_{14}).$
Then 
\be
\left\lVert\triv_{n,k}\right\rVert\le
\left(\frac{M_{n}}{n^2}\right)^k\binom{n^2}{k}
=M_{n}^{k}\cfrac{\binom{n^2}{k}}{n^{2k}}\le
\frac{M_{n}^{k}}{k!}
\ee
for all $k\in\{0,1,\dots,n^2\}.$
For the second claim, we note that $\ds\lim_{n\to\infty}M_{n}$
converges to a finite value due to 
the arguments preceding the proof of Proposition 
\ref{prop:trivnconverges} in this appendix since 
the sequence of such terms converges, upon refinement, 
to a term of the form $\un{\a_{\triv(\g_{s,t})}}\big(B(s,t)\big).$ 
Set
\be
\tilde{M}:=\max_{(s,t)}\left\lVert\un{\a_{\triv(\g_{s,t})}}\big(B(s,t)\big)\right\rVert,
\ee
which is finite and well-defined by compactness and smoothness of all inputs. 
For each $\e>0,$ let $\tilde{N}_{\e}\in\N$ be large enough so that 
$|M_{n}-\tilde{M}|<\e$ for all $n\ge\tilde{N}_{\e}.$ 
Choose any $\e>0$ and set 
$M:=\tilde{M}+\e$
and 
$N:=\max\left\{\tilde{N}_{\e},\left\lfloor\sqrt{2k}\right\rfloor+1\right\}.$ 
From this it follows that 
\be
\left\lVert\triv_{n,k}\right\rVert\le
\left(\frac{M}{n^2}\right)^k\binom{n^2}{k}
=M^{k}\cfrac{\binom{n^2}{k}}{n^{2k}}\le
\frac{M^{k}}{k!}
\ee
for all $k\in\{0,1,\dots,n^2\}$ for all $n\ge N.$ 
The final claim follows from the triangle inequality and 
the fact that $\e$ can be taken to be arbitrarily small. 
\eprf

\section{Proof of configurations Lemma}
\label{app:config}
This appendix serves to give a rigorous proof of Lemma
\ref{lem:twoboxeslimit}. 
For the proof of this Lemma, it is useful to rewrite $S_{n,k}$ as
\be
\begin{split}
S_{n,k}&=\frac{1}{k!}\sum_{2n-1\ge i_{k}\ne i_{k-1}\ne\cdots\ne i_{1}\ge1}
l_{n}(i_1)\cdots l_{n}(i_{k}) \\
&=\frac{1}{k!}\sum_{i_{k}=1}^{2n-1}l_{n}(i_{k})
\sum_{\substack{i_{k-1}=1\\ i_{k-1}\ne i_{k}}}^{2n-1}l_{n}(i_{k-1})
\cdots
\sum_{\substack{i_{2}=1\\i_{2}\ne i_{3}\\ \vdots\\ i_{2}\ne i_{k}}}^{2n-1}l_{n}(i_{2})
\sum_{\substack{i_{1}=1\\i_{1}\ne i_{2}\\ \vdots\\ i_{1}\ne i_{k}}}^{2n-1}l_{n}(i_{1}),
\end{split}
\ee
where it is understood that any sum operation on the left acts on
everything to the right. Before working out this
summation to obtain a more explicit formula, for each $n\in\Z^{+},$
define the function 
\be
\begin{split}
\{1,2,\dots,2n-1\}&\xrightarrow{\phi_{n}}\Z\\
p&\mapsto\phi_{n}(p):=\sum_{i=1}^{2n-1}l_{n}(i)^{p}.
\end{split}
\ee
Explicitly, this can be calculated as follows
\cite{We}.
\be
\label{eq:phinp}
\phi_{n}(p)
=2\sum_{q=1}^{n}q^{p}-n^{p}
=\frac{2}{p+1}\left(\sum_{q=1}^{p+1}(-1)^{\de_{qp}}
\binom{p+1}{q}B_{p+1-q}n^{q}\right)-n^{p},
\ee
where $\de_{qp}$ is the Kronecker delta function and 
$B_{r}$ is the Bernoulli number defined, for instance, by the
power series expansion (thought of as a formal power series in the
variable $x$)
\be
\frac{x}{e^{x}-1}=\sum_{r=0}^{\infty}\frac{B_{r}x^{r}}{r!}.
\ee
The first few of these Bernoulli numbers are
\be
\begin{split}
B_{0}&=1\\
B_{1}&=-\frac{1}{2}\\
B_{2}&=\frac{1}{6}\\
B_{3}&=0\\
B_{4}&=-\frac{1}{30}
\end{split}
\ee
while the first few $\phi_{n}$ are
\be
\begin{split}
\phi_{n}(1)&=n^2\\
\phi_{n}(2)&=\frac{n(n^2+1)}{3}\\
\phi_{n}(3)&=\frac{n^2(n^2+1)}{2}\\
\phi_{n}(4)&=\frac{n(6n^4+10n^2-1)}{15}.
\end{split}
\ee
Examining $\phi_{n}(p)$ a little more, one immediately notices the 
crucial result
\be
\label{eq:phinpasymptotics}
\lim_{n\to\infty}\frac{\phi_{n}(p)}{n^{2p}}=0
\qquad\mbox{ for } p\ge2.
\ee
Now, $S_{n,k}$ can be written as a polynomial in the $\phi_{n}$'s 
\be
\begin{split}
S_{n,k}&=
\frac{1}{k!}\sum_{i_{k}=1}^{2n-1}l_{n}(i_{k})
\sum_{\substack{i_{k-1}=1\\ i_{k-1}\ne i_{k}}}^{2n-1}l_{n}(i_{k-1})
\cdots
\sum_{\substack{i_{2}=1\\i_{2}\ne i_{3}\\ \vdots\\ i_{2}\ne i_{k}}}^{2n-1}l_{n}(i_{2})
\left[\phi_{n}(1)-\sum_{j_{1}=2}^{k}l_{n}(i_{j_{1}})\right]\\
&=\frac{1}{k!}\sum_{i_{k}=1}^{2n-1}l_{n}(i_{k})
\cdots
\sum_{\substack{i_{3}=1\\i_{3}\ne i_{4}\\ \vdots\\ i_{3}\ne i_{k}}}^{2n-1}l_{n}(i_{3})
\left[
\phi_{n}(1)^2-\phi_{n}(2)-2\phi_{n}(1)\sum_{j_{1}=3}^{k}l_{n}(i_{j_{1}})
\right.\\
&\qquad\qquad\qquad\qquad\qquad\qquad\left.
+\sum_{j_{2}=3}^{k}l_{n}(i_{j_2})^{2}
+\sum_{j_{2}=3}^{k}\sum_{j_{1}=3}^{k}l_{n}(i_{j_2})l_{n}(i_{j_1})
\right]\\
&=\frac{1}{k!}\sum_{i_{k}=1}^{2n-1}l_{n}(i_{k})
\cdots
\sum_{\substack{i_{4}=1\\i_{4}\ne i_{5}\\ \vdots\\ i_{4}\ne i_{k}}}^{2n-1}l_{n}(i_{4})
\left[
*_{n,4}
\right],
\end{split}
\ee
where
\be
\begin{split}
*_{n,4}&=
\phi_{n}(1)^{3}-3\phi_{n}(1)\phi_{n}(2)+2\phi_{n}(3)
+3\Big(\phi_{n}(2)-\phi_{n}(1)^{2}\Big)
\sum_{j_{1}=4}^{k}l_{n}(i_{j_{1}})\\
&+3\phi_{n}(1)\left(\
\sum_{j_{1}=4}^{k}l_{n}(i_{j_1})^{2}
+\sum_{j_2=4}^{k}\sum_{j_1=4}^{k}l_{n}(i_{j_1})l_{n}(i_{j_2})
\right)\\
&-2\sum_{j_{1}=4}^{k}l_{n}(i_{j_{1}})^{3}
-3\sum_{j_2=4}^{k}\sum_{j_1=4}^{k}l_{n}(i_{j_1})^{2}l_{n}(i_{j_2})
-\sum_{j_3=4}^{k}\sum_{j_2=4}^{k}\sum_{j_1=4}^{k}l_{n}(i_{j_1})l_{n}(i_{j_2})l_{n}(i_{j_3}),
\end{split}
\ee
and so on (a more explicit formula will be given momentarily). 
For example, one obtains the following
expressions for small values of $k$:
\be
\begin{split}
S_{n,1}&=\phi_{n}(1)\\
S_{n,2}&=\frac{1}{2!}\Big(\phi_{n}(1)^{2}-\phi_{n}(2)\Big)\\
S_{n,3}&=\frac{1}{3!}\Big(\phi_{n}(1)^{3}-3\phi_{n}(1)\phi_{n}(2)
+2\phi_{n}(3)\Big)\\
S_{n,4}&=\frac{1}{4!}\Big(\phi_{n}(1)^{4}-6\phi_{n}(1)^{2}\phi_{n}(2)
+8\phi_{n}(1)\phi_{n}(3)+3\phi_{n}(2)^{2}
-6\phi_{n}(4)\Big)\\
S_{n,5}&=\frac{1}{5!}\Big(\phi_{n}(1)^{5}+
10\phi_{n}(1)^{3}\phi_{n}(2)+20\phi_{n}(1)^{2}\phi_{n}(3)
+15\phi_{n}(1)\phi_{n}(2)^{2}\\
&\qquad\qquad -30\phi_{n}(1)\phi_{n}(4)
-20\phi_{n}(2)\phi_{n}(3)+24\phi_{n}(5)\Big)
\end{split}
\ee
Looking back at the expressions for $S_{n,k},$ one
sees that there is a recursion relation for $S_{n,k}.$
Setting $S_{n,0}:=1,$ this recursion relation reads
\be
S_{n,k}=\frac{1}{k}\sum_{j=1}^{k}(-1)^{j+1}S_{n,k-j}\phi_{n}(j).
\ee
This recursion relation can be used
to express $S_{n,k}$ purely in terms of the $\phi_{n}$'s
and is given by 
\be
S_{n,k}=\sum_{j_{1}=1}^{k}\sum_{j_2=1}^{k=j_{1}}
\sum_{j_3=1}^{k-j_1-j_2}\cdots
\hspace{-2mm}
\sum_{j_k=1}^{k-j_1-j_2-\cdots-j_{k-1}}
\hspace{-5mm}
\frac{(-1)^{k+j_1+j_2+\cdots+j_k}\phi_{n}(j_1)\phi_{n}(j_2)\cdots\phi_{n}(j_k)}{k(k-j_1)(k-j_1-j_2)\cdots(k-j_1-j_2-\cdots-j_{k-1})}
,
\ee
where it is understood that the sum terminates earlier if
any of the $j$'s are larger than $1.$ For example, if there
are $s$ of them, then
\be
\sum_{r=1}^{s}j_{r}=k.
\ee
Therefore, fixing $k,$ one obtains
\be
\label{eq:configproduct}
\lim_{n\to\infty}\cfrac{\prod_{r=1}^{s}\phi_{n}(j_{r})}{\phi_{n}(1)^{k}}
=\begin{cases}
0 &\mbox{ if } s < k\\
1 &\mbox { if } s=k
\end{cases}
.
\ee
To see this, first notice that $j_{r}=1$ for all $r\in\{1,\dots,s\}$ when $s=k$
in which case the denominator and numerator in 
(\ref{eq:configproduct}) are equal and the limit is $1.$ 
However, when $s<k,$ by the
formula for $\phi_{n}(p)$ in (\ref{eq:phinp}) and the
asymptotics of this given in (\ref{eq:phinpasymptotics}),
\be
\begin{split}
\lim_{n\to\infty}\cfrac{\prod_{r=1}^{s}\phi_{n}(j_{r})}{\phi_{n}(1)^{k}}
&=\lim_{n\to\infty}\cfrac{\prod_{r=1}^{s}\phi_{n}(j_{r})}{n^{2k}}\\
&=\lim_{n\to\infty}\cfrac{\prod_{r=1}^{s}\phi_{n}(j_{r})}{n^{2(j_{1}+\cdots+j_{s})}}\\
&=\lim_{n\to\infty}\prod_{r=1}^{s}\frac{\phi_{n}(j_{r})}{n^{2j_{r}}}\\
&=0.
\end{split}
\ee
Hence, 
\be
\lim_{n\to\infty}\frac{k!S_{n,k}}{n^{2k}}=1.
\ee
Finally going back to $R_{n,k}$ and using this fact gives
\be
\begin{split}
\lim_{n\to\infty}R_{n,k}
&=\lim_{n\to\infty}
\frac{k!S_{n,k}}{n^2(n^2-1)\cdots(n^2-k+1)}\\
&=\lim_{n\to\infty}
\frac{k!S_{n,k}}{n^{2k}\left(1-\frac{1}{n^2}\right)\cdots\left(1-\frac{k-1}{n^2}\right)}\\
&=1.
\end{split}
\ee

\end{appendices}

\pagebreak
\section*{Index of notation}
\addcontentsline{toc}{section}{\numberline{}Index of notation}
\vspace{-2mm}
\
\begin{longtable}{c|c|c|c}
\hline
Notation & Name/description & Location & Page \\
\hline
$\mC$ & category/2-category & Def \ref{defn:1dcat}/\ref{defn:2dcat} 
& \pageref{defn:1dcat}/\pageref{defn:2dcat} \\
\hline
$G$ & group & Def \ref{ex:BG} & \pageref{ex:BG} \\
\hline
$\B G$ & one-object groupoid & Ex \ref{ex:BG} & \pageref{ex:BG} \\
\hline
$\s:F\Rightarrow G$ & \begin{tabular}{c}natural transformation\\from functor $F$ to $G$\end{tabular} & Def \ref{defn:nattrans1cat}/\ref{defn:nattransf} & \pageref{defn:nattrans1cat}/\pageref{defn:nattransf} \\
\hline
$\mathcal{G}\equiv(H,G,\t,\a)$ & crossed module & Def \ref{defn:crossedmodule2dalg} & \pageref{defn:crossedmodule2dalg} \\
\hline
$\B \mathcal{G}$ & one-object 2-groupoid & Ex \ref{ex:2groups} & \pageref{ex:2groups} \\
\hline
$A$ & 1-form potential & Eq (\ref{eq:Agrp}) & \pageref{eq:Agrp} \\
\hline
$M$ & smooth manifold & Sec \ref{sec:2dalgebra1dgt} & \pageref{sec:2dalgebra1dgt} \\
\hline
$\mathds{1}$ & identity matrix & Eq (\ref{eq:grpproductptexpansion}) & \pageref{eq:grpproductptexpansion} \\
\hline
$\triv(\g)$ & local transport along a path $\g$ & Eq (\ref{eq:grouppt}) & \pageref{eq:grouppt} \\
\hline
$\mathcal{P}^{1}M$ & path groupoid of $M$ & After Eq \ref{eq:trivgrouphomo} & \pageref{eq:trivgrouphomo} \\
\hline
$g$ & (thin) gauge transformation & Eq (\ref{eq:usualgaugetransformation}) & \pageref{eq:usualgaugetransformation} \\
\hline
$\mathcal{P}^{2}M$ & path 2-groupoid of $M$ & Before Def \ref{defn:bigon} & \pageref{defn:bigon} \\
\hline
$\triv(\S)$ & local transport along a bigon $\S$ & Eq (\ref{eq:trivbigons})\&(\ref{eq:surfaceiteratedintegral}) & \pageref{eq:trivbigons}\&\pageref{eq:surfaceiteratedintegral} \\
\hline
$B$ & 2-form potential & Eq (\ref{eq:bij}) & \pageref{eq:bij} \\
\hline
$F$ & 2-form curvature of $A$ & Eq (\ref{eq:curvature2form}) & \pageref{eq:curvature2form} \\
\hline
$\vf$ & fat gauge transformation & Eq (\ref{eq:fatgtvfform}) & \pageref{eq:fatgtvfform} \\
\hline
$H$ & 3-form curvature of $(A,B)$ & Eq (\ref{eq:3formcurvH}) & \pageref{eq:3formcurvH} \\
\hline
$(\mathfrak{h},\mathfrak{g},\un\t,\un\a)$ & differential crossed module & Appendix \ref{app:dlcm} & \pageref{eq:differentialcrossedmoduleidentity1} \\
\hline
\end{longtable}

\addcontentsline{toc}{section}{\numberline{}Bibliography}

\end{document}